\documentclass[10pt]{article} 
\usepackage{graphicx}
\usepackage{tikz}
\usepackage{longtable}
\usepackage[top=1in, left=0.95in, bottom=1.1in, right=0.95in]{geometry}
\usepackage[colorlinks=true,
linkcolor=blue,
urlcolor=red,
citecolor=red]{hyperref}
\usepackage[toc]{appendix}
\usepackage{amssymb, amsmath,mathrsfs}
\usepackage{times} % Loads the Times Roman Fonts
\usepackage{mathptmx} % Loads the Times Roman Math Fonts
\usepackage{upgreek}
\usepackage{multicol}
\usepackage{float}
\usepackage{caption}
\usepackage{multirow}
\usepackage{ytableau}
\usepackage[numbers,sort&compress]{natbib}
\usepackage{slashed}
\usepackage{bbm}
\usepackage{comment}
\usepackage{tikz-cd}
\usepackage{array}

\numberwithin{equation}{section}

\begin{document}

\begin{center}
{\Large \textbf{Complete Operator Basis for the modular invariant SMEFT}}\\[10mm]

Luo-Jia Kang$^{a,b,c}$\footnote{kangluojia23@mails.ucas.ac.cn},Hao Sun$^{a, b}$\footnote{sunhao@itp.ac.cn}, Jiang-Hao Yu$^{a, b, c, d}$\footnote{jhyu@itp.ac.cn}\\[10mm]

\noindent 
$^a${\em \small School of Fundamental Physics and Mathematical Sciences, Hangzhou Institute for Advanced Study, UCAS, Hangzhou 310024, China}  \\
$^b${\em \small School of Physical Sciences, University of Chinese Academy of Sciences,   Beijing 100049, P.R. China}   \\
$^c${\em \small Institute of Theoretical Physics, Chinese Academy of Sciences,   Beijing 100190, P. R. China}\\
$^d${ \small International Centre for Theoretical Physics Asia-Pacific, Beijing/Hangzhou, China}\\[10mm]

\date{\today}   
\end{center}

\begin{abstract}

We implement modular flavor symmetries within the Standard Model Effective Field Theory (SMEFT) framework, using the flavor group $A_4^{(q)} \times A_4^{(e)}$ with distinct moduli $\tau_q$ and $\tau_e$, and assigning different modular weights to right-handed quarks using simplest weight assignment. By treating the moduli as non-dynamical spurions,  adopting the MFV-like assumption, and neglecting effects associated with $\mathrm{Im}\,\tau$, we systematically construct a finite set of independent modular-invariant higher-dimensional operators via the Hilbert-series techniques. In the holomorphic $A_4$ scenario, where all modular forms derive from the weight-2 triplet $Y^{(2)}_{\mathbf{3}}$, we present two equivalent Hilbert-series bases. This establishes that higher-dimensional operators can be formally organized as $[Y{\mathbf{r}}^{(k_Y)},{Y_{\mathbf{r}'}^{(k_Y')}}^{*},\mathcal{O}]_{\mathbf{1}}$ singlets. We subsequently enumerate all independent operators up to dimension 7 under this assumption and provide explicit constructions for all dimension-5 operators as well as baryon- and lepton-number conserving dimension-6 operators. Relaxing holomorphicity to the non-holomorphic case of polyharmonic Maa\ss\ forms, considering that non-holomorphic modular forms are not closed under multiplication, adopting the holomorphic organizing idea would generically lead to an infinite proliferation of modular-invariant structures. To retain a finite and complete operator basis, we therefore impose the same minimal formal organizing principle, which reproduces the benchmark Weinberg operator and the corresponding dimension-$6$ operators.

\end{abstract}

\newpage

\tableofcontents
\newpage

% ====== 正文章节开始 ======
\section{Introduction} 
The Standard Model (SM) is widely regarded as the low-energy limit of a more fundamental theory at higher energy scales.
Despite its remarkable success, the absence of new-physics signals at the LHC suggests a significant mass gap between the SM particles and possible new degrees of freedom.
In this context, effective field theories provide a powerful and model-independent framework for describing physics beyond the SM.
The Standard Model Effective Field Theory (SMEFT) extends the SM by adding higher-dimensional operators constructed from SM fields while preserving the SM gauge symmetry,
$SU(3)_C\otimes SU(2)_L\otimes U(1)_Y$.
These operators encode the possible effects of heavy new particles above a cutoff scale $\Lambda$ in low-energy processes.
In particular, the unique dimension-5 operator was written down by Weinberg~\cite{Weinberg:1979sa};
a complete and independent basis of dimension-6 operators was presented in~\cite{Buchmuller:1985jz,Grzadkowski:2010es};
and classifications of dimension-7 operators were given in~\cite{Lehman_2014}.
Based on the Young-tensor method~\cite{Li:2020gnx,Li:2020xlh,Li:2022tec}, the dimension-8 and dimension-9 operators have also been constructed.
Alternatively, the number of SMEFT operators can be counted using Hilbert-series methods~\cite{Lehman:2015via,Lehman:2015coa,Henning:2015daa,Henning:2017fpj,Marinissen:2020jmb,Henning:2015alf,Sun:2022aag,Kondo:2022wcw}.
%Moreover, adding the spurion field to construct EFT operators can also use this method~\cite{Sun:2022aag,Kondo:2022wcw}.
Recently, Hilbert-series techniques have begun to be extended to theories involving discrete flavor symmetries~\cite{Cal__2023}.
The Hilbert series is a powerful algebraic tool for systematically counting independent operators that are invariant under specified symmetries.
One of its main advantages is that it can account for redundancies arising from integration by parts (IBP) and equations of motion (EOM), thereby providing a universal way to determine the number of operators at each mass dimension.
Although the Hilbert series does not yield explicit operator expressions or index structures, it offers a robust cross-check of the completeness and consistency of any proposed operator basis.

%A well-known challenge in the Standard Model is the flavor problem. A common strategy is to assume that flavor and CP violation are governed by the SM Yukawa couplings themselves, promoted to spurion fields transforming under a global flavor symmetry such as $U(3)^5$ or $U(2)^5$. This idea, known as Minimal Flavor Violation (MFV)~\cite{Chivukula:1987py,DAmbrosio:2002vsn,Bonnefoy:2020yee,Aoude:2020dwv,Bruggisser:2021duo,Kobayashi:2021uam,Bruggisser:2022rhb,Bartocci:2023nvp,Bartocci:2024fgj}, ensures that any flavor-violating effects beyond the SM are aligned with the observed structure of the SM Yukawas. In the MFV paradigm, higher-dimensional operators are systematically classified according to their transformation properties under the flavor group and spurion insertions~\cite{sun2025flavorcpsymmetriesstandard}. 

A well-known open problem in the Standard Model concerns the origin of flavor. A common strategy is to assume that flavor and CP violation are governed by the SM Yukawa couplings, treated as spurions transforming under a global flavor symmetry, e.g.\ $U(3)^5$ or $U(2)^5$. This idea, known as Minimal Flavor Violation (MFV)~\cite{Chivukula:1987py,DAmbrosio:2002vsn,Bonnefoy:2020yee,Aoude:2020dwv,Bruggisser:2021duo,Kobayashi:2021uam,Bruggisser:2022rhb,Bartocci:2023nvp,Bartocci:2024fgj}, ensures that any flavor-violating effects beyond the SM are aligned with the observed hierarchies and mixing encoded in the SM Yukawa matrices. In the MFV paradigm, higher-dimensional operators are systematically classified according to their transformation properties under the flavor group and the required spurion insertions~\cite{sun2025flavorcpsymmetriesstandard}.
%The MFV hypothesis shows that the SM Yukawa matrices $Y_u$, $Y_d$, $Y_e$ are the only sources of flavor symmetry breaking, and the Yukawa couplings can be thought of as spurion fields that break this symmetry in well-defined directions~\cite{sun2025flavorcpsymmetriesstandard}.
The MFV hypothesis shows that the SM Yukawa matrices $Y_u$, $Y_d$, $Y_e$ are the only sources of flavor symmetry breaking, and the Yukawa matrices are assigned spurion transformation properties under the flavor group so that any flavor violation can be written in terms of Yukawa insertions~\cite{sun2025flavorcpsymmetriesstandard}.
%Besides the MFV approach, as a bottom-up thinking, non-Abelian discrete flavor symmetries have been proposed as alternative solutions to the flavor problem~\cite{Ding_2025,Altarelli_2010,bartlett2005categoricalaspectstopologicalquantum,Altarelli_2005,ma2004nonabeliandiscretefamilysymmetries,Ishimori_2010,King_2014}. 
%These symmetries were initially introduced to account for the tri-bimaximal mixing pattern observed in the lepton sector~\cite{Harrison_2002,Harrison1_2002,Xing_2002,harrison2004statustribimaximalneutrinomixing,Harrison_2004, Babu_2003}.
Complementary to MFV, and from a bottom-up perspective, non-Abelian discrete flavor symmetries have been proposed as alternative solutions to the flavor problem~\cite{Ding_2025,Altarelli_2010,bartlett2005categoricalaspectstopologicalquantum,Altarelli_2005,ma2004nonabeliandiscretefamilysymmetries,Ishimori_2010,King_2014}. These symmetries were originally introduced to reproduce the tri-bimaximal pattern of lepton mixing~\cite{Harrison_2002,Harrison1_2002,Xing_2002,harrison2004statustribimaximalneutrinomixing,Harrison_2004,Babu_2003}.
%In the eigenbasis of the charged-lepton mass matrix $M_e$, and assuming that neutrinos are Majorana particles, the PMNS matrix can, as an experimental approximation, be taken to have the tri-bimaximal form~\cite{Harrison_2002}, which can be written as 
%\begin{equation}
%    U_{\text{PMNS}} = U_{\text{TB}} = 
%    \begin{pmatrix}
%        \sqrt{2/3} & 1/\sqrt{3} & 0 \\
%        -1/\sqrt{6} & 1/\sqrt{3} & -1/\sqrt{2} \\
%        -1/\sqrt{6} & 1/\sqrt{3} & +1/\sqrt{2}.
%        \label{eq:TB}
%        \end{pmatrix}.
%\end{equation}
Working in the basis where the charged-lepton mass matrix is diagonal, and assuming Majorana neutrinos, the PMNS matrix is, to a good approximation, given by the tri-bimaximal form~\cite{Harrison_2002},
\begin{equation}
    U_{\rm PMNS} \simeq U_{\rm TB} =
    \begin{pmatrix}
        \sqrt{2/3} & 1/\sqrt{3} & 0 \\
        -1/\sqrt{6} & 1/\sqrt{3} & -1/\sqrt{2} \\
        -1/\sqrt{6} & 1/\sqrt{3} & +1/\sqrt{2}
    \end{pmatrix},
    \label{eq:TB}
\end{equation}
where $U_{\rm TB}$ denotes the tri-bimaximal mixing matrix. 
%Before electroweak and flavor symmetry breaking, we assume that the lepton sector is well approximated by tri-bimaximal (TB) mixing. In this limit, the effective Majorana neutrino mass matrix is diagonalized by the TB mixing matrix \(U_{\rm TB}\),
%\begin{equation}
%M_\nu = U_{\rm TB}\,\mathrm{diag}(m_1,m_2,m_3)\,U_{\rm TB}^{T},
%\end{equation}
%with a hierarchical spectrum \(m_3 \gg m_2 \gg m_1\). On the same weak basis, the charged-lepton mass matrix is taken to be diagonal as \(M_\ell=\mathrm{diag}(m_e,m_\mu,m_\tau)\), so that all lepton mixing originates from the neutrino sector. 
In the flavor-symmetry breaking, we assume that the lepton sector is well described by TB mixing. In this limit, the effective Majorana neutrino mass matrix is diagonalized by $U_{\rm TB}$,
\begin{equation}
    M_\nu = U_{\rm TB}\,\mathrm{diag}(m_1,m_2,m_3)\,U_{\rm TB}^{T},
\end{equation}
with a hierarchical spectrum $m_3 \gg m_2 \gg m_1$. In the same weak basis, the charged-lepton mass matrix is taken to be diagonal, $M_\ell=\mathrm{diag}(m_e,m_\mu,m_\tau)$, so that all lepton mixing originates from the neutrino sector.
%A useful way to characterize this situation is through residual flavor symmetries. The diagonal charged-lepton structure is left invariant by a transformation \(Z_3\) generated by \(T=\mathrm{diag}(1,\omega,\omega^2)\), where \(\omega=e^{2\pi i/3}\). The TB neutrino mass matrix, on the other hand, is invariant under a \(Z_2\) transformation generated by the standard matrix \(S\) in the triplet representation. Taking \(S\) and \(T\) as generators then yields the full \(A_4\) flavor symmetry of the Lagrangian\footnote{Also, the neutrino mass matrix can also remain invariant under $P = T^{2} S T^{2}$, which corresponds to an invariance under $\mu$--$\tau$ reflection (i.e.\ $\mu$--$\tau$ mass symmetry). If $P$ is also taken to be an element of the flavor group, the associated flavor symmetry is then promoted to the $S_4$ symmetry. }. 
A useful way to characterize this situation is in terms of residual flavor symmetries. The diagonal charged-lepton structure is invariant under a residual $Z_3$ generated by $T=\mathrm{diag}(1,\omega,\omega^2)$, where $\omega=e^{2\pi i/3}$. The TB neutrino mass matrix, on the other hand, is invariant under a $Z_2$ generated by the usual matrix $S$ in the triplet representation. Taking $S$ and $T$ as generators yields the full $A_4$ flavor symmetry of the Lagrangian%
\footnote{The neutrino mass matrix may also be invariant under $P=T^{2}ST^{2}$, corresponding to $\mu$--$\tau$ reflection (i.e.\ $\mu$--$\tau$ mass symmetry). If $P$ is included as an element of the flavor group, the symmetry is promoted to $S_4$.}.
%An effective choice for Yukawa terms for getting the leptonic sector with cutoff scale $\Lambda$ is~\cite{Altarelli_2005}
%\begin{equation}
%    \mathcal{L}
%= y_e\, e^{c}\,(\varphi_T\, l)\,\frac{h_d}{\Lambda}
%+ y_\mu\, \mu^{c}\,(\varphi_T\, l)^{\prime}\,\frac{h_d}{\Lambda}
%+ y_\tau\, \tau^{c}\,(\varphi_T\, l)^{\prime\prime}\,\frac{h_d}{\Lambda}
%+ \frac{x_a}{\Lambda^{2}}\,\xi\,(l h_u\, l h_u)
%+ \frac{x_b}{\Lambda^{2}}\,(\varphi_S\, l h_u\, l h_u)
%+ \text{h.c.} + \ldots,
%\end{equation} 
%with the leptonic and scalar fields transforming under $A_4$ as $\{e^c,h_u,h_d\} \sim 1,\; \mu^c\sim 1^{\prime\prime},\; \tau^c\sim 1^\prime,\; l\sim 3$, and the three flavons \(\phi_T\) and \(\phi_S\) transforming as triplets, and $\xi$ transforming as a trivial singlet under \(A_4\). The relevant models in extra dimension and 4-dimensional SUSY were first proposed by Guido Altarelli and Ferruccio Feruglio in~\cite{Altarelli_2005} and~\cite{Altarelli_2006}.  
As an illustrative example, the effective lepton Yukawa Lagrangian reads~\cite{Altarelli_2005}
\begin{equation}
    \mathcal{L}
= y_e\, e^{c}\,(\varphi_T\, l)\,\frac{h_d}{\Lambda}
+ y_\mu\, \mu^{c}\,(\varphi_T\, l)^{\prime}\,\frac{h_d}{\Lambda}
+ y_\tau\, \tau^{c}\,(\varphi_T\, l)^{\prime\prime}\,\frac{h_d}{\Lambda}
+ \frac{x_a}{\Lambda^{2}}\,\xi\,(l h_u\, l h_u)
+ \frac{x_b}{\Lambda^{2}}\,(\varphi_S\, l h_u\, l h_u)
+ \text{h.c.} + \ldots ,
\end{equation}
with $\Lambda$ the cutoff scale and $A_4$ assignments
$\{e^c,h_u,h_d\}\sim 1$, $\mu^c\sim 1^{\prime\prime}$, $\tau^c\sim 1^\prime$, $l\sim 3$,
$\varphi_T\sim 3$, $\varphi_S\sim 3$, $\xi\sim 1$.
Concrete realizations in extra dimensions and in four-dimensional SUSY were first proposed in~\cite{Altarelli_2005} and ~\cite{Altarelli_2006}.
%Nevertheless, these theoretical constructions still have several shortcomings. First, the framework typically requires the introduction of multiple flavon fields and relies on specific vacuum alignments of different flavon fields to generate the mass matrices. Second, if one demands that the quark sector follow an analogous structure to that of the leptonic sector, the CKM matrix becomes the identity at leading order, but its subleading corrections are far too small to account for the experimentally observed mixing. Finally, all Yukawa interactions arise from non-renormalizable operators with mass dimension greater than 4. One approach that inherits the philosophy of discrete symmetry is modular symmetry.
Nevertheless, these constructions still have several limitations. First, they typically introduce several flavon fields and rely on specific vacuum alignments to generate the desired mass matrices. Second, if one imposes an analogous structure in the quark sector, the CKM matrix is the identity at leading order, while subleading corrections are generically too small to reproduce the observed mixing. Finally, all Yukawa interactions arise from non-renormalizable operators of mass dimension greater than four. An alternative that retains the spirit of discrete symmetries is provided by modular symmetry.

%Modular symmetry is a symmetry that makes the torus $T^2$ which is defined by a lattice $\Lambda=\mathbb{Z} \omega_1+\mathbb{Z}\omega_2$ invariant~\cite{Ding:2023htn}.
%Such lattice-preserving transformations can be realized as elements of the group $\mathrm{SL}(2,\mathbb{Z})$.
%Since we are not concerned with isomorphic transformations of the torus, we may reparametrize the lattice by defining $\tau = \omega_1 / \omega_2$, which uniquely fixes the geometry of the torus.
%Such a reparametrization introduces an overall rescaling factor $(c\tau + d)$.
%For functions defined on the lattice, it is natural to assume homogeneity
%of degree $k$ with respect to this rescaling.
%By further requiring holomorphicity and appropriate convergence properties,
%one is then led to define modular forms as functions satisfying, under
%$\mathrm{SL}(2,\mathbb{Z})$,
%\begin{equation}
%    f(\gamma \tau) = (c\tau + d)^k f(\tau) \, .
%\end{equation}
Modular symmetry is the group of transformations that preserve the lattice $\Lambda=\mathbb{Z}\,\omega_1+\mathbb{Z}\,\omega_2$ associated with the torus $T^2$~\cite{Ding:2023htn}. Such lattice-preserving transformations are realized by elements of $\mathrm{SL}(2,\mathbb{Z})$. Since an overall rescaling of the lattice does not change the complex structure, it is convenient to parameterize the lattice by the modulus $\tau=\omega_1/\omega_2$, which fixes the geometry of the torus. Under $\gamma=\begin{pmatrix}a&b\\ c&d\end{pmatrix}\in \mathrm{SL}(2,\mathbb{Z})$, this reparameterization induces a fractional linear transformation $\tau\to \gamma\tau=(a\tau+b)/(c\tau+d)$, together with an overall rescaling of the lattice by $(c\tau+d)$. For functions on the lattice, it is then natural to consider functions that are homogeneous of degree $k$ under this rescaling. Imposing holomorphicity, especially at the cusps, and appropriate growth conditions leads to modular forms $f(\tau)$ of weight $k$, which satisfy
\begin{equation}
    f(\gamma \tau) = (c\tau + d)^k f(\tau)\, .
\end{equation}

%Moreover, not for original modular group $SL(2,\mathbb{Z})$, the principal congruence subgroups $\Gamma(N)$ are always considered.
%Because the modular forms defined in $\Gamma(N)$ has more inequivalent functional forms, such forms can serve as representation vectors of the group in $\Gamma_N\equiv SL(2,\mathbb{Z})/\left\{ \pm \Gamma(N) \right\}$, which is a discrete group. 
In modular-invariant model building, one typically works with the finite modular group
$\Gamma_N \equiv \mathrm{SL}(2,\mathbb{Z})/\{\pm \Gamma(N)\}$, obtained by quotienting the modular group by the principal congruence subgroup $\Gamma(N)$. Modular forms of level $N$ can then be organized into multiplets transforming in representations of $\Gamma_N$.
%In SUSY modular-invariant theories, the superpotential is holomorphic function of modulus $\tau$ and other chiral supermultiplets $\phi^{(I)}$ is transformed under modular symmetry as
%\begin{equation}
%    \phi^{(I)} (\gamma \tau) = (c\tau + d)^{-k} \, \rho(\gamma) \phi^{(I)}(\tau).
%\end{equation}
%Here the $\rho(\gamma)$ is representation matrix of the quotient group $\Gamma_N$. 
%Therefore, the superpotential is naturally constructed from holomorphic modular forms. However, since the Kähler potential is not required to be holomorphic,
%its explicit form cannot be fully determined by imposing modular symmetry alone.
%The essential reason is that non-holomorphic operators transform with factors
%of the form $(c\tau + d)^{k_1}(c\overline{\tau}+d)^{k_2}$, which can be combined with different
%powers of $\mathrm{Im}\,\tau$ with proper weight modular forms to build Kähler-invariant structures, with $\mathrm{Im}\,\tau$ transforming as $\operatorname{Im}(\gamma\tau)= \frac{\operatorname{Im}\,\tau}{(c\tau+d)\,(c\bar{\tau}+d)}\,.$
%As a consequence, a large number of modular-invariant terms are allowed,
%and the resulting parameter space becomes highly redundant~\cite{Chen_2020}.
In SUSY modular-invariant theories, the superpotential is a holomorphic function of the modulus $\tau$ and chiral supermultiplets $\phi^{(I)}$, which transform under modular symmetry as
\begin{equation}
    \phi^{(I)} (\gamma \tau) = (c\tau + d)^{-k_I}\,\rho(\gamma)\,\phi^{(I)}(\tau)\, .
\end{equation}
Here $\rho(\gamma)$ denotes a representation matrix of the finite modular group $\Gamma_N$. Consequently, the superpotential can be built from holomorphic modular forms. 
The K\"ahler potential, however, is generically non-holomorphic and therefore cannot be constructed from holomorphic modular forms alone. In particular, non-holomorphic operators transform with factors of the form $(c\tau + d)^{k_1}(c\bar{\tau}+d)^{k_2}$. To build modular-invariant combinations, such factors are typically compensated by appropriate powers of $\mathrm{Im}\,\tau$, and always together with modular forms of suitable weight. 
Indeed, $\mathrm{Im}\,\tau$ transforms as
\begin{equation}
    \operatorname{Im}(\gamma\tau)=\frac{\operatorname{Im}\,\tau}{(c\tau+d)(c\bar{\tau}+d)}\, .
\end{equation}
As a result, many modular-invariant K\"ahler terms are allowed, and the corresponding parameter space can become highly redundant~\cite{Chen_2020}.
%Up to now, many models with modular symmetry have been considered, like $\Gamma_2 \simeq S_3$~\cite{Liu:2020akv,Yao:2020qyy,Behera:2025tpj,Behera:2024ark,Meloni:2023aru,Marciano_2024,behera2024neutrinophenomenologymodulars3,Meloni_2023,mishra2020neutrinomixingleptogenesismodular,Okada_2019,Kobayashi_2019}, $\Gamma_3 \simeq A_4$~\cite{nomura2023texturezerosrealizationthreeloop,Kumar_2024,gogoi2023leptogenesisdarkmatterminimal,mishra2023exploringmodelsmodularsymmetry,Ding:2022bzs,Chen:2021zty,nomura2023quarkleptonmodelflavor,kim2023fermilatgevexcessmuon,devi2023retrievingtexturezeros31,Ding:2021eva,dasgupta2021diracradiativeneutrinomass,Nomura_2020,pathak2025neutrinomassgenesisscotoinverse,Wang_2020,nomura2024novelapproachradiativelinear,Kashav_2021,Kashav_2023,Kobayashi_2020,nomura2019inverseseesawmodela4modular,behera2020modulara4symmetricscotogenic,Kobayashi1_2018,Kobayashi_2018}, $\Gamma_4 \simeq S_4$~\cite{Ding:2019gof,Penedo2_2019,Novichkov2_2019,Kobayashi2_2019,Liu_2021,varzielas2023quarksmodulars4cusp,King_2020,Ding:2021zbg,Ding:2024inn,Penedo_2019}, and $\Gamma_5 \simeq A_5$~\cite{Ding:2019xna,Yao:2020zml,Novichkov_2019}.
To date, a wide variety of modular-symmetric models have been studied, for example based on $\Gamma_2 \simeq S_3$~\cite{Liu:2020akv,Yao:2020qyy,Behera:2025tpj,Behera:2024ark,Meloni:2023aru,Marciano_2024,behera2024neutrinophenomenologymodulars3,Meloni_2023,mishra2020neutrinomixingleptogenesismodular,Okada_2019,Kobayashi_2019}, $\Gamma_3 \simeq A_4$~\cite{nomura2023texturezerosrealizationthreeloop,Kumar_2024,gogoi2023leptogenesisdarkmatterminimal,mishra2023exploringmodelsmodularsymmetry,Ding:2022bzs,Chen:2021zty,nomura2023quarkleptonmodelflavor,kim2023fermilatgevexcessmuon,devi2023retrievingtexturezeros31,Ding:2021eva,dasgupta2021diracradiativeneutrinomass,Nomura_2020,pathak2025neutrinomassgenesisscotoinverse,Wang_2020,nomura2024novelapproachradiativelinear,Kashav_2021,Kashav_2023,Kobayashi_2020,nomura2019inverseseesawmodela4modular,behera2020modulara4symmetricscotogenic,Kobayashi1_2018,Kobayashi_2018}, $\Gamma_4 \simeq S_4$~\cite{Ding:2019gof,Penedo2_2019,Novichkov2_2019,Kobayashi2_2019,Liu_2021,varzielas2023quarksmodulars4cusp,King_2020,Ding:2021zbg,Ding:2024inn,Penedo_2019}, and $\Gamma_5 \simeq A_5$~\cite{Ding:2019xna,Yao:2020zml,Novichkov_2019}.

Although the superpotential is holomorphic, so that its interaction terms are expressed in terms of holomorphic modular forms, higher-dimensional operators in the non-supersymmetric Standard Model are generically non-holomorphic. Nevertheless, the holomorphic sector of modular forms is mathematically well structured, and we impose modular symmetry in the SM while continuing to build Yukawa couplings from holomorphic modular forms. This setup may be viewed as a minimal way to realize the SM as a low-energy effective theory descending from SUSY. 
%We assume that all flavor structures arise solely from the leading-order Yukawa sector; accordingly, and treat $\tau$ as a background parameter, i.e. it does not vary with the coordinates. $\tau$ is treated as a non-dynamical spurion and we do not introduce explicit $\mathrm{Im}\,\tau$--dependent structures. Flavor-invariant operators are then constructed by inserting suitable modular forms.
We assume that all flavor structures arise solely from the leading-order Yukawa sector; accordingly, we treat $\tau$ as a background parameter that does not vary with the coordinates. So that we can take $\tau$ to be a non-dynamical spurion and do not introduce explicit $\mathrm{Im}\,\tau$--dependent structures. Flavor-invariant operators are then constructed by inserting appropriate modular forms.
For the holomorphic theory at level $N=3$, it suffices to take the weight--2 triplet $Y^{(2)}_{\mathbf{3}}$ as the fundamental spurion, since all higher-weight modular forms can be obtained from its symmetric tensor powers. In this way, flavor symmetry breaking in higher-dimensional operators is entirely controlled by modulus $\tau$, especially by the modular forms, and does not require additional sources. More generally, for any level $N$, whenever modular forms of arbitrary weight can be generated via tensor products of a finite set of seed forms, these seeds may be taken as the basic building blocks. This yields an MFV-like construction: the only flavor-symmetry-breaking sources are the modular forms themselves, and the flavor symmetry is broken only once $\tau$ acquires a vacuum expectation value (VEV). Using the Hilbert series, we can then count, at each operator dimension, the number of independent singlet structures that can be formed.

%\textbf{This assumption in fact includes the coupling ansatz used in Ref.~\cite{Kobayashi_2022} for constructing SMEFT operators, namely that the four-point couplings of matter fields are built from three-point couplings of matter fields. At the same time, in our work we indeed reproduce part of their results, especiall for the bilinear dimension-6 operators they consider, while some quadratic dimension-6 operators written in this way naturally introduce a lot of redundancy, and this approach cannot account for the redundancy implied by the holomorphic modular-form constraints.One example is the dimension-5 case, namely the Weinberg operator. Using this construction, we find that if we countthe operators directly, some of them are actually linearly related, and cannot take into account the constraint $Y^{(4)}_\mathbf{3} = 0$ among the modular forms. But in our approach, using the Hilbert series, we can naturally remove these redundancies and obtain both the number and the explicit form of independent operators.}
This assumption subsumes the coupling ansatz adopted in Ref.~\cite{Kobayashi_2022} for constructing SMEFT operators, namely that four-point couplings of matter fields can be assembled from three-point couplings. Within our framework, we reproduce part of their results, in particular for the bilinear dimension-6 operators considered there. However, for some quadratic dimension-6 operators their construction is highly redundant and misses the relations imposed by holomorphic modular-form constraints.
A simple illustration already appears at dimension five, i.e.\ for the Weinberg operator. If one counts operators directly within the above construction, some of the resulting structures are in fact linearly dependent; moreover, the counting does not properly incorporate modular-form constraints. By contrast, our Hilbert-series approach removes these redundancies in a systematic way and yields both the number and an explicit basis of independent operators.

While holomorphic modular forms have useful mathematical properties, the Standard Model does not require the Lagrangian to be constructed solely from holomorphic modular forms. It is therefore important to consider the more general non-holomorphic case~\cite{qu2024nonholomorphicmodularflavorsymmetry}.
%For non-holomorphic modular forms, the main difference from the holomorphic case is that they do not satisfy holomorphy but instead obey a Laplacian condition.This condition means that, unlike in the holomorphic case, under minimal assumptions we cannot find a single special modular form to use as a building block to generate all modular forms.Therefore, when we consider higher-dimensional operators, even without introducing $\operatorname{Im}\tau$, we stillhave to include many additional combinations of modular forms, which leads to a large number of operators.So, as a reasonable and structured operator construction, inspired by the holomorphic case, we assume that the non-holomorphic case also follows the same operator structure as in the holomorphic case, namely that
$
\bigl[Y_\mathbf{r}^{(k_Y)} \, {Y_{\mathbf{r}'}^{(k_Y')}}^{*} \, \mathcal{O}\bigr]_{\mathbf{1}}
$
%serves as a minimal-assumption construction for the non-holomorphic case to avoid large number of operators, \textbf{which means any comformal factor compensation be provided entirely by independent modular forms.} With benchmark lepton model considered in~\cite{qu2024nonholomorphicmodularflavorsymmetry}, we construct dimension-5 Weinberg operators and dimension-6 operators in lepton sector.
For non-holomorphic modular forms, the key difference from the holomorphic case is that they are not holomorphic; instead, they obey a Laplacian condition. As a consequence, under our minimal assumptions in holomorphic case we cannot identify a single special modular form to use as a building block to generate all modular forms. When constructing higher-dimensional operators, even if we do not introduce $\operatorname{Im}\tau$ as building block, we still have to include infinite additional combinations of modular forms, which leads to a large number of operators.
Motivated by the holomorphic construction, we therefore adopt the following structured ansatz for the non-holomorphic case:
\[
\bigl[\,Y_{\mathbf{r}}^{(k_Y)} \, (Y_{\mathbf{r}'}^{(k_Y')})^{*} \, \mathcal{O}\,\bigr]_{\mathbf{1}}\,,
\]
which provides a minimal-assumption prescription to avoid an excessive number of operators; in particular, any modular weight compensation is supplied entirely by independent modular forms. Using the benchmark lepton model of Ref.~\cite{qu2024nonholomorphicmodularflavorsymmetry}, we then construct the dimension-5 Weinberg operator and the dimension-6 operators in the lepton sector.

%Meanwhile, in our modular symmetry SM model, we assume that the quark andlepton sectors transform under distinct flavor symmetries, so that theoverall flavor group is taken to be $A_{4}^{(q)} \times A_{4}^{(e)}$, withtwo independent moduli $\tau_q$ and $\tau_e$ to fit models with different fix points in quark and lepton sectors. As for the lepton sector, we assume that neutrinos are Majorana particles.Furthermore, for the right-handed up- and down-type quarks, we assign lowest properweights that fit the observed values at the model-building level. This choice is mainly motivated by the desire to compare our construction with that of Ref.~\cite{Kobayashi_2022}. Anyway, their construction does not take into account the redundancy introduced by the constarin in $A_4$ modular forms.In the leading-order Lagrangian, we neglect the kinetic terms for $\tau$,so that all modular forms can be treated as spurions. With our minimal assumption, we do not includehigher-order structures built from $\mathrm{Im}\,\tau$, in order to avoidintroducing an excessive number of additional parameters.Finally, by means of the Hilbert series, we determine both the number and the explicit structure of SMEFT singlet operators that are invariant under the imposed modular flavor symmetry, and we present explicit constructions for these singlet combinations.
In our modular invariant SM model, we assume that the quark and lepton sectors transform under distinct flavor symmetries, such that the overall flavor group is taken to be $A_{4}^{(q)} \times A_{4}^{(e)}$, with two independent moduli $\tau_q$ and $\tau_e$ to accommodate models with different fixed points in the quark and lepton sectors. As for the lepton sector, we assume that neutrinos are Majorana particles. Furthermore, for the right-handed up- and down-type quarks, we assign the lowest admissible weights that fit the observed values at the model-building level. 
%This choice is mainly motivated by the desire to compare our construction with that of Ref.~\cite{Kobayashi_2022}. However, their construction does not take into account the redundancy introduced by the constraint in $A_4$ modular forms. 
At leading order, we are only interested in $\tau$ as a background parameter; accordingly, we treat the basic modular forms $Y^{(2)}_{\mathbf{3}}(\tau_q)$ and $Y^{(2)}_{\mathbf{3}}(\tau_e)$ as spurions. Under our minimal assumption, we do not include higher-order structures built from $\mathrm{Im}\,\tau$, in order to avoid introducing an excessive number of additional parameters. Finally, using the Hilbert series, we determine both the number and the explicit structure of SMEFT singlet operators that are invariant under the imposed modular flavor symmetry, and we present explicit constructions of these singlet combinations.

The paper is organized as follows. In Section~\ref{ModularSM}, In Section~\ref{ModularSM}, we review the Standard Model flavor structure and the MFV flavor pattern. We then construct a holomorphic modular version of the Standard Model, and review the MFV implementation in the SM. Adopting a specific modular $A_4$ realization of the Standard Model, we introduce a modular-invariant SMEFT and list all fields together with their representations.
In Section~\ref{sec:hilbert}, we review the construction of the Hilbert series for continuous symmetry groups, using the SMEFT as an illustrative example. In addition, we present the Hilbert-series method for discrete symmetries. In particular, we spell out the building blocks of our construction, generate higher-order modular invariants, and validate the resulting operators in two complementary ways: (i) by explicitly constructing them from the weight-2 triplet \(Y^{(2)}_{\mathbf{3}}\) and its conjugate; and (ii) by forming the tensor product \(Y_{\mathbf{r}}^{(k_Y)}\,{Y_{\mathbf{r}'}^{(k_Y')}}^{*}\,\mathcal{O}\) and projecting it onto the group-invariant singlet.
In Section~\ref{sec:smeft}, we build the effective operators relevant for our setup, focusing on the explicit dimension-5 and dimension-6 cases with baryon- and lepton-number-conserving operators. Section~\ref{sec:Maabeta_construction} turns to the non-holomorphic case. We outline what is meant by a non-holomorphic modular form, clarify how it differs from the holomorphic case, and highlight the obstruction that arises when the requirement of building $YY^* \mathcal{O}\rightarrow \mathbf{1}$ is relaxed. Finally, we summarize our conclusions in Section~\ref{sec:conclusion}.          % 对应 \section{Introduction}

\section{Standard Model and modular symmetry}
\label{ModularSM}
%In this section, we review the SM flavor structure and flavor-symmetry patterns, recall the MFV implementation in SMEFT,  which will serve as a useful point of reference for our modular construction, and introduce a model combining modular flavor symmetry and SM. These results will be used in the next section to construct SMEFT operators consistent with modular symmetry.

In this section, we review the SM flavor structure and the associated patterns of flavor symmetries, and we recall the implementation of MFV in the SMEFT, which will serve as a useful reference for our modular construction. We introduce a model that combines modular flavor symmetry with the SM. These results will be used in the next section to construct SMEFT operators consistent with modular symmetry.

\subsection{Standard Model flavor structure and flavor symmetry patterns}
\begin{comment}
The Standard Model (SM) is a renormalizable gauge theory based on the symmetry group
$SU(3)_C \otimes SU(2)_{L} \otimes U(1)_Y$, which governs strong, weak, and hypercharge
interactions.
The SM Lagrangian can be decomposed into four parts,
\begin{equation}
\mathcal{L}_{\rm SM}
= \mathcal{L}_{\rm gauge}
+ \mathcal{L}_{\rm kinetic}
+ \mathcal{L}_{\rm Higgs}
+ \mathcal{L}_{\rm Yuk}.
\label{eq:LSM}
\end{equation}
The gauge part contains the kinetic terms of the gauge fields,
\begin{equation}
\mathcal{L}_{\rm gauge}
= -\frac14\, G_{\mu\nu} G^{\mu\nu}
  -\frac14\, W_{\mu\nu} W^{\mu\nu}
  -\frac14\, B_{\mu\nu} B^{\mu\nu}.
\end{equation}
Here $G_\mu$ are the gluon fields of $SU(3)_C$, $W_\mu$ are the weak gauge bosons of $SU(2)_{L}$, and $B_\mu$ is the hypercharge gauge boson of $U(1)_Y$.
\end{comment}

The Standard Model (SM) is a renormalizable gauge theory based on the symmetry group
$SU(3)_C \otimes SU(2)_{L} \otimes U(1)_Y$, which governs strong, weak, and
$U(1)_Y$ (hypercharge) interactions.
The SM Lagrangian can be decomposed into four parts,
\begin{equation}
\mathcal{L}_{\rm SM}
= \mathcal{L}_{\rm gauge}
+ \mathcal{L}_{\rm kinetic}
+ \mathcal{L}_{\rm Higgs}
+ \mathcal{L}_{\rm Yuk}.
\label{eq:LSM}
\end{equation}
The gauge part contains the kinetic terms of the gauge fields,
\begin{equation}
\mathcal{L}_{\rm gauge}
= -\frac14\, G_{\mu\nu} G^{\mu\nu}
  -\frac14\, W_{\mu\nu} W^{\mu\nu}
  -\frac14\, B_{\mu\nu} B^{\mu\nu}.
\end{equation}
Here $G_\mu$ are the gluon fields of $SU(3)_C$, $W_\mu$ are the weak gauge bosons of
$SU(2)_{L}$, and $B_\mu$ is the hypercharge gauge boson of $U(1)_Y$.

\begin{comment}
The kinetic terms for SM fermions are
\begin{equation}
\mathcal{L}_{\rm kinetic}
= \sum_{i=1}^3\Big[
\overline{Q_L^i}i\slashed{D}Q_L^i
+\overline{u_R^i}i\slashed{D}u_R^i
+\overline{d_R^i}i\slashed{D}d_R^i
+\overline{L_L^i}i\slashed{D}L_L^i
+\overline{e_R^i}i\slashed{D}e_R^i
\Big].
\end{equation}
The fields
$Q_L^i=(u_L^i,d_L^i)^T$ and $L_L^i=(\nu_L^i,e_L^i)^T$ transform as doublets under
$SU(2)_{L}$, while $u_R^i,d_R^i,e_R^i$ are singlets and index
$i=1,2,3$ denotes the flavor index of SM fermions..
\end{comment}
The kinetic terms for SM fermions are
\begin{equation}
\mathcal{L}_{\rm kinetic}
= \sum_{i=1}^3\Big[
\overline{Q_L^i}i\slashed{D}Q_L^i
+\overline{u_R^i}i\slashed{D}u_R^i
+\overline{d_R^i}i\slashed{D}d_R^i
+\overline{L_L^i}i\slashed{D}L_L^i
+\overline{e_R^i}i\slashed{D}e_R^i
\Big].
\end{equation}
The fields
$Q_L^i=(u_L^i,d_L^i)^T$ and $L_L^i=(\nu_L^i,e_L^i)^T$ transform as doublets under
$SU(2)_{L}$, while $u_R^i,d_R^i,e_R^i$ are singlets. The index
$i=1,2,3$ denotes the flavor index of SM fermions.

The Higgs part is
\begin{equation}
\mathcal{L}_{\rm Higgs}
=(D_\mu H)^\dagger(D^\mu H)-V(H),
\qquad
V(H)= -\mu^2 H^\dagger H + \lambda (H^\dagger H)^2,
\end{equation}
and the Higgs doublet $H$ is an $SU(2)_{L}$ doublet with hypercharge $Y=1/2$.
When the Higgs field acquires a vacuum expectation value (VEV), the electroweak symmetry
$SU(2)_{L}\otimes U(1)_Y$ is spontaneously broken to $U(1)_{\rm em}$, generating masses
for the $W^\pm$ and $Z$ bosons. Moreover, this mechanism also generates fermion masses
through the Yukawa terms,
\begin{equation}
\mathcal{L}_{\rm Yuk}
= \,u_R^{c\,j}\, (Y_u)_{ji}\, \tilde H^\dagger\, Q_L^i
   +\,d_R^{c\,j}\, (Y_d)_{ji}\, H^\dagger\, Q_L^i
   +\,e_R^{c\,j}\, (Y_e)_{ji}\, H^\dagger\, L_L^i
+ {\rm h.c.},
\end{equation}
with $\tilde H = i\sigma_2 H^\ast$ and $Y_{u,d,e}$ being $3\times 3$ complex Yukawa matrices.
After electroweak symmetry breaking, the Yukawa interactions generate fermion mass matrices.
In the quark sector,
\begin{equation}
\mathcal{L}_{\rm Yuk} \supset
\,u_R^{c\,i}(M_u)_{ij}u_L^j
+\,d_R^{c\,i}(M_d)_{ij}d_L^j
+{\rm h.c.},
\end{equation}
which are in general non-diagonal in flavor space. They are diagonalized by the transformations,
\begin{equation}
U_R^{u\dagger} M_u U_L^u = M_u^{\rm diag},
\qquad
U_R^{d\dagger} M_d U_L^d = M_d^{\rm diag}.
\end{equation}
Rotating quark fields to their mass eigenstates, the charged-current interaction becomes
\begin{equation}
\mathcal{L}_{CC}=
\frac{g}{\sqrt2}\,\bar u_L^i \gamma^\mu \left(V_{\rm CKM}\right)_{ij}\, d_L^j\, W_\mu^{+}
+{\rm h.c.},
\qquad
V_{\rm CKM}=U_L^{u\dagger}U_L^{d},
\end{equation}
so the CKM matrix arises from the mismatch between the left-handed rotations that diagonalize
$M_u$ and $M_d$. In the lepton sector, the charged-lepton masses are generated analogously,
\begin{equation}
M_\ell=\frac{v}{\sqrt2}\,Y_e,
\qquad
U_R^{\ell\dagger} M_\ell U_L^\ell = M_\ell^{\rm diag}\,.
\end{equation}
Neutrinos are massless in the renormalizable SM. However, neutrino masses can be included
via non-renormalizable operators, for instance the dimension-five Weinberg operator.
After the Higgs acquires a VEV, the neutrino mass term is
\begin{equation}
\mathcal{L}_{\nu\,\text{mass}}
= \frac12\, \nu_L^{T}\,C^{-1} M_\nu\, \nu_L \;+\; {\rm h.c.}\;.
\end{equation}
This mass matrix is diagonalized by a unitary matrix $U_\nu$,
\begin{equation}
U_\nu^{T} M_\nu U_\nu = M_\nu^{\rm diag}.
\end{equation}
Upon transforming to the mass basis, the leptonic charged-current interaction reads
\begin{equation}
\mathcal{L}_{CC}=
\frac{g}{\sqrt2}\,\bar \nu_L^i \gamma^\mu \left(U_{\rm PMNS}\right)_{ij}\, e_L^j\, W_\mu^{+}
+{\rm h.c.},
\qquad
U_{\rm PMNS}=U_L^{\ell\dagger}U_\nu.
\end{equation}
The PMNS matrix~\cite{Pontecorvo:1957qd,Maki:1962mu} is the counterpart of the CKM matrix in the leptonic sector.

It is worth noting that, in the limit $Y_{u,d,e}\to 0$, the fermion kinetic terms possess
the largest global symmetry,
\begin{equation}
\label{eq:u35}
G_F \;=\; U(3)^5 \;=\; U(3)_q \times U(3)_u \times U(3)_d \times U(3)_\ell \times U(3)_e\,,
\end{equation}
which commutes with the gauge interactions. Turning on the Yukawa couplings $Y_u$, $Y_d$,
and $Y_e$ explicitly breaks $G_F$. In the SM, these Yukawas are the unique
sources that distinguish different fermion flavors and induce CP violation.

Neglecting the Weinberg operator, the three complex Yukawa matrices
$Y_u, Y_d,$ and $Y_e$ contain in total $54$ real parameters.
Within the flavor symmetry $G_F$, there are four global $U(1)$ symmetries
that do not affect $Y_u, Y_d,$ or $Y_e$. These form the rephasing group
\begin{equation}
\label{eq:rephasing}
U(1)^4 \;=\; U(1)_{e-\mu}\times U(1)_{\mu-\tau}\times U(1)_B\times U(1)_L \, .
\end{equation}
Therefore, the final independent parameters consist of
13 quantities: the six quark masses, the four CKM parameters, and
the three charged-lepton masses. When the Weinberg operator is taken into account,
\begin{equation}
\mathcal{L}_{\rm Weinberg}
= \,\frac{1}{2\Lambda}\,
\kappa_{ij}\,
\bigl(\overline{L_{L i}^{\,c}}\,\tilde H^\ast\bigr)
\bigl(L_{L j}\,\tilde H^\ast\bigr)
+ {\rm h.c.},
\end{equation}
the three lepton-sector $U(1)$ symmetries are broken, while the operator
introduces 12 additional independent parameters. As a result, the
total number of flavor parameters becomes 22: the original 13 parameters, plus
3 neutrino masses, 3 PMNS mixing angles, and 3 CP-violating phases in
the PMNS matrix.

Minimal Flavor Violation (MFV) assumes that this structure remains the only source in the presence
of new physics, so that any additional flavor and CP violation is aligned with the SM
Yukawa matrices. Technically, one treats the Yukawas as spurions transforming under $G_F$ so that the Yukawa Lagrangian is formally invariant. Restricting to the
non-Abelian subgroup $SU(3)^5$, one assigns
\begin{equation}
Y_u \sim (\bar 3, 3,1,1,1)\,, \qquad
Y_d \sim (\bar 3,1,3,1,1)\,, \qquad
Y_e \sim (1,1,1,\bar 3,3)\,,
\end{equation}
which ensures covariance under flavor rotations. Higher-dimensional operators and possible new-physics couplings are then constructed from SM fields and these spurions in a $G_F$-invariant manner.

Another line of thought follows a bottom--up approach, in which the flavor symmetry is
assumed to originate from certain discrete groups. In particular, when neutrinos are
Majorana fermions, the lepton mass sector can preserve different residual symmetries in
the charged-lepton and neutrino sectors, and these residual symmetries may generate the
full discrete flavor group. In this viewpoint, the observed lepton mixing originates from
a mismatch of residual flavor symmetries: $G_\ell$ is preserved in the charged-lepton
sector and $G_\nu$ is preserved in the neutrino sector after the breaking of an underlying
flavor group $G_f$.

Residual flavor symmetries are defined as transformations that leave the corresponding
mass matrices invariant. For the charged leptons we have
\begin{equation}
M_\ell^\dagger M_\ell \;\to\; G_\ell^\dagger\, M_\ell^\dagger M_\ell\, G_\ell
= M_\ell^\dagger M_\ell ,
\label{eq:residualGl}
\end{equation}
while for Majorana neutrinos the invariance condition reads
\begin{equation}
M_\nu \;\to\; G_\nu^{T}\, M_\nu\, G_\nu
= M_\nu .
\label{eq:residualGnu}
\end{equation}
Equations \eqref{eq:residualGl}--\eqref{eq:residualGnu} imply that the residual generators
can be expressed in terms of the diagonalization matrices as
\begin{equation}
G_\ell = U_\ell\, \mathrm{diag}\!\left(e^{i\alpha_e},e^{i\alpha_\mu},e^{i\alpha_\tau}\right)
\,U_\ell^\dagger ,
\qquad
G_\nu = U_\nu\, \mathrm{diag}(\pm1,\pm1,\pm1)\,U_\nu^\dagger .
\end{equation}
Hence $G_\ell$ is typically a rephasing symmetry, whose minimal choice is often taken to
be $Z_3$, whereas for Majorana neutrinos the allowed residual symmetry leads to a Klein symmetry
$G_\nu\simeq Z_2\times Z_2$.
Since $G_\ell$ ($G_\nu$) shares eigenvectors with $M_\ell^\dagger M_\ell$ ($M_\nu$), the
diagonalization matrices of the residual symmetries coincide with those of the
corresponding mass matrices. Therefore, the mismatch between $G_\ell$ and $G_\nu$ directly
reflects the misalignment of the charged-lepton and neutrino mass bases, yielding a
highly constrained mixing matrix $U_{\rm PMNS}=U_\ell^\dagger U_\nu$. In this framework
the full flavor group is reconstructed as the group generated by the two residual
symmetries,
\begin{equation}
G_f = \langle\, G_\ell,\, G_\nu \,\rangle ,
\end{equation}
and a minimal realization is to choose
$G_\ell=Z_3=\{1,T,T^2\}$ and $G_\nu=Z_2\times Z_2=\{1,S,TST^2,T^2ST\}$ with the semidirect product
$G_f\simeq G_\ell \rtimes G_\nu$,
which generates $G_f=A_4$ with generators $S$ and $T$ satisfying
$S^2=(ST)^3=T^3=1$.

In the minimal residual-symmetry setup, corresponding to the standard choice
$G_\ell=Z_3$ and $G_\nu=Z_2\times Z_2$ discussed above, one
possible realization of this symmetry is the tri-bimaximal mixing (TBM) pattern~\cite{Harrison_2002}. It is
convenient to work in the weak basis where the charged-lepton mass matrix is already
diagonal, $M_\ell=M_\ell^{\rm diag}$, so that $U_\ell = I$ and all lepton
mixing originates from the neutrino sector. In this limit the PMNS matrix is therefore of
TBM form,
\begin{equation}
U_{\rm PMNS}\simeq U_{\rm TB},
\end{equation}
with $U_{\rm TB}$ as an experimentally motivated ansatz given in
Eq.~(\ref{eq:TB}). Equivalently, the effective Majorana neutrino mass matrix prior to
electroweak and flavor symmetry breaking is diagonalized by $U_{\rm TB}$,
\begin{equation}
M_\nu = U_{\rm TB}\,M_\nu^{\rm diag}\,U_{\rm TB}^{T},
\qquad (m_3\gg m_2\gg m_1).
\end{equation}

Concretely, the charged-lepton residual symmetry is generated by
\begin{equation}
T^\dagger M_\ell^\dagger M_\ell\, T = M_\ell^\dagger M_\ell,
\qquad
T=\mathrm{diag}(1,\omega,\omega^2),\ \ \omega=e^{2\pi i/3},
\end{equation}
so that $G_\ell=\langle T\rangle\simeq Z_3$, while the TBM neutrino mass matrix satisfies
\begin{equation}
S^T M_\nu S = M_\nu,
\qquad
S=\frac{1}{3}
\begin{pmatrix}
-1&2&2\\
2&-1&2\\
2&2&-1
\end{pmatrix}.
\end{equation}
For Majorana neutrinos this $Z_2$ invariance is naturally promoted to the full Klein
symmetry $G_\nu\simeq Z_2\times Z_2$, whose diagonalization fixes $U_\nu= U_{\rm TB}$.
If the neutrino sector additionally preserves the $\mu$--$\tau$ reflection-type generator
$P=T^2ST^2$, the residual symmetry is enlarged and the generated group is promoted to
$S_4$, giving alternative mixing textures.

However, in tri-bimaximal models one inevitably introduces
non-renormalizable Yukawa operators including many flavons. The desired
separated residual symmetries, typically $G_\ell\simeq Z_3$ and
$G_\nu\simeq Z_2\times Z_2$, are enforced by assuming multiple flavons with specific vacuum
alignments in different mass sectors. This not only enlarges the parameter space, but
also leads to a nontrivial and often model-dependent alignment potential, thereby
reducing predictivity and obscuring the origin of flavor breaking.

A compelling way to overcome these drawbacks is provided by modular flavor symmetry. In
this framework the discrete flavor group is identified with a finite modular group
\begin{equation}
\Gamma_N \;\equiv\; PSL(2,\mathbb Z)/\Gamma(N),
\end{equation}
generated by the same elements $S$ and $T$ satisfying $S^2=(ST)^3=T^N=1$. Consequently,
\begin{equation}
\Gamma_3\simeq A_4,\qquad \Gamma_4\simeq S_4,
\end{equation}
and, for $N=4$, the element $P=T^2ST^2$ naturally appears as the $\mu$--$\tau$ exchange
generator.

Crucially, Yukawa couplings are no longer generated by flavon VEVs,
but by modular forms $Y^{(k)}_{\mathbf r}(\tau)$, holomorphic functions of the modulus
$\tau$ transforming in irreducible representations $\mathbf r$ of $\Gamma_N$ with modular
weight $k$. Renormalizable Yukawa interactions can then be written directly in terms of
SM fields and modular forms, so that flavor breaking is controlled solely by the
VEV $\langle\tau\rangle$ rather than by flavon alignments.
From the bottom--up residual-symmetry viewpoint, the separated residual subgroups can be understood as residual modular symmetries, \emph{i.e.} the stabilizer subgroups of $\langle\tau\rangle$, which arise when $\langle\tau\rangle$ sits at special fixed points in the fundamental domain. Once $\tau$ is fixed at such a point (or in its vicinity), the explicit forms of the mass matrices follow from the values of the relevant modular forms evaluated at $\langle\tau\rangle$, thereby constraining the allowed textures of the mass matrices. In a minimal setup with a single modulus, the residual modular symmetry is therefore common to all sectors.
In this sense,
modular symmetry offers a predictive and effective way to realize the residual
flavor-symmetry paradigm, with significantly fewer free parameters and without the need
for many auxiliary flavons.

\subsection{Modular flavor symmetry in SM}
Defining $\mathcal{H}$ as the upper half-plane of the complex plane, the modulus $\tau$ is a point in $\mathcal{H}$.
Furthermore, let $SL(2,\mathbb{Z})$ denote the group consisting of matrices
$\left(\begin{smallmatrix} a & b \\ c & d \end{smallmatrix}\right)$ with $a,b,c,d\in \mathbb{Z}$ and $ad-bc=1$.
More generally, for $\gamma=\left(\begin{smallmatrix} a & b \\ c & d \end{smallmatrix}\right)\in SL(2,\mathbb{R})$,
the action on $\tau$ is defined by
\begin{equation}
\gamma\tau=\frac{a\tau+b}{c\tau+d}\in \mathcal{H},\qquad \forall\,\tau\in\mathcal{H},
\end{equation}
which implies that $\mathcal{H}$ is stable under $SL(2,\mathbb{R})$.

Note that $-I=\bigl(\begin{smallmatrix}-1&0\\[2pt]0&-1\end{smallmatrix}\bigr)$ acts trivially on $\mathcal H$.
Hence the action of $SL(2,\mathbb R)$ on $\mathcal H$ has kernel $\{\pm I\}$ and factors through
\begin{equation}
PSL(2,\mathbb R)=SL(2,\mathbb R)/\{\pm I\}.
\end{equation}

The modular group $SL(2,\mathbb{Z})$ is the subgroup of $SL(2,\mathbb{R})$ generated by
$S=\left(\begin{smallmatrix} 0 & -1 \\ 1 & 0 \end{smallmatrix}\right)$ and
$T=\left(\begin{smallmatrix} 1 & 1 \\ 0 & 1 \end{smallmatrix}\right)$.
A modular form of weight $k$ for $SL(2,\mathbb{Z})$ is a holomorphic function $f(\tau)$ on $\mathcal{H}$, which is also holomorphic at the cusp as $\mathrm{Im}(\tau)\to\infty$ (equivalently, it has a $q$-expansion with no negative powers),
and it satisfies the covariance condition
\begin{equation}
f(\gamma \tau)=(c\tau +d)^k f(\tau),\qquad \forall\,\gamma\in SL(2,\mathbb{Z}).
\end{equation}
If it additionally vanishes at cusp, it is called a cusp form.
Under the generators, $\tau$ transforms as
\begin{equation}
S\tau=-\frac{1}{\tau},\qquad T\tau=\tau+1.
\end{equation}
Under the action of $-\mathbb{I}$, the modular form transforms as
\begin{equation}
f(-\mathbb{I}\tau)=f(\tau)=(-1)^k f(\tau),
\end{equation}
which implies that the modular weight $k$ must be even; otherwise $f(\tau)=0$.
Accordingly, the modular group can be redefined as: 
\begin{equation}
\overline{\Gamma}\equiv PSL(2,\mathbb{Z})=SL(2,\mathbb{Z})/\{\pm I\}.
\end{equation}

In general, we consider the principal congruence subgroups $\Gamma(N)$, defined as
\begin{equation}
\Gamma(N)=\left\{\gamma=
\begin{pmatrix}
a & b \\
c & d
\end{pmatrix}\in SL(2,\mathbb{Z})\;\middle|\;
\gamma \equiv
\begin{pmatrix}
1 & 0 \\
0 & 1
\end{pmatrix}
\ \mathrm{mod}\ N\right\}.
\end{equation}
The finite modular group is defined as $\Gamma_N\equiv\overline{\Gamma}/\overline{\Gamma}(N)$, where
$\overline{\Gamma}(N)$ is a normal subgroup of $\overline{\Gamma}$.
Specifically, for $N=2$ one has $\overline{\Gamma}(2)=\Gamma(2)/\{\pm I\}$,
while for $N>2$ it simplifies to $\overline{\Gamma}(N)=\Gamma(N)$.
For $N=2,3,4,5,\dots$, these groups are isomorphic to well-known permutation groups,
$\Gamma_2\simeq S_3$, $\Gamma_3\simeq A_4$, $\Gamma_4\simeq S_4$, $\Gamma_5\simeq A_5$, \dots\;.

The modular forms of level $N$ and weight $k$ form a linear space $\mathcal{M}_{k}(\Gamma(N))$
with dimension $d_k(\Gamma(N))$.
For instance, for $N=3$ and even $k$, one has $d_k(\Gamma(3))=k+1$.
This space furnishes representations of the finite modular group $\Gamma_N$, and it can be decomposed into
a direct sum of irreducible representations. On a suitable basis, the modular forms transform as
\begin{equation}
f_i(\gamma \tau) = (c\tau + d)^k \, \rho_{ij}(\gamma)\, f_j(\tau),
\qquad \forall\,\gamma\in \Gamma_N.
\end{equation}

Although holomorphy is not required by the SM at leading order, we still assume that the Yukawa couplings are given by holomorphic modular forms $Y(\tau)$. This Lagrangian can be regarded as the low-energy effective description of an underlying supersymmetric (SUSY) ultraviolet completion.
we thus write a generic leading-order Yukawa interaction as
\begin{equation}
\mathcal{L}^{D}_{\rm Yuk}=Y(\tau)\,\psi_R^c\,\Phi^\dagger\,\psi_L\,+{\rm h.c.}\; ,
\end{equation}
Here, $\psi_L$ is a left-handed $SU(2)_{L}$ doublet, while $\psi_R^{c}$ denotes the charge-conjugated field of a right-handed fermion. 
The $SU(2)_L$ doublet $\Phi$ is chosen as $\Phi=H$ or $\Phi=\tilde H\equiv i\sigma_2 H^*$, such that $\Phi^\dagger\psi_L$ is gauge invariant.
These fields transform under the $\Gamma_N$ modular symmetry as
\begin{equation}
{\psi_L}_i \;\longrightarrow\; (c\tau + d)^{-k_L}\,
\bigl[\rho_{\mathbf{r}}(\gamma)\bigr]_{ij}\,{\psi_L}_j \, ,
\label{eq:L}
\end{equation}
\begin{equation}
{\psi_R^c}_i \;\longrightarrow\; (c\tau + d)^{-k_R}\,
\bigl[\rho_{\mathbf{r}}(\gamma)\bigr]_{ij}\,{\psi_R^c}_j \, ,
\label{eq:R}
\end{equation}
where $\rho_{\mathbf r}$ denotes the representation matrix acting on $\psi_L$ (and $\psi_R^c$), and $-k_L$ ($-k_R$) is the corresponding weight.
For scalar fields, we assume that they are invariant under the modular symmetry.

In this case, when treating $\tau$ as a dynamical field, the modular-invariant kinetic terms are modified, and they can be written explicitly as~\cite{feruglio2017neutrinomassesmodularforms}
\begin{equation}
\mathcal{L}_{\rm kinetic}
\supset
\frac{h}{\left\langle-i\tau + i\bar{\tau}\right\rangle^2}\,
D'_\mu \bar{\tau}\,D'^{\mu}\tau
+
\sum_{\psi}
\frac{1}{\left\langle i(\overline{\tau} - \tau)\right\rangle^{k_\psi}}\,
\big[ i\,{\psi}^\dagger \,\slashed{D}' \psi \big]_{\mathbf{1}}\,,
\label{eq:kinetic}
\end{equation}
where $h$ is a mass square scale introduced to canonically normalize the kinetic term of the modulus $\tau$.
The subscript $[\cdots]_{\mathbf{1}}$ denotes the contraction to the trivial singlet, which is invariant under the finite modular group.
The covariant derivative $D'$ is modified compared to the standard SM derivative, and when acting on a field of modular weight $k$ it takes the form
\begin{equation}
D'_\mu \;=\; D_\mu \;+\; \frac{k\pi i}{6}\,E_2(\tau)\, ,
\label{eq:d}
\end{equation}
where $E_2(\tau)$ is the Eisenstein series,
\begin{equation}
E_2(\tau)=1-24\sum_{n=1}^{\infty}\sigma_1(n)\,q^n,
\qquad q=e^{2\pi i\tau},
\end{equation}
and $\sigma_1(n)=\sum_{d\mid n} d$ is the divisor function.

If we focus on the impact of $\tau$ on the mass structures, we regard $\tau$ as a background parameter and neglect its dynamical term.
When $\tau$ acquires a VEV $\langle\tau\rangle$, the fermion kinetic terms in Eq.~\eqref{eq:kinetic} reduce to
\begin{equation}
\mathcal{L}_{\rm kinetic}
\supset
\sum_{\psi}
\frac{1}{\left\langle i\overline{\tau} - i\tau \right\rangle^{k_\psi}}\,
\big[ i\,{\psi}^\dagger \,\slashed{D}' \psi \big]_{\mathbf{1}}\,.
\label{eq:kinetic2}
\end{equation}
%The factor $\left\langle i\overline{\tau} - i\tau \right\rangle^{k}$ can be absorbed by a rescaling of the fields to match the canonical normalization, \emph{i.e.}
%\begin{equation}
%\psi\rightarrow \left\langle i\overline{\tau} - i\tau %\right\rangle^{k/2}\psi.
%\label{eq:redefine}
%\end{equation}

Because $\mathrm{Im}\,\tau$ transforms covariantly under modular transformations,
kinetic terms allow insertions of $\mathrm{Im}\,\tau$ together with modular forms of appropriate weights to compensate the covariant factor, which can be written explicitly as~\cite{Chen_2020}.
\begin{equation}
    \Delta \mathcal{L}_{\rm kinetic}= \sum_{\psi,k,\mathbf{r},\mathbf{r'}} \alpha_{\psi,k,\mathbf{r},\mathbf{r'}}\left\langle i\overline{\tau} - i\tau \right\rangle^{k-k_{\psi}}\big[\big(Y^{(k)}{Y^*}^{(k)}\big)_{\mathbf{r}} \big( i\,{\psi}^\dagger \,\slashed{D}' \psi \big)_{\mathbf{r'}}\big]_{\mathbf{1}}\,,
\end{equation}
with $k$ running over all even integers with $k\ge 2$.
The introduction of such additional kinetic terms implies that canonical normalization of the kinetic terms reshuffles the original Yukawa matrices, thereby modifying the mixing angles and CP phases inferred from them. 
At the same time, the additional terms typically introduces additional free parameters, which in turn reduces the model's predictive power. 
Therefore, we require that, in the renormalizable modular-symmetric SM Lagrangian, all flavor structures reside entirely in the Yukawa terms; consequently, we take the kinetic sector in its minimal form, i.e.~we keep only the structure in Eq.~\eqref{eq:kinetic2}. 
%And we assume that, for the Yukawa sector considered in our paper, the corresponding kinetic terms have already been canonically normalized. 
At the same time, higher-dimensional operators generally involve non-holomorphic structures. Since allowing insertions of $\mathrm{Im}\,\tau$ would lead to an infinite tower of operators, we follow an approach analogous to MFV that all flavor structures be fully encoded in the leading-order Yukawa sector. In modular-invariant Yukawa terms, the contributions to the mass matrices are provided by modular forms rather than by $\mathrm{Im}\,\tau$. 
%Moreover, $\mathrm{Im}\,\tau$ carries no non-trivial $\Gamma_N$ representation and by itself cannot generate non-trivial flavor textures; its role is mainly to compensate modular weights under modular transformations rather than to produce masses.
Therefore, as a minimal choice, we consider only the contributions of modular forms to the higher-dimensional operators, and we will no longer take the effects of $\mathrm{Im}\,\tau$ into account.

\begin{comment}
In modular-symmetry models, mass matrices are most often studied nearby fixed points.
In the modular group $PSL(2,\mathbb{Z})$, a point $\tau\in\mathcal H$ is called a fixed point if $\gamma\tau=\tau$.
Up to modular transformations, there are three inequivalent classes of fixed
points:
\begin{align}
  \tau_S &= i\,, & S\,\tau_S &= \tau_S\,,\\
  \tau_{ST} &= \omega \equiv -\frac12 + \frac{\sqrt3}{2}\,i
            = e^{2\pi i/3}\,, &
            ST\,\tau_{ST} &= \tau_{ST}\,,\\[2pt]
  \tau_T &= i\infty\,, & T\,\tau_T &= \tau_T\,,
\end{align}
where $\tau_T$ denotes the cusp at infinity.

Any point in the $PSL(2,\mathbb Z)$ orbit of
$\tau_S$, $\tau_{ST}$ or $\tau_T$ is again a fixed point, with a conjugated
residual symmetry.  For example, consider $\tau_0=T\tau_S=i+1$.
Since $S\tau_S=\tau_S$, we obtain
\begin{equation}
TST^{-1}\,\tau_0 = TS\tau_S = T\tau_S = \tau_0\,,
\end{equation}
so $\tau_0$ is fixed by the element $TST^{-1}$.
More generally, if $\tau_0=\gamma\tau_S$ for some $\gamma\in PSL(2,\mathbb Z)$,
the residual symmetry at $\tau_0$ is generated by
\begin{equation}
G_{\rm res}(\tau_0)=\{\,1,\;\gamma S\gamma^{-1}\,\}\,,
\end{equation}
and analogous statements hold for $\tau_{ST}$ and $\tau_T$.
Thus, at the level of the infinite modular group there are only three
inequivalent fixed points, associated with the subgroups generated by $S$,
$ST$ and $T$.
\end{comment}
In modular-symmetry models, mass matrices are most often studied in the vicinity of fixed points of the modulus.
For the modular group $PSL(2,\mathbb{Z})$, a point $\tau\in\mathcal H$ is called a fixed point (of a non-trivial element $\gamma$) if $\gamma\tau=\tau$.
Up to modular transformations, there are three inequivalent classes of fixed points (including the cusp):
\begin{align}
  \tau_S &= i\,, & S\,\tau_S &= \tau_S\,,\\
  \tau_{ST} &= \omega \equiv -\frac12 + \frac{\sqrt3}{2}\,i
            = e^{2\pi i/3}\,, &
            ST\,\tau_{ST} &= \tau_{ST}\,,\\[2pt]
  \tau_T &= i\infty\,, & T\,\tau_T &= \tau_T\,,
\end{align}
where $\tau_T$ denotes the cusp at infinity.

Any point in the $PSL(2,\mathbb Z)$ orbit of $\tau_S$, $\tau_{ST}$, or $\tau_T$ is again a fixed point, with a conjugate residual symmetry.
For example, consider $\tau_0=T\tau_S=i+1$.
Since $S\tau_S=\tau_S$, we obtain
\begin{equation}
TST^{-1}\,\tau_0 = TS\tau_S = T\tau_S = \tau_0\,,
\end{equation}
so $\tau_0$ is fixed by the element $TST^{-1}$.
More generally, if $\tau_0=\gamma\tau_S$ for some $\gamma\in PSL(2,\mathbb Z)$,
the residual symmetry at $\tau_0$ is generated by
\begin{equation}
G_{\rm res}(\tau_0)=\{\,1,\;\gamma S\gamma^{-1}\,\}\,,
\end{equation}
and analogous statements hold for $\tau_{ST}$ and $\tau_T$.
Thus, at the level of the infinite modular group there are only three inequivalent fixed-point orbits, associated with the subgroups generated by $S$, $ST$, and $T$.

After electroweak symmetry breaking, the Yukawa couplings generate Dirac and Majorana mass terms. In a modular-invariant theory, they can be written as
\begin{equation}
\mathcal L_{\rm mass} \supset
\psi^{c}_R\,M_D(\tau)\,\psi_L
\;+\;
\frac12\,\psi_L^{T} C^{-1} M_\nu(\tau)\,\psi_L
\;+\;{\rm h.c.},
\end{equation}
where $M_D$ and $M_\nu$ denote the Dirac and Majorana mass matrices, respectively.
Requiring the Dirac mass term to be invariant under a modular transformation $\gamma$ leads to
\begin{equation}
M_D(\gamma\tau)
=
(c\tau+d)^{k_c+k}\,
\rho^{c}(\gamma)^{*}\,
M_D(\tau)\,
\rho(\gamma)^{\dagger}.
\end{equation}
Similarly, invariance of the Majorana term implies
\begin{equation}
M_\nu(\gamma\tau)
=
(c\tau+d)^{2k}\,
\rho(\gamma)^{*}\,
M_\nu(\tau)\,
\rho(\gamma)^{\dagger}.
\end{equation}

Now consider a fixed point $\tau_0$ with residual symmetry $G_{\rm res}$, \emph{i.e.}\ $g\tau_0=\tau_0$ for all $g\in G_{\rm res}$.
At this point the Dirac mass term must still be invariant, so that
\begin{equation}
M_D(\tau_0)
=
\rho^{c}(g)^{*}\,
M_D(\tau_0)\,
\rho(g)^{\dagger},
\qquad g\in G_{\rm res}.
\label{eq:Dirac-res}
\end{equation}
The overall factor $(c\tau_0+d)^{k_c+k}$ can be absorbed by the transformation factors in
Eqs.~\eqref{eq:L} and \eqref{eq:R}, and hence the non-trivial constraint at the fixed point is the residual symmetry condition \eqref{eq:Dirac-res}.
Unlike in the usual minimal flavor violation framework, where Yukawa matrices explicitly break the flavor symmetry $U(3)^5$, the modular symmetry is broken solely by fixing $\tau$ at a specific value $\tau_0$.
The residual subgroup $G_{\rm res}$ remains unbroken and imposes strong constraints on the structure of the mass matrices.

For the Dirac sector we study the Hermitian combination $M_D^\dagger M_D$. From Eq.~\eqref{eq:Dirac-res} we obtain
\begin{equation}
\rho(g)\,M_D^\dagger(\tau_0)M_D(\tau_0)\,\rho(g)^{\dagger}
=
M_D^\dagger(\tau_0)M_D(\tau_0)\,,
\qquad g\in G_{\rm res},
\end{equation}
or equivalently
\begin{equation}
\bigl[\,M_D^\dagger M_D,\;\rho(g)\,\bigr]=0\,.
\end{equation}
Hence $M_D^\dagger M_D$ and the representation matrices $\rho(g)$ can be diagonalized simultaneously: there exists a unitary matrix $V$ such that
\begin{equation}
V^\dagger \rho(g) V = \rho_{\rm diag}(g)\,,\qquad
V^\dagger M_D^\dagger M_D V
= \mathrm{diag}(m_1^2,m_2^2,m_3^2)\,.
\end{equation}
The matrices $\rho^{c}(g)$ do not enter this condition, so the weights of the conjugated fields do not affect the structure of $M_D^\dagger M_D$.

The Majorana mass matrix at the fixed point satisfies the analogous residual symmetry condition
\begin{equation}
M_\nu(\tau_0)
=
\rho(g)^{*}\,M_\nu(\tau_0)\,\rho(g)^{\dagger},
\qquad g\in G_{\rm res}.
\label{eq:Majorana-res}
\end{equation}
Let $M_\nu$ be diagonalized by a unitary matrix $V$,
\begin{equation}
D_\nu = V^{T} M_\nu(\tau_0) V
      = \mathrm{diag}(m_1,m_2,m_3)\,.
\end{equation}
Then Eq.~\eqref{eq:Majorana-res} implies
\begin{equation}
V^{T}\rho(g)^{*}V^{*}\,D_\nu\,V^{\dagger}\rho(g)^{\dagger}V = D_\nu\,.
\end{equation}
Defining
\begin{equation}
\rho_m(g) \equiv V^{\dagger}\rho(g)V\,,
\end{equation}
we obtain
\begin{equation}
\rho_m(g)^{*} D_\nu \rho_m(g)^{\dagger} = D_\nu\,.
\end{equation}
For non-degenerate eigenvalues $m_i$ this condition forces $\rho_m(g)$ to be diagonal with entries $\pm1$, \emph{i.e.}
\begin{equation}
\rho_m(g)=\mathrm{diag}(\pm1,\pm1,\pm1)\,,
\end{equation}
which corresponds to a residual $Z_2$ symmetry. In this case $M_\nu(\tau_0)$ and $\rho(g)$ can be diagonalized simultaneously.

For the other fixed points, such as $\tau_{ST}=\omega$ and $\tau_T=i\infty$, the residual symmetries are generated by $ST$ and $T$, respectively.
In particular, in the finite modular group $\Gamma_N$ the $T$-residual is $Z_N$ (and thus $Z_3$ for $N=3$).
In these cases $\rho(g)$ and $M_\nu(\tau_0)$ are in general not simultaneously diagonalizable, and their misalignment leads to non-trivial lepton mixing.
When $\tau$ is fixed at $\omega$ or $i\infty$, one must therefore diagonalize $M_\nu$ independently, subject to the residual symmetry constraint \eqref{eq:Majorana-res}.
Since the mass matrices constructed at fixed values of $\tau$ are often too strongly constrained, one sometimes considers values of $\tau$ near the fixed points in order to obtain a better fit to the data.

In this work, we focus on the case $\Gamma_3 \cong A_4$, which is the minimal finite modular group admitting a triplet irreducible representation.
For $A_4$, modular forms can be constructed from the Dedekind eta function, defined as~\cite{feruglio2017neutrinomassesmodularforms}
\begin{equation}
\eta(\tau) = q^{1/24} \prod_{n=1}^\infty (1 - q^n), \qquad q = e^{2\pi i \tau},
\end{equation}
where $\eta(\tau)$ transforms as a modular form of weight $1/2$ up to a multiplier system.

In the complex basis, the modular generators $S$ and $T$ are represented by
\begin{equation}
\rho(S) = \frac{1}{3}
\begin{pmatrix}
-1 & 2 & 2 \\
2 & -1 & 2 \\
2 & 2 & -1
\end{pmatrix}, \qquad
\rho(T) =
\begin{pmatrix}
1 & 0 & 0 \\
0 & \omega & 0 \\
0 & 0 & \omega^2
\end{pmatrix}, \qquad \omega = e^{2\pi i/3}.
\end{equation}
In this basis, the weight-2 modular form triplet can be written as
\begin{equation}
Y^{(2)}_{\mathbf{3}}(\tau) =
\begin{pmatrix}
Y_1(\tau) \\
Y_2(\tau) \\
Y_3(\tau)
\end{pmatrix}.
\end{equation}
Its explicit $q$-expansions read
\begin{equation}
\begin{aligned}
Y_1(\tau) &= 1 + 12q + 36q^2 + 12q^3 + \cdots, \\
Y_2(\tau) &= -6q^{1/3}\bigl(1 + 7q + 8q^2 + \cdots\bigr), \\
Y_3(\tau) &= -18q^{2/3}\bigl(12q + 5q^2 + \cdots\bigr),
\end{aligned}
\end{equation}
subject to the constraint
\begin{equation}
Y_2^2 + 2Y_1Y_3 = 0.
\label{eq:modular_constrain}
\end{equation}
Its complex conjugate takes the form
\begin{equation}
{Y^{(2)}_{\mathbf{3}}}^*(\overline{\tau}) =
\begin{pmatrix}
Y_1(\overline{\tau}) \\
Y_3(\overline{\tau}) \\
Y_2(\overline{\tau})
\end{pmatrix}.
\end{equation}

For the group $A_4$, all modular forms of weight $k>2$ can be generated from the triplet $Y_{\mathbf{3}}^{(2)}$ by taking $A_4$ tensor products.
The explicit product rules are summarized in Appendix~\ref{app:tensor_product}.
These higher-weight objects arise from the symmetric tensor powers of $Y_{\mathbf{3}}^{(2)}$.
For a weight $k=2n$ modular form, the number of independent symmetric monomials is
$\binom{n+3-1}{3-1}=\frac{(n+2)(n+1)}{2}$.
Upon imposing the constraint in Eq.~\eqref{eq:modular_constrain}, this reduces to $k+1$~\cite{feruglio2017neutrinomassesmodularforms}.

Furthermore, by explicitly constructing forms of arbitrary weight $k>2$ (see Appendix~\ref{app:forms}),
we show that, under Eq.~\eqref{eq:modular_constrain}, the structure of the constrained symmetric tensor product
$\bigl(Y_{\mathbf{3}}^{(2)}\otimes \dots \otimes Y_{\mathbf{3}}^{(2)}\bigr)_{{sym}^n}'$
is determined by the recursion relation
\begin{equation}
\bigl(Y_{\mathbf{3}}^{(2)}\otimes\cdots\otimes Y_{\mathbf{3}}^{(2)}\bigr)_{\mathrm{Sym}^n}
=
\bigl(Y_{\mathbf{3}}^{(2)}\otimes\cdots\otimes Y_{\mathbf{3}}^{(2)}\bigr)_{\mathrm{Sym}^n}^{\prime}
\;\oplus\;
\bigl(Y_{\mathbf{3}}^{(2)}\otimes\cdots\otimes Y_{\mathbf{3}}^{(2)}\bigr)_{\mathrm{Sym}^{\,n-2}}
\otimes Y_{\mathbf{1^{\prime\prime}}}^{(4)} \, .
\label{tensor_rules}
\end{equation}
This relation allows us to write down the Hilbert series of modular forms, which we then use to obtain a complete basis of modular-invariant SMEFT operators.

\subsection{SMEFT and flavor symmetry}

When considering physics beyond the Standard Model (BSM), a primary strategy is to employ
effective field theory (EFT) to parameterize the impact of higher-energy effects by the Wilson coefficients of the effective operators. 
Regarding the Lagrangian in Eq.~\eqref{eq:LSM} as the leading Lagrangian, the SM is extended to the standard model effective field theory (SMEFT). The effective Lagrangian takes the form,
\begin{equation}
\mathcal{L}_{\rm SMEFT}
\;=\;
\mathcal{L}_{\rm SM}
\;+\;
\sum_{d>4}\,\sum_i
\frac{C_i^{(d)}}{\Lambda^{\,d-4}}\,
\mathcal{O}_i^{(d)} \, ,
\label{eq:SMEFT}
\end{equation}
where $\Lambda$ denotes the new-physics scale, $\mathcal{O}_i^{(d)}$ are independent 
gauge-invariant operators of mass dimension $d$, and $C_i^{(d)}$ are the corresponding
Wilson coefficients.

Considering the flavor structures, we suppose the flavor number of the fermions are $n_f\geq 1$, the effective operators involving $N$ fermions are $N$-rank flavor tensors that
\begin{equation}
\label{eq:flavortensor}
    \mathcal{O}[\psi]_{pr\dots s} = \psi_{1p}\psi_{2r}\dots \psi_{Ns}\,,
\end{equation}
where the Lorentz and gauge structures are ignored. The flavor structures make the operator basis complicated. If the $N$ fields are distinct, such a tensor counts $n_f^N$ operators. Nevertheless, if there are repeated fields, they should be of specific permutation symmetries, which means the flavor tensors in Eq.~\eqref{eq:flavortensor} should be decomposed into irreducible representations of the symmetric group~\cite{Li:2020gnx,Li:2020xlh,Li:2022tec}. The irreducible representations are characterized by the standard Young diagrams, and can be projected by the associated Young stabilizer. For example, the 2 irreducible representations of the $S_2$ group correspond to the Young diagrams,
\begin{equation}
    [2] = \ydiagram{2}\,,\quad [1^2] = \ydiagram{1,1}\,.
\end{equation}
and the Young stabilizers are the symmetrizer and antisymmetrizer,
\begin{equation}
\ytableausetup{smalltableaux}
    \mathcal{Y}\left(\ydiagram{2}\right) = \frac{1}{2}\left[1 + (12)\right]\,,\quad 
    \mathcal{Y}\left(\ydiagram{1,1}\right) = \frac{1}{2}\left[1 - (12)\right]\,,
\end{equation}
For example, the dimension-6 operator
\begin{equation}
    \left[\mathcal{O}_{quqd}^{(1)}\right]_{prst} = (\overline{q}^{j}_pu_r)\epsilon_{jk}(\overline{q}^{k}_sd_t)\,,
\end{equation}
contains 2 repeated $\overline{q}$. It can be decomposed as 
\begin{equation}
\label{eq:quqd}
    \left[\mathcal{O}_{quqd}^{(1)}\right]_{prst} = \mathcal{Y}\left(\ytableaushort{ps}\right)(\overline{q}^{j}_pu_r)\epsilon_{jk}(\overline{q}^{k}_sd_t) + \mathcal{Y}\left(\ytableaushort{p,s}\right)(\overline{q}^{j}_pu_r)\epsilon_{jk}(\overline{q}^{k}_sd_t)\,.
\end{equation}

Specifying $n_f=3$, although the flavor is not conserved, if the Yukawa interactions are switched off in the SM Lagrangian in Eq.~\eqref{eq:LSM}, the fermion kinetic and gauge sectors are invariant under the independent unitary rotations in the flavor space for each fermion species. This defines the maximal global flavor symmetry compatible with the SM gauge group~\cite{Chivukula:1987py,Gerard1983FermionMS}, $G_F \;=\; U(3)^5$ in Eq.~\eqref{eq:u35}.
Under $G_F$ the fields transform as
\begin{equation}
q_L \to V_q\, q_L,\quad
u_R \to V_u\, u_R,\quad
d_R \to V_d\, d_R,\quad
\ell_L \to V_\ell\, \ell_L,\quad
e_R \to V_e\, e_R,
\qquad V_f\in U(3)_f .
\end{equation}
These flavor rotations correspond to changes in the weak basis for each
fermion triplet, leaving the kinetic and gauge interactions unchanged while
reparametrizing the Yukawa or mass matrices.
At leading order, turning on the Yukawa couplings breaks the $U(3)^5$ symmetry to the rephasing symmetry in Eq.~\eqref{eq:rephasing}.
\iffalse
\begin{equation}
U(3)^5 \;\longrightarrow\; U(1)^4 = U(1)_{e-\mu}\times U(1)_{\mu-\tau}\times U(1)_B\times U(1)_L .
\end{equation}
\fi
$U(1)_B$ and $U(1)_L$ correspond to baryon- and lepton-number conservation,
while $U(1)_{e-\mu}$ and $U(1)_{\mu-\tau}$ are two independent
family-dependent rephasings in the charged-lepton sector that survive in the
$Y_e^{\rm diag}$ basis~\cite{sun2025flavorcpsymmetriesstandard}.

Extending to the high-dimensional operators, the rephasing symmetry could be violated further. This leads to flavor-violating patterns different from the Yukawa matrices, which, nevertheless, are not observed. For example, the experiments of the flavor-changed-charged-current (FCNC) are consistent with the SM prediction~\cite{ParticleDataGroup:2024cfk}.
This implies the flavor violations due to new physics are highly suppressed, which pushes the new-physics scale considerably high, $\Lambda\sim 10^5\text{ TeV}$~\cite{Isidori:2010kg}. A solution to this puzzle is the minimal flavor violation (MFV) hypothesis~\cite{Chivukula:1987py,DAmbrosio:2002vsn,Bonnefoy:2020yee,Aoude:2020dwv,Bruggisser:2021duo,Kobayashi:2021uam,Bruggisser:2022rhb,Bartocci:2023nvp,Bartocci:2024fgj}, which states that the Yukawa matrices are the only sources of flavor and CP violation. Technically, one promotes $Y_{u,d,e}$ to spurions
transforming under $G_F$ 
\begin{equation}
    Y_u \rightarrow V_q Y_u V_u^\dagger\,,\quad Y_d \rightarrow V_q Y_d V_d^\dagger\,,\quad Y_e \rightarrow V_l Y_e V_e^\dagger\,,\quad V_f \in U(3)_f\,, 
\end{equation}
so that the Yukawa Lagrangian is formally invariant.

Combining the MFV hypothesis with the SMEFT further assumes that all flavor and CP violation in the higher-dimensional operators is fully governed by the SM Yukawa matrices. 
Treating the Yukawa couplings as spurions, we impose that the higher-dimensional operators are also invariant under the flavor symmetry $G_F$
\begin{equation}
    \mathcal{L}_{\text{SMEFT}}\supset \frac{C^{(d)}}{\Lambda^{d-4}}f(Y_u,Y_d,Y_e)\cdot \mathcal{O}^{(d)}[\psi]\,,
\end{equation}
where $f$ is a polynomial of the Yukawa matrices, $\mathcal{O}$ contains the dynamic fields, and $\cdot$ means the $f$ and $\mathcal{O}$ should form $G_F$ invariant.
%. As a result, their flavor-breaking
%structures can only be built from the Yukawa spurions, 
This reduces the number of independent Wilson coefficients compared to the flavor-general SMEFT and
thereby leads to a more predictive framework.
%When constructing high-dimensional operators, 
%since in MFV the spurions $Y_u$, $Y_d$, etc.\ are assigned in chiral-paired flavor representations. For instance, $Y_u$ must be contracted with ${Q_L}$ and $u_R^c$
%to form a flavor singlet at the group-theory level. 
In particular, the MFV-SMEFT
operators are necessarily baryon- and lepton-number conserving since the Yukawa matrices are of no baryon- and lepton-number.

For the three-flavor SMEFT, the dimension-five sector contains a single operator class (the Weinberg operator), which counts 6 for $n_f=3$.
Considering the conjugate terms, there are 12 independent operators, i.e. $12$ real parameters. At dimension six, one finds $3045$ independent real operator
structures in total, among which $2499$ conserve baryon number, while the remaining
$546$ violate baryon number. By contrast, in the three-flavor MFV-SMEFT, the
baryon-number-conserving dimension-six sector is reduced to $894$ operators~\cite{Faroughy:2020ina,sun2025flavorcpsymmetriesstandard}. 
%%%%%%%%%%%%%%%%%%%%%%%%%%%%%%%%%%%%%%%%%%%
However, MFV also has a clear limitation: it does not predict the PMNS matrix nor the charged-lepton and neutrino mass spectrum. In other words, the spurions are taken as experimental inputs. In the SM flavor sector, this amounts to $22$ independent physical parameters to be fixed by data (six quark masses, three charged-lepton masses, three neutrino masses, four CKM parameters, and six PMNS parameters for Majorana neutrinos). 
%Therefore, the introduction of modular symmetry becomes highly desirable.
Besides, the MFV hypothesis can be generalized to other cases of various flavor symmetries. For example, the SMEFT with modular symmetry considered in this paper is formulated under an MFV-like hypothesis. It states that all the flavor coefficients are modular forms composed of the ones with the lowest weight, which serve as the Yukawa matrices in the SM.

\subsection{Modular invariant SMEFT}
\label{sub_M}
\begin{comment}
Here, our model is constructed in a way analogous to the form in~\cite{Kobayashi_2022} and~\cite{okada2021unificationquarkleptonflavors}, for convenience to make a clearer comparison between their construction and ours in some relevant operators.
\end{comment}
Here, our model is constructed in a way analogous to Refs.~\cite{Kobayashi_2022,okada2021unificationquarkleptonflavors}, which facilitates a direct comparison between their setup and ours for the operators relevant to our construction.
Our concrete field content and symmetry assignments are summarized in Table~\ref{tab:SM_model}.
Motivated by the fact that the modulus is expected to lie close to a fixed point, this model choice corresponds to the simplest possible assignment of modular weights. In the lepton sector the modular weights are already taken to be minimal, whereas in the quark sector a fully low-weight assignment for each flavor, \emph{e.g.}\ $(0,0,0)$, leaves too few independent parameters to reproduce the observed CKM matrix. In practice, when fitting the CKM parameters this setup tends to drive the fitted value of $\tau_q$ towards the nearby fixed point $\tau_0=i$, where the $A_4$ modular symmetry is broken to the residual subgroup $\mathbb{Z}_2^{S}$~\cite{Kobayashi_2022}.

Moreover, we take the full flavor modular group to be $A_{4}^{(q)} \otimes A_{4}^{(e)}$, \emph{i.e.}\ the modular symmetry is implemented separately in the quark and lepton sectors.
This choice is made to allow for the possibility that the quark and lepton sectors are associated with different fixed points (and hence different moduli $\tau_q$ and $\tau_e$).
For simplicity, the Higgs doublet is taken to carry modular weight zero.
Left-handed fermions are placed in the triplet representation of $A_4$, whereas each right-handed fermion transforms as an $A_4$ singlet, with different flavors assigned to inequivalent singlet representations.
In the last column, $Y_{\mathbf{r}_q}^{(k_{qY})}(\tau_q)$ and $Y_{\mathbf{r}_e}^{(k_{eY})}(\tau_e)$ denote modular forms of weights $k_{qY}$ and $k_{eY}$, transforming in the $A_4$ representations $\mathbf{r}_q$ and $\mathbf{r}_e$ associated with the quark and lepton sectors, and depending on the corresponding moduli $\tau_q$ and $\tau_e$.

%At this fixed point the modular group $A_4$ will break into modular subgroups as $Z_2^S=\{1 ,S %\}$. Near a symmetric point with a residual symmetry,
%many entries of the fermion mass matrices vanish. When $\tau$ slightly moves away from %$\tau=i$, it will generate a hierarchical structure in the mass
%matrices.

\begin{table}[H]
    \centering
    \begin{tabular}{ |c|c|c|c|c|c|c|c| }
    \hline
    & $Q_L$ &
      $(u_R^{c},\, c_R^{c},\, t_R^{c})$ &
      $(d_R^{c},\, s_R^{c},\, b_R^{c})$ &
      $L_L$ &
      $(e_R^{c},\, \mu_R^{c},\, \tau_R^{c})$ &
      $H$ &
      $Y_{\mathbf{r}_{q}}^{(k_{qY})}(\tau_q),\, Y_{\mathbf{r}_{e}}^{(k_{eY})}(\tau_e)$ \\
    \hline
    $SU(2)_L$ & $2$ & $1$ & $1$ & $2$ & $1$ & $2$ & $1$ \\
    \hline
    $A_4$ & $\mathbf{3}$ &
      $(\mathbf{1},\, \mathbf{1}^{\prime\prime},\, \mathbf{1}^{\prime})$ &
      $(\mathbf{1},\, \mathbf{1}^{\prime\prime},\, \mathbf{1}^{\prime})$ &
      $\mathbf{3}$ &
      $(\mathbf{1},\, \mathbf{1}^{\prime\prime},\, \mathbf{1}^{\prime})$ &
      $\mathbf{1}$ &
      $\mathbf{3}$ \\
    \hline
    $k$ & $2$ & $(4,4,4)$ & $(0,0,0)$ & $2$ & $(0,0,0)$ & $0$ &
      $k_{qY},\, k_{eY}$ \\
    \hline
    \end{tabular}
    \caption{Field content and symmetry assignments in the quark and lepton
    sectors. Here $Y_{\mathbf{r}_q}^{(k_{qY})}(\tau_q)$
    and $Y_{\mathbf{r}_e}^{(k_{eY})}(\tau_e)$ denote modular forms of weights
    $k_{qY}$ and $k_{eY}$ transforming in $A_4$ representations
    $\mathbf{r}_q$ and $\mathbf{r}_e$ in the quark and lepton sectors.}
    \label{tab:SM_model}
\end{table}

Under a modular transformation with
$\gamma = \begin{pmatrix} a & b \\ c & d \end{pmatrix} \in A_4$,
the modular forms $Y_{\mathbf{r}_{q}}^{(k_{qY})}(\tau_q)$ and
$Y_{\mathbf{r}_{e}}^{(k_{eY})}(\tau_e)$ appearing in the last column of
Table~\ref{tab:SM_model} transform as
\begin{equation}
\begin{aligned}
Y_{\mathbf{r}_{q}}^{(k_{qY})}(\tau_q) &\;\longrightarrow\;
(c\tau_q + d)^{k_{qY}}\,
\rho_{\mathbf{r}_q}(\gamma)\,
Y_{\mathbf{r}_{q}}^{(k_{qY})}(\tau_q) \,,
\\[4pt]
Y_{\mathbf{r}_{e}}^{(k_{eY})}(\tau_e) &\;\longrightarrow\;
(c\tau_e + d)^{k_{eY}}\,
\rho_{\mathbf{r}_e}(\gamma)\,
Y_{\mathbf{r}_{e}}^{(k_{eY})}(\tau_e) \,.
\end{aligned}
\end{equation}

Modular invariance of the Lagrangian requires that, for each interaction term,
the tensor product of the $A_4$ representations carried by the matter fields
and by the corresponding modular form contains the trivial representation.
Concretely, in the quark sector one demands
\begin{equation}
\rho_{\mathbf{r}_{q_1}} \otimes \rho_{\mathbf{r}_{q_2}} \otimes \cdots
\otimes \rho_{\mathbf{r}_{q_n}} \otimes \rho_{\mathbf{r}_q}
\supset \mathbf{1} \,,
\end{equation}
where $\rho_{\mathbf{r}_{q_i}}$ denotes the $A_4$ representation matrix of the quark field
$q_i$ appearing in the operator. Similarly, in the lepton sector one requires
\begin{equation}
\rho_{\mathbf{r}_{\ell_1}} \otimes \rho_{\mathbf{r}_{\ell_2}} \otimes \cdots
\otimes \rho_{\mathbf{r}_{\ell_m}} \otimes \rho_{\mathbf{r}_e}
\supset \mathbf{1} \,,
\end{equation}
where $\rho_{\mathbf{r}_{\ell_j}}$ denotes the $A_4$ representation matrix of the lepton field
$\ell_j$ appearing in the operator.

The modular weights are constrained in the same way: for each quark-sector operator,
the sum of the modular weights of the quark fields must match the weight of the associated modular form,
\begin{equation}
\sum_{i \in \text{quarks}} k_{q_i} = k_{qY} \,,
\end{equation}
while in the lepton sector one has
\begin{equation}
\sum_{j \in \text{leptons}} k_{\ell_j} = k_{eY} \,.
\end{equation}
These conditions ensure that all Yukawa terms constructed from the field content in
Table~\ref{tab:SM_model} are invariant under the modular symmetry.

For the quark sector, the left-handed doublets are assembled into an $A_4$ triplet,
\begin{equation}
Q_L=\begin{pmatrix} Q_1 \\[2pt] Q_2 \\[2pt] Q_3 \end{pmatrix}, \qquad
\overline{Q}_L=\begin{pmatrix} \overline{Q}_1 \\[2pt] \overline{Q}_3 \\[2pt] \overline{Q}_2 \end{pmatrix}, \qquad
Q_1=\begin{pmatrix} u_L \\ d_L \end{pmatrix},\quad
Q_2=\begin{pmatrix} c_L \\ s_L \end{pmatrix},\quad
Q_3=\begin{pmatrix} t_L \\ b_L \end{pmatrix}.
\end{equation}
The right-handed quarks are taken to be $A_4$ singlets with assignments
\begin{equation}
(u_R^c,\, c_R^c,\, t_R^c)=(\mathbf{1},\, \mathbf{1''},\, \mathbf{1'}),\quad
(u_R,\, c_R,\, t_R)=(\mathbf{1},\, \mathbf{1'},\, \mathbf{1''}),
\end{equation}
\begin{equation}
(d_R^c,\, s_R^c,\, b_R^c)=(\mathbf{1},\, \mathbf{1''},\, \mathbf{1'}),\quad
(d_R,\, s_R,\, b_R)=(\mathbf{1},\, \mathbf{1'},\, \mathbf{1''}) \, .
\end{equation}
Analogously, in the lepton sector,
\begin{equation}
L_L=\begin{pmatrix} L_1 \\[2pt] L_2 \\[2pt] L_3 \end{pmatrix}, \qquad
\overline{L}_L=\begin{pmatrix} \overline{L}_1 \\[2pt] \overline{L}_3 \\[2pt] \overline{L}_2 \end{pmatrix}, \qquad
L_1=\begin{pmatrix} \nu_{eL} \\ e_L \end{pmatrix},\quad
L_2=\begin{pmatrix} \nu_{\mu L} \\ \mu_L \end{pmatrix},\quad
L_3=\begin{pmatrix} \nu_{\tau L} \\ \tau_L \end{pmatrix},
\end{equation}
and
\begin{equation}
(e_R^c,\, \mu_R^c,\, \tau_R^c)=(\mathbf{1},\, \mathbf{1''},\, \mathbf{1'}),\qquad
(e_R,\, \mu_R,\, \tau_R)=(\mathbf{1},\, \mathbf{1'},\, \mathbf{1''}) \, .
\end{equation}
Within this model, the Yukawa sector can be written explicitly as
%\begin{align}
%\mathcal{L}_u &= \alpha_u\,u^{c} H^* \mathbf{Y}_{\mathbf{3a}}^{(6)}%(\tau_q) Q
%      + \alpha'_u\,u^{c} H^* \mathbf{Y}_{\mathbf{3b}}^{(6)}(\tau_q) Q
%      + \beta_u\,c^{c} H^* \mathbf{Y}_{\mathbf{3a}}^{(6)}(\tau_q) Q
%      + \beta'_u\,c^{c} H^* \mathbf{Y}_{\mathbf{3b}}^{(6)}(\tau_q) Q  %\notag\\
%      &\quad
%      + \gamma_u\,t^{c} H^* \mathbf{Y}_{\mathbf{3a}}^{(6)}(\tau_q) Q
%      + \gamma'_u\,t^{c} H^* \mathbf{Y}_{\mathbf{3b}}^{(6)}(\tau_q) Q + {\rm h.c.}, \label{eq:Lu}\\[4pt]
%\mathcal{L}_d &= \alpha_d\,d^{c} H^* \mathbf{Y}_{\mathbf{3}}^{(2)}(\tau_q) Q
%      + \beta_d\,s^{c} H^* \mathbf{Y}_{\mathbf{3}}^{(2)}(\tau_q) Q
%      + \gamma_d\,b^{c} H^* \mathbf{Y}_{\mathbf{3}}^{(2)}(\tau_q) Q + {\rm h.c.}\,, \label{eq:Ld}\\
%\mathcal{L}_e &= \alpha_e\, e^{c} H^* \mathbf{Y}_{\mathbf{3}}^{(2)}(\tau_e) L
%    + \beta_e\, \mu^{c} H^* \mathbf{Y}_{\mathbf{3}}^{(2)}(\tau_e) L
%    + \gamma_e\, \tau^{c} H^* \mathbf{Y}_{\mathbf{3}}^{(2)}(\tau_e) L + {\rm h.c.} \, . \label{eq:Le}
%\end{align}
\begin{align}
\mathcal{L}_u
&=
\alpha_u\,u^{c}\,\tilde H^\dagger\,\mathbf{Y}_{\mathbf{3a}}^{(6)}(\tau_q)\,Q
+\alpha'_u\,u^{c}\,\tilde H^\dagger\,\mathbf{Y}_{\mathbf{3b}}^{(6)}(\tau_q)\,Q
+\beta_u\,c^{c}\,\tilde H^\dagger\,\mathbf{Y}_{\mathbf{3a}}^{(6)}(\tau_q)\,Q
+\beta'_u\,c^{c}\,\tilde H^\dagger\,\mathbf{Y}_{\mathbf{3b}}^{(6)}(\tau_q)\,Q \notag\\
&\quad
+\gamma_u\,t^{c}\,\tilde H^\dagger\,\mathbf{Y}_{\mathbf{3a}}^{(6)}(\tau_q)\,Q
+\gamma'_u\,t^{c}\,\tilde H^\dagger\,\mathbf{Y}_{\mathbf{3b}}^{(6)}(\tau_q)\,Q
+{\rm h.c.},
\label{eq:Lu}
\\[4pt]
\mathcal{L}_d
&=
\alpha_d\,d^{c}\,H^\dagger\,\mathbf{Y}_{\mathbf{3}}^{(2)}(\tau_q)\,Q
+\beta_d\,s^{c}\,H^\dagger\,\mathbf{Y}_{\mathbf{3}}^{(2)}(\tau_q)\,Q
+\gamma_d\,b^{c}\,H^\dagger\,\mathbf{Y}_{\mathbf{3}}^{(2)}(\tau_q)\,Q
+{\rm h.c.},
\label{eq:Ld}
\\
\mathcal{L}_e
&=
\alpha_e\,e^{c}\,H^\dagger\,\mathbf{Y}_{\mathbf{3}}^{(2)}(\tau_e)\,L
+\beta_e\,\mu^{c}\,H^\dagger\,\mathbf{Y}_{\mathbf{3}}^{(2)}(\tau_e)\,L
+\gamma_e\,\tau^{c}\,H^\dagger\,\mathbf{Y}_{\mathbf{3}}^{(2)}(\tau_e)\,L
+{\rm h.c.}\,.
\label{eq:Le}
\end{align}

In this Yukawa setup, there are in total $28$ real free parameters.
They consist of $12$ complex Yukawa coefficients,
$\alpha_u,\alpha'_u,\beta_u,\beta'_u,\gamma_u,\gamma'_u$ in the up-quark sector,
$\alpha_d,\beta_d,\gamma_d$ in the down-quark sector, and
$\alpha_e,\beta_e,\gamma_e$ in the lepton sector, together with two complex moduli
$\tau_q$ and $\tau_e$, which contribute $4$ additional real parameters.

Once the fields and their representations are fixed as in Table~\ref{tab:SM_model}, the original basis symmetry $U(3)^5$ is no longer preserved.
In particular, for the left-handed multiplets the $U(3)$ symmetry is reduced to a single $U(1)$ phase rotation, while for the right-handed fields the $U(3)$ acting on the three generations is broken down to independent $U(1)$ symmetries for each flavor.
Consequently, the initial $U(3)^5$ symmetry is reduced to a product of $11$ $U(1)$ factors, so that there are $11$ independent $U(1)$ phases in total.

\begin{figure}[h]
  \centering
  \begin{equation}
    \begin{array}{c}
      U(3)_q \\
      U(3)_\ell \\
      U(3)_u \\
      U(3)_d \\
      U(3)_e
    \end{array}
    \xrightarrow{\;\;\;\;\;\;A_4 \;\;\text{modular symmetry}\;\;\;\;\;\;}
    \begin{array}{c}
      U(1)_q \\
      U(1)_\ell \\
      U(1)_u \times U(1)_c \times U(1)_t \\
      U(1)_d \times U(1)_s \times U(1)_b \\
      U(1)_e \times U(1)_\mu \times U(1)_\tau
    \end{array}
  \end{equation}
  \caption{$U(3)^5$ is broken down to $U(1)^{11}$ by the $A_4$ modular symmetry.}
\end{figure}

Among these, the combinations corresponding to baryon number $U(1)_B$ and lepton number $U(1)_L$ still leave the leading-order Yukawa sector invariant, and thus do not reduce the number of parameters.
Consequently, there are $11-2=9$ independent rephasings that can be used to remove unphysical phases, leaving
\begin{equation}
28 - 9 = 19
\end{equation}
independent real parameters in the modular Yukawa sector.
More concretely, the quark sector contains $14$ independent parameters, while the lepton sector contains $5$.

Under our minimal assumption that the leading-order Yukawa sector is the sole origin of flavor structure, we do not include the kinetic term of $\tau$ nor the effects of $\mathrm{Im}\,\tau$ on higher-dimensional operators; instead, we treat all Yukawa-sector modular forms as building blocks.
Since any holomorphic modular form can be generated from the triplet $Y_{\mathbf{3}}^{(2)}$ via symmetric tensor powers, it is sufficient to take $Y_{\mathbf{3}}^{(2)}(\tau_q)$ and $Y_{\mathbf{3}}^{(2)}(\tau_e)$ as the elementary building blocks.
When the modular form $Y^{(2)}_{\mathbf{3}}(\tau)$ is treated as a spurion field and used as a building block, the flavor-breaking sources of higher-dimensional operators are correspondingly restricted.
In this way, the flavor symmetry is broken uniquely by two moduli $\tau_q$ and $\tau_e$ when considering $A_{4}^{(q)} \otimes A_{4}^{(e)}$.
This procedure is therefore MFV-like.
Building on the modular-symmetric Standard Model together with the $A_4$ tensor algebra in Appendix~\ref{app:tensor_product}, we then employ the Hilbert series method to construct general SMEFT operators that are invariant under the $A_4$ modular symmetry.

\section{Hilbert series method for \texorpdfstring{$A_4$}{A4} modular symmetry}
\label{sec:hilbert}
In this section, we introduce the Hilbert series. We first discuss the Hilbert series for compact continuous groups, and then present the corresponding structure for discrete groups. Moreover, taking the fundamental modular form $Y^{(2)}_{\mathbf{3}}$ as the basic building block, we derive the Hilbert series that determines the number of independent operators.

\subsection{Hilbert series for continuous group}
Given a set of multiplets \(q_i\) transforming in representations
\(R_i\) of a (compact) group \(G\), the Hilbert series is defined as
the generating function
\begin{equation}
  H(q_i)\;=\;\sum_{k_1,\dots,k_N=0}^{\infty}
  c_{k_1,\dots,k_N}\,
  q_1^{k_1}\cdots q_N^{k_N},
\end{equation}
where the coefficients \(c_{k_1,\dots,k_N}\) are the multiplicities of the trivial representation in the tensor product $\bigotimes_{i=1}^N \mathrm{Sym}^{k_i} R_i$ with \(\mathrm{Sym}^{k_i} R_i\) denoting the \(k_i\)-order symmetric tensor product of \(R_i\).

For a compact group $G$, the Hilbert series can be obtained using the integration formula,
\begin{equation}
    H(q_i) = \oint d\mu(z)\mathrm{PE}\!\left[q_i\,\chi_{q_i}(z)\right]
\end{equation}
where $z$ is the maximal torus variable of the group $G$, and $d\mu(z)$, $\chi_{q_i}(z)$ are the integral measure and character expressed by them, respectively. The symmetrized products of the characters are generated by the plethystic exponential (PE)
\begin{equation}
  \mathrm{PE}\!\left[q_i\,\chi_{q_i}(z)\right]
  \;\equiv\;
  \exp\!\left[
    \sum_{n=1}^{\infty}
    \frac{q_i^{\,n}}{n}\,
    \chi_{q_i}(z^n)
  \right]
  =
  \sum_{n=0}^{\infty}
  q_i^{\,n}\,
  \chi_{\mathrm{Sym}^n(q_i)}(z)\,,
\end{equation}
Because of the orthogonality of the group characters, 
such an integral extracts the multiplicity of the trivial representation, which counts the number of operators.
In effective field theory (EFT), we treat all fields \(\phi_i\) and the covariant derivative \(D_\mu\) as group multiplets. 
Truncating at a certain mass order, the Hilbert series presents both the number and field contents of the operators.
%By expanding
%the corresponding Hilbert series in the total mass dimension (mass
%order), we obtain both the number and the structure
%of independent invariant operators at each given mass order.

The derivative is a special building block since it raises two redundancies of the operators, equation of motion (EOM) and integration by part (IBP). To eliminate these redundancies, the Lorentz group should be extended to the conformal group. In the conformal group, the derivative is a lower operator of its Cartan sub-algebra, so that for each field \(\phi\), its conformal representation takes the form
\begin{equation}
  \phi \rightarrow R_\phi
  =
  \begin{pmatrix}
    \phi\\[2pt]
    D_\mu\phi\\[2pt]
    D_{(\mu_1}D_{\mu_2)}\phi\\[2pt]
    \vdots\\[2pt]
    D_{(\mu_1\cdots\mu_n)}\phi\\[2pt]
    \vdots
  \end{pmatrix},
\end{equation}
where \(D_{(\mu_1\cdots\mu_n)}\) denotes the totally symmetrized combination of derivatives. 
This defines the single-particle module
(SPM) $R_\phi$~\cite{Henning:2017fpj}.
%Antisymmetric combinations of derivativescan be put into the field strength \(F_{\mu\nu}\) andare therefore not counted independently. 
The SPM character is defined as
\begin{equation}
  \chi_{\phi}(D,z)
  =
  \sum_{n=0}^{\infty}
  D^n\,
  \chi_{\mathrm{sym}^n(D)}(z)\,
  \chi_\phi(z),
\end{equation}
where \(D\) is the spurion corresponding to the  covariant derivative formal, \(\chi_\phi(g)\) is the character of the Lorentz representation of $\phi$, and \(\chi_{\mathrm{sym}^n(D)}(z)\) denotes the character of the rank-\(n\) symmetric tensor representation of the derivatives.

For the multi-particle sector, we can introduce generating functions for symmetric and antisymmetric tensor products. For a bosonic field
\(\phi\), the \(n\)-particle character corresponds to the symmetric tensor product \(\mathrm{Sym}^n(\phi)\) of the SPM. While for the fermionic field, the $n$-particle character corresponds to the antisymmetric tensor product. Thus, the PE is generalized to~\cite{Feng:2007ur} 
\begin{equation}
\begin{aligned}
    \text{boson:}\quad  \mathrm{PE}\!\left[\phi\,\chi_{\phi}(D,z)\right]&
  \;\equiv\;
  \exp\!\left[
    \sum_{n=1}^{\infty}
    \frac{\phi^{\,n}}{n}\,
    \chi_{\phi}(D^n,z^n)
  \right]
  =
  \sum_{n=0}^{\infty}
  \phi^{\,n}\,
  \chi_{\mathrm{Sym}^n(\phi)}(D,z)\,,\\
  \text{fermion:}\quad \mathrm{PEF}\!\left[\phi\,\chi_{\phi}(D,z)\right]&
  \;\equiv\;
  \exp\!\left[
    -\sum_{n=1}^{\infty}
    \frac{(-\phi)^{n}}{n}\,
    \chi_{\phi}(D^n,z^n)
  \right]
  =
  \sum_{n=0}^{\infty}
  \phi^{\,n}\,
  \chi_{\wedge^n(\phi)}(D,z)\,,
\end{aligned}
\end{equation}
where $\chi_{\wedge^n(\phi)}(D,z)$ denotes the antisymmetric product of the characters.
\iffalse
For a fermionic field \(\phi\), the \(n\)-particle character instead
corresponds to the antisymmetric tensor product \(\wedge^n(\phi)\),
and its generating function is given by the fermionic plethystic
exponential \(\mathrm{PEF}\),
\begin{equation}
  \mathrm{PEF}\!\left[\phi\,\chi_{\phi}(D,g)\right]
  \;\equiv\;
  \exp\!\left[
    -\sum_{n=1}^{\infty}
    \frac{(-\phi)^{n}}{n}\,
    \chi_{\phi}(D^n,g^n)
  \right]
  =
  \sum_{n=0}^{\infty}
  \phi^{\,n}\,
  \chi_{\wedge^n(\phi)}(D,g)\,.
\end{equation}
\fi
For a theory with several distinct fields, the multi-particle generating function is obtained by multiplying together the PEs and PEFs of each field.

%For a compact continuous group \(G\), the inner product of
%two characters is given by

%where \(d\mu(g)\) is the normalized Haar measure on \(G\). The the multiplicities
%of the trivial representation
%in a representation \(R\) with character
%\(\chi_R(g)\) can be extracted as
\iffalse
\begin{equation}
  c_R
  =\langle\chi_R,1\rangle
  =\int_G d\mu(g)\,\chi_R(g).
\end{equation}
Applied to the Hilbert series, this inner product projects onto the
trivial singlet of the full symmetry group. 

For a single field
\(\phi\), we therefore obtain
\begin{equation}
  H(D,\phi)
  =\int_G d\mu(g)\,
  \begin{cases}
    \mathrm{PE}\!\big[\phi\,\chi_{\phi}(D,g)\big], & \phi\
    \text{bosonic},\\[4pt]
    \mathrm{PEF}\!\big[\phi\,\chi_{\phi}(D,g)\big], & \phi\
    \text{fermionic},
  \end{cases}
\end{equation}
and for multiple fields \(\{\phi_i\}\) the Hilbert series generalizes
to
\begin{equation}
  H(D,\{\phi_i\})
  =\int_G d\mu(g)\,
  \prod_i
  \begin{cases}
    \mathrm{PE}\!\big[\phi_i\,\chi_{\phi_i}(D,g)\big], & \phi_i\
    \text{bosonic},\\[4pt]
    \mathrm{PEF}\!\big[\phi_i\,\chi_{\phi_i}(D,g)\big], & \phi_i\
    \text{fermionic}.
  \end{cases}
\end{equation}
\fi
When considering the EOM, the SPM $\phi$ for a scalar field should be reduced by removing all components that
are proportional to the scalar EOM $(D^2+m^2)\phi=0$. Under this constraint, the character can be written as 
\begin{equation}
  \chi_{\phi}(D,z)
  \rightarrow \chi_\phi(D,z)-D^2\chi(D,z)
  = (1 - D^{2})\,P(D,z)\,\chi_{(0,0)}(z) \, .
\end{equation}
Here, $\chi_{(0,0)}$ denote the character of the Lorentz representation, and $P(D,z)$ denotes the character of the symmetrical derivatives tensor product as 
\begin{equation}
P(D,z) \equiv 
\frac{1}{(1 - D z_{1})(1 - D/z_{1})(1 - D z_{2})(1 - D/z_{2})} \, .
\end{equation}
For a left-handed fermion, the decomposition of $\mathrm{sym}^n(D)\psi_L$ takes the form
\begin{equation}
\mathrm{sym}^n\!\left(\tfrac12,\tfrac12\right)\otimes\left(\tfrac12,0\right)
  = \left(\tfrac{n+1}{2},\,\tfrac{n}{2}\right)\oplus\cdots .
\end{equation}
Here only the highest–weight component $\left(\tfrac{n+1}{2},\,\tfrac{n}{2}\right)$ is kept. Taking such constraint into account, the character of the fermionic module
$R_{\psi_L}$ becomes
\begin{equation}
  \chi_{{\psi_L}}(D,z)
  = \Bigl(\chi_{(\tfrac12,0)}(z)
        - D\,\chi_{(0,\tfrac12)}(z)\Bigr)\,P(D,z)\, .
\end{equation}
For the left-handed field strength $F_L$ with $n$ derivatives, the redundancies
\begin{equation}
    D_\mu F_L^{\mu\nu} = D_\mu D_\nu F^{\mu\nu}_L=0\,,
\end{equation}
imply the decomposition
\iffalse
\begin{equation}
  \mathrm{sym}^n\!\left(\tfrac12,\tfrac12\right)\otimes (1,0)
  = \left(\tfrac{n+2}{2},\,\tfrac12\right)\oplus\cdots .
  \label{eq:symn-FL}
\end{equation}
By keeping only the highest-weight component $\bigl(\tfrac{n+2}{2},\tfrac12\bigr)$, the
equation-of-motion redundancy
$
  D_\mu F_L^{\mu\nu} = J^\nu
$
is removed. And the character of the left-handed field-strength module
$R_{F_L}$ becomes
\fi
\begin{equation}
  \chi_{{F_L}}(D,z)
  = \Bigl(\chi_{(1,0)}(D,z)
          - D\,\chi_{(\tfrac12,\tfrac12)}(D,z)
          + D^{2}\chi_{(0,0)}(D,z)\Bigr)\,P(D,z) .
\end{equation}
IBP redundancy means that if one operator differs from another only by a total derivative, then these two operators are physically equivalent, because the total derivative becomes a boundary term and does not change the action after integrating the Lagrangian over space–time. By considering this redundancy, the Hilbert series can finally be represented as:
\begin{equation}
    H(D,\phi)
= \int d\mu_L(z)\,\frac{1}{P(D,z)}\,
   \mathrm{PE}\!\left(\phi\,\chi_{\phi}(D,z)\right)
   + \Delta H\, ,
\end{equation}
where $\Delta H$ only receives contributions from operators with
mass-dimension below four and will therefore be ignored, and $d\mu_L$ denotes
the Haar measure of the Lorentz group $SO(4,\mathbb{C})$.

In an EFT, the symmetry group is typically taken to be
\begin{equation}
  G = SO(4)\times I,
\end{equation}
where \(SO(4)\) is the Lorentz group and \(I\) is the
internal symmetry group. 
In this case, the full Hilbert series can
be written in the form
\begin{equation}
  H(D,\{\phi_i\})
  =
  \int d\mu_{SO(4)}(z)\,
       d\mu_{I}(z)\,
  \frac{1}{P(D,z)}\,
  \prod_i
  \begin{cases}
    \mathrm{PE}\!\big[\phi_i\,\chi_{\phi_i}(D,z,a)\big], & \phi_i\
    \text{bosonic},\\[4pt]
    \mathrm{PEF}\!\big[\phi_i\,\chi_{\phi_i}(D,z,a)\big], & \phi_i\
    \text{fermionic}.
  \end{cases}
\end{equation}

As an example, in SMEFT, the symmetry group is
\begin{equation}
  G_{\text{SMEFT}} \;=\;
  SO(4)\times SU(3)_c\times SU(2)_L\times U(1)_Y\,,
\end{equation}
where \(SO(4)\) is the Euclidean Lorentz group and the remaining
factors form the Standard Model gauge group. We use maximal-torus
variables \(z=(z_1,z_2)\) for \(SO(4)\), and collectively denote by
\(a\) the maximal-torus variables of the internal gauge group
\(SU(3)_c\times SU(2)_L\times U(1)_Y\).
The SMEFT field content includes the Higgs doublet \(H\), the
left-handed quark and lepton doublets \(Q\) and \(L\), the
right-handed singlets \(u^c, d^c, e^c\), and the gauge field strengths
\(G_{\mu\nu}, W_{\mu\nu}, B_{\mu\nu}\). For each of these fields
\(\phi_i\), we construct an EOM-reduced single-particle module
\(R_{\phi_i}\) as in the general discussion, and define its
single-particle character
\begin{equation}
  \chi_{\phi_i}(D,z,a)\,,
\end{equation}
which factorizes into a Lorentz part (built from the appropriate
scalar, spinor, or field-strength SPM with EOM already removed) and an
internal part encoding the corresponding
\(\big(SU(3)_c,SU(2)_L,U(1)_Y\big)\) representation of \(\phi_i\).

The SMEFT Hilbert series follows by applying the general formula for
\(G = SO(4)\times I\) to the SMEFT group. Explicitly,
\begin{equation}
  H_{\text{SMEFT}}(D,\{\phi_i\})
  =
  \int d\mu_{SO(4)}(z)\,
       d\mu_{SU(3)_c\times SU(2)_L\times U(1)_Y}(a)\,
  \frac{1}{P(D,z)}\,
  \prod_i
  \begin{cases}
    \mathrm{PE}\!\big[\phi_i\,\chi_{{\phi_i}}(D,z,a)\big],
      & \phi_i\ \text{bosonic},\\[4pt]
    \mathrm{PEF}\!\big[\phi_i\,\chi_{{\phi_i}}(D,z,a)\big],
      & \phi_i\ \text{fermionic}.
  \end{cases}
\end{equation}
To obtain the spectrum of SMEFT operators at fixed mass dimension
\(d\), we rescale
\begin{equation}
  D \;\to\; \varepsilon\,D\,, \qquad
  \phi_i \;\to\; \varepsilon^{d_{\phi_i}}\,\phi_i\,,
\end{equation}
where \(d_{\phi_i}\) is the canonical mass dimension of the field
\(\phi_i\), and expand the SMEFT Hilbert series in powers of
\(\varepsilon\). The coefficient of \(\varepsilon^d\) then gives the
number and field contents of independent SMEFT operators at mass dimension
\(d\), with the IBP and EOM removed.

\subsection{Hilbert series for discrete group}

%In the analysis of higher-order effects within effective field theory, it's necessary to ensure the minimality of the subleading operators set, which is to avoid double-counting contributions from the integrated-out physics. And the Hilbert series method gives us an efficient and systematic way to solve this issue.

%The Hilbert series applies not only to the compact Lie group but also to the discrete group. Suppose a finite group $G$ of $|G|$ elements, the character $\chi$ is a function of the conjugacy classes $C_i\subset G$, $\chi(C_i)$. If there are $n_i$ group elements in the class $C_i$, the orthogonality of the characters is
The Hilbert series applies not only to compact Lie groups but also to finite discrete groups. Consider a finite group $G$ with $|G|$ elements. The character $\chi$ is a class function, which is constant on each conjugacy class $C_i\subset G$ and can be written as $\chi(C_i)$. If the class $C_i$ contains $n_i$ elements, the orthogonality relation for characters reads
\begin{equation}
    \langle \chi_1, \chi_2 \rangle = \frac{1}{|G|} \sum_i n_i \, \chi_1(C_i) \, \chi_2^*(C_i).
\end{equation}
%As an example, we consider two multiplets $q_1$ and $q_2$, which transform under representations $R_1$ and $R_2$ of a finite group $G$. For any 
%conjugacy class $C_i\in G$ that contains $n_i$ group elements, we denote the corresponding characters as $\chi_1(C_i)$ and $\chi_2(C_i)$. The inner product of characters is defined as:
Based on the inner product, we define the multiplicity n of the irreducible representation $R$ as:
\begin{equation}
    n=\langle \chi, \chi_{R} \rangle.
\end{equation}
If $\chi_R=1$, then the multiplicity $n$ gives the number of independent operators.
%Therefore, the Hilbert series can tell us its number.
\begin{table}[t]
  \centering
  \[
  \begin{array}{c|cccc}
          & C_1      & C_2          & C_3            & C_4          \\ \hline
  \text{class rep.}
          & e        & {T^2}        & {T}          & {S}     \\
  \text{class size}
          & 1        & 4            & 4              & 3            \\ \hline
  \chi^{(\mathbf{1})}      
          & 1        & 1            & 1              & 1            \\
  \chi^{(\mathbf{1'})}     
          & 1        & \omega       & \omega^2       & 1            \\
  \chi^{(\mathbf{1''})}    
          & 1        & \omega^2     & \omega         & 1            \\
  \chi^{(\mathbf{3})}      
          & 3        & 0            & 0              & -1           
  \end{array}
  \]
    \caption{Character table of the alternating group $A_4$. $C_1$ is the identity class, $C_2$ and $C_3$ are the two conjugacy classes of $3$-cycles, and $C_4$ is the class of double transpositions. The irreducible representations are three singlets $\mathbf{1}$, $\mathbf{1'}$, $\mathbf{1''}$ and one triplet $\mathbf{3}$. Here $\omega = e^{2\pi i/3}$.}
        \label{tab:A4-character-table}
  \vspace{4pt}
  \raggedright\footnotesize
\end{table}
\begin{comment}
For the tensor-product representation $q_1\otimes q_2$, the character satisfies
\begin{equation}
    \chi_{q_1 \otimes q_2}(C_i) = \chi_1(C_i) \cdot \chi_2(C_i),
\end{equation}
which can be separated into the symmetric and anti-symmetric parts. For instance, taking 3-dimension representation with three eigenvalues $\lambda_1$, $\lambda_2$, $\lambda_3$, the symmetric and anti-symmetry characters can be written as:
\begin{equation}
    \chi_{{\text{sym}}^{\otimes 2}}=\lambda_1^2+\lambda_2^2+\lambda_3^2+\lambda_1\lambda_2+\lambda_1\lambda_3+\lambda_2\lambda_3,
\end{equation}
\begin{equation}
    \chi_{{\text{asym}}^{\otimes 2}}=\lambda_1\lambda_2+\lambda_1\lambda_3+\lambda_2\lambda_3.
\end{equation}
\end{comment}
For the tensor-product representation $q_1\otimes q_2$, the character satisfies
\begin{equation}
\chi_{q_1 \otimes q_2}(C_i) = \chi_{q_1}(C_i)\,\chi_{q_2}(C_i)\,.
\end{equation}
When $q_1=q_2\equiv q$, the product representation $q\otimes q$ can be decomposed into its symmetric and antisymmetric parts.
For instance, for a three-dimensional representation with eigenvalues $\lambda_1$, $\lambda_2$, $\lambda_3$ of a representative group element in the class $C_i$, the symmetric and antisymmetric characters can be written as
\begin{equation}
\chi_{\mathrm{Sym}^2}(C_i)
=\lambda_1^2+\lambda_2^2+\lambda_3^2+\lambda_1\lambda_2+\lambda_1\lambda_3+\lambda_2\lambda_3\,,
\end{equation}
\begin{equation}
\chi_{\wedge^2}(C_i)
=\lambda_1\lambda_2+\lambda_1\lambda_3+\lambda_2\lambda_3\,.
\end{equation}

\begin{comment}
In general, if there are $n$ identical terms, and for each term the eigenvalues are $\lambda_1,...,\lambda_k$, the corresponding generating functions are given by plethystic exponentials. The symmetric and anti-symmetric polynomials are respectively:
\begin{equation}
    \text{PE}(q;\lambda(g))=\text{exp}(\sum_{i=1}^{k}\sum_{r=1}^{\infty}\frac{q^r\lambda_i^r(g)}{r}),
\end{equation}
\begin{equation}
    \text{PEF}(q;\lambda(g))=\text{exp}(-\sum_{i=1}^{k}\sum_{r=1}^{\infty}\frac{(-q)^r\lambda_i^r(g)}{r}).
\end{equation}
For each $q^m$ term expanded by generating functions corresponds to the character of the (anti)symmetric representation at order $m$.
For a single variable \(q\), the Hilbert series can be written as
\begin{equation}
    H(q)=\frac{1}{|G|} \sum_{g\in G}
    \begin{cases}
        \text{PE}(q;\lambda(g)), & \text{if } q \text{ is bosonic},\\
        \text{PEF}(q;\lambda(g)), & \text{if } q \text{ is fermionic}.
    \end{cases}
\end{equation}
For several multiplets \(q_i\), the Hilbert series naturally takes the form
\begin{equation}
    H(q_1,\dots,q_i)=\frac{1}{|G|} \sum_{g\in G}\prod _i
    \begin{cases}
        \text{PE}(q_i;\lambda(g)), & \text{if } q_i \text{ is bosonic},\\
        \text{PEF}(q_i;\lambda(g)), & \text{if } q_i \text{ is fermionic}.
    \end{cases}
    \label{eq:HFS}
\end{equation}
\end{comment}
In general, suppose there are $n$ identical terms, and for each term the eigenvalues are $\lambda_1,\dots,\lambda_k$.
The corresponding generating functions are given by plethystic exponentials.
The symmetric and antisymmetric generating functions are
\begin{equation}
\text{PE}(q;\lambda(g))=\exp\!\left(\sum_{i=1}^{k}\sum_{r=1}^{\infty}\frac{q^r\,\lambda_i^r(g)}{r}\right),
\end{equation}
\begin{equation}
\text{PEF}(q;\lambda(g))=\exp\!\left(-\sum_{i=1}^{k}\sum_{r=1}^{\infty}\frac{(-q)^r\,\lambda_i^r(g)}{r}\right).
\end{equation}
Each $q^m$ term in the expansion of these generating functions corresponds to the character of the (anti)symmetric representation at order $m$.

For a single variable $q$, the Hilbert series can be written as
\begin{equation}
H(q)=\frac{1}{|G|} \sum_{g\in G}
\begin{cases}
\text{PE}(q;\lambda(g)), & \text{if } q \text{ is bosonic},\\
\text{PEF}(q;\lambda(g)), & \text{if } q \text{ is fermionic}.
\end{cases}
\end{equation}
For several multiplets $q_i$, the Hilbert series naturally takes the form
\begin{equation}
H(q_1,\dots,q_i)=\frac{1}{|G|} \sum_{g\in G}\prod _i
\begin{cases}
\text{PE}(q_i;\lambda(g)), & \text{if } q_i \text{ is bosonic},\\
\text{PEF}(q_i;\lambda(g)), & \text{if } q_i \text{ is fermionic}.
\end{cases}
\label{eq:HFS}
\end{equation}

As an example, we consider two distinct $A_4$ triplets, denoted by $T_1$ and $T_2$.
The relevant $A_4$ representations are given in Table~\ref{tab:A4-character-table}.
Using the three-dimensional matrix representation of the $A_4$ group elements summarized in Appendix~\ref{app:tensor_product}, we write the triplet character as
\begin{equation}
\chi_{\mathbf{3}}(g)
= \sum_{k=1}^{3}\lambda_k(g),
\qquad g\in A_4,
\end{equation}
where $\lambda_k(g)$ are the eigenvalues of the representation matrix.
%where $\lambda_k(g)$ are the eigenvalues of a representative $A_4$ group element $g$.
%If the class $C_i$ contains $n_i$ elements, then $n_i$ enters the Molien average through the sum over conjugacy classes.
Then, we can then write the corresponding plethystic exponential as
\begin{equation}
\text{PE}(T_1,T_2;\lambda(g))
=\exp\!\left(
\sum_{i=1}^{3}\sum_{r=1}^{\infty}\frac{T_1^{r}\lambda_i^{r}(g)}{r}
+\sum_{i=1}^{3}\sum_{r=1}^{\infty}\frac{T_2^{r}\lambda_i^{r}(g)}{r}
\right).
\end{equation}
Using Eq.~\eqref{eq:HFS}, we obtain the Hilbert series
\begin{equation}
H(T_1,T_2)=
\frac{
T_1^4 \left(T_2^4 - T_2^2 + 1\right)
+ T_1^3 T_2 \left(T_2^2 + T_2 + 1\right)
- T_1^2 (T_2+1)^2 \left(T_2^2 - 3 T_2 + 1\right)
+ T_1 T_2 \left(T_2^2 + T_2 + 1\right)
+ T_2^4 - T_2^2 + 1
}{
(1-T_1^2)^2(1-T_2^2)^2 (1-T_1^3)(1-T_2^3)
}\, .
\end{equation}
Expanding in the total degree of $T_1$ and $T_2$, we obtain the multiplicity of invariant singlets at each tensor-product order,
\begin{equation}
\begin{aligned}
H(T_1,T_2)=1
&+ \bigl(T_1^2 + T_1 T_2 + T_2^2\bigr) \\
&+ \bigl(T_1^3 + T_1^2 T_2 + T_1 T_2^2 + T_2^3\bigr) \\
&+ \bigl(2 T_1^4 + 3 T_1^3 T_2 + 4 T_1^2 T_2^2 + 3 T_1 T_2^3 + 2 T_2^4\bigr) \\
&+ \bigl(T_1^5 + 3 T_1^4 T_2 + 4 T_1^3 T_2^2 + 4 T_1^2 T_2^3 + 3 T_1 T_2^4 + T_2^5\bigr)
+ O\!\left(6\right).
\end{aligned}
\end{equation}

As a quick check, we consider the trivial singlet structures in $T_1 \otimes T_2$.
Comparing with the tensor-product decomposition in Appendix~\ref{app:tensor_product}, we find that the number of trivial singlets is equal to one, which corresponds to the term $T_1T_2$.

We also examine the example of flavons in the renormalizable scalar potential of Ref.~\cite{Altarelli_2005}.
In that model two flavons $\phi$ and $\phi'$ are introduced, both transforming as $A_4$ triplets.
This setup is directly analogous to the present case.
The only difference is that $\phi'$ changes sign under a $Z_4$ transformation, so the full symmetry is $A_4\times Z_4$.
In the Hilbert series this is implemented by removing all terms that contain an odd power of $\phi'$.
Taking the flavons to have mass dimension one, the Hilbert series for renormalizable operators is
\begin{equation}
HS = 1 + \phi^{2} + \phi'^{2} + \phi^{3} + \phi \phi'^{2}
     + 2 \phi^{4} + 4 \phi^{2} \phi'^{2} + 2 \phi'^{4} \, .
\end{equation}
We find that there are $12$ invariant operators in total, and the polynomial structure of each term matches exactly the $12$ renormalizable structures given in Ref.~\cite{Altarelli_2005}.

%For any field $\phi$, the corresponding generating function for bosons is given by
%\begin{equation}
%    \text{PE}(\phi;\chi(D,g))=\text{exp}(\sum_{r=1}^{\infty}\frac{\phi^r\chi(D^r,g^r)}{r}),
%\end{equation}
%\begin{equation}
%    \text{PEF}(\phi;\chi(D,g))=\text{exp}(-\sum_{r=1}^{\infty}\frac{(-\phi)^r\chi(D^r,g^r)}{r}).
%\end{equation}%

%从model如何引入到SMEFT 
%算符的形式引入

\subsection{Computing Hilbert series with modular forms}
%Because all the modular forms can be constructed by the $Y_{\mathbf{3}}^{(2)}$, which means it can be considered as a building block. 
%When the Yukawa couplings are constructed from holomorphic modular forms, modular forms of arbitrary weight can be built from the fundamental modular form $Y^{(2)}_{\mathbf{3}}$ through tensor products. Because the tensor product of any two modular forms of weights \( k \) and \( \widetilde{k} \) results in a modular form of weight \( k + \widetilde{k} \), which itself can be expressed as a tensor product of $Y^{(2)}_{\mathbf{3}}$, therefore $Y^{(2)}_{\mathbf{3}}$ should be regarded as a spurion field that acts as a fundamental building block.
When the Yukawa couplings are constructed from holomorphic modular forms in $A_4$ modular symmetry,  modular forms of arbitrary weight can be built from the fundamental modular form $Y^{(2)}_{\mathbf{3}}$ through tensor products.
Since the tensor product of two modular forms of weights $k$ and $\widetilde{k}$ has weight $k+\widetilde{k}$, any higher-weight modular form can ultimately be expressed in terms of tensor products of $Y^{(2)}_{\mathbf{3}}$.
Therefore, $Y^{(2)}_{\mathbf{3}}$ can be regarded as a spurion field and taken as the fundamental building block of the flavor structures.

For convenience, we use $Y$ to denote $Y^{(2)}_{\mathbf{3}}$.
The plethystic exponential (PE) involving modular forms can be written as
\begin{equation}
\text{PE}[D,\phi,Y,Y^*] = \text{PE}[D,\phi]\cdot \text{PE}[Y]\cdot \text{PE}[Y^*] \, .
\end{equation}
Here, the generating function for $Y$ takes the form
\begin{equation}
\text{PE}[Y]=\exp\!\left(\sum_{r=1}^{\infty}\frac{Y^r\,\chi(g^r)}{r}\right).
\end{equation}
Since modular forms are trivial singlets under both the Lorentz group and the gauge group, their characters depend only on the representation of the modular group.

With the $A_4$ constraint in Eq.~\eqref{eq:modular_constrain} imposed, the symmetric tensor product at weight $n$ is constrained by removing the components proportional to $Y_{\mathbf{1''}}^{(4)}$ arising from the product of a weight-$(n-2)$ symmetric tensor with $Y_{\mathbf{1''}}^{(4)}$.
Taking this effect into account, we define the modified generating function
\begin{equation}
\text{PE}'[Y]=\exp\!\left(\sum_{r=1}^{\infty}\frac{Y^r\,\chi(g^r)}{r}-\sum_{r=1}^{\infty}\frac{Y^{2r}\,\chi_{\mathbf{1''}}(g^r)}{r}\right),
\label{eq:remove}
\end{equation}
where $\chi_{\mathbf{1''}}$ denotes the character of the $A_4$ singlet representation $\mathbf{1''}$.

For the model considered here, the generating function for the two modular forms $Y(\tau_q)$ and $Y(\tau_e)$ takes the form
\begin{equation}
\text{PE}'[Y(\tau_q),Y(\tau_e)]
=\exp\!\left(\sum_{r=1}^{\infty}\left[\frac{Y^r(\tau_q)\,\chi_q(g^r)}{r}+\frac{Y^r(\tau_e)\,\chi_e(g^r)}{r}-\frac{Y^{2r}(\tau_q)\,\chi_{q\mathbf{1''}}(g^r)}{r}-\frac{Y^{2r}(\tau_e)\,\chi_{e\mathbf{1''}}(g^r)}{r}\right]\right).
\end{equation}
Including the anti-holomorphic modular forms, the complete Hilbert series can be formulated as
\begin{equation}
\begin{aligned}
H(D,\phi,Y,Y^*)
&= \frac{1}{|G_d|}
\sum_{g\in G_d}
\int_{G_c} d\mu_{G_c}\,
\frac{1}{P(D,z)}\,
\text{PE}'\bigl[Y(\tau_q),Y(\tau_e)\bigr]\,
\text{PE}'\bigl[Y^*(\tau_q),Y^*(\tau_e)\bigr] \\
&\quad\times
\prod_i
\begin{cases}
\text{PE}\bigl(\phi_i;\chi(D,z,a,g)\bigr),
& \text{if } \phi_i \text{ is boson},\\[2pt]
\text{PEF}\bigl(\phi_i;\chi(D,z,a,g)\bigr),
& \text{if } \phi_i \text{ is fermion},
\end{cases}
\end{aligned}
\label{Hilbert_series}
\end{equation}
with
\begin{equation}
G_d=A_{4q}\otimes A_{4e},\qquad
G_c=G_{SO_4}\otimes G_{\rm gauge}\otimes U(1)_q\otimes U(1)_{qd}\otimes U(1)_e\otimes U(1)_{ed}.
\end{equation}

Meanwhile, a more direct approach is to follow the leading-order Yukawa structure and construct holomorphic modular-invariant higher-dimensional operators in the schematic form
\begin{equation}
Y_\mathbf{r}^{(k_Y)} \, {Y_\mathbf{r'}^{(k_Y')}}^* \, \mathcal{O},
\label{eq:operator}
\end{equation}
which is a combination of a holomorphic modular form, an anti-holomorphic modular form, and a normal field operator.
This can be expressed by the tensor-product relation
\begin{equation}
\begin{aligned}
\left[\mathcal{O}(D,\phi,Y,Y^*)\right]_{\mathbf{1}}
&=
\bigl[(Y\otimes \dots \otimes Y)_{\mathrm{Sym}^{k_Y/2}}
\otimes (Y^*\otimes \dots \otimes Y^*)_{\mathrm{Sym}^{k_Y'/2}}
\otimes \mathcal{O}(D,\phi)\bigr]_{\mathbf{1}} \\
&\equiv
\bigl[Y^{(k_Y)}_{s}\otimes {Y^*_{s}}^{(k_Y')} \otimes \mathcal{O}(D,\phi)\bigr]_{\mathbf{1}} \, ,
\end{aligned}
\label{tensor_construction}
\end{equation}
where $Y^{(k_Y)}_{s}$ and ${Y^*_{s}}^{(k_Y')}$ denote the corresponding symmetric tensor products of $Y$ and $Y^*$.

This relation implies separate weight counting and tensor-product constraints for holomorphic and anti-holomorphic factors,
\begin{equation}
k_Y=\sum_{i}k_i,\quad k'_Y=\sum_{j}k'_j,\quad
\rho\otimes {\rho}^*\otimes\rho^{(1)}\otimes\cdots\otimes\rho^{(n)}\otimes{{\rho}^{(n+1)}}^*\otimes\cdots\otimes{{\rho}^{(n+m)}}^*
\supset \mathbf{1},
\label{eq:invariant_condition}
\end{equation}
for an operator containing $n$ non-conjugated factors and $m$ conjugated factors.
In addition, the equivalence between this tensor-product construction and the Hilbert-series counting can be verified explicitly.

Equivalently, one may treat all modular forms (e.g.\ $Y^{(2)}_{\mathbf{3}}(\tau_q)$, $Y^{(2)}_{\mathbf{3}}(\tau_e)$, $Y^{(4)}_{\mathbf{1}}(\tau_q)$, $Y^{(4)}_{\mathbf{1}}(\tau_e)$, $\dots$) as building blocks, with the restriction that each distinct modular form appears at most once in a given term.
The corresponding Hilbert series can then be written as
\begin{equation}
\begin{aligned}
H\!\left(D,\phi,Y_{\mathbf{r}}^{(k_Y)}, {Y_{\mathbf{r'}}^{(k_Y')}}^{*}\right)
&=\frac{1}{|G_d|}\sum_{g\in G_d}\int_{G_c} d\mu_{G_c}\,
\frac{1}{P(D,z)} \\
&\quad\times
\left(\prod_k \mathrm{PE}\!\left[Y_{\mathbf{r}}^{(k_Y)}, {Y_{\mathbf{r'}}^{(k_Y')}}^{*}\right]\right)
\prod_i
\begin{cases}
\mathrm{PE}\!\bigl(\phi_i;\chi(D,z,a,g)\bigr), & \text{if } \phi_i \text{ is boson},\\
\mathrm{PEF}\!\bigl(\phi_i;\chi(D,z,a,g)\bigr), & \text{if } \phi_i \text{ is fermion}.
\end{cases}
\end{aligned}
\label{Hilbert_series2}
\end{equation}

Here,
\begin{equation}
\prod_k \text{PE}[Y_{\mathbf{r}}^{(k_Y)},{Y_{\mathbf{r'}}^{(k_Y')}}^*]
=
\prod_k
\text{PE}[\{Y_{\mathbf{rq}}^{(k_{qY})}\}]
\text{PE}[\{Y_{\mathbf{re}}^{(k_{eY})}\}]
\text{PE}[\{{Y_{\mathbf{rq'}}^{(k_{qY}')}}^*\}]
\text{PE}[\{{Y_{\mathbf{re'}}^{(k_{eY}')}}^*\}]\,,
\end{equation}
with $\mathbf{rq}$, $\mathbf{rq'}$ and $\mathbf{re}$, $\mathbf{re'}$ denoting representations of $A_{4q}$ and $A_{4e}$, respectively.
Because each type of modular form appears at most once per term, the factor $\prod_k\text{PE}[\{Y^{(k_{qY})}_{\mathbf{rq}}(\tau_q)\}]$ reduces to a finite polynomial; schematically,
\begin{equation}
\prod_k\text{PE}[\{Y_{\mathbf{rq}}^{(k_{qY})}\}]
\simeq
\left(1+Y^{(2)}_{\mathbf{3}}(\tau_q)\chi_{q3}(g)+Y^{(4)}_{\mathbf{1}}(\tau_q)\chi_{q1}(g)+Y^{(4)}_{\mathbf{1''}}(\tau_q)\chi_{q\mathbf{1''}}(g)+\dots\right).
\end{equation}

By explicitly computing the Hilbert series in Eqs.~\eqref{Hilbert_series} and~\eqref{Hilbert_series2}, we find that each term obtained from~\eqref{Hilbert_series2} matches the corresponding contribution in Eq.~\eqref{Hilbert_series}.
For example, consider the dimension-six SMEFT operator $\mathcal{O}_{uu}$ with flavor $u$ and $\overline{u}$.
Equation~\eqref{Hilbert_series} yields the term $7\,u^2 {u^\dagger}^2 {Y^*}^4 Y^4$, while Eq.~\eqref{Hilbert_series2} resolves it into the following seven structures,
\begin{equation}
\begin{aligned}
u^2& {u^\dagger}^2 \, Y^{(8)}_{\mathbf{1}}(\tau_q)\; {Y^{(8)}_{\mathbf{1}}}^*(\tau_q),\\
u^2& {u^\dagger}^2 \, Y^{(8)}_{\mathbf{1''}}(\tau_q)\; {Y^{(8)}_{\mathbf{1'}}}^*(\tau_q),\\
u^2& {u^\dagger}^2 \, Y^{(8)}_{\mathbf{1'}}(\tau_q)\; {Y^{(8)}_{\mathbf{1''}}}^*(\tau_q),\\
u^2& {u^\dagger}^2 \, Y^{(8)}_{\mathbf{3}a}(\tau_q)\; {Y^{(8)}_{\mathbf{3}b}}^*(\tau_q),\\
u^2& {u^\dagger}^2 \, Y^{(8)}_{\mathbf{3}b}(\tau_q)\; {Y^{(8)}_{\mathbf{3}a}}^*(\tau_q),\\
u^2& {u^\dagger}^2 \, Y^{(8)}_{\mathbf{3}a}(\tau_q)\; {Y^{(8)}_{\mathbf{3}a}}^*(\tau_q),\\
u^2& {u^\dagger}^2 \, Y^{(8)}_{\mathbf{3}b}(\tau_q)\; {Y^{(8)}_{\mathbf{3}b}}^*(\tau_q).
\end{aligned}
\end{equation}
Here $a$ and $b$ label the two independent triplet modular forms of $A_4$ at weight $8$.
Furthermore, both Hilbert series yield the same total number of operators: $2961$ at mass dimension $6$ and $360$ at mass dimension $7$.

For (anti-)holomorphic modular forms, constructing operators via tensor products of modular forms is fully consistent with the alternative procedure of enumerating all modular forms at a given weight and then building operators.
The two approaches therefore provide a cross-check of each other.
To make the operator structures more transparent, we will base our discussion on the terms appearing in the Hilbert series in Eq.~\eqref{Hilbert_series2}.

\section{\texorpdfstring{$A_4$}{A4} modular invariant SMEFT operators}
\label{sec:smeft}

%dimension-5详细结构

%In this section, we employ the Hilbert series framework to systematically determine the operator flavor structures and the corresponding multiplicities. 
In this section, we employ the Hilbert-series framework to systematically determine the operator flavor structures and their corresponding multiplicities.
%\textbf{Meanwhile, this construction realises the SM as the low-energy limit of supersymmetry.}
%The coefficients of the series directly yield the number of independent operators associated with each structure.
The coefficients of the series directly give the number of independent operators associated with each structure.

%As a first application, we analyze the dimension-5 Weinberg operator and identify all flavor-invariant structures consistent with the $A_{4}^{(q)} \otimes A_{4}^{(e)}$ symmetry. At dimension six, we proceed analogously and construct the explicit forms of all baryon- and lepton-number conserving operators.
As a first application, we analyze the dimension-five Weinberg operator and identify all flavor-invariant structures consistent with the $A_{4}^{(q)} \otimes A_{4}^{(e)}$ modualr symmetry.
At dimension six, we proceed analogously and construct the explicit forms of all baryon- and lepton-number conserving operators.

In this way, we obtain a systematic classification of all relevant operators
considered in this work.
\begin{table}[!htbp]
  \centering
  \small
  \setlength{\tabcolsep}{6pt}      % 扩大列间距
  \renewcommand{\arraystretch}{1.2} % 放大行高
  \begin{tabular}{|c|c|c|c|}
    \hline
    \textbf{class} 
      & \textbf{relevant dim-5 operators} 
      & \textbf{A$_4$ invariant comb.} 
      & \textbf{number} \\
    \hline
    \multirow{1}{*}{$\psi^2 H^2$}
    & $\mathcal{O}_5=(LH)(LH)+\mathrm{h.c.}$ 
    & $Y^{(4)}_\mathbf{re}[(LH)(LH)]+\mathrm{h.c.}$ 
    & 6  \\ 
    \hline
  \end{tabular}
  \caption{Dimension-5 operators with $A_4$ modular invariants.}
  \label{tab:dim5_A4}
\end{table}

\begin{table}[!htbp]
  \centering
  \small
  \setlength{\tabcolsep}{6pt}      % 扩大列间距
  \renewcommand{\arraystretch}{1.2} % 放大行高
  \begin{tabular}{|c|c|c|c|}
    \hline
    \textbf{class} 
      & \textbf{relevant dim-6 operators} 
      & \textbf{A$_4$ invariant comb.} 
      & \textbf{number} \\
    \hline
    \multirow{1}{*}{bosonic sector}
    &  &  & 15  \\ 
     \hline
    % 1: ψ²H³ + \mathrm{h.c.}
    \multirow{3}{*}{$\psi^2H^3$+\text{h.c.}}
      & $\mathcal{O}_{eH}+ \mathrm{h.c.}$ & ${Y^{(2)}_\mathbf{re'}}^*[\overline{L}_L \Gamma E_R]+ \mathrm{h.c.}$ & 6 \\ \cline{2-4}
      & $\mathcal{O}_{uH}+ \mathrm{h.c.}$ & ${Y^{(6)}_\mathbf{rq'}}^*[\overline{Q}_L \Gamma U_R]+ \mathrm{h.c.}$ & 12 \\ \cline{2-4}
      & $\mathcal{O}_{dH}+ \mathrm{h.c.}$ & ${Y^{(2)}_\mathbf{rq'}}^*[\overline{Q}_L \Gamma D_R]+ \mathrm{h.c.}$ & 6\\ 
    \hline

    % 2: ψ²XH + \mathrm{h.c.}
    \multirow{8}{*}{$\psi^2XH$+\text{h.c.}}
      & $\mathcal{O}_{eW}+ \mathrm{h.c.}$ & ${Y^{(2)}_\mathbf{re'}}^*[\overline{L}_L \Gamma E_R]+ \mathrm{h.c.}$ & 6 \\ \cline{2-4}
      & $\mathcal{O}_{eB}+ \mathrm{h.c.}$ & ${Y^{(2)}_\mathbf{re'}}^*[\overline{L}_L \Gamma E_R]+ \mathrm{h.c.}$ & 6 \\ \cline{2-4}
      & $\mathcal{O}_{uG}+ \mathrm{h.c.}$ & ${Y^{(6)}_\mathbf{rq'}}^*[\overline{Q}_L \Gamma U_R]+ \mathrm{h.c.}$ & 12 \\ \cline{2-4}
      & $\mathcal{O}_{uW}+ \mathrm{h.c.}$ & ${Y^{(6)}_\mathbf{rq'}}^*[\overline{Q}_L \Gamma U_R]+ \mathrm{h.c.}$ & 12 \\ \cline{2-4}
      & $\mathcal{O}_{uB}+ \mathrm{h.c.}$ & ${Y^{(6)}_\mathbf{rq'}}^*[\overline{Q}_L \Gamma U_R]+ \mathrm{h.c.}$ & 12 \\ \cline{2-4}
      & $\mathcal{O}_{dG}+ \mathrm{h.c.}$ & ${Y^{(2)}_\mathbf{rq'}}^*[\overline{Q}_L \Gamma D_R]+ \mathrm{h.c.}$ & 6 \\ \cline{2-4}
      & $\mathcal{O}_{dW}+ \mathrm{h.c.}$ & ${Y^{(2)}_\mathbf{rq'}}^*[\overline{Q}_L \Gamma D_R]+ \mathrm{h.c.}$ & 6 \\ \cline{2-4}
      & $\mathcal{O}_{dB}+ \mathrm{h.c.}$ & ${Y^{(2)}_\mathbf{rq'}}^*[\overline{Q}_L \Gamma D_R]+ \mathrm{h.c.}$ & 6 \\ 
    \hline

    % 3: ψ²H²D
    \multirow{8}{*}{$\psi^2H^2D$}
      & $\mathcal{O}_{Hl}^{(1)}$ & ${Y^{(2)}_\mathbf{re}}{Y^{(2)}_\mathbf{re'}}^*[\overline{L}_L \Gamma L_L]$ & 7 \\ \cline{2-4}
      & $\mathcal{O}_{Hl}^{(3)}$ & ${Y^{(2)}_\mathbf{re}}{Y^{(2)}_\mathbf{re'}}^*[\overline{L}_L \Gamma L_L]$ & 7 \\ \cline{2-4}
      & $\mathcal{O}_{He}$       & $[\overline{E}_R \Gamma E_R]$ & 3 \\ \cline{2-4}
      & $\mathcal{O}_{Hq}^{(1)}$ & ${Y^{(2)}_\mathbf{rq}}{Y^{(2)}_\mathbf{rq'}}^*[\overline{Q}_L \Gamma Q_L]$ & 7 \\ \cline{2-4}
      & $\mathcal{O}_{Hq}^{(3)}$ & ${Y^{(2)}_\mathbf{rq}}{Y^{(2)}_\mathbf{rq'}}^*[\overline{Q}_L \Gamma Q_L]$ & 7 \\ \cline{2-4}
      & $\mathcal{O}_{Hu}$       & ${Y^{(4)}_\mathbf{rq}}{Y^{(4)}_\mathbf{rq'}}^*[\overline{U}_R \Gamma U_R]$ & 21 \\ \cline{2-4}
      & $\mathcal{O}_{Hd}$       & $[\overline{D}_R \Gamma D_R]$ & 3 \\ \cline{2-4}
      & $\mathcal{O}_{Hud}+ \mathrm{h.c.}$      & ${Y^{(4)}_\mathbf{rq}}[\overline{U}_R \Gamma D_R]+ \mathrm{h.c.}$ & 12 \\ 
    \hline
  \end{tabular}
\caption{Dimension-6 bosonic and bilinear SMEFT operators with $A_4$ modular invariants. Here $\Gamma$ denotes a generic combination of Dirac matrices, color generators, and $SU(2)$ generators. $U_R$, $D_R$, and $E_R$ represent the three types of right-handed fermions in $A_4$ singlets. For example, $E_R$ can denote $e_R$, $\mu_R$, or $\tau_R$.}
  \label{tab:dim6_boson_bilinear}
\end{table}

\clearpage

\begin{table}[!htb]
  \centering
  \small
  \setlength{\tabcolsep}{4pt}      % 列间距
  \renewcommand{\arraystretch}{1.1} % 行高
  \begin{tabular}{|c|c|c|c|}
    \hline
    \textbf{class} 
      & \textbf{relevant dim-6 operators} 
      & \textbf{$A_4$ invariant comb.} 
      & \textbf{number} \\
    \hline

    % 4: (LL)(LL)
    \multirow{5}{*}{$(\overline{L}L)(\overline{L}L)$}
      & $\mathcal{O}_{ll}$       & ${Y^{(4)}_\mathbf{re}}{Y^{(4)}_\mathbf{re}}^*[\overline{L}_L\Gamma L_L][\overline{L}_L\Gamma L_L]$ & 95 \\ \cline{2-4}
      & $\mathcal{O}_{qq}^{(1)}$ & ${Y^{(4)}_\mathbf{rq}}{Y^{(4)}_\mathbf{rq'}}^*[\overline{Q}_L\Gamma Q_L][\overline{Q}_L\Gamma Q_L]$ & 95 \\ \cline{2-4}
      & $\mathcal{O}_{qq}^{(3)}$ & ${Y^{(4)}_\mathbf{rq}}{Y^{(4)}_\mathbf{rq'}}^*[\overline{Q}_L\Gamma Q_L][\overline{Q}_L\Gamma Q_L]$ & 95 \\ \cline{2-4}
      & $\mathcal{O}_{lq}^{(1)}$ & ${Y^{(2)}_\mathbf{re}}{Y^{(2)}_\mathbf{re'}}^*{Y^{(2)}_\mathbf{rq}}{Y^{(2)}_\mathbf{rq'}}^*[\overline{L}_L\Gamma L_L][\overline{Q}_L\Gamma Q_L]$ & 49 \\ \cline{2-4}
      & $\mathcal{O}_{lq}^{(3)}$ & ${Y^{(2)}_\mathbf{re}}{Y^{(2)}_\mathbf{re'}}^*{Y^{(2)}_\mathbf{rq}}{Y^{(2)}_\mathbf{rq'}}^*[\overline{L}_L\Gamma L_L][\overline{Q}_L\Gamma Q_L]$ & 49 \\
    \hline

    % 5: (RR)(RR)
    \multirow{7}{*}{$(\overline{R}R)(\overline{R}R)$}
      & $\mathcal{O}_{ee}$       & $[\overline{E}_R\Gamma E_R][\overline{E}_R\Gamma E_R]$ & 12 \\ \cline{2-4}
      & $\mathcal{O}_{uu}$       & ${Y^{(8)}_\mathbf{rq}}{Y^{(8)}_\mathbf{rq'}}^*[\overline{U}_R\Gamma U_R][\overline{U}_R\Gamma U_R]$ & 315 \\ \cline{2-4}
      & $\mathcal{O}_{dd}$       & $[\overline{D}_R\Gamma D_R][\overline{D}_R\Gamma D_R]$ & 15 \\ \cline{2-4}
      & $\mathcal{O}_{eu}$       & ${Y^{(4)}_\mathbf{rq}}{Y^{(4)}_\mathbf{rq'}}^*[\overline{E}_R\Gamma E_R][\overline{U}_R\Gamma U_R]$ & 63 \\ \cline{2-4}
      & $\mathcal{O}_{ed}$       & $[\overline{E}_R\Gamma E_R][\overline{D}_R\Gamma D_R]$ & 9 \\ \cline{2-4}
      & $\mathcal{O}_{ud}^{(1)}$ & ${Y^{(4)}_\mathbf{rq}}{Y^{(4)}_\mathbf{rq'}}^*[\overline{U}_R\Gamma U_R][\overline{D}_R\Gamma D_R]$ & 189 \\ \cline{2-4}
      & $\mathcal{O}_{ud}^{(8)}$ & ${Y^{(4)}_\mathbf{rq}}{Y^{(4)}_\mathbf{rq'}}^*[\overline{U}_R\Gamma U_R][\overline{D}_R\Gamma D_R]$ & 189 \\
    \hline

    % 6: (LL)(RR)
    \multirow{8}{*}{$(\overline{L}L)(\overline{R}R)$}
      & $\mathcal{O}_{le}$       & ${Y^{(2)}_\mathbf{re}}{Y^{(2)}_\mathbf{re'}}^*[\overline{L}_L\Gamma L_L][\overline{E}_R\Gamma E_R]$ & 63 \\ \cline{2-4}
      & $\mathcal{O}_{lu}$       & ${Y^{(4)}_\mathbf{rq}}{Y^{(4)}_\mathbf{rq'}}^*{Y^{(2)}_\mathbf{re}}{Y^{(2)}_\mathbf{re'}}^*[\overline{L}_L\Gamma L_L][\overline{U}_R\Gamma U_R]$ & 147 \\ \cline{2-4}
      & $\mathcal{O}_{ld}$       & ${Y^{(2)}_\mathbf{re}}{Y^{(2)}_\mathbf{re'}}^*[\overline{L}_L\Gamma L_L][\overline{D}_R\Gamma D_R]$ & 21 \\ \cline{2-4}
      & $\mathcal{O}_{qe}$       & ${Y^{(2)}_\mathbf{rq}}{Y^{(2)}_\mathbf{rq'}}^*[\overline{Q}_L\Gamma Q_L][\overline{E}_R\Gamma E_R]$ & 21 \\ \cline{2-4}
      & $\mathcal{O}_{qu}^{(1)}$ & ${Y^{(6)}_\mathbf{rq}}{Y^{(6)}_\mathbf{rq'}}^*[\overline{Q}_L\Gamma Q_L][\overline{U}_R\Gamma U_R]$ & 333 \\ \cline{2-4}
      & $\mathcal{O}_{qu}^{(8)}$ & ${Y^{(6)}_\mathbf{rq}}{Y^{(6)}_\mathbf{rq'}}^*[\overline{Q}_L\Gamma Q_L][\overline{U}_R\Gamma U_R]$ & 333 \\ \cline{2-4}
      & $\mathcal{O}_{qd}^{(1)}$ & ${Y^{(2)}_\mathbf{rq}}{Y^{(2)}_\mathbf{rq'}}^*[\overline{Q}_L\Gamma Q_L][\overline{D}_R\Gamma D_R]$ & 63 \\ \cline{2-4}
      & $\mathcal{O}_{qd}^{(8)}$ & ${Y^{(2)}_\mathbf{rq}}{Y^{(2)}_\mathbf{rq'}}^*[\overline{Q}_L\Gamma Q_L][\overline{D}_R\Gamma D_R]$ & 63 \\
    \hline

    % 7: (LR)(RL) + \mathrm{h.c.}
    \multirow{1}{*}{$(\overline{L}R)(\overline{R}L)$+\text{h.c.}}
      & $\mathcal{O}_{ledq}+ \mathrm{h.c.}$     & ${Y^{(2)}_\mathbf{rq}}{Y^{(2)}_\mathbf{re'}}^*[\overline{D}_R\Gamma Q_L][\overline{L}_L\Gamma E_R]+ \mathrm{h.c.}$ & 18 \\
    \hline

    % 8: (LR)(LR) + \mathrm{h.c.}
    \multirow{4}{*}{$(\overline{L}R)(\overline{L}R)$+\text{h.c.}}
      & $\mathcal{O}_{quqd}^{(1)}+ \mathrm{h.c.}$ & ${Y^{(8)}_\mathbf{rq'}}^*[\overline{Q}_L\Gamma U_R][\overline{Q}_L\Gamma D_R]+ \mathrm{h.c.}$ & 126 \\ \cline{2-4}
      & $\mathcal{O}_{quqd}^{(8)}+ \mathrm{h.c.}$ & ${Y^{(8)}_\mathbf{rq'}}^*[\overline{Q}_L\Gamma U_R][\overline{Q}_L\Gamma D_R]+ \mathrm{h.c.}$ & 126 \\ \cline{2-4}
      & $\mathcal{O}_{lequ}^{(1)}+ \mathrm{h.c.}$ & ${Y^{(6)}_\mathbf{rq'}}^*{Y^{(2)}_\mathbf{re'}}^*[\overline{Q}_L\Gamma U_R][\overline{L}_L\Gamma E_R]+ \mathrm{h.c.}$ & 36 \\ \cline{2-4}
      & $\mathcal{O}_{lequ}^{(3)}+ \mathrm{h.c.}$ & ${Y^{(6)}_\mathbf{rq'}}^*{Y^{(2)}_\mathbf{re'}}^*[\overline{Q}_L\Gamma U_R][\overline{L}_L\Gamma E_R]+ \mathrm{h.c.}$ & 36 \\
    \hline

    % 9: B-violating
    \multirow{2}{*}{$B$-violating}
      & $Q_{duq}+ \mathrm{h.c.}$ & ${Y^{(2)}_\mathbf{rq}}{Y^{(4)}_\mathbf{rq'}}^*{Y^{(2)}_\mathbf{re}}[D_R^T\Gamma U_R][Q_L^T\Gamma L_L]+ \mathrm{h.c.}$ & 72 \\ \cline{2-4}
      & $Q_{qqu}+ \mathrm{h.c.}$ & ${Y^{(4)}_\mathbf{rq}}{Y^{(4)}_\mathbf{rq'}}^*[Q_L^T\Gamma Q_L][U_R^T\Gamma E_R]+ \mathrm{h.c.}$ & 78 \\ \cline{2-4}
      & $Q_{qqq}+ \mathrm{h.c.}$ & ${Y^{(6)}_\mathbf{rq}}{Y^{(2)}_\mathbf{re}}[Q_L^T\Gamma Q_L][Q_L^T\Gamma L_L]+ \mathrm{h.c.}$ & 24 \\ \cline{2-4}
      & $Q_{duu}+ \mathrm{h.c.}$ & ${Y^{(8)}_\mathbf{rq'}}^*[D_R^T\Gamma U_R][U_R^T\Gamma E_R]+ \mathrm{h.c.}$ & 54 \\
    \hline

  \end{tabular}
  \caption{Dimension-6 SMEFT four-fermion operators with $A_4$ modular invariants.}
  \label{tab:classes4-9}
\end{table}

\begin{table}[!htb]
  \centering
  \small
  \setlength{\tabcolsep}{6pt}
  \renewcommand{\arraystretch}{1.2}
  \begin{tabular}{|l|r|}
    \hline
    \textbf{Class}                & \textbf{number} \\ 
    \hline
    bosonic sectors               & 15   \\ 
    $\psi^2H^3 + \mathrm{h.c.}$   & 24   \\ 
    $\psi^2XH + \mathrm{h.c.}$    & 66  \\ 
    $\psi^2H^2D$                  & 67   \\ 
    $\psi^4$                      & 2789 \\ 
    \hline
    \textbf{Total}                & 2961 \\ 
    \hline
  \end{tabular}
\caption{Total number of dimension-6 SMEFT operators with $A_4$ modular invariants by class.}
  \label{tab:dim6_total_number}
\end{table}

\subsection{Dimension-5 case}
\label{dimension-W}
At dimension five, the only operator is the Weinberg operator, whose explicit form is
\begin{equation}
\mathcal{O}_5 = (L H)(L H)+\text{h.c.}\, .
\end{equation}
A detailed discussion of modular-invariant SMEFT operators at this order can be found in Ref.~\cite{Okada_2021}.
Before inserting the vacuum expectation value (VEV) of the Higgs field, there are three possible structures:
\begin{equation}
\begin{aligned}
  &\left[Y^{(4)}_{\mathbf{3}}\big[(L H)(L H)\big]_{\mathbf{3}}\right]_{\mathbf{1}},\\
  &\left[Y^{(4)}_{\mathbf{1}}\big[(L H)(L H)\big]_{\mathbf{1}}\right]_{\mathbf{1}},\\
  &\left[Y^{(4)}_{\mathbf{1'}}\big[(L H)(L H)\big]_{\mathbf{1''}}\right]_{\mathbf{1}}.
\end{aligned}
\label{eq:Weinberg}
\end{equation}

Using the tensor-product rules summarized in Appendix~\ref{app:tensor_product}, these three terms can be written explicitly as
\begin{equation}
\begin{aligned}
\left[Y^{(4)}_{\mathbf{3}}\big[(L H)(L H)\big]_{\mathbf{3}}\right]_{\mathbf{1}}
={}&
Y^{(4)}_{1}\big[2(L_1H)(L_1H)-(L_2H)(L_3H)-(L_3H)(L_2H)\big]\\
&+Y^{(4)}_{2}\big[2(L_3H)(L_3H)-(L_1H)(L_2H)-(L_2H)(L_1H)\big]\\
&+Y^{(4)}_{3}\big[2(L_2H)(L_2H)-(L_1H)(L_3H)-(L_3H)(L_1H)\big]\,,
\end{aligned}
\end{equation}
\begin{equation}
\left[Y^{(4)}_{\mathbf{1}}\big[(L H)(L H)\big]_{\mathbf{1}}\right]_{\mathbf{1}}
=
Y^{(4)}_{\mathbf{1}}\big[(L_1H)(L_1H)+(L_2H)(L_3H)+(L_3H)(L_2H)\big]\,,
\end{equation}
\begin{equation}
\left[Y^{(4)}_{\mathbf{1'}}\big[(L H)(L H)\big]_{\mathbf{1''}}\right]_{\mathbf{1}}
=
Y^{(4)}_{\mathbf{1'}}\big[(L_2H)(L_2H)+(L_1H)(L_3H)+(L_3H)(L_1H)\big]\,.
\end{equation}
Here the weight-4 triplet modular form is written as
\begin{equation}
Y^{(4)}_{\mathbf{3}}=\begin{pmatrix} Y^{(4)}_{1} \\ Y^{(4)}_{2} \\ Y^{(4)}_{3} \end{pmatrix}.
\end{equation}

For the explicit tensor products of fields, the contractions into
$[\cdots]_{\mathbf{1}}$, $[\cdots]_{\mathbf{1'}}$, $[\cdots]_{\mathbf{1''}}$,
$[\cdots]_{\mathbf{3}_{\mathrm{s}}}$, $[\cdots]_{\mathbf{3}_{\mathrm{a}}}$
are given by
\begin{equation}
[(LH)(LH)]_{\mathbf{1}}
=
(L_1H)(L_1H)+(L_2H)(L_3H)+(L_3H)(L_2H)\,,
\end{equation}
\begin{equation}
[(LH)(LH)]_{\mathbf{1'}}
=
(L_3H)(L_3H)+(L_1H)(L_2H)+(L_2H)(L_1H)\,,
\end{equation}
\begin{equation}
[(LH)(LH)]_{\mathbf{1''}}
=
(L_2H)(L_2H)+(L_1H)(L_3H)+(L_3H)(L_1H)\,,
\end{equation}
\begin{equation}
[(LH)(LH)]_{\mathbf{3}}
=[(LH)(LH)]_{\mathbf{3}_{\mathrm{s}}}
=
\begin{pmatrix}
2(L_1H)(L_1H)-(L_2H)(L_3H)-(L_3H)(L_2H)\\
2(L_3H)(L_3H)-(L_1H)(L_2H)-(L_2H)(L_1H)\\
2(L_2H)(L_2H)-(L_1H)(L_3H)-(L_3H)(L_1H)
\end{pmatrix}\!,
\end{equation}
and
\begin{equation}
[(LH)(LH)]_{\mathbf{3}_{\mathrm{a}}}
=
\begin{pmatrix}
(L_2H)(L_3H)-(L_3H)(L_2H)\\
(L_1H)(L_2H)-(L_2H)(L_1H)\\
(L_1H)(L_3H)-(L_3H)(L_1H)
\end{pmatrix}\!.
\end{equation}

Within the Hilbert-series framework, and applying series~\eqref{Hilbert_series2}, the dimension-five contribution can be written as
\begin{equation}
\text{Hilbert5} =
{Y^{(4)}_{\mathbf{3}}}^*(\tau_e) \, {h^\dagger}^2 {l^\dagger}^2
+ {Y^{(4)}_{\mathbf{1'}}}^*(\tau_e) \, {h^\dagger}^2 {l^\dagger}^2
+ {Y^{(4)}_{\mathbf{1}}}^*(\tau_e) \, {h^\dagger}^2 {l^\dagger}^2
+ Y^{(4)}_{\mathbf{3}}(\tau_e) \, h^2 l^2
+ Y^{(4)}_{\mathbf{1'}}(\tau_e) \, h^2 l^2
+ Y^{(4)}_{\mathbf{1}}(\tau_e) \, h^2 l^2 \, .
\label{eq:MSSMHS5}
\end{equation}
From this expression we see that the Weinberg operator and its conjugate together generate three distinct operator types and their conjugates.
As shown in Eq.~\eqref{eq:Weinberg}, these provide the most elementary realization.
Moreover, each term in the Hilbert series can be put in one-to-one correspondence with the cases discussed in Ref.~\cite{Okada_2021}.

In this case, the dimension-five Lagrangian can be written as
\begin{equation}
\mathcal{L}_W
= \frac{1}{\Lambda}\Big(
\big[ Y^{(4)}_{\mathbf{3}}\big[(L H)(L H)\big]_{\mathbf{3}}\big]_{\mathbf{1}}
+ g_1 \big[ Y^{(4)}_{\mathbf{1}}\big[(L H)(L H)\big]_{\mathbf{1}}\big]_{\mathbf{1}}
+ g_2 \big[ Y^{(4)}_{\mathbf{1'}}\big[(L H)(L H)\big]_{\mathbf{1''}}\big]_{\mathbf{1}}
\Big) + \text{h.c.}\, .
\end{equation}

An alternative modular-invariant construction of SMEFT operators is presented in~\cite{Kobayashi_2022}.
In that framework, all higher-dimensional operators are generated from three-point Yukawa interactions by integrating out heavy fields.
Using this construction, the Weinberg operator can be written as
\begin{equation}
(Y^{(2)}_{\mathbf{3}}LH)_{\mathbf{1},\mathbf{1'},\mathbf{1''},\mathbf{3}_{\mathrm{s}},\mathbf{3}_{\mathrm{a}}}
\otimes
(Y^{(2)}_{\mathbf{3}}LH)_{\mathbf{1},\mathbf{1'},\mathbf{1''},\mathbf{3}_{\mathrm{s}},\mathbf{3}_{\mathrm{a}}}\, .
\label{wein}
\end{equation}
After contracting the Lorentz and $SU(2)_L$ indices, the fully contracted operator is symmetric under $L_i \leftrightarrow L_j$.
So in this construction, there are five singlet contractions:
\begin{align}
\mathcal{C}_1 &\equiv \mathbf{1}\otimes \mathbf{1}, \\
\mathcal{C}_2 &\equiv \mathbf{1'}\otimes \mathbf{1''}, \\
\mathcal{C}_3 &\equiv (\mathbf{3}_{\mathrm{s}}\otimes \mathbf{3}_{\mathrm{s}})_{\mathbf{1}}, \\
\mathcal{C}_4 &\equiv (\mathbf{3}_{\mathrm{a}}\otimes \mathbf{3}_{\mathrm{a}})_{\mathbf{1}}, \\
\mathcal{C}_5 &\equiv (\mathbf{3}_{\mathrm{a}}\otimes \mathbf{3}_{\mathrm{s}})_{\mathbf{1}}.
\end{align}
However, these contractions are not all independent, and they satisfy the linear relation
\begin{equation}
\mathcal{C}_3 \;=\; 4\big(\mathcal{C}_1-\mathcal{C}_2\big)\;-\;3\,\mathcal{C}_4\,.
\label{redundent}
\end{equation}
Therefore, only four contractions are independent.

Another alternative construction is
\begin{equation}
\big(Y^{(2)}_{\mathbf{3}} Y^{(2)}_{\mathbf{3}}\big)_{\mathbf{1},\mathbf{1}',\mathbf{1}'',\mathbf{3}}
\;\otimes\;
\big[(LH)(LH)\big]_{\mathbf{1},\mathbf{1}',\mathbf{1}'',\mathbf{3}}\,.
\end{equation}
Compared to Eq.~\eqref{wein}, this is simply a different choice of basis.
The four singlet contractions in this basis can be chosen as
\begin{align}
\widetilde{\mathcal{C}}_1 &\equiv \big(Y^{(2)}_{\mathbf{3}} Y^{(2)}_{\mathbf{3}}\big)_{\mathbf{1}}\;
\big[(LH)(LH)\big]_{\mathbf{1}},\\
\widetilde{\mathcal{C}}_2 &\equiv \big(Y^{(2)}_{\mathbf{3}} Y^{(2)}_{\mathbf{3}}\big)_{\mathbf{1}'}\;
\big[(LH)(LH)\big]_{\mathbf{1}''},\\
\widetilde{\mathcal{C}}_3 &\equiv \big(Y^{(2)}_{\mathbf{3}} Y^{(2)}_{\mathbf{3}}\big)_{\mathbf{1}''}\;
\big[(LH)(LH)\big]_{\mathbf{1}'},\\
\widetilde{\mathcal{C}}_4 &\equiv
\Big(\big(Y^{(2)}_{\mathbf{3}} Y^{(2)}_{\mathbf{3}}\big)_{\mathbf{3}}
\otimes \big[(LH)(LH)\big]_{\mathbf{3}}\Big)_{\mathbf{1}}.
\end{align}
This choice of basis removes the redundancy in Eq.~\eqref{redundent}.
Moreover, imposing the constraint in Eq.~\eqref{eq:modular_constrain} yields
\begin{equation}
\big(Y^{(2)}_{\mathbf{3}} Y^{(2)}_{\mathbf{3}}\big)_{\mathbf{1}''}=Y^{(4)}_{\mathbf{1''}}=0\,,
\end{equation}
which implies $\widetilde{\mathcal{C}}_3=0$.
Hence, only three independent terms remain, in agreement with the result obtained from the Hilbert series.
Moreover, in the
holomorphic case they consider all pairwise tensor products between
modular forms of weights $k$ and $k'$, i.e.\ they work in the direct sum
$\bigoplus_{k,k'} \mathcal{M}_k \otimes \mathcal{M}_{k'}$. The natural
multiplication map
\begin{equation}
\mu_{k,k'}:\; \mathcal{M}_k \otimes \mathcal{M}_{k'} \longrightarrow
\mathcal{M}_{k+k'}
\end{equation}
is generically not injective, and typically
$\dim(\mathcal{M}_k)\,\dim(\mathcal{M}_{k'}) >
\dim(\mathcal{M}_{k+k'})$. Consequently, many product tensors lie in
$\ker \mu_{k,k'}$ and can be rewritten as linear combinations of a basis
of $\mathcal{M}_{k+k'}$ (i.e., modular forms of weight $k+k'$). Keeping both the original products and those
linear combinations leads to an overcomplete set and thus introduces
redundancy.

\subsection{Dimension-6 case}
\label{sec:dim-6}
\begin{comment}
In the classification of dimension-6 SMEFT operators~\cite{Grzadkowski:2010es},
we organize the operators according to the number of fermion fields they
contain, and present the corresponding results for each class. All modular
forms with weight $k>2$ that are required in our analysis are listed in
Appendix~\ref{app:forms}.
\end{comment}
In the classification of dimension-6 SMEFT operators in Ref.~\cite{Grzadkowski:2010es}, we organize the operators according to the number of fermion fields they contain and present the corresponding results for each class.
All modular forms with weight $k>2$ required in our analysis are listed in Appendix~\ref{app:forms}.

\begin{itemize}
    \item \textbf{Bilinear structures}

    \begin{enumerate}
\item $\overline{L}R + \text{h.c.}$:
\begin{comment}
There are three independent flavor structures of this type, which we
denote by $(\overline{L}e)_{\mathbf{1}}$, $(\overline{Q}u)_{\mathbf{1}}$
and $(\overline{Q}d)_{\mathbf{1}}$. Here, “type’’ labels the contraction
of left- and right-handed fermions into an $A_4$ singlet. The second
column of Table~\ref{tab:LR_bilinears} shows the schematic field content
of each operator, while the third column gives the modular weight assigned to
the corresponding modular form. The subscripts “ed” and “qd” indicate
that the modular form belongs to the lepton ($A_{4e}$) or quark ($A_{4q}$)
sector, respectively. The last column displays the generic $A_4$-invariant
completion, written in terms of modular forms and the corresponding fields constructions. 
In the following we show explicitly how the Hilbert series result
for each flavor structure is matched to these $A_4$-invariant operators.
\end{comment}
There are three independent flavor structures of this type, which we denote by $(\overline{L}e)_{\mathbf{1}}$, $(\overline{Q}u)_{\mathbf{1}}$ and $(\overline{Q}d)_{\mathbf{1}}$.
Here, ``type'' labels the contraction of left- and right-handed fermions into an $A_4$ singlet.
The second column of Table~\ref{tab:LR_bilinears} shows the schematic field content of each operator, while the third column gives the modular weight assigned to the corresponding modular form.
The subscripts ``ed'' and ``qd'' indicate that the modular form belongs to the lepton ($A_{4e}$) or quark ($A_{4q}$) sector, respectively.
The last column displays the generic $A_4$-invariant completion, written in terms of modular forms and the corresponding field contractions.
In the following, we show explicitly how the Hilbert series result for each flavor structure is matched to these $A_4$-invariant operators.

\begin{table}[H]
\begin{center}
    \begin{tabular}{|l|l|l|l|}
        \hline
        type &  & weight & $A_4$ invariant \\ \hline
        $(\overline{L}e)_{\mathbf{1}}$ &
        $\overline{L}_L\,\Gamma\,E_R$ &
        $2_{ed}$ &
        {${Y^{(2)}_{\mathbf{r}'_e}}^{*}\,[\overline{L}_L\,\Gamma\,E_R]$} \\ \hline
        $(\overline{Q}u)_{\mathbf{1}}$ &
        $\overline{Q}_L\,\Gamma\,U_R$ &
        $6_{qd}$ &
        {${Y^{(6)}_{\mathbf{r}'_q}}^{*}\,[\overline{Q}_L\,\Gamma\,U_R]$} \\ \hline
        $(\overline{Q}d)_{\mathbf{1}}$ &
        $\overline{Q}_L\,\Gamma\,D_R$ &
        $2_{qd}$ &
        {${Y^{(2)}_{\mathbf{r}'_q}}^{*}\,[\overline{Q}_L\,\Gamma\,D_R]$} \\ \hline
    \end{tabular}
    \caption{Bilinear SMEFT operators of the form $\overline{L}R$,
    together with their modular weights and schematic $A_4$-invariant
    completions.}
    \label{tab:LR_bilinears}
\end{center}
\end{table}

\begin{enumerate}
    \item $(\overline{\ell}e)_\mathbf{1}+\text{h.c.}$:
    For this flavor structure, the Hilbert series yields the following combination of monomials in the lepton sector:
    \begin{equation}
        \begin{aligned}
        & e \, h^2 \, {h^\dagger} \, {\ell^\dagger} \, {Y^{(2)}_{\mathbf{3}}}^*(\tau_e)
        + B \, e \, h \, {\ell^\dagger} \, {Y^{(2)}_{\mathbf{3}}}^*(\tau_e)
        + e \, h \, {\ell^\dagger} \, W \, {Y^{(2)}_{\mathbf{3}}}^*(\tau_e) \\
        & + h^2 \, {h^\dagger} \, {\ell^\dagger} \, \mu \, {Y^{(2)}_{\mathbf{3}}}^*(\tau_e)
        + B \, h \, {\ell^\dagger} \, \mu \, {Y^{(2)}_{\mathbf{3}}}^*(\tau_e)
        + h \, {\ell^\dagger} \, \mu \, W \, {Y^{(2)}_{\mathbf{3}}}^*(\tau_e) \\
        & + h^2 \, {h^\dagger} \, {\ell^\dagger} \, \tau \, {Y^{(2)}_{\mathbf{3}}}^*(\tau_e)
        + B \, h \, {\ell^\dagger} \, \tau \, {Y^{(2)}_{\mathbf{3}}}^*(\tau_e)
        + h \, {\ell^\dagger} \, \tau \, W \, {Y^{(2)}_{\mathbf{3}}}^*(\tau_e)
        + \text{h.c.}\; .
        \end{aligned}
    \end{equation}
    Grouping these monomials by their gauge and Higgs content, one identifies
    the three SMEFT operators $\mathcal{O}_{eH}$, $\mathcal{O}_{eW}$ and
    $\mathcal{O}_{eB}$. Since these operators only involve leptons, they are trivial
    singlets under $A_{4q}$ by construction and must be promoted to invariants
    of the full $A_{4e}\times A_{4q}$ flavor symmetry by an appropriate
    contraction in the $A_{4e}$ sector.

In our setup $L_L$ and $\overline{L}_L$ transform as an $A_{4e}$ triplet, while $e_R$, $\mu_R$ and $\tau_R$ are $A_{4e}$ singlets.
Using the $A_4$ tensor-product rules collected in Appendix~\ref{app:tensor_product}, we can combine the triplet modular form $Y^{(2)}_{\mathbf{3}}(\tau_e)$ with $\overline{L}_L$ into the three inequivalent $A_{4e}$ singlets $\mathbf{1}$, $\mathbf{1'}$ and $\mathbf{1''}$:
\begin{align}
\left({Y^{(2)}_{\mathbf{3}}}^*(\tau_e)\,\overline{L}_L\right)_{\mathbf{1}}\Gamma e_R 
+ \text{h.c.}\;, \label{eq:Le-type-1}\\
\left({Y^{(2)}_{\mathbf{3}}}^*(\tau_e)\,\overline{L}_L\right)_{\mathbf{1''}}\Gamma \mu_R 
+ \text{h.c.}\;, \label{eq:Le-type-1pp}\\
\left({Y^{(2)}_{\mathbf{3}}}^*(\tau_e)\,\overline{L}_L\right)_{\mathbf{1'}}\Gamma \tau_R 
+ \text{h.c.}\;. \label{eq:Le-type-1p}
\end{align}
Each of these structures multiplies the operators $\mathcal{O}_{eH}$, $\mathcal{O}_{eW}$ and $\mathcal{O}_{eB}$, thereby reproducing the three $A_{4e}$-invariant combinations counted by the Hilbert series.

    \item $(\overline{q}u)_\mathbf{1}+\text{h.c.}$:
    For this type, the Hilbert series in the up-quark sector gives
    \begin{equation}
        \begin{aligned}
        & B \, {h^\dagger} \, {q^\dagger} \, u \, {Y^{(6)}_{\mathbf{3}a}}^*(\tau_q)
        + G \, {h^\dagger} \, {q^\dagger} \, u \, {Y^{(6)}_{\mathbf{3}a}}^*(\tau_q)
        + {h^\dagger} \, {q^\dagger} \, u \, W \, {Y^{(6)}_{\mathbf{3}a}}^*(\tau_q) \\
        & + B \, {h^\dagger} \, {q^\dagger} \, u \, {Y^{(6)}_{\mathbf{3}b}}^*(\tau_q)
        + G \, {h^\dagger} \, {q^\dagger} \, u \, {Y^{(6)}_{\mathbf{3}b}}^*(\tau_q)
        + {h^\dagger} \, {q^\dagger} \, u \, W \, {Y^{(6)}_{\mathbf{3}b}}^*(\tau_q) \\
        & + h \, {h^\dagger}^2 \, {q^\dagger} \, u \, {Y^{(6)}_{\mathbf{3}a}}^*(\tau_q)
        + h \, {h^\dagger}^2 \, {q^\dagger} \, u \, {Y^{(6)}_{\mathbf{3}b}}^*(\tau_q) \\
        & + B \, c \, {h^\dagger} \, {q^\dagger} \, {Y^{(6)}_{\mathbf{3}a}}^*(\tau_q)
        + c \, G \, {h^\dagger} \, {q^\dagger} \, {Y^{(6)}_{\mathbf{3}a}}^*(\tau_q)
        + c \, {h^\dagger} \, {q^\dagger} \, W \, {Y^{(6)}_{\mathbf{3}a}}^*(\tau_q) \\
        & + B \, c \, {h^\dagger} \, {q^\dagger} \, {Y^{(6)}_{\mathbf{3}b}}^*(\tau_q)
        + c \, G \, {h^\dagger} \, {q^\dagger} \, {Y^{(6)}_{\mathbf{3}b}}^*(\tau_q)
        + c \, {h^\dagger} \, {q^\dagger} \, W \, {Y^{(6)}_{\mathbf{3}b}}^*(\tau_q) \\
        & + c \, h \, {h^\dagger}^2 \, {q^\dagger} \, {Y^{(6)}_{\mathbf{3}a}}^*(\tau_q)
        + c \, h \, {h^\dagger}^2 \, {q^\dagger} \, {Y^{(6)}_{\mathbf{3}b}}^*(\tau_q) \\
        & + B \, {h^\dagger} \, {q^\dagger} \, t \, {Y^{(6)}_{\mathbf{3}a}}^*(\tau_q)
        + G \, {h^\dagger} \, {q^\dagger} \, t \, {Y^{(6)}_{\mathbf{3}a}}^*(\tau_q)
        + {h^\dagger} \, {q^\dagger} \, t \, W \, {Y^{(6)}_{\mathbf{3}a}}^*(\tau_q) \\
        & + B \, {h^\dagger} \, {q^\dagger} \, t \, {Y^{(6)}_{\mathbf{3}b}}^*(\tau_q)
        + G \, {h^\dagger} \, {q^\dagger} \, t \, {Y^{(6)}_{\mathbf{3}b}}^*(\tau_q)
        + {h^\dagger} \, {q^\dagger} \, t \, W \, {Y^{(6)}_{\mathbf{3}b}}^*(\tau_q) \\
        & + h \, {h^\dagger}^2 \, {q^\dagger} \, t \, {Y^{(6)}_{\mathbf{3}a}}^*(\tau_q)
        + h \, {h^\dagger}^2 \, {q^\dagger} \, t \, {Y^{(6)}_{\mathbf{3}b}}^*(\tau_q)
        + \mathrm{h.c.} \;.
        \end{aligned}
    \end{equation}
These monomials can be organized into the SMEFT operators $\mathcal{O}_{uH}$, $\mathcal{O}_{uG}$, $\mathcal{O}_{uW}$ and $\mathcal{O}_{uB}$.
Since they involve only quark fields, they are singlets under $A_{4e}$, and the non-trivial flavor structure resides in the $A_{4q}$ sector.

Here, $Q_L$ and $\overline{Q}_L$ transform as $A_{4q}$ triplets, while $u_R$, $c_R$ and $t_R$ are $A_{4q}$ singlets.
The Hilbert series tells us that the modular forms entering this sector have weight $6$ and transform as two independent triplets $Y^{(6)}_{\mathbf{3}a}(\tau_q)$ and $Y^{(6)}_{\mathbf{3}b}(\tau_q)$.
Using the $A_4$ tensor products, we can construct three $A_{4q}$ singlets:
\begin{equation}
\begin{aligned}
& \left({Y^{(6)}_{\mathbf{3}a,\mathbf{3}b}}^*(\tau_q)\,\overline{Q}_L\right)_{\mathbf{1}}\Gamma u_R 
+ \text{h.c.}\;, \\
& \left({Y^{(6)}_{\mathbf{3}a,\mathbf{3}b}}^*(\tau_q)\,\overline{Q}_L\right)_{\mathbf{1''}}\Gamma c_R 
+ \text{h.c.}\;, \\
& \left({Y^{(6)}_{\mathbf{3}a,\mathbf{3}b}}^*(\tau_q)\,\overline{Q}_L\right)_{\mathbf{1'}}\Gamma t_R 
+ \text{h.c.}\;.
\end{aligned}
\end{equation}
Dressing $\mathcal{O}_{uH}$, $\mathcal{O}_{uG}$, $\mathcal{O}_{uW}$ and $\mathcal{O}_{uB}$ with these singlet combinations yields precisely the $A_{4q}$-invariant operators counted by the Hilbert series.

    \item $(\overline{q}d)_\mathbf{1}+\text{h.c.}$:
    For the down-quark case, the Hilbert series result reads
    \begin{equation}
        \begin{aligned}
        & d \, h^2 \, {h^\dagger} \, {q^\dagger} \, {Y^{(2)}_{\mathbf{3}}}^*(\tau_q)
        + B \, d \, h \, {q^\dagger} \, {Y^{(2)}_{\mathbf{3}}}^*(\tau_q)
        + d \, G \, h \, {q^\dagger} \, {Y^{(2)}_{\mathbf{3}}}^*(\tau_q)
        + d \, h \, {q^\dagger} \, W \, {Y^{(2)}_{\mathbf{3}}}^*(\tau_q) \\
        & + h^2 \, {h^\dagger}  \, {q^\dagger} \, s \, {Y^{(2)}_{\mathbf{3}}}^*(\tau_q)
        + B \, h \, {q^\dagger} \, s \, {Y^{(2)}_{\mathbf{3}}}^*(\tau_q)
        + G \, h \, {q^\dagger} \, s \, {Y^{(2)}_{\mathbf{3}}}^*(\tau_q)
        + h \, {q^\dagger} \, s \, W \, {Y^{(2)}_{\mathbf{3}}}^*(\tau_q) \\
        & + b \, h^2 \, {h^\dagger}  \, {q^\dagger} \, {Y^{(2)}_{\mathbf{3}}}^*(\tau_q)
        + b \, B \, h \, {q^\dagger} \, {Y^{(2)}_{\mathbf{3}}}^*(\tau_q)
        + b \, G \, h \, {q^\dagger} \, {Y^{(2)}_{\mathbf{3}}}^*(\tau_q)
        + b \, h \, {q^\dagger} \, W \, {Y^{(2)}_{\mathbf{3}}}^*(\tau_q)
        + \text{h.c.}\;.
        \end{aligned}
    \end{equation}
These monomials correspond to the SMEFT operators $\mathcal{O}_{dH}$, $\mathcal{O}_{dG}$, $\mathcal{O}_{dW}$ and $\mathcal{O}_{dB}$.
As in the up-sector case, the operators are singlets under $A_{4e}$, and the non-trivial flavor structure is in the $A_{4q}$ sector.

Using again that $Q_L$ and $\overline{Q}_L$ are $A_{4q}$ triplets, while $d_R$, $s_R$ and $b_R$ are $A_{4q}$ singlets, and that the relevant modular forms in this sector have weight $2$ and transform as a triplet $Y^{(2)}_{\mathbf{3}}(\tau_q)$, we can construct the three $A_{4q}$ singlet combinations
\begin{equation}
\begin{aligned}
& \left({Y^{(2)}_{\mathbf{3}}}^*(\tau_q)\,\overline{Q}_L\right)_{\mathbf{1}} \Gamma d_R 
+ \text{h.c.}\;, \\
& \left({Y^{(2)}_{\mathbf{3}}}^*(\tau_q)\,\overline{Q}_L\right)_{\mathbf{1''}} \Gamma s_R 
+ \text{h.c.}\;, \\
& \left({Y^{(2)}_{\mathbf{3}}}^*(\tau_q)\,\overline{Q}_L\right)_{\mathbf{1'}} \Gamma b_R 
+ \text{h.c.}\;.
\end{aligned}
\end{equation}
Dressing $\mathcal{O}_{dH}$, $\mathcal{O}_{dG}$, $\mathcal{O}_{dW}$ and $\mathcal{O}_{dB}$ with these singlet structures gives the three $A_{4q}$-invariant operators, in one-to-one correspondence with the multiplicities obtained from the Hilbert series.
\end{enumerate}

\item $\overline{L}L$:
There are two classes of operators of this type, distinguished by whether
the associated modular invariance resides in the lepton or quark sector:
\begin{itemize}
    \item Operators invariant under $A_{4e}$:
    \begin{equation}
        (\overline{\ell}\,\ell)_{\mathbf{1}} \,,
    \end{equation}
    \item Operator invariant under $A_{4q}$:
    \begin{equation}
        (\overline{q}\,q)_{\mathbf{1}} \,.
    \end{equation}
\end{itemize}

\begin{center}
    \begin{tabular}{|l|l|l|l|}
        \hline
        type &  & weight & $A_4$ invariant \\ \hline
        $(\overline{\ell}\,\ell)_\mathbf{1}$ &
        $\overline{L}_L\,\Gamma\,L_L$ &
        $2_q + 2_{qd}$ &
        $\bigl({Y^{(2)}_{\mathbf{r}_e'}}^{*} Y^{(2)}_{\mathbf{r}_e}\bigr)\,[\overline{L}_L\,\Gamma\,L_L]$ \\ \hline
        $(\overline{q}q)_\mathbf{1}$ &
        $\overline{Q}_L\,\Gamma\,Q_L$ &
        $2_q + 2_{qd}$ &
        $\bigl({Y^{(2)}_{\mathbf{r}_q'}}^{*} Y^{(2)}_{\mathbf{r}_q}\bigr)\,[\overline{Q}_L\,\Gamma\,Q_L]$ \\ \hline
    \end{tabular}
\end{center}

\begin{enumerate}
    \item $(\overline{\ell}\,\ell)_\mathbf{1}$:
    For this flavor structure the Hilbert series yields
    \begin{equation}
        14 \, h \, h^\dagger \, \ell \, \ell^\dagger \, D \,
        Y^{(2)}_{\mathbf{3}}(\tau_e) \, {Y^{(2)}_{\mathbf{3}}}^{*}(\tau_e)\,.
    \end{equation}
    Thus there are 14 independent invariants with the schematic field content
    $\overline{L}_L \Gamma L_L$ dressed by the modular forms
    $Y^{(2)}_{\mathbf{3}}(\tau_e)\,{Y^{(2)}_{\mathbf{3}}}^{*}(\tau_e)$.
    The corresponding SMEFT operators are $\mathcal{O}_{Hl}^{(1)}$ and
    $\mathcal{O}_{Hl}^{(3)}$, so each of them is accompanied by 7 distinct
    $A_{4e}$-invariant flavor structures.

    Using the $A_4$ tensor products in Appendix~\ref{app:tensor_product},
    the product of two triplet modular forms decomposes as
    $Y^{(2)}_{\mathbf{3}} \otimes {Y^{(2)}_{\mathbf{3}}}^{*}
     \supset \mathbf{1} \oplus \mathbf{1'} \oplus \mathbf{1''} \oplus
     \mathbf{3}_s \oplus \mathbf{3}_a$,
    and similarly for the bilinear $\overline{L}_L \Gamma L_L$.
    The seven independent $A_{4e}$ singlets can then be written as
    \begin{equation}
        \begin{aligned}
        & \bigl({Y^{(2)}_{\mathbf{3}}}^{*}(\tau_e)\,Y^{(2)}_{\mathbf{3}}(\tau_e)\bigr)_{\mathbf{1}}
          \,\bigl(\overline{L}_L \Gamma L_L\bigr)_{\mathbf{1}}, \\
        & \bigl({Y^{(2)}_{\mathbf{3}}}^{*}(\tau_e)\,Y^{(2)}_{\mathbf{3}}(\tau_e)\bigr)_{\mathbf{1'}}
          \,\bigl(\overline{L}_L \Gamma L_L\bigr)_{\mathbf{1''}}, \\
        & \bigl({Y^{(2)}_{\mathbf{3}}}^{*}(\tau_e)\,Y^{(2)}_{\mathbf{3}}(\tau_e)\bigr)_{\mathbf{1''}}
          \,\bigl(\overline{L}_L \Gamma L_L\bigr)_{\mathbf{1'}}, \\
        & \bigl[\bigl({Y^{(2)}_{\mathbf{3}}}^{*}(\tau_e)\,Y^{(2)}_{\mathbf{3}}(\tau_e)\bigr)_{\mathbf{3}_s}
          \,\bigl(\overline{L}_L \Gamma L_L\bigr)_{\mathbf{3}_s}\bigr]_{\mathbf{1}}, \\
        & \bigl[\bigl({Y^{(2)}_{\mathbf{3}}}^{*}(\tau_e)\,Y^{(2)}_{\mathbf{3}}(\tau_e)\bigr)_{\mathbf{3}_s}
          \,\bigl(\overline{L}_L \Gamma L_L\bigr)_{\mathbf{3}_a}\bigr]_{\mathbf{1}}, \\
        & \bigl[\bigl({Y^{(2)}_{\mathbf{3}}}^{*}(\tau_e)\,Y^{(2)}_{\mathbf{3}}(\tau_e)\bigr)_{\mathbf{3}_a}
          \,\bigl(\overline{L}_L \Gamma L_L\bigr)_{\mathbf{3}_s}\bigr]_{\mathbf{1}}, \\
        & \bigl[\bigl({Y^{(2)}_{\mathbf{3}}}^{*}(\tau_e)\,Y^{(2)}_{\mathbf{3}}(\tau_e)\bigr)_{\mathbf{3}_a}
          \,\bigl(\overline{L}_L \Gamma L_L\bigr)_{\mathbf{3}_a}\bigr]_{\mathbf{1}}.
        \end{aligned}
    \end{equation}
    Here, the labels $\mathbf{3}_s$ and $\mathbf{3}_a$ denote, respectively,
    the symmetric and antisymmetric triplet components in the tensor product
    $\mathbf{3}\otimes\mathbf{3}$.

    \item $(\overline{q}q)_\mathbf{1}$:
    The analysis in the quark sector proceeds in complete analogy with the
    lepton case. The Hilbert series now gives
    \begin{equation}
        14 \, h \, h^\dagger \, D \, q \, q^\dagger \,
        Y^{(2)}_{\mathbf{3}}(\tau_q)\, {Y^{(2)}_{\mathbf{3}}}^{*}(\tau_q)\,,
    \end{equation}
    corresponding to the operators $\mathcal{O}_{Hq}^{(1)}$ and
    $\mathcal{O}_{Hq}^{(3)}$. Since these operators involve only quark
    fields, they are singlets under $A_{4e}$ and must be completed to
    $A_{4q}$ invariants.

    Using that $Q_L$ transforms as an $A_{4q}$ triplet and following the same
    tensor-product decomposition as above, we obtain the seven
    $A_{4q}$ singlet structures:
    \begin{equation}
        \begin{aligned}
        & \bigl({Y^{(2)}_{\mathbf{3}}}^{*}(\tau_q)\,Y^{(2)}_{\mathbf{3}}(\tau_q)\bigr)_{\mathbf{1}}
          \,\bigl(\overline{Q}_L \Gamma Q_L\bigr)_{\mathbf{1}}, \\
        & \bigl({Y^{(2)}_{\mathbf{3}}}^{*}(\tau_q)\,Y^{(2)}_{\mathbf{3}}(\tau_q)\bigr)_{\mathbf{1'}}
          \,\bigl(\overline{Q}_L \Gamma Q_L\bigr)_{\mathbf{1''}}, \\
        & \bigl({Y^{(2)}_{\mathbf{3}}}^{*}(\tau_q)\,Y^{(2)}_{\mathbf{3}}(\tau_q)\bigr)_{\mathbf{1''}}
          \,\bigl(\overline{Q}_L \Gamma Q_L\bigr)_{\mathbf{1'}}, \\
        & \bigl[\bigl({Y^{(2)}_{\mathbf{3}}}^{*}(\tau_q)\,Y^{(2)}_{\mathbf{3}}(\tau_q)\bigr)_{\mathbf{3}_s}
          \,\bigl(\overline{Q}_L \Gamma Q_L\bigr)_{\mathbf{3}_s}\bigr]_{\mathbf{1}}, \\
        & \bigl[\bigl({Y^{(2)}_{\mathbf{3}}}^{*}(\tau_q)\,Y^{(2)}_{\mathbf{3}}(\tau_q)\bigr)_{\mathbf{3}_s}
          \,\bigl(\overline{Q}_L \Gamma Q_L\bigr)_{\mathbf{3}_a}\bigr]_{\mathbf{1}}, \\
        & \bigl[\bigl({Y^{(2)}_{\mathbf{3}}}^{*}(\tau_q)\,Y^{(2)}_{\mathbf{3}}(\tau_q)\bigr)_{\mathbf{3}_a}
          \,\bigl(\overline{Q}_L \Gamma Q_L\bigr)_{\mathbf{3}_s}\bigr]_{\mathbf{1}}, \\
        & \bigl[\bigl({Y^{(2)}_{\mathbf{3}}}^{*}(\tau_q)\,Y^{(2)}_{\mathbf{3}}(\tau_q)\bigr)_{\mathbf{3}_a}
          \,\bigl(\overline{Q}_L \Gamma Q_L\bigr)_{\mathbf{3}_a}\bigr]_{\mathbf{1}}.
        \end{aligned}
    \end{equation}
    These seven structures for each of $\mathcal{O}_{Hq}^{(1)}$ and
    $\mathcal{O}_{Hq}^{(3)}$ reproduce the total of 14
    $A_{4q}$-invariant operators predicted by the Hilbert series.
\end{enumerate}

    \item $\overline{R}R $: There are three types of operators, classified according to their associated modular symmetry invariance as follows:
    \begin{itemize}
        \item Operators invariant under $A_{4e}$:
        \begin{equation}
        (\overline{e}e)_\mathbf{1}.
        \end{equation}
    
        \item Operators invariant under $A_{4q}$:
        \begin{equation}
        \begin{aligned}
        & (\overline{u}u)_\mathbf{1}, \\
        & (\overline{d}d)_\mathbf{1}.
        \end{aligned}
        \end{equation}
    \end{itemize}    
    \begin{center}
        \begin{tabular}{|l|l|l|l|}
            \hline
            type &  & weight& $A_4$ invariant  \\ \hline
            $(\overline{e}e)_\mathbf{1}$ & $\overline{E}_R\Gamma E_R$ & $0$& $\overline{E}_R\Gamma E_R$  \\ \hline
            $(\overline{u}u)_\mathbf{1} $ & $\overline{U}_R\Gamma U_R$ & $4_q+4_{qd}$& $({Y^{(4)}_\mathbf{rq}}{Y^{(4)}_\mathbf{rq'}}^*)[\overline{U}_R\Gamma U_R]$  \\ \hline
            $(\overline{d}d)_\mathbf{1}$ & $\overline{D}_R\Gamma D_R$ & $0$& $\overline{D}_R\Gamma D_R$  \\ \hline
        \end{tabular}
    \end{center}
\begin{enumerate}
    \item $(\overline{e}e)_\mathbf{1}$: For this type, the result given by the Hilbert series is
    \begin{equation}
        e \, e^{\dagger} \, h \, {h^\dagger}  \, D 
        + h \, {h^\dagger}  \, \mu \, {\mu^\dagger}  \, D
        + h \, {h^\dagger}  \, D \, \tau \, {\tau^\dagger} .
        \end{equation}                           
    The corresponding SMEFT operators is $\mathcal{O}_{He}$. Because all field components are $A_{4e}$ singlets with weight $0$, we can easily find these structures as:
    \begin{equation}
        \begin{aligned}
        & \overline{e}_R \Gamma e_R, \\
        & \overline{\mu}_R \Gamma \mu_R, \\
        & \overline{\tau}_R \Gamma \tau_R.
        \end{aligned}
    \end{equation}
    \item $(\overline{u}u)_\mathbf{1} $: For this type, the result given by the Hilbert series is
    \begin{equation}
        \begin{aligned}
        & h \, {h^\dagger}  \, D \, u \, {u^\dagger}  \, Y^{(4)}_\mathbf{1}(\tau_q) \, {Y^{(4)}_\mathbf{1}}^*(\tau_q)
        + h \, {h^\dagger}  \, D \, u \, {u^\dagger}  \, Y^{(4)}_{\mathbf{3}}(\tau_q) \, {Y^{(4)}_{\mathbf{3}}}^*(\tau_q) \\
        & + c \, h \, {h^\dagger}  \, D \, {u^\dagger}  \, Y^{(4)}_{\mathbf{1'}}(\tau_q) \, {Y^{(4)}_{\mathbf{1'}}}^*(\tau_q)
        + c \, h \, {h^\dagger}  \, D \, {u^\dagger}  \, Y^{(4)}_{\mathbf{3}}(\tau_q) \, {Y^{(4)}_{\mathbf{3}}}^*(\tau_q) \\
        & + h \, {h^\dagger}  \, D \, t \, {u^\dagger}  \, Y^{(4)}_{\mathbf{1'}}(\tau_q) \, {Y^{(4)}_\mathbf{1}}^*(\tau_q)
        + h \, {h^\dagger}  \, D \, t \, {u^\dagger}  \, Y^{(4)}_\mathbf{1}(\tau_q) \, {Y^{(4)}_{\mathbf{1'}}}^*(\tau_q) \\
        & + h \, {h^\dagger}  \, D \, t \, {u^\dagger}  \, Y^{(4)}_{\mathbf{3}}(\tau_q) \, {Y^{(4)}_{\mathbf{3}}}^*(\tau_q) \\
        & + {c^\dagger}  \, h \, {h^\dagger}  \, D \, u \, Y^{(4)}_{\mathbf{1'}}(\tau_q) \, {Y^{(4)}_\mathbf{1}}^*(\tau_q)
        + {c^\dagger} \, h \, {h^\dagger}  \, D \, u \, Y^{(4)}_\mathbf{1}(\tau_q) \, {Y^{(4)}_{\mathbf{1'}}}^*(\tau_q) \\
        & + {c^\dagger} \, h \, {h^\dagger}  \, D \, u \, Y^{(4)}_{\mathbf{3}}(\tau_q) \, {Y^{(4)}_{\mathbf{3}}}^*(\tau_q) \\
        & + c \, {c^\dagger} \, h \, {h^\dagger}  \, D \, Y^{(4)}_\mathbf{1}(\tau_q) \, {Y^{(4)}_\mathbf{1}}^*(\tau_q)
        + c \, {c^\dagger} \, h \, {h^\dagger}  \, D \, Y^{(4)}_{\mathbf{3}}(\tau_q) \, {Y^{(4)}_{\mathbf{3}}}^*(\tau_q) \\
        & + {c^\dagger} \, h \, {h^\dagger}  \, D \, t \, Y^{(4)}_{\mathbf{1'}}(\tau_q) \, {Y^{(4)}_{\mathbf{1'}}}^*(\tau_q)
        + {c^\dagger} \, h \, {h^\dagger}  \, D \, t \, Y^{(4)}_{\mathbf{3}}(\tau_q) \, {Y^{(4)}_{\mathbf{3}}}^*(\tau_q) \\
        & + h \, {h^\dagger}  \, D \, {t^\dagger} \, u \, Y^{(4)}_{\mathbf{1'}}(\tau_q) \, {Y^{(4)}_{\mathbf{1'}}}^*(\tau_q)
        + h \, {h^\dagger}  \, D \, {t^\dagger} \, u \, Y^{(4)}_{\mathbf{3}}(\tau_q) \, {Y^{(4)}_{\mathbf{3}}}^*(\tau_q) \\
        & + c \, h \, {h^\dagger}  \, D \, {t^\dagger} \, Y^{(4)}_{\mathbf{1'}}(\tau_q) \, {Y^{(4)}_\mathbf{1}}^*(\tau_q)
        + c \, h \, {h^\dagger}  \, D \, {t^\dagger} \, Y^{(4)}_\mathbf{1}(\tau_q) \, {Y^{(4)}_{\mathbf{1'}}}^*(\tau_q) \\
        & + c \, h \, {h^\dagger}  \, D \, {t^\dagger} \, Y^{(4)}_{\mathbf{3}}(\tau_q) \, {Y^{(4)}_{\mathbf{3}}}^*(\tau_q) \\
        & + h \, {h^\dagger}  \, D \, t \, {t^\dagger} \, Y^{(4)}_\mathbf{1}(\tau_q) \, {Y^{(4)}_\mathbf{1}}^*(\tau_q)
        + h \, {h^\dagger}  \, D \, t \, {t^\dagger} \, Y^{(4)}_{\mathbf{3}}(\tau_q) \, {Y^{(4)}_{\mathbf{3}}}^*(\tau_q)
        \end{aligned}
        \end{equation}        
    %The corresponding SMEFT operators is $\mathcal{O}_{Hu}$. The $U_R$ and $\overline{U}_R$ are $A_{4q}$ singlets with weight 4, we should consider the effect of modular forms with weight 4. More specifically, we can write them as:
    The corresponding SMEFT operator is $\mathcal{O}_{Hu}$. The $U_R$ and $\overline{U}_R$ are $A_{4q}$ singlets with weight $4$, so we must include modular forms with weight $4$. More specifically, we can write them as:
    \begin{equation}
        \begin{aligned}
        & \left({Y^{(4)}_{\mathbf{1}}}(\tau_q) {Y^{(4)}_{\mathbf{1}}}^*(\tau_q)\right)_{\mathbf{1}} \left(\overline{u}_R \Gamma u_R\right)_{\mathbf{1}}, \left({Y^{(4)}_{\mathbf{3}}}(\tau_q)  {Y^{(4)}_{\mathbf{3}}}^*(\tau_q)\right)_{\mathbf{1}} \left(\overline{u}_R \Gamma u_R\right)_{\mathbf{1}}, \\
        & \left({Y^{(4)}_{\mathbf{1'}}}(\tau_q) {Y^{(4)}_{\mathbf{1'}}}^*(\tau_q)\right)_{\mathbf{1''}} \left(\overline{u}_R \Gamma c_R\right)_{\mathbf{1'}}, \left({Y^{(4)}_{\mathbf{3}}}(\tau_q)   {Y^{(4)}_{\mathbf{3}}}^*(\tau_q)\right)_{\mathbf{1''}} \left(\overline{u}_R \Gamma c_R\right)_{\mathbf{1'}}, \\
        & \left({Y^{(4)}_{\mathbf{1}}}(\tau_q) {Y^{(4)}_{\mathbf{1'}}}^*(\tau_q)\right)_{\mathbf{1'}} \left(\overline{u}_R \Gamma t_R\right)_{\mathbf{1''}},  \left({Y^{(4)}_{\mathbf{1'}}}(\tau_q) {Y^{(4)}_{\mathbf{1}}}^*(\tau_q)\right)_{\mathbf{1'}} \left(\overline{u}_R \Gamma t_R\right)_{\mathbf{1''}}, \\
        & \left({Y^{(4)}_{\mathbf{3}}}(\tau_q)  {Y^{(4)}_{\mathbf{3}}}^*(\tau_q)\right)_{\mathbf{1'}} \left(\overline{u}_R \Gamma t_R\right)_{\mathbf{1''}}.
        \end{aligned}
        \end{equation}
        \begin{equation}
            \begin{aligned}
            & \left({Y^{(4)}_{\mathbf{1}}}(\tau_q) {Y^{(4)}_{\mathbf{1}}}^*(\tau_q)\right)_{\mathbf{1}} \left(\overline{c}_R \Gamma c_R\right)_{\mathbf{1}}, \left({Y^{(4)}_{\mathbf{3}}}(\tau_q)  {Y^{(4)}_{\mathbf{3}}}^*(\tau_q)\right)_{\mathbf{1}} \left(\overline{c}_R \Gamma c_R\right)_{\mathbf{1}}, \\
            & \left({Y^{(4)}_{\mathbf{1'}}}(\tau_q) {Y^{(4)}_{\mathbf{1'}}}^*(\tau_q)\right)_{\mathbf{1''}} \left(\overline{c}_R \Gamma t_R\right)_{\mathbf{1'}}, \left({Y^{(4)}_{\mathbf{3}}}(\tau_q)  {Y^{(4)}_{\mathbf{3}}}^*(\tau_q)\right)_{\mathbf{1''}} \left(\overline{c}_R \Gamma t_R\right)_{\mathbf{1'}}, \\
            & \left({Y^{(4)}_{\mathbf{1}}}(\tau_q) {Y^{(4)}_{\mathbf{1'}}}^*(\tau_q)\right)_{\mathbf{1'}} \left(\overline{c}_R \Gamma u_R\right)_{\mathbf{1''}}, \left({Y^{(4)}_{\mathbf{1'}}}(\tau_q) {Y^{(4)}_{\mathbf{1}}}^*(\tau_q)\right)_{\mathbf{1'}} \left(\overline{c}_R \Gamma u_R\right)_{\mathbf{1''}},\\
            & \left({Y^{(4)}_{\mathbf{3}}}(\tau_q)  {Y^{(4)}_{\mathbf{3}}}^*(\tau_q)\right)_{\mathbf{1'}} \left(\overline{c}_R \Gamma u_R\right)_{\mathbf{1''}}.
            \end{aligned}
        \end{equation}
    And
    \begin{equation}
        \begin{aligned}
                & \left({Y^{(4)}_{\mathbf{1}}}(\tau_q) {Y^{(4)}_{\mathbf{1}}}^*(\tau_q)\right)_{\mathbf{1}} \left(\overline{t}_R \Gamma t_R\right)_{\mathbf{1}}, \left({Y^{(4)}_{\mathbf{1}}}(\tau_q) {Y^{(4)}_{\mathbf{1}}}^*(\tau_q)\right)_{\mathbf{1}} \left(\overline{t}_R \Gamma t_R\right)_{\mathbf{1}}, \\
                & \left({Y^{(4)}_{\mathbf{1'}}}(\tau_q) {Y^{(4)}_{\mathbf{1'}}}^*(\tau_q)\right)_{\mathbf{1''}} \left(\overline{t}_R \Gamma u_R\right)_{\mathbf{1'}}, \left({Y^{(4)}_{\mathbf{1'}}}(\tau_q) {Y^{(4)}_{\mathbf{1'}}}^*(\tau_q)\right)_{\mathbf{1''}} \left(\overline{t}_R \Gamma u_R\right)_{\mathbf{1'}}, \\
                & \left({Y^{(4)}_{\mathbf{1}}}(\tau_q) {Y^{(4)}_{\mathbf{1'}}}^*(\tau_q)\right)_{\mathbf{1'}} \left(\overline{t}_R \Gamma c_R\right)_{\mathbf{1''}}, \left({Y^{(4)}_{\mathbf{1'}}}(\tau_q) {Y^{(4)}_{\mathbf{1}}}^*(\tau_q)\right)_{\mathbf{1'}} \left(\overline{t}_R \Gamma c_R\right)_{\mathbf{1''}}, \\
                & \left({Y^{(4)}_{\mathbf{3}}}(\tau_q)  {Y^{(4)}_{\mathbf{3}}}^*(\tau_q)\right)_{\mathbf{1'}} \left(\overline{t}_R \Gamma c_R\right)_{\mathbf{1''}}.
        \end{aligned}
    \end{equation}                
    \item $(\overline{d}d)_\mathbf{1}$: For this type, the result given by the Hilbert series is
    \begin{equation}
        d \, {d^\dagger} \, h \, {h^\dagger} \, D
        + h \, {h^\dagger} \, D \, s \, {s^\dagger}
        + b \, {b^\dagger} \, h \, {h^\dagger} \, D.
        \end{equation}        
    The corresponding SMEFT operator is $\mathcal{O}_{Hd}$. The construction rules are identical to $(\overline{e}e)_{\mathbf{1}}$:

    \begin{equation}
        \begin{aligned}
        & \overline{d}_R \Gamma d_R, \\
        & \overline{s_R} \Gamma s_R, \\
        & \overline{b_R} \Gamma b_R.
        \end{aligned}
        \end{equation}        
\end{enumerate}

\item $\overline{R}R' + \text{h.c.}$:
In this class there is a single flavor structure, whose non-trivial modular
invariance resides in the quark sector:
\begin{itemize}
    \item Operator invariant under $A_{4q}$:
    \begin{equation}
        (\overline{u}d)_\mathbf{1}\,.
    \end{equation}
\end{itemize}

\begin{center}
    \begin{tabular}{|l|l|l|l|}
        \hline
        type &  & weight & $A_4$ invariant \\ \hline
        $(\overline{u}d)_\mathbf{1}$ &
        $\overline{U}_R\,\Gamma\,D_R$ &
        $4_q$ &
        $Y^{(4)}_{\mathbf{r}_q}\,[\overline{U}_R\,\Gamma\,D_R]$ \\ \hline
    \end{tabular}
\end{center}

For this flavor structure the Hilbert series gives
\begin{equation}
    \begin{aligned}
        & d \, h^2 \, D \, {u^\dagger} \, Y^{(4)}_\mathbf{1}(\tau_q)
        + {c^\dagger} \, d \, h^2 \, D \, Y^{(4)}_{\mathbf{1'}}(\tau_q)
        + {c^\dagger} \, h^2 \, D \, s \, Y^{(4)}_\mathbf{1}(\tau_q) \\
        & + h^2 \, D \, s \, {t^\dagger} \, Y^{(4)}_{\mathbf{1'}}(\tau_q)
        + b \, h^2 \, D \, {t^\dagger} \, Y^{(4)}_{\mathbf{1'}}(\tau_q)
        + b \, h^2 \, D \, {t^\dagger} \, Y^{(4)}_\mathbf{1}(\tau_q)
        + \text{h.c.}\; .
    \end{aligned}
\end{equation}
This corresponds to 12 independent invariants built from the SMEFT operator
$\mathcal{O}_{Hud}+\text{h.c.}$, with different flavor contractions in the
up- and down-type quark fields. In terms of the modular forms $Y^{(4)}_{\mathbf{1}}$
and $Y^{(4)}_{\mathbf{1'}}$, the $A_{4q}$-invariant combinations can be written as:
\begin{equation}
    \begin{aligned}
        & Y^{(4)}_{\mathbf{1}}(\tau_q)\,\bigl(\overline{u}_R \Gamma d_R\bigr)_{\mathbf{1}}
          + \text{h.c.}\;, \\
        & Y^{(4)}_{\mathbf{1}}(\tau_q)\,\bigl(\overline{c}_R \Gamma s_R\bigr)_{\mathbf{1}}
          + \text{h.c.}\;, \\
        & Y^{(4)}_{\mathbf{1}}(\tau_q)\,\bigl(\overline{t}_R \Gamma b_R\bigr)_{\mathbf{1}}
          + \text{h.c.}\;, \\
        & Y^{(4)}_{\mathbf{1'}}(\tau_q)\,\bigl(\overline{c}_R \Gamma d_R\bigr)_{\mathbf{1''}}
          + \text{h.c.}\;, \\
        & Y^{(4)}_{\mathbf{1'}}(\tau_q)\,\bigl(\overline{t}_R \Gamma s_R\bigr)_{\mathbf{1''}}
          + \text{h.c.}\;, \\
        & Y^{(4)}_{\mathbf{1'}}(\tau_q)\,\bigl(\overline{u}_R \Gamma b_R\bigr)_{\mathbf{1''}}
          + \text{h.c.}\;.
    \end{aligned}
\end{equation}
In each case the product of the modular singlet representation
($\mathbf{1}$ or $\mathbf{1'}$) with the corresponding bilinear
representation ($\mathbf{1}$ or $\mathbf{1''}$) gives an $A_{4q}$ singlet,
so that all six structures contribute to $\mathcal{O}_{Hud}$ in a way that
matches the Hilbert-series counting.

\end{enumerate}

\item \textbf{Quadrilinear Structure:}
\begin{enumerate}
\item $(\overline{L}L)(\overline{L}L)$:
There are three classes of four-fermion operators of this type, which can be
classified according to their $A_4$ transformation properties:
\begin{itemize}
    \item Operators invariant under $A_{4e}\otimes A_{4q}$:
    \begin{equation}
        \bigl[(\overline{\ell}\,\ell)_\mathbf{1}\,(\overline{q}q)_\mathbf{1}\bigr]_{\mathbf{1}}\,.
    \end{equation}

    \item Operators invariant under $A_{4e}$:
    \begin{equation}
        \bigl[(\overline{\ell}\,\ell)(\overline{\ell}\,\ell)\bigr]_{\mathbf{1}}\,.
    \end{equation}

    \item Operators invariant under $A_{4q}$:
    \begin{equation}
        \bigl[(\overline{q}q)(\overline{q}q)\bigr]_{\mathbf{1}}\,.
    \end{equation}
\end{itemize}

\begin{enumerate}
    \item $\bigl[(\overline{\ell}\,\ell)_\mathbf{1}(\overline{q}q)_\mathbf{1}\bigr]_{\mathbf{1}}$:
    For this type, the Hilbert series gives
    \begin{equation}
        98 \,\ell \,\ell^\dagger \, q \, q^\dagger \,
        Y^{(2)}_{\mathbf{3}}(\tau_e)\,Y^{(2)}_{\mathbf{3}}(\tau_q)\,
        {Y^{(2)}_{\mathbf{3}}}^{*}(\tau_e)\,{Y^{(2)}_{\mathbf{3}}}^{*}(\tau_q)\,.
    \end{equation}
    The corresponding SMEFT operators are $\mathcal{O}_{lq}^{(1)}$ and
    $\mathcal{O}_{lq}^{(3)}$. Thus, for each of these operators there are
    $49$ independent $A_4$-invariant flavor structures.

    \begin{center}
        \begin{tabular}{|l|l|l|l|}
            \hline
            type &  & weight & $A_4$ invariant \\ \hline
            $(\overline{\ell}\,\ell)_\mathbf{1}(\overline{q}q)_\mathbf{1}$ &
            $[\overline{L}_L\Gamma L_L][\overline{Q}_L\Gamma Q_L]$ &
            $2_q+2_{qd}+2_e+2_{ed}$ &
            $\bigl({Y^{(2)}_{\mathbf{r}_e}}{Y^{(2)}_{\mathbf{r}'_e}}^{*}\bigr)
             \bigl({Y^{(2)}_{\mathbf{r}_q}}{Y^{(2)}_{\mathbf{r}'_q}}^{*}\bigr)
             [\overline{L}_L\Gamma L_L][\overline{Q}_L\Gamma Q_L]$ \\ \hline
        \end{tabular}
    \end{center}
    The $A_4$-invariant completion factorizes into an $A_{4e}$ singlet times
    an $A_{4q}$ singlet and can be written schematically as
    \begin{equation}
        \bigl[({Y^{(2)}_{\mathbf{r}_e}}{Y^{(2)}_{\mathbf{r}'_e}}^{*})
              (\overline{L}_L\Gamma L_L)\bigr]_{\mathbf{1}}\,
        \bigl[({Y^{(2)}_{\mathbf{r}_q}}{Y^{(2)}_{\mathbf{r}'_q}}^{*})
              (\overline{Q}_L\Gamma Q_L)\bigr]_{\mathbf{1}}\,,
    \end{equation}
i.e.\ it is simply obtained by combining the bilinear structures of
type $(\overline{\ell}\,\ell)_{\mathbf{1}}$ and $(\overline{q}q)_{\mathbf{1}}$.

    \item $\bigl[(\overline{\ell}\,\ell)(\overline{\ell}\,\ell)\bigr]_{\mathbf{1}}$:
    For this type, the Hilbert series yields
    \begin{equation}
        \begin{aligned}
        & 5 \,\ell^2 \, {\ell^\dagger}^2 \, Y^{(4)}_\mathbf{1}(\tau_e) \,
          {Y^{(4)}_\mathbf{1}}^*(\tau_e)
        + 5 \,\ell^2 \, {\ell^\dagger}^2 \, Y^{(4)}_{\mathbf{1'}}(\tau_e) \,
          {Y^{(4)}_\mathbf{1}}^*(\tau_e) \\
        & + 10 \,\ell^2 \, {\ell^\dagger}^2 \, Y^{(4)}_{\mathbf{3}}(\tau_e) \,
          {Y^{(4)}_\mathbf{1}}^*(\tau_e)
        + 5 \,\ell^2 \, {\ell^\dagger}^2 \, Y^{(4)}_\mathbf{1}(\tau_e) \,
          {Y^{(4)}_{\mathbf{1'}}}^*(\tau_e) \\
        & + 5 \,\ell^2 \, {\ell^\dagger}^2 \, Y^{(4)}_{\mathbf{1'}}(\tau_e) \,
          {Y^{(4)}_{\mathbf{1'}}}^*(\tau_e)
        + 10 \,\ell^2 \, {\ell^\dagger}^2 \, Y^{(4)}_{\mathbf{3}}(\tau_e) \,
          {Y^{(4)}_{\mathbf{1'}}}^*(\tau_e) \\
        & + 10 \,\ell^2 \, {\ell^\dagger}^2 \, Y^{(4)}_\mathbf{1}(\tau_e) \,
          {Y^{(4)}_{\mathbf{3}}}^*(\tau_e)
        + 10 \,\ell^2 \, {\ell^\dagger}^2 \, Y^{(4)}_{\mathbf{1'}}(\tau_e) \,
          {Y^{(4)}_{\mathbf{3}}}^*(\tau_e) \\
        & + 35 \,\ell^2 \, {\ell^\dagger}^2 \, Y^{(4)}_{\mathbf{3}}(\tau_e) \,
          {Y^{(4)}_{\mathbf{3}}}^*(\tau_e)\,.
        \end{aligned}
        \label{barllcc}
    \end{equation}
    The corresponding SMEFT operator is $\mathcal{O}_{ll}$.

    \begin{center}
        \begin{tabular}{|l|l|l|l|}
            \hline
            type &  & weight & $A_4$ invariant \\ \hline
            $\bigl[(\overline{\ell}\,\ell)(\overline{\ell}\,\ell)\bigr]_{\mathbf{1}}$ &
            $[\overline{L}_L\Gamma L_L][\overline{L}_L\Gamma L_L]$ &
            $4_e+4_{ed}$ &
            $\bigl({Y^{(4)}_{\mathbf{r}_e}}{Y^{(4)}_{\mathbf{r}'_e}}^{*}\bigr)
             [\overline{L}_L\Gamma L_L][\overline{L}_L\Gamma L_L]$ \\ \hline
        \end{tabular}
    \end{center}
    The $A_4$-invariant completion is obtained from
    \begin{equation}
        \Bigl[\bigl({Y^{(4)}_{\mathbf{r}_e}}{Y^{(4)}_{\mathbf{r}'_e}}^{*}\bigr)
              [\overline{L}_L\Gamma L_L][\overline{L}_L\Gamma L_L]\Bigr]_{\mathbf{1}}\,.
    \end{equation}
    We now construct the $A_4$-invariant terms explicitly.

    For the modular forms, there are six possible triplet combinations,
    namely:
    \begin{equation}
        \begin{aligned}
        & \bigl(Y^{(4)}_{\mathbf{3}}(\tau_e)\,{Y^{(4)}_{\mathbf{3}}}^{*}(\tau_e)\bigr)_{\mathbf{3}_s}, \\
        & \bigl(Y^{(4)}_{\mathbf{3}}(\tau_e)\,{Y^{(4)}_{\mathbf{3}}}^{*}(\tau_e)\bigr)_{\mathbf{3}_a}, \\
        & Y^{(4)}_{\mathbf{1}}(\tau_e)\,{Y^{(4)}_{\mathbf{3}}}^{*}(\tau_e), \\
        & Y^{(4)}_{\mathbf{1'}}(\tau_e)\,{Y^{(4)}_{\mathbf{3}}}^{*}(\tau_e), \\
        & Y^{(4)}_{\mathbf{3}}(\tau_e)\,{Y^{(4)}_{\mathbf{1}}}^{*}(\tau_e), \\
        & Y^{(4)}_{\mathbf{3}}(\tau_e)\,{Y^{(4)}_{\mathbf{1'}}}^{*}(\tau_e)\,.
        \end{aligned}
    \end{equation}
    To obtain singlets, these triplets must be combined with triplets built
    from $[\overline{L}_L\Gamma L_L][\overline{L}_L\Gamma L_L]$. There are
    ten such triplet structures:
    \begin{equation}
        \begin{aligned}
        & [\overline{L}_L \Gamma L_L]_{\mathbf{1}}\,[\overline{L}_L \Gamma L_L]_{\mathbf{3}_s}, \\
        & [\overline{L}_L \Gamma L_L]_{\mathbf{1}}\,[\overline{L}_L \Gamma L_L]_{\mathbf{3}_a}, \\
        & [\overline{L}_L \Gamma L_L]_{\mathbf{1'}}\,[\overline{L}_L \Gamma L_L]_{\mathbf{3}_s}, \\
        & [\overline{L}_L \Gamma L_L]_{\mathbf{1'}}\,[\overline{L}_L \Gamma L_L]_{\mathbf{3}_a}, \\
        & [\overline{L}_L \Gamma L_L]_{\mathbf{1''}}\,[\overline{L}_L \Gamma L_L]_{\mathbf{3}_s}, \\
        & [\overline{L}_L \Gamma L_L]_{\mathbf{1''}}\,[\overline{L}_L \Gamma L_L]_{\mathbf{3}_a}, \\
        & \bigl([\overline{L}_L \Gamma L_L]_{\mathbf{3}_s}
                [\overline{L}_L \Gamma L_L]_{\mathbf{3}_a}\bigr)_{\mathbf{3}_s}, \\
        & \bigl([\overline{L}_L \Gamma L_L]_{\mathbf{3}_s}
                [\overline{L}_L \Gamma L_L]_{\mathbf{3}_a}\bigr)_{\mathbf{3}_a}, \\
        & \bigl([\overline{L}_L \Gamma L_L]_{\mathbf{3}_s}
                [\overline{L}_L \Gamma L_L]_{\mathbf{3}_s}\bigr)_{\mathbf{3}}, \\
        & \bigl([\overline{L}_L \Gamma L_L]_{\mathbf{3}_a}
                [\overline{L}_L \Gamma L_L]_{\mathbf{3}_a}\bigr)_{\mathbf{3}}.
        \end{aligned}
    \end{equation}

    In addition, for the modular forms there are seven independent singlet
    combinations:
    \begin{equation}
        \begin{aligned}
        & \bigl(Y^{(4)}_{\mathbf{3}}(\tau_e)\,{Y^{(4)}_{\mathbf{3}}}^{*}(\tau_e)\bigr)_{\mathbf{1}}, \\
        & \bigl(Y^{(4)}_{\mathbf{3}}(\tau_e)\,{Y^{(4)}_{\mathbf{3}}}^{*}(\tau_e)\bigr)_{\mathbf{1'}}, \\
        & \bigl(Y^{(4)}_{\mathbf{3}}(\tau_e)\,{Y^{(4)}_{\mathbf{3}}}^{*}(\tau_e)\bigr)_{\mathbf{1''}}, \\
        & \bigl(Y^{(4)}_{\mathbf{1}}(\tau_e)\,{Y^{(4)}_{\mathbf{1}}}^{*}(\tau_e)\bigr)_{\mathbf{1}}, \\
        & \bigl(Y^{(4)}_{\mathbf{1}}(\tau_e)\,{Y^{(4)}_{\mathbf{1'}}}^{*}(\tau_e)\bigr)_{\mathbf{1'}}, \\
        & \bigl(Y^{(4)}_{\mathbf{1'}}(\tau_e)\,{Y^{(4)}_{\mathbf{1}}}^{*}(\tau_e)\bigr)_{\mathbf{1'}}, \\
        & \bigl(Y^{(4)}_{\mathbf{1'}}(\tau_e)\,{Y^{(4)}_{\mathbf{1'}}}^{*}(\tau_e)\bigr)_{\mathbf{1''}}.
        \end{aligned}
    \end{equation}
    For each singlet irrep of $A_4$ (trivial or non-trivial), there are five
    possible contractions built from the factors
    $[\overline{L}_L \Gamma L_L][\overline{L}_L \Gamma L_L]$. We classify
    them according to the resulting singlet representation:

    \begin{itemize}
        \item \textbf{$\mathbf{1}$ case:}
        \begin{equation}
            \begin{aligned}
            & [\overline{L}_L \Gamma L_L]_{\mathbf{1}}\,
              [\overline{L}_L \Gamma L_L]_{\mathbf{1}}, \\
            & [\overline{L}_L \Gamma L_L]_{\mathbf{1'}}\,
              [\overline{L}_L \Gamma L_L]_{\mathbf{1''}}, \\
            & \bigl([\overline{L}_L \Gamma L_L]_{\mathbf{3}_s}
                    [\overline{L}_L \Gamma L_L]_{\mathbf{3}_s}\bigr)_{\mathbf{1}}, \\
            & \bigl([\overline{L}_L \Gamma L_L]_{\mathbf{3}_s}
                    [\overline{L}_L \Gamma L_L]_{\mathbf{3}_a}\bigr)_{\mathbf{1}}, \\
            & \bigl([\overline{L}_L \Gamma L_L]_{\mathbf{3}_a}
                    [\overline{L}_L \Gamma L_L]_{\mathbf{3}_a}\bigr)_{\mathbf{1}}.
            \end{aligned}
        \end{equation}

        \item \textbf{$\mathbf{1'}$ case:}
        \begin{equation}
            \begin{aligned}
            & [\overline{L}_L \Gamma L_L]_{\mathbf{1}}\,
              [\overline{L}_L \Gamma L_L]_{\mathbf{1'}}, \\
            & [\overline{L}_L \Gamma L_L]_{\mathbf{1''}}\,
              [\overline{L}_L \Gamma L_L]_{\mathbf{1''}}, \\
            & \bigl([\overline{L}_L \Gamma L_L]_{\mathbf{3}_s}
                    [\overline{L}_L \Gamma L_L]_{\mathbf{3}_s}\bigr)_{\mathbf{1'}}, \\
            & \bigl([\overline{L}_L \Gamma L_L]_{\mathbf{3}_s}
                    [\overline{L}_L \Gamma L_L]_{\mathbf{3}_a}\bigr)_{\mathbf{1'}}, \\
            & \bigl([\overline{L}_L \Gamma L_L]_{\mathbf{3}_a}
                    [\overline{L}_L \Gamma L_L]_{\mathbf{3}_a}\bigr)_{\mathbf{1'}}.
            \end{aligned}
        \end{equation}

        \item \textbf{$\mathbf{1''}$ case:}
        \begin{equation}
            \begin{aligned}
            & [\overline{L}_L \Gamma L_L]_{\mathbf{1}}\,
              [\overline{L}_L \Gamma L_L]_{\mathbf{1''}}, \\
            & [\overline{L}_L \Gamma L_L]_{\mathbf{1'}}\,
              [\overline{L}_L \Gamma L_L]_{\mathbf{1'}}, \\
            & \bigl([\overline{L}_L \Gamma L_L]_{\mathbf{3}_s}
                    [\overline{L}_L \Gamma L_L]_{\mathbf{3}_s}\bigr)_{\mathbf{1''}}, \\
            & \bigl([\overline{L}_L \Gamma L_L]_{\mathbf{3}_s}
                    [\overline{L}_L \Gamma L_L]_{\mathbf{3}_a}\bigr)_{\mathbf{1''}}, \\
            & \bigl([\overline{L}_L \Gamma L_L]_{\mathbf{3}_a}
                    [\overline{L}_L \Gamma L_L]_{\mathbf{3}_a}\bigr)_{\mathbf{1''}}.
            \end{aligned}
        \end{equation}
    \end{itemize}

Summing up, we find
\begin{equation}
6 \times 10 \; (\text{triplet--triplet contractions})
\;+\; 7 \times 5 \; (\text{singlet--singlet contractions})
\;=\; 95
\end{equation}
possible $A_{4e}$-invariant constructions. The coefficient of the
corresponding Hilbert-series term differs from that in
Ref.~\cite{Kobayashi_2022}.
The origin of this redundancy is the same as that discussed in Section~\ref{dimension-W}.

    \item $\bigl[(\overline{q}q)(\overline{q}q)\bigr]_{\mathbf{1}}$:
    For this type, the Hilbert series gives
    \begin{equation}
        \begin{aligned}
        & 10 \, q^2 \, {q^\dagger}^2 \, Y^{(4)}_\mathbf{1}(\tau_q) \,
          {Y^{(4)}_\mathbf{1}}^*(\tau_q)
        + 10 \, q^2 \, {q^\dagger}^2 \, Y^{(4)}_{\mathbf{1'}}(\tau_q) \,
          {Y^{(4)}_\mathbf{1}}^*(\tau_q) \\
        & + 20 \, q^2 \, {q^\dagger}^2 \, Y^{(4)}_{\mathbf{3}}(\tau_q) \,
          {Y^{(4)}_\mathbf{1}}^*(\tau_q)
        + 10 \, q^2 \, {q^\dagger}^2 \, Y^{(4)}_\mathbf{1}(\tau_q) \,
          {Y^{(4)}_{\mathbf{1'}}}^*(\tau_q) \\
        & + 10 \, q^2 \, {q^\dagger}^2 \, Y^{(4)}_{\mathbf{1'}}(\tau_q) \,
          {Y^{(4)}_{\mathbf{1'}}}^*(\tau_q)
        + 20 \, q^2 \, {q^\dagger}^2 \, Y^{(4)}_{\mathbf{3}}(\tau_q) \,
          {Y^{(4)}_{\mathbf{1'}}}^*(\tau_q) \\
        & + 20 \, q^2 \, {q^\dagger}^2 \, Y^{(4)}_\mathbf{1}(\tau_q) \,
          {Y^{(4)}_{\mathbf{3}}}^*(\tau_q)
        + 20 \, q^2 \, {q^\dagger}^2 \, Y^{(4)}_{\mathbf{1'}}(\tau_q) \,
          {Y^{(4)}_{\mathbf{3}}}^*(\tau_q) \\
        & + 70 \, q^2 \, {q^\dagger}^2 \, Y^{(4)}_{\mathbf{3}}(\tau_q) \,
          {Y^{(4)}_{\mathbf{3}}}^*(\tau_q)\,.
        \end{aligned}
        \label{barqqcc}
    \end{equation}
    The corresponding SMEFT operators are $\mathcal{O}_{qq}^{(1)}$ and
    $\mathcal{O}_{qq}^{(3)}$.

    \begin{center}
        \begin{tabular}{|l|l|l|l|}
            \hline
            type &  & weight & $A_4$ invariant \\ \hline
            $\bigl[(\overline{q}q)(\overline{q}q)\bigr]_{\mathbf{1}}$ &
            $[\overline{Q}_L\Gamma Q_L][\overline{Q}_L\Gamma Q_L]$ &
            $4_q+4_{qd}$ &
            $\bigl({Y^{(4)}_{\mathbf{r}_q}}{Y^{(4)}_{\mathbf{r}'_q}}^{*}\bigr)
             [\overline{Q}_L\Gamma Q_L][\overline{Q}_L\Gamma Q_L]$ \\ \hline
        \end{tabular}
    \end{center}
    The $A_4$-invariant completion is
    \begin{equation}
        \Bigl[\bigl({Y^{(4)}_{\mathbf{r}_q}}{Y^{(4)}_{\mathbf{r}'_q}}^{*}\bigr)
              [\overline{Q}_L\Gamma Q_L][\overline{Q}_L\Gamma Q_L]\Bigr]_{\mathbf{1}}\,.
    \end{equation}
    The structure of the contractions is completely analogous to the
    leptonic case $\bigl[(\overline{\ell}\,\ell)(\overline{\ell}\,\ell)\bigr]_{\mathbf{1}}$,
    with the only difference that the invariants are now built in the
    $A_{4q}$ sector rather than $A_{4e}$.
\end{enumerate}

\item $(\overline{R}R)(\overline{R}R)$: 
The structure $(\overline{R}R)(\overline{R}R)$ gives rise to six types of
operators, which can be classified as follows:
\begin{itemize}
    \item Operators invariant under $A_{4e} \otimes A_{4q}$:
    \begin{equation}
        \begin{aligned}
        & \bigl[(\overline{e}e)_\mathbf{1} (\overline{u}u)_\mathbf{1}\bigr]_{\mathbf{1}}, \\
        & \bigl[(\overline{e}e)_\mathbf{1} (\overline{d}d)_\mathbf{1}\bigr]_{\mathbf{1}}.
        \end{aligned}
    \end{equation}

    \item Operators invariant under $A_{4e}$:
    \begin{equation}
        \bigl[(\overline{e}e)(\overline{e}e)\bigr]_{\mathbf{1}}.
    \end{equation}

    \item Operators invariant under $A_{4q}$:
    \begin{equation}
        \begin{aligned}
        & \bigl[(\overline{u}u) (\overline{u}u)\bigr]_{\mathbf{1}}, \\
        & \bigl[(\overline{d}d) (\overline{d}d)\bigr]_{\mathbf{1}}, \\
        & \bigl[(\overline{u}u) (\overline{d}d)\bigr]_{\mathbf{1}}.
        \end{aligned}
    \end{equation}
\end{itemize}

\begin{enumerate}
\item $\bigl[(\overline{e}e)_\mathbf{1}(\overline{u}u)_\mathbf{1}\bigr]_{\mathbf{1}}$: 
The corresponding SMEFT operator is $\mathcal{O}_{eu}$.
\begin{center}
    \begin{tabular}{|l|l|l|l|}
        \hline
        type &  & weight & $A_4$ invariant  \\ \hline
        $(\overline{e}e)_\mathbf{1}(\overline{u}u)_\mathbf{1}$ &
        $[\overline{E}_R\Gamma E_R][\overline{U}_R\Gamma U_R]$ &
        $4_q+4_{qd}$ &
        $\bigl(Y_{\mathbf{r}_e}^{(4)} {Y^{(4)}_{\mathbf{r}'_q}}^{*}\bigr)
        [\overline{E}_R\Gamma E_R][\overline{U}_R\Gamma U_R]$ \\ \hline
    \end{tabular}
\end{center}
We directly give the construction types. As for the bilinear structures
$(\overline{e}e)_{\mathbf{1}}$ and $(\overline{u}u)_{\mathbf{1}}$ discussed
above, in the $(\overline{e}e)_{\mathbf{1}}$ sector there are three possible
terms, and in the $(\overline{u}u)_{\mathbf{1}}$ sector there are $21$. The
number of invariant combinations is therefore
\begin{equation}
3\;(\text{num.\ of }(\overline{e}e)_{\mathbf{1}})\times21\;(\text{num.\ of }(\overline{u}u)_{\mathbf{1}}) = 63\,.
\end{equation}

\item $\bigl[(\overline{e}e)_\mathbf{1}(\overline{d}d)_\mathbf{1}\bigr]_{\mathbf{1}}$: 
The corresponding SMEFT operator is $\mathcal{O}_{ed}$.
\begin{center}
    \begin{tabular}{|l|l|l|l|}
        \hline
        type &  & weight & $A_4$ invariant  \\ \hline
        $(\overline{e}e)_\mathbf{1}(\overline{d}d)_\mathbf{1}$ &
        $[\overline{E}_R\Gamma E_R][\overline{D}_R\Gamma D_R]$ &
        $0$ &
        $[\overline{E}_R\Gamma E_R]_{\mathbf{1}}
         [\overline{D}_R\Gamma D_R]_{\mathbf{1}}$  \\ \hline
    \end{tabular}
\end{center}
This case is simply the product of the bilinear sectors
$(\overline{e}e)_{\mathbf{1}}$ and $(\overline{d}d)_{\mathbf{1}}$, with modular
weight zero. Each sector contributes three singlet structures, so the
number of invariant operators is
\begin{equation}
N_{ed}=3\;(\text{num.\ of }(\overline{e}e)_{\mathbf{1}})\times3\;(\text{num.\ of }(\overline{d}d)_{\mathbf{1}}) = 9\,.
\end{equation}

\item $\bigl[(\overline{u}u)(\overline{u}u)\bigr]_{\mathbf{1}}$:  
The corresponding SMEFT operator is $\mathcal{O}_{uu}$.
\begin{center}
    \begin{tabular}{|l|l|l|l|}
        \hline
        type &  & weight & $A_4$ invariant  \\ \hline
        $[(\overline{u}u)(\overline{u}u)]_{\mathbf{1}}$ &
        $[\overline{U}_R\Gamma U_R][\overline{U}_R\Gamma U_R]$ &
        $8_q+8_{qd}$ &
        $\bigl(Y_{\mathbf{r}_q}^{(8)} {Y_{\mathbf{r}'_q}^{(8)}}^{*}\bigr)
        [\overline{U}_R\Gamma U_R][\overline{U}_R\Gamma U_R]$  \\ \hline
    \end{tabular}
\end{center}
The construction can be separated into two parts: one involving the modular
forms and the other involving the SMEFT operators.

\paragraph{Modular forms:}

The possible contractions of modular forms at weight $8$ are classified
according to the resulting one-dimensional $A_4$ representation:

\begin{itemize}
    \item Tensor products of $A_{4}$ modular forms (singlet
    $\mathbf{1}$):
    \begin{equation}
    \begin{aligned}
    & Y^{(8)}_{\mathbf{1}}(\tau_q)\,{Y^{(8)}_{\mathbf{1}}}^{*}(\tau_q), \\
    & Y^{(8)}_{\mathbf{1'}}(\tau_q)\,{Y^{(8)}_{\mathbf{1''}}}^{*}(\tau_q), \\
    & Y^{(8)}_{\mathbf{1''}}(\tau_q)\,{Y^{(8)}_{\mathbf{1'}}}^{*}(\tau_q), \\
    & \bigl[Y^{(8)}_{\mathbf{3}a}(\tau_q)\,{Y^{(8)}_{\mathbf{3}a}}^{*}(\tau_q)\bigr]_{\mathbf{1}}, \\
    & \bigl[Y^{(8)}_{\mathbf{3}a}(\tau_q)\,{Y^{(8)}_{\mathbf{3}b}}^{*}(\tau_q)\bigr]_{\mathbf{1}}, \\
    & \bigl[Y^{(8)}_{\mathbf{3}b}(\tau_q)\,{Y^{(8)}_{\mathbf{3}a}}^{*}(\tau_q)\bigr]_{\mathbf{1}}, \\
    & \bigl[Y^{(8)}_{\mathbf{3}b}(\tau_q)\,{Y^{(8)}_{\mathbf{3}b}}^{*}(\tau_q)\bigr]_{\mathbf{1}}.
    \end{aligned}
    \end{equation}

    \item Tensor products of $A_{4}$ modular forms (non-trivial singlet
    $\mathbf{1'}$):
    \begin{equation}
    \begin{aligned}
    & Y^{(8)}_{\mathbf{1}}(\tau_q)\,{Y^{(8)}_{\mathbf{1'}}}^{*}(\tau_q), \\
    & Y^{(8)}_{\mathbf{1'}}(\tau_q)\,{Y^{(8)}_{\mathbf{1}}}^{*}(\tau_q), \\
    & Y^{(8)}_{\mathbf{1''}}(\tau_q)\,{Y^{(8)}_{\mathbf{1''}}}^{*}(\tau_q), \\
    & \bigl[Y^{(8)}_{\mathbf{3}a}(\tau_q)\,{Y^{(8)}_{\mathbf{3}a}}^{*}(\tau_q)\bigr]_{\mathbf{1'}}, \\
    & \bigl[Y^{(8)}_{\mathbf{3}a}(\tau_q)\,{Y^{(8)}_{\mathbf{3}b}}^{*}(\tau_q)\bigr]_{\mathbf{1'}}, \\
    & \bigl[Y^{(8)}_{\mathbf{3}b}(\tau_q)\,{Y^{(8)}_{\mathbf{3}a}}^{*}(\tau_q)\bigr]_{\mathbf{1'}}, \\
    & \bigl[Y^{(8)}_{\mathbf{3}b}(\tau_q)\,{Y^{(8)}_{\mathbf{3}b}}^{*}(\tau_q)\bigr]_{\mathbf{1'}}.
    \end{aligned}
    \end{equation}

    \item Tensor products of $A_{4}$ modular forms (non-trivial singlet
    $\mathbf{1''}$):
    \begin{equation}
    \begin{aligned}
    & Y^{(8)}_{\mathbf{1}}(\tau_q)\,{Y^{(8)}_{\mathbf{1''}}}^{*}(\tau_q), \\
    & Y^{(8)}_{\mathbf{1''}}(\tau_q)\,{Y^{(8)}_{\mathbf{1}}}^{*}(\tau_q), \\
    & Y^{(8)}_{\mathbf{1'}}(\tau_q)\,{Y^{(8)}_{\mathbf{1'}}}^{*}(\tau_q), \\
    & \bigl[Y^{(8)}_{\mathbf{3}a}(\tau_q)\,{Y^{(8)}_{\mathbf{3}a}}^{*}(\tau_q)\bigr]_{\mathbf{1''}}, \\
    & \bigl[Y^{(8)}_{\mathbf{3}a}(\tau_q)\,{Y^{(8)}_{\mathbf{3}b}}^{*}(\tau_q)\bigr]_{\mathbf{1''}}, \\
    & \bigl[Y^{(8)}_{\mathbf{3}b}(\tau_q)\,{Y^{(8)}_{\mathbf{3}a}}^{*}(\tau_q)\bigr]_{\mathbf{1''}}, \\
    & \bigl[Y^{(8)}_{\mathbf{3}b}(\tau_q)\,{Y^{(8)}_{\mathbf{3}b}}^{*}(\tau_q)\bigr]_{\mathbf{1''}}.
    \end{aligned}
    \end{equation}
\end{itemize}
Thus, for each one-dimensional representation $\mathbf{1}$, $\mathbf{1'}$ and $\mathbf{1''}$, there are seven independent modular contractions at weight $8$.

\paragraph{SMEFT operators:}

The basic structures for the up-type right-handed bilinears are
\begin{equation}
\begin{aligned}
& (\overline{u}_R \Gamma u_R)_\mathbf{1}, \quad
  (\overline{u}_R \Gamma c_R)_{\mathbf{1'}}, \quad
  (\overline{u}_R \Gamma t_R)_{\mathbf{1''}}, \\
& (\overline{c}_R \Gamma c_R)_\mathbf{1}, \quad
  (\overline{c}_R \Gamma t_R)_{\mathbf{1'}}, \quad
  (\overline{c}_R \Gamma u_R)_{\mathbf{1''}}, \\
& (\overline{t}_R \Gamma t_R)_\mathbf{1}, \quad
  (\overline{t}_R \Gamma u_R)_{\mathbf{1'}}, \quad
  (\overline{t}_R \Gamma c_R)_{\mathbf{1''}}.
\end{aligned}
\end{equation}
The tensor products of two such bilinears can be classified according to
their flavor structure:

\begin{itemize}
    \item Type1: \textbf{Four identical flavors.}  \\
    In this class both bilinears carry the same flavor, schematically
    paired as $(\overline{f}f)_\rho(\overline{f}f)_{\rho'}$ with $f=u,c,t$, and
$\rho,\rho' \in \{\mathbf{1},\mathbf{1'},\mathbf{1''}\}$ denote the
one-dimensional representations of $A_{4q}$. We have
    ${u^\dagger}^2 u^2$, ${c^\dagger}^2 c^2$ and ${t^\dagger}^2 t^2$, giving
    the three SMEFT structures:
    \begin{equation}
    \begin{aligned}
    & (\overline{u}_R \Gamma u_R)_\mathbf{1}
      (\overline{u}_R \Gamma u_R)_\mathbf{1},\\
    & (\overline{c}_R \Gamma c_R)_\mathbf{1}
      (\overline{c}_R \Gamma c_R)_\mathbf{1},\\
    & (\overline{t}_R \Gamma t_R)_\mathbf{1}
      (\overline{t}_R \Gamma t_R)_\mathbf{1}.
    \end{aligned}
    \end{equation}
    These operators are hermitian by themselves.

    \item Type2: \textbf{Three identical and one different flavor.}  \\
    Here the four external fields contain three copies of one flavor and one
    copy of a different flavor; schematically the bilinears can be thought of
    as $(\overline{f}f)_{\rho}(\overline{f}f')_{\rho'}$ with $f\neq f'$. The six possible
    flavor patterns are
    \begin{equation}
        \begin{aligned}
            & {u^\dagger}^2 u\,c + \text{h.c.},\quad
              {u^\dagger}^2 u\,t + \text{h.c.},\\
            & {c^\dagger}^2 c\,u + \text{h.c.},\quad
              {c^\dagger}^2 c\,t + \text{h.c.},\\
            & {t^\dagger}^2 t\,u + \text{h.c.},\quad
              {t^\dagger}^2 t\,c + \text{h.c.}\;.
        \end{aligned}
    \end{equation}
    For instance, for ${c^\dagger}^2 c\,u$ we can write
    \begin{equation}
        \bigl[(\overline{c}_R \Gamma c_R)_{\mathbf{1}}
              (\overline{c}_R \Gamma u_R)_{\mathbf{1''}}\bigr]_{\mathbf{1''}},
    \end{equation}
    while the conjugate structure is
    \begin{equation}
        \bigl[(\overline{u}_R \Gamma c_R)_{\mathbf{1'}}
              (\overline{c}_R \Gamma c_R)_{\mathbf{1}}\bigr]_{\mathbf{1'}}.
    \end{equation}
    Counting the operator and its hermitian conjugate separately, this class
    contributes $6\times 2$ flavor structures.

    \item Type3: \textbf{Two identical bilinears with different internal flavors.}  
    In this class the two bilinears are identical, and each bilinear involves
    two different flavors; schematically they are paired as
    $(\overline{f}f')_{\rho}(\overline{f}f')_{\rho'}$. The three flavor
    patterns are
    \begin{equation}
        {u^\dagger}^2 c^2 + \text{h.c.},\quad
        {u^\dagger}^2 t^2 + \text{h.c.},\quad
        {t^\dagger}^2 c^2 + \text{h.c.}\;.
    \end{equation}
    More explicitly, the relevant SMEFT operators are
    \begin{equation}
        \begin{aligned}
            & \bigl[(\overline{u}_R \Gamma c_R)_{\mathbf{1'}}
                    (\overline{u}_R \Gamma c_R)_{\mathbf{1'}}\bigr]_{\mathbf{1''}}
              + \text{h.c.}\;,\\
            & \bigl[(\overline{u}_R \Gamma t_R)_{\mathbf{1''}}
                    (\overline{u}_R \Gamma t_R)_{\mathbf{1''}}\bigr]_{\mathbf{1'}}
              + \text{h.c.}\;,\\
            & \bigl[(\overline{c}_R \Gamma t_R)_{\mathbf{1'}}
                    (\overline{c}_R \Gamma t_R)_{\mathbf{1'}}\bigr]_{\mathbf{1''}}
              + \text{h.c.}\;.
        \end{aligned}
    \end{equation}
    This gives $3\times 2 = 6$ flavor structures (including their conjugates).

    \item Type4: \textbf{Two different flavors paired across two bilinears.}  \\
    In this class the two bilinears involve the same pair of flavors, paired
    as $(\overline{f}f)_{\rho}(\overline{f'}f')_{\rho'}$. The corresponding flavor patterns are
    \begin{equation}
        \begin{aligned}
            & {u^\dagger}u\,{t^\dagger}t,\\
            & {u^\dagger}u\,{c^\dagger}c,\\
            & {c^\dagger}c\,{t^\dagger}t.
        \end{aligned}
    \end{equation}
    For each flavor pattern there are two independent $A_4$ singlet
    contractions. For example, for ${u^\dagger}u\,{t^\dagger}t$ we have
    \begin{equation}
        \bigl[(\overline{u}_R \Gamma u_R)_{\mathbf{1}}
              (\overline{t}_R \Gamma t_R)_{\mathbf{1}}\bigr]_{\mathbf{1}},
    \end{equation}
    and
    \begin{equation}
        \bigl[(\overline{u}_R \Gamma t_R)_{\mathbf{1''}}
              (\overline{t}_R \Gamma u_R)_{\mathbf{1'}}\bigr]_{\mathbf{1}}.
    \end{equation}
    Altogether this class contributes $3\times 2 = 6$ flavor structures.

    \item Type5: \textbf{Two identical and two distinct flavors.}  
    This class contains flavor patterns with two copies of one flavor and two
    different additional flavors. Schematically the operators can be grouped
    as $(\overline{f}f)_{\rho}(\overline{f'}f'')_{\rho'}$ or
    $(\overline{f}f')_{\rho}(\overline{f}f'')_{\rho'}$ with $f,f',f''$ all different. It is
    convenient to distinguish two subcases.

    \medskip
    \emph{(i) Two identical antifields in the same bilinear.}  
    Here the two identical flavors sit in the same bilinear,
    $(\overline{f}f')(\overline{f}f'')$ with $f',f''\neq f$ and
    $f'\neq f''$. The three flavor
    patterns are
    \begin{equation}
        {u^\dagger}^2 c\,t + \text{h.c.},\quad
        {c^\dagger}^2 u\,t + \text{h.c.},\quad
        {t^\dagger}^2 u\,c + \text{h.c.}\;.
    \end{equation}
    For example, for ${u^\dagger}^2 c\,t$ we can write
    \begin{equation}
        \bigl[(\overline{u}_R \Gamma t_R)_{\mathbf{1''}}
              (\overline{u}_R \Gamma c_R)_{\mathbf{1'}}\bigr]_{\mathbf{1}},
    \end{equation}
    together with its hermitian conjugate. This subcase therefore gives
    $3\times 2 = 6$ flavor structures.

    \medskip
    \emph{(ii) Two identical flavors distributed over two bilinears.}  
    In this subcase the identical flavor is split between the two bilinears,
    schematically $(\overline{f}f)_{\rho}(\overline{f'}f'')_{\rho'}$. The remaining three patterns are
    \begin{equation}
        {u^\dagger}u\,{t^\dagger}c+\text{h.c.},\quad
        {c^\dagger}c\,{u^\dagger}t+\text{h.c.},\quad
        {t^\dagger}t\,{u^\dagger}c+\text{h.c.}.
    \end{equation}
    For each of them there are two distinct $A_4$ singlet contractions. For
    instance, for ${u^\dagger}u\,{t^\dagger}c$ we obtain
    \begin{equation}
        \begin{aligned}
            & \bigl[(\overline{u}_R \Gamma c_R)_{\mathbf{1'}}
                    (\overline{t}_R \Gamma u_R)_{\mathbf{1'}}\bigr]_{\mathbf{1''}},\\
            & \bigl[(\overline{u}_R \Gamma u_R)_{\mathbf{1}}
                    (\overline{t}_R \Gamma c_R)_{\mathbf{1''}}\bigr]_{\mathbf{1''}}.
        \end{aligned}
    \end{equation}
    Each of these operators also comes with its hermitian conjugate, so this
    subcase contributes $3\times 2\times 2 = 12$ flavor structures.
\end{itemize}

Collecting all classes, the number of distinct flavor structures (counting
each operator and its hermitian conjugate separately when appropriate) is
\begin{equation}
N_{uu}=3\;(\text{Type1}) \;+\; 12\;(\text{Type2}) \;+\; 6\;(\text{Type3}) \;+\; 6\;(\text{Type4}) \;+\; 6\;(\text{Type5\;i}) \;+\; 12\;(\text{Type5\;ii})
= 45\,.
\end{equation}
For each one-dimensional representation of $A_{4q}$ there are seven possible
weight-8 modular singlet contractions, so the total number of
$A_4$-invariant SMEFT operators of this type is
\begin{equation}
45 \times 7 = 315\,.
\end{equation}

\item $\bigl[(\overline{d}d)(\overline{d}d)\bigr]_{\mathbf{1}}$: 
The corresponding SMEFT operator is $\mathcal{O}_{dd}$.
\begin{center}
    \begin{tabular}{|l|l|l|l|}
        \hline
        type &  & weight & $A_4$ invariant  \\ \hline
        $[(\overline{d}d)(\overline{d}d)]_{\mathbf{1}}$ &
        $[\overline{D}_R\Gamma D_R][\overline{D}_R\Gamma D_R]$ &
        $0$ &
        $[\overline{D}_R\Gamma D_R][\overline{D}_R\Gamma D_R]$  \\ \hline
    \end{tabular}
\end{center}
Since all these terms have modular weight zero, they do not need to be
combined with modular forms. Therefore, the possible structures are
analogous to those in the SMEFT operator constructions where only $A_{4q}$
singlets are considered in the type
$[(\overline{u}u)(\overline{u}u)]_{\mathbf{1}}$. The complete set of
flavor structures can be organized as follows (with
$f,f',f''\in\{d,s,b\}$):

\begin{itemize}
    \item Type1: \textbf{Four identical flavors.} \\ 
In this class both bilinears carry the same flavor, schematically
paired as $(\overline{f}f)_{\rho}(\overline{f}f)_{\rho'}$. Explicitly,
\begin{equation}
\begin{aligned}
& (\overline{d}_R \Gamma d_R)_{\mathbf{1}}
  (\overline{d}_R \Gamma d_R)_{\mathbf{1}},\\
& (\overline{s}_R \Gamma s_R)_{\mathbf{1}}
  (\overline{s}_R \Gamma s_R)_{\mathbf{1}},\\
& (\overline{b}_R \Gamma b_R)_{\mathbf{1}}
  (\overline{b}_R \Gamma b_R)_{\mathbf{1}}.
\end{aligned}
\end{equation}
Each of these operators is hermitian by itself, giving $3$ flavor
structures.

\item Type2: \textbf{Two different flavors paired across two bilinears.}  \\
Here the two bilinears involve the same pair of different flavors,
paired as $(\overline{f}f)_{\rho}(\overline{f'}f')_{\rho'}$ with $f\neq f'$. The three
flavor pairs are $(d,s)$, $(d,b)$ and $(s,b)$, and for each pair there
are two independent $A_4$ singlet contractions:
\begin{equation}
\begin{aligned}
& (\overline{d}_R \Gamma d_R)_{\mathbf{1}}
  (\overline{s}_R \Gamma s_R)_{\mathbf{1}},\quad
  \bigl[(\overline{d}_R \Gamma s_R)_{\mathbf{1'}}
        (\overline{s}_R \Gamma d_R)_{\mathbf{1''}}\bigr]_{\mathbf{1}},\\
& (\overline{d}_R \Gamma d_R)_{\mathbf{1}}
  (\overline{b}_R \Gamma b_R)_{\mathbf{1}},\quad
  \bigl[(\overline{d}_R \Gamma b_R)_{\mathbf{1''}}
        (\overline{b}_R \Gamma d_R)_{\mathbf{1'}}\bigr]_{\mathbf{1}},\\
& (\overline{s}_R \Gamma s_R)_{\mathbf{1}}
  (\overline{b}_R \Gamma b_R)_{\mathbf{1}},\quad
  \bigl[(\overline{s}_R \Gamma b_R)_{\mathbf{1'}}
        (\overline{b}_R \Gamma s_R)_{\mathbf{1''}}\bigr]_{\mathbf{1}}.
\end{aligned}
\end{equation}
These combinations can be chosen hermitian, so no explicit
$\text{h.c.}$ is required. Altogether this class contributes
$3\times 2 = 6$ flavor structures.

    \item Type3: \textbf{Two identical and two distinct flavors.}  \\
    In this class there are two copies of one flavor and two different
    additional flavors, schematically paired as
    $(\overline{f}f')_{\rho}(\overline{f}f'')_{\rho'}$ with $f',f''\neq f$ and
    $f'\neq f''$. The three independent patterns are
    \begin{equation}
    \begin{aligned}
    & \bigl[(\overline{d}_R \Gamma s_R)_{\mathbf{1'}}
            (\overline{d}_R \Gamma b_R)_{\mathbf{1''}}\bigr]_{\mathbf{1}}
      + \text{h.c.},\\
    & \bigl[(\overline{s}_R \Gamma d_R)_{\mathbf{1''}}
            (\overline{s}_R \Gamma b_R)_{\mathbf{1'}}\bigr]_{\mathbf{1}}
      + \text{h.c.},\\
    & \bigl[(\overline{b}_R \Gamma d_R)_{\mathbf{1'}}
            (\overline{b}_R \Gamma s_R)_{\mathbf{1''}}\bigr]_{\mathbf{1}}
      + \text{h.c.}.
    \end{aligned}
    \end{equation}
    Each of these comes together with its hermitian conjugate, so this class
    contributes $3\times 2 = 6$ flavor structures.
\end{itemize}
In summary, there are
\begin{equation}
N_{dd}=3\;(\text{Type1}) + 6\;(\text{Type2}) + 6\;(\text{Type3}) = 15
\end{equation}
possible $A_4$-singlet constructions for this operator class.

\item $\bigl[(\overline{u}u)(\overline{d}d)\bigr]_{\mathbf{1}}$: 
The corresponding SMEFT operators are $\mathcal{O}_{ud}^{(1)}$ and
$\mathcal{O}_{ud}^{(8)}$.
\begin{center}
    \begin{tabular}{|l|l|l|l|}
        \hline
        type &  & weight & $A_4$ invariant  \\ \hline
        $[(\overline{u}u)(\overline{d}d)]_{\mathbf{1}}$ &
        $[\overline{U}_R\Gamma U_R][\overline{D}_R\Gamma D_R]$ &
        $4_q+4_{qd}$ &
        $\bigl(Y_{\mathbf{r}_q}^{(4)} {Y^{(4)}_{\mathbf{r}'_q}}^{*}\bigr)
        [\overline{U}_R\Gamma U_R][\overline{D}_R\Gamma D_R]$  \\ \hline
    \end{tabular}
\end{center}
The construction is divided into two parts: one involving the modular forms
and the other involving the SMEFT operators.

%\paragraph{Modular forms:}

%The possible contractions of modular forms at weight~4 are classified
%according to the resulting $A_4$ one-dimensional representation:

%\begin{itemize}
%    \item Tensor products of $A_{4}$ modular forms (singlet
%    $\mathbf{1}$):
%    \begin{equation}
%    \begin{aligned}
%    & Y^{(4)}_{\mathbf{1}}(\tau_q)\,{Y^{(4)}_{\mathbf{1}}}^{*}(\tau_q), \\
%    & \bigl[Y^{(4)}_{\mathbf{3}}(\tau_q)\,{Y^{(4)}_{\mathbf{3}}}^{*}(\tau_q)\bigr]_{\mathbf{1}}.
%    \end{aligned}
%    \end{equation}

%    \item Tensor products of $A_{4}$ modular forms (non-trivial singlet
%    $\mathbf{1'}$):
%    \begin{equation}
%    \begin{aligned}
%    & Y^{(4)}_{\mathbf{1}}(\tau_q)\,{Y^{(4)}_{\mathbf{1'}}}^{*}(\tau_q), \\
%    & Y^{(4)}_{\mathbf{1'}}(\tau_q)\,{Y^{(4)}_{\mathbf{1}}}^{*}(\tau_q), \\
%    & \bigl[Y^{(4)}_{\mathbf{3}}(\tau_q)\,{Y^{(4)}_{\mathbf{3}}}^{*}(\tau_q)\bigr]_{\mathbf{1'}}.
%    \end{aligned}
%    \end{equation}

%    \item Tensor products of $A_{4}$ modular forms (non-trivial singlet
%    $\mathbf{1''}$):
%    \begin{equation}
%    \begin{aligned}
%    & Y^{(4)}_{\mathbf{1'}}(\tau_q)\,{Y^{(4)}_{\mathbf{1'}}}^{*}(\tau_q), \\
%    & \bigl[Y^{(4)}_{\mathbf{3}}(\tau_q)\,{Y^{(4)}_{\mathbf{3}}}^{*}(\tau_q)\bigr]_{\mathbf{1''}}.
%    \end{aligned}
%    \end{equation}
%\end{itemize}
%In total there are
%\[
%2\;(\text{for }\mathbf{1})\;+\;
%3\;(\text{for }\mathbf{1'})\;+\;
%2\;(\text{for }\mathbf{1''}) \;=\; 7
%\]
%independent singlet contractions of the modular forms at weight~4.
\paragraph{Modular forms:}

The possible contractions of modular forms at weight $4$ are classified
according to the resulting $A_4$ one-dimensional representation:

\begin{itemize}
    \item Tensor products of $A_{4}$ modular forms (singlet
    $\mathbf{1}$):
    \begin{equation}
    \begin{aligned}
    & Y^{(4)}_{\mathbf{1}}(\tau_q)\,{Y^{(4)}_{\mathbf{1}}}^{*}(\tau_q), \\
    & \bigl[Y^{(4)}_{\mathbf{3}}(\tau_q)\,{Y^{(4)}_{\mathbf{3}}}^{*}(\tau_q)\bigr]_{\mathbf{1}}.
    \end{aligned}
    \end{equation}

    \item Tensor products of $A_{4}$ modular forms (non-trivial singlet
    $\mathbf{1'}$):
    \begin{equation}
    \begin{aligned}
    & Y^{(4)}_{\mathbf{1}}(\tau_q)\,{Y^{(4)}_{\mathbf{1'}}}^{*}(\tau_q), \\
    & Y^{(4)}_{\mathbf{1'}}(\tau_q)\,{Y^{(4)}_{\mathbf{1}}}^{*}(\tau_q), \\
    & \bigl[Y^{(4)}_{\mathbf{3}}(\tau_q)\,{Y^{(4)}_{\mathbf{3}}}^{*}(\tau_q)\bigr]_{\mathbf{1'}}.
    \end{aligned}
    \end{equation}

    \item Tensor products of $A_{4}$ modular forms (non-trivial singlet
    $\mathbf{1''}$):
    \begin{equation}
    \begin{aligned}
    & Y^{(4)}_{\mathbf{1'}}(\tau_q)\,{Y^{(4)}_{\mathbf{1'}}}^{*}(\tau_q), \\
    & \bigl[Y^{(4)}_{\mathbf{3}}(\tau_q)\,{Y^{(4)}_{\mathbf{3}}}^{*}(\tau_q)\bigr]_{\mathbf{1''}}.
    \end{aligned}
    \end{equation}
\end{itemize}
In total there are
\begin{equation}
2\;(\text{for }\mathbf{1})\;+\;
3\;(\text{for }\mathbf{1'})\;+\;
2\;(\text{for }\mathbf{1''}) \;=\; 7
\end{equation}
independent singlet contractions of the modular forms at weight $4$.

\paragraph{SMEFT operators:}

The basic up-type right-handed bilinears are
\begin{equation}
\begin{aligned}
& (\overline{u}_R \Gamma u_R)_{\mathbf{1}}, \quad
  (\overline{u}_R \Gamma c_R)_{\mathbf{1'}}, \quad
  (\overline{u}_R \Gamma t_R)_{\mathbf{1''}}, \\
& (\overline{c}_R \Gamma c_R)_{\mathbf{1}}, \quad
  (\overline{c}_R \Gamma t_R)_{\mathbf{1'}}, \quad
  (\overline{c}_R \Gamma u_R)_{\mathbf{1''}}, \\
& (\overline{t}_R \Gamma t_R)_{\mathbf{1}}, \quad
  (\overline{t}_R \Gamma u_R)_{\mathbf{1'}}, \quad
  (\overline{t}_R \Gamma c_R)_{\mathbf{1''}}.
\end{aligned}
\end{equation}
and the down-type right-handed bilinears are
\begin{equation}
\begin{aligned}
& (\overline{d}_R \Gamma d_R)_{\mathbf{1}}, \quad
  (\overline{d}_R \Gamma s_R)_{\mathbf{1'}}, \quad
  (\overline{d}_R \Gamma b_R)_{\mathbf{1''}}, \\
& (\overline{s}_R \Gamma s_R)_{\mathbf{1}}, \quad
  (\overline{s}_R \Gamma b_R)_{\mathbf{1'}}, \quad
  (\overline{s}_R \Gamma d_R)_{\mathbf{1''}}, \\
& (\overline{b}_R \Gamma b_R)_{\mathbf{1}}, \quad
  (\overline{b}_R \Gamma d_R)_{\mathbf{1'}}, \quad
  (\overline{b}_R \Gamma s_R)_{\mathbf{1''}}.
\end{aligned}
\end{equation}
A generic flavor structure in this sector can be written schematically as
\begin{equation}
(\overline{f_u} \Gamma f_u)_{\rho}\,
(\overline{f_d} \Gamma f_d)_{\rho'}\,,
\end{equation}
where $f_u \in \{u,c,t\}$ and $f_d \in \{d,s,b\}$ label the up- and
down-type quark flavors, while
$\rho,\rho' \in \{\mathbf{1},\mathbf{1'},\mathbf{1''}\}$ denote the
one-dimensional irreps of $A_{4q}$.

To make the counting explicit, let us first fix the up-type bilinear to
transform as the singlet $\mathbf{1}$, i.e.\ we restrict to
\begin{equation}
(\overline{u}_R \Gamma u_R)_{\mathbf{1}},\quad
(\overline{c}_R \Gamma c_R)_{\mathbf{1}},\quad
(\overline{t}_R \Gamma t_R)_{\mathbf{1}}.
\end{equation}
In this case the $A_{4q}$ irrep of the four-quark fermion operator is
entirely determined by the irrep of the down-type bilinear.

\medskip
\noindent
\textbf{Fermionic operators in the singlet $\mathbf{1}$.}  
Here we choose down-type bilinears transforming as $\mathbf{1}$:
\begin{equation}
(\overline{d}_R \Gamma d_R)_{\mathbf{1}},\quad
(\overline{s}_R \Gamma s_R)_{\mathbf{1}},\quad
(\overline{b}_R \Gamma b_R)_{\mathbf{1}}.
\end{equation}
Combining them with the three up-type bilinears in $\mathbf{1}$ gives the
nine flavor structures
\begin{equation}
\begin{aligned}
& (\overline{u}_R \Gamma u_R)_{\mathbf{1}}
  (\overline{d}_R \Gamma d_R)_{\mathbf{1}},\quad
  (\overline{u}_R \Gamma u_R)_{\mathbf{1}}
  (\overline{s}_R \Gamma s_R)_{\mathbf{1}},\quad
  (\overline{u}_R \Gamma u_R)_{\mathbf{1}}
  (\overline{b}_R \Gamma b_R)_{\mathbf{1}},\\
& (\overline{c}_R \Gamma c_R)_{\mathbf{1}}
  (\overline{d}_R \Gamma d_R)_{\mathbf{1}},\quad
  (\overline{c}_R \Gamma c_R)_{\mathbf{1}}
  (\overline{s}_R \Gamma s_R)_{\mathbf{1}},\quad
  (\overline{c}_R \Gamma c_R)_{\mathbf{1}}
  (\overline{b}_R \Gamma b_R)_{\mathbf{1}},\\
& (\overline{t}_R \Gamma t_R)_{\mathbf{1}}
  (\overline{d}_R \Gamma d_R)_{\mathbf{1}},\quad
  (\overline{t}_R \Gamma t_R)_{\mathbf{1}}
  (\overline{s}_R \Gamma s_R)_{\mathbf{1}},\quad
  (\overline{t}_R \Gamma t_R)_{\mathbf{1}}
  (\overline{b}_R \Gamma b_R)_{\mathbf{1}}.
\end{aligned}
\end{equation}
Each of these can be dressed with any of the \emph{two} modular singlet
structures transforming as $\mathbf{1}$, so this class yields
$2\times 9$ independent operator structures.

\medskip
\noindent
\textbf{Fermionic operators in the non-trivial singlet $\mathbf{1'}$.}  
Next we choose down-type bilinears in the irrep $\mathbf{1'}$:
\begin{equation}
(\overline{d}_R \Gamma s_R)_{\mathbf{1'}},\quad
(\overline{s}_R \Gamma b_R)_{\mathbf{1'}},\quad
(\overline{b}_R \Gamma d_R)_{\mathbf{1'}}.
\end{equation}
Combining them with the three up-type bilinears in $\mathbf{1}$ again gives
nine flavor structures, now transforming as $\mathbf{1'}$:
\begin{equation}
\begin{aligned}
& (\overline{u}_R \Gamma u_R)_{\mathbf{1}}
  (\overline{d}_R \Gamma s_R)_{\mathbf{1'}},\quad
  (\overline{u}_R \Gamma u_R)_{\mathbf{1}}
  (\overline{s}_R \Gamma b_R)_{\mathbf{1'}},\quad
  (\overline{u}_R \Gamma u_R)_{\mathbf{1}}
  (\overline{b}_R \Gamma d_R)_{\mathbf{1'}},\\
& (\overline{c}_R \Gamma c_R)_{\mathbf{1}}
  (\overline{d}_R \Gamma s_R)_{\mathbf{1'}},\quad
  (\overline{c}_R \Gamma c_R)_{\mathbf{1}}
  (\overline{s}_R \Gamma b_R)_{\mathbf{1'}},\quad
  (\overline{c}_R \Gamma c_R)_{\mathbf{1}}
  (\overline{b}_R \Gamma d_R)_{\mathbf{1'}},\\
& (\overline{t}_R \Gamma t_R)_{\mathbf{1}}
  (\overline{d}_R \Gamma s_R)_{\mathbf{1'}},\quad
  (\overline{t}_R \Gamma t_R)_{\mathbf{1}}
  (\overline{s}_R \Gamma b_R)_{\mathbf{1'}},\quad
  (\overline{t}_R \Gamma t_R)_{\mathbf{1}}
  (\overline{b}_R \Gamma d_R)_{\mathbf{1'}}.
\end{aligned}
\end{equation}
On the modular side there are \emph{three} independent singlet contractions
transforming as $\mathbf{1'}$, so this class contributes
$3\times 9$ operator structures.

\medskip
\noindent
\textbf{Fermionic operators in the non-trivial singlet $\mathbf{1''}$.}  
Finally we take down-type bilinears in the irrep $\mathbf{1''}$:
\begin{equation}
(\overline{d}_R \Gamma b_R)_{\mathbf{1''}},\quad
(\overline{s}_R \Gamma d_R)_{\mathbf{1''}},\quad
(\overline{b}_R \Gamma s_R)_{\mathbf{1''}}.
\end{equation}
Combining them with the three up-type bilinears in $\mathbf{1}$ yields nine
flavor structures transforming as $\mathbf{1''}$:
\begin{equation}
\begin{aligned}
& (\overline{u}_R \Gamma u_R)_{\mathbf{1}}
  (\overline{d}_R \Gamma b_R)_{\mathbf{1''}},\quad
  (\overline{u}_R \Gamma u_R)_{\mathbf{1}}
  (\overline{s}_R \Gamma d_R)_{\mathbf{1''}},\quad
  (\overline{u}_R \Gamma u_R)_{\mathbf{1}}
  (\overline{b}_R \Gamma s_R)_{\mathbf{1''}},\\
& (\overline{c}_R \Gamma c_R)_{\mathbf{1}}
  (\overline{d}_R \Gamma b_R)_{\mathbf{1''}},\quad
  (\overline{c}_R \Gamma c_R)_{\mathbf{1}}
  (\overline{s}_R \Gamma d_R)_{\mathbf{1''}},\quad
  (\overline{c}_R \Gamma c_R)_{\mathbf{1}}
  (\overline{b}_R \Gamma s_R)_{\mathbf{1''}},\\
& (\overline{t}_R \Gamma t_R)_{\mathbf{1}}
  (\overline{d}_R \Gamma b_R)_{\mathbf{1''}},\quad
  (\overline{t}_R \Gamma t_R)_{\mathbf{1}}
  (\overline{s}_R \Gamma d_R)_{\mathbf{1''}},\quad
  (\overline{t}_R \Gamma t_R)_{\mathbf{1}}
  (\overline{b}_R \Gamma s_R)_{\mathbf{1''}}.
\end{aligned}
\end{equation}
There are \emph{two} singlet contractions of the modular forms in the
irrep $\mathbf{1''}$, so this class produces $2\times 9$ operator
structures.

\medskip

Putting these three cases together, and still keeping the up-type bilinear
in the singlet $\mathbf{1}$, we obtain
\begin{equation}
2\times 9 \;(\text{from }\mathbf{1})
\;+\; 3\times 9 \;(\text{from }\mathbf{1'})
\;+\; 2\times 9 \;(\text{from }\mathbf{1''})
\;=\; (2+3+2)\times 9\,,
\end{equation}
independent $A_4$-invariant operators. Here the factor $(2+3+2)$ counts
the singlet contractions of the modular forms, while the factor $9$ counts
the flavor combinations of up- and down-type bilinears.

By symmetry of the three one-dimensional irreps
$\mathbf{1},\mathbf{1'},\mathbf{1''}$ of $A_{4q}$, the same counting
applies when the up-type bilinear transforms as $\mathbf{1'}$ or
$\mathbf{1''}$ instead of $\mathbf{1}$. Therefore, there are three such
sectors, and the total number of $A_4$-invariant SMEFT operators of this
type is
\begin{equation}
N_{ud} \;=\; 3 \times (2+3+2)\times 9 \;=\; 189.
\end{equation}

\item $\bigl[(\overline{e}e)(\overline{e}e)\bigr]_{\mathbf{1}}$: 
The corresponding SMEFT operator is $\mathcal{O}_{ee}$.
\begin{center}
    \begin{tabular}{|l|l|l|l|}
        \hline
        type &  & weight & $A_4$ invariant  \\ \hline
        $[(\overline{e}e)(\overline{e}e)]_{\mathbf{1}}$ &
        $[\overline{E}_R\Gamma E_R][\overline{E}_R\Gamma E_R]$ &
        $0$ &
        $[\overline{E}_R\Gamma E_R][\overline{E}_R\Gamma E_R]$  \\ \hline
    \end{tabular}
\end{center}
Since all these terms have modular weight zero, they do not need to be
combined with modular forms. As in the up-type quark case, the flavor
structures can be classified according to their pattern of lepton flavors.
However, the Hilbert series indicates that in the purely leptonic sector
some of the structures that are distinct for quarks are no longer
independent. In particular, for a given pair of flavors, say $(e,\mu)$, a
term of the schematic form $e\,e^\dagger\,\mu\,\mu^\dagger$ can only
represent one independent construction, which may be written either as
\begin{equation}
    \bigl[(\overline{e}e)_{\mathbf{1}} (\overline{\mu}\mu)_{\mathbf{1}}\bigr]_{\mathbf{1}},
\end{equation}
or as
\begin{equation}
    \bigl[(\overline{e}\mu)_{\mathbf{1'}} (\overline{\mu}e)_{\mathbf{1''}}\bigr]_{\mathbf{1}}.
\end{equation}
Thus these two contractions are not independent and must be counted only
once. Taking this into account, the relevant $A_4$-singlet constructions are:

\begin{itemize}
    \item Type 1: \textbf{Four identical flavors.}  \\
    Both bilinears involve the same lepton flavor, paired as
    $(\overline{f}f)_{\rho}(\overline{f}f)_{\rho'}$ with $f \in \{e,\mu,\tau\}$:
    \begin{equation}
    \begin{aligned}
    & (\overline{e}_R \Gamma e_R)_{\mathbf{1}}
      (\overline{e}_R \Gamma e_R)_{\mathbf{1}},\\
    & (\overline{\mu}_R \Gamma \mu_R)_{\mathbf{1}}
      (\overline{\mu}_R \Gamma \mu_R)_{\mathbf{1}},\\
    & (\overline{\tau}_R \Gamma \tau_R)_{\mathbf{1}}
      (\overline{\tau}_R \Gamma \tau_R)_{\mathbf{1}}.
    \end{aligned}
    \end{equation}
    These operators are hermitian by themselves.

    \item Type 2: \textbf{Two different flavors paired across two bilinears.}  \\
    Here the two bilinears involve two distinct flavors $f\neq f'$,
    paired as $(\overline{f}f)_{\rho}(\overline{f'}f')_{\rho'}$ with
    $f,f'\in\{e,\mu,\tau\}$:
    \begin{equation}
    \begin{aligned}
    & (\overline{e}_R \Gamma e_R)_{\mathbf{1}}
      (\overline{\mu}_R \Gamma \mu_R)_{\mathbf{1}},\\
    & (\overline{e}_R \Gamma e_R)_{\mathbf{1}}
      (\overline{\tau}_R \Gamma \tau_R)_{\mathbf{1}},\\
    & (\overline{\mu}_R \Gamma \mu_R)_{\mathbf{1}}
      (\overline{\tau}_R \Gamma \tau_R)_{\mathbf{1}}.
    \end{aligned}
    \end{equation}
    As explained above, the alternative contractions with off-diagonal
    bilinears,
    \begin{equation}
      \notag
      \bigl[(\overline{f}f')_{\mathbf{1'}}
            (\overline{f'}f)_{\mathbf{1''}}\bigr]_{\mathbf{1}},
    \end{equation}
    are not independent of the diagonal products and are therefore not
    counted separately.

    \item Type 3: \textbf{Two identical and two distinct flavors.}  \\
    In this class the two bilinears contain two copies of one flavor and
    one copy each of the remaining two flavors, e.g.\ $f,f',f''$ with
    $f\neq f'\neq f''$. The independent structures can be chosen as
    \begin{equation}
    \begin{aligned}
    & \bigl[(\overline{e}_R \Gamma \mu_R)_{\mathbf{1'}}
            (\overline{e}_R \Gamma \tau_R)_{\mathbf{1''}}\bigr]_{\mathbf{1}}
      + \text{h.c.},\\
    & \bigl[(\overline{\mu}_R \Gamma e_R)_{\mathbf{1''}}
            (\overline{\mu}_R \Gamma \tau_R)_{\mathbf{1'}}\bigr]_{\mathbf{1}}
      + \text{h.c.},\\
    & \bigl[(\overline{\tau}_R \Gamma e_R)_{\mathbf{1'}}
            (\overline{\tau}_R \Gamma \mu_R)_{\mathbf{1''}}\bigr]_{\mathbf{1}}
      + \text{h.c.}.
    \end{aligned}
    \end{equation}
    Each flavor pattern comes with its hermitian conjugate and is counted
    once.
\end{itemize}
In summary, there are
\begin{equation}
\notag
N_{ee}=3\;(\text{Type 1})
\;+\;3\;(\text{Type 2})
\;+\;3\times 2\;(\text{Type 3 with h.c.})
\;=\;12
\end{equation}
independent $A_4$-singlet SMEFT flavor structures of the type
$\bigl[(\overline{e}e)(\overline{e}e)\bigr]_{\mathbf{1}}$.
\end{enumerate}

\item $(\overline{L}L)(\overline{R}R)$: The structure
$(\overline{L}L)(\overline{R}R)$ gives rise to six types of operators, which
can be classified as follows:
    \begin{itemize}
        \item Operators invariant under $A_{4e} \otimes A_{4q}$:
        \begin{equation}
        \begin{aligned}
        & \bigl[(\overline{\ell}\ell)_\mathbf{1} (\overline{u}u)_\mathbf{1}\bigr]_{\mathbf{1}}, \\
        & \bigl[(\overline{\ell}\ell)_\mathbf{1} (\overline{d}d)_\mathbf{1}\bigr]_{\mathbf{1}}, \\
        & \bigl[(\overline{q}q)_\mathbf{1} (\overline{e}e)_\mathbf{1}\bigr]_{\mathbf{1}}.
        \end{aligned}
        \end{equation}

        \item Operators invariant under $A_{4e}$:
        \begin{equation}
        \bigl[(\overline{\ell}\ell)(\overline{e}e)\bigr]_{\mathbf{1}}.
        \end{equation}

        \item Operators invariant under $A_{4q}$:
        \begin{equation}
        \begin{aligned}
        & \bigl[(\overline{q}q) (\overline{u}u)\bigr]_{\mathbf{1}}, \\
        & \bigl[(\overline{q}q) (\overline{d}d)\bigr]_{\mathbf{1}}.
        \end{aligned}
        \end{equation}
    \end{itemize}

\begin{enumerate}
\item $\bigl[(\overline{\ell}\ell)_\mathbf{1}(\overline{u}u)_\mathbf{1}\bigr]_{\mathbf{1}}$: 
The corresponding SMEFT operator is $\mathcal{O}_{\ell u}$.
\begin{center}
    \begin{tabular}{|l|l|l|l|}
        \hline
        type &  & weight & $A_4$ invariant  \\ \hline
        $(\overline{\ell}\ell)_\mathbf{1}(\overline{u}u)_\mathbf{1}$ &
        $[\overline{L}_L\Gamma L_L][\overline{U}_R\Gamma U_R]$ &
        $4_q+4_{qd}+2_e+2_{ed}$ &
        $(Y^{(2)}_{\mathbf{r}_e}{Y^{(2)}_{\mathbf{r}'_e}}^*)
         (Y^{(4)}_{\mathbf{r}_q}{Y^{(4)}_{\mathbf{r}'_q}}^*)
         [\overline{L}_L\Gamma L_L][\overline{U}_R\Gamma U_R]$  \\ \hline
    \end{tabular}
\end{center}
The $A_4$-invariant coefficient factorizes into a lepton and a quark part,
\begin{equation}
\Bigl[(Y^{(2)}_{\mathbf{r}_e}{Y^{(2)}_{\mathbf{r}'_e}}^*)
      (\overline{L}_L\Gamma L_L)\Bigr]_{\mathbf{1}}\,
\Bigl[(Y^{(4)}_{\mathbf{r}_q}{Y^{(4)}_{\mathbf{r}'_q}}^*)
      (\overline{U}_R\Gamma U_R)\Bigr]_{\mathbf{1}}.
\end{equation}

For the lepton sector, this is identical to the bilinear type
$(\overline{\ell}\ell)_\mathbf{1}$: combining the weight-2 modular triplet
$Y^{(2)}_{\mathbf{3}}(\tau_e)$ yields
\begin{equation}
N_\ell = 7
\end{equation}
independent $A_{4e}$ singlet contractions.

For the up-quark sector, we reuse the classification of
$(\overline{u}u)_\mathbf{1}$ at weight $4$. As in the
$\bigl[(\overline{u}u)(\overline{d}d)\bigr]_{\mathbf{1}}$ case, there are
\begin{equation}
(2+3+2)=7
\end{equation}
independent $A_{4q}$ singlet contractions of the weight-$4$ modular forms
(two of type $\mathbf{1}$, three of type non-trivial singlet $\mathbf{1'}$ and two of type
non-trivial singlet $\mathbf{1''}$). Each of these seven modular singlets can be combined with
the three up-type flavors $u,c,t$, giving
\begin{equation}
N_u = 7 \times 3 = 21
\end{equation}
independent quark structures of the form
$\bigl[(\overline{f_u}\Gamma f_u)_\mathbf{1}Y^{(4)}Y^{(4)\,*}\bigr]_{\mathbf{1}}$.

Multiplying the lepton and quark sectors, the total number of
$A_4$-invariant structures of type $\mathcal{O}_{\ell u}$ is
\begin{equation}
N_{\ell u} = N_\ell \times N_u = 7 \times 21 = 147.
\end{equation}

%----------------------------------------------------------
\begin{comment}
\item $\bigl[(\overline{\ell}\ell)_\mathbf{1}(\overline{d}d)_\mathbf{1}\bigr]_{\mathbf{1}}$: 
The corresponding SMEFT operator is $\mathcal{O}_{\ell d}$.
\begin{center}
    \begin{tabular}{|l|l|l|l|}
        \hline
        type &  & weight & $A_4$ invariant  \\ \hline
        $(\overline{\ell}\ell)_\mathbf{1}(\overline{d}d)_\mathbf{1}$ &
        $[\overline{L}_L\Gamma L_L][\overline{D}_R\Gamma D_R]$ &
        $2_e+2_{ed}$ &
        $(Y^{(2)}_{\mathbf{r}_e}{Y^{(2)}_{\mathbf{r}'_e}}^*)
        [\overline{L}_L\Gamma L_L][\overline{D}_R\Gamma D_R]$  \\ \hline
    \end{tabular}
\end{center}
The $A_4$-invariant combination can be written as
\begin{equation}
\Bigl[(Y^{(2)}_{\mathbf{r}_e}{Y^{(2)}_{\mathbf{r}'_e}}^*)
      (\overline{L}_L\Gamma L_L)\Bigr]_{\mathbf{1}}\,
\Bigl[(\overline{D}_R\Gamma D_R)\Bigr]_{\mathbf{1}}.
\end{equation}
The lepton sector contributes $N_\ell = 7$ singlet structures as above. The
down-quark sector has modular weight zero and is already an $A_{4q}$
singlet; there are three flavor-diagonal bilinears,
\[
(\overline{d}_R\Gamma d_R)_\mathbf{1},\quad
(\overline{s}_R\Gamma s_R)_\mathbf{1},\quad
(\overline{b}_R\Gamma b_R)_\mathbf{1},
\]
so
\[
N_d = 3.
\]
Thus the total number of $A_4$-invariant operators of type $\mathcal{O}_{\ell d}$ is
\[
N_{\ell d} = N_\ell \times N_d = 7 \times 3 = 21.
\]
\end{comment}
\item $\bigl[(\overline{\ell}\ell)_\mathbf{1}(\overline{d}d)_\mathbf{1}\bigr]_{\mathbf{1}}$: 
The corresponding SMEFT operator is $\mathcal{O}_{\ell d}$.
\begin{center}
    \begin{tabular}{|l|l|l|l|}
        \hline
        type &  & weight & $A_4$ invariant  \\ \hline
        $(\overline{\ell}\ell)_\mathbf{1}(\overline{d}d)_\mathbf{1}$ &
        $[\overline{L}_L\Gamma L_L][\overline{D}_R\Gamma D_R]$ &
        $2_e+2_{ed}$ &
        $(Y^{(2)}_{\mathbf{r}_e}{Y^{(2)}_{\mathbf{r}'_e}}^*)
        [\overline{L}_L\Gamma L_L][\overline{D}_R\Gamma D_R]$  \\ \hline
    \end{tabular}
\end{center}
The $A_4$-invariant combination can be written as
\begin{equation}
\Bigl[(Y^{(2)}_{\mathbf{r}_e}{Y^{(2)}_{\mathbf{r}'_e}}^*)
      (\overline{L}_L\Gamma L_L)\Bigr]_{\mathbf{1}}\,
\Bigl[(\overline{D}_R\Gamma D_R)\Bigr]_{\mathbf{1}}.
\end{equation}
The lepton sector contributes $N_\ell = 7$ singlet structures as above. The
down-quark sector has modular weight zero and is already an $A_{4q}$
singlet; there are three flavor-diagonal bilinears,
\begin{equation}
(\overline{d}_R\Gamma d_R)_\mathbf{1},\quad
(\overline{s}_R\Gamma s_R)_\mathbf{1},\quad
(\overline{b}_R\Gamma b_R)_\mathbf{1},
\end{equation}
so
\begin{equation}
N_d = 3.
\end{equation}
Thus the total number of $A_4$-invariant operators of type $\mathcal{O}_{\ell d}$ is
\begin{equation}
N_{\ell d} = N_\ell \times N_d = 7 \times 3 = 21.
\end{equation}

%----------------------------------------------------------
\begin{comment}
\item $\bigl[(\overline{q}q)_\mathbf{1}(\overline{e}e)_\mathbf{1}\bigr]_{\mathbf{1}}$:
The corresponding SMEFT operator is $\mathcal{O}_{qe}$.
\begin{center}
    \begin{tabular}{|l|l|l|l|}
        \hline
        type &  & weight & $A_4$ invariant  \\ \hline
        $(\overline{q}q)_\mathbf{1}(\overline{e}e)_\mathbf{1}$ &
        $[\overline{Q}_L\Gamma Q_L][\overline{E}_R\Gamma E_R]$ &
        $2_q+2_{qd}$ &
        $(Y^{(2)}_{\mathbf{r}_q}{Y^{(2)}_{\mathbf{r}'_q}}^*)
        [\overline{Q}_L\Gamma Q_L][\overline{E}_R\Gamma E_R]$  \\ \hline
    \end{tabular}
\end{center}
The $A_4$-invariant piece is
\begin{equation}
\Bigl[(Y^{(2)}_{\mathbf{r}_q}{Y^{(2)}_{\mathbf{r}'_q}}^*)
      (\overline{Q}_L\Gamma Q_L)\Bigr]_{\mathbf{1}}\,
\Bigl[(\overline{E}_R\Gamma E_R)\Bigr]_{\mathbf{1}}.
\end{equation}
This is completely analogous to the $\mathcal{O}_{\ell d}$ case, with
$(Q_L,E_R,Y^{(2)}_{\mathbf{3}}(\tau_q))$ replacing
$(L_L,D_R,Y^{(2)}_{\mathbf{3}}(\tau_e))$. The quark sector at weight~2 provides
\[
N_q = 7
\]
independent modular–bilinear singlets,
while the charged-lepton sector supplies three flavor-diagonal singlets,
\[
(\overline{e}_R\Gamma e_R)_\mathbf{1},\quad
(\overline{\mu}_R\Gamma \mu_R)_\mathbf{1},\quad
(\overline{\tau}_R\Gamma \tau_R)_\mathbf{1},
\]
so
\[
N_e = 3.
\]
Therefore,
\[
N_{qe} = N_q \times N_e = 7 \times 3 = 21.
\]
\end{comment}
\item $\bigl[(\overline{q}q)_\mathbf{1}(\overline{e}e)_\mathbf{1}\bigr]_{\mathbf{1}}$:
The corresponding SMEFT operator is $\mathcal{O}_{qe}$.
\begin{center}
    \begin{tabular}{|l|l|l|l|}
        \hline
        type &  & weight & $A_4$ invariant  \\ \hline
        $(\overline{q}q)_\mathbf{1}(\overline{e}e)_\mathbf{1}$ &
        $[\overline{Q}_L\Gamma Q_L][\overline{E}_R\Gamma E_R]$ &
        $2_q+2_{qd}$ &
        $(Y^{(2)}_{\mathbf{r}_q}{Y^{(2)}_{\mathbf{r}'_q}}^*)
        [\overline{Q}_L\Gamma Q_L][\overline{E}_R\Gamma E_R]$  \\ \hline
    \end{tabular}
\end{center}
The $A_4$-invariant piece is
\begin{equation}
\Bigl[(Y^{(2)}_{\mathbf{r}_q}{Y^{(2)}_{\mathbf{r}'_q}}^*)
      (\overline{Q}_L\Gamma Q_L)\Bigr]_{\mathbf{1}}\,
\Bigl[(\overline{E}_R\Gamma E_R)\Bigr]_{\mathbf{1}}.
\end{equation}
This is completely analogous to the $\mathcal{O}_{\ell d}$ case, with
$(Q_L,E_R,Y^{(2)}_{\mathbf{3}}(\tau_q))$ replacing
$(L_L,D_R,Y^{(2)}_{\mathbf{3}}(\tau_e))$. The quark sector at weight $2$ provides
\begin{equation}
N_q = 7
\end{equation}
independent modular--bilinear singlets,
while the charged-lepton sector supplies three flavor-diagonal singlets,
\begin{equation}
(\overline{e}_R\Gamma e_R)_\mathbf{1},\quad
(\overline{\mu}_R\Gamma \mu_R)_\mathbf{1},\quad
(\overline{\tau}_R\Gamma \tau_R)_\mathbf{1},
\end{equation}
so
\begin{equation}
N_e = 3.
\end{equation}
Therefore,
\begin{equation}
N_{qe} = N_q \times N_e = 7 \times 3 = 21.
\end{equation}

\item $\bigl[(\overline{q}q)(\overline{u}u)\bigr]_{\mathbf{1}}$: 
The corresponding SMEFT operators are $\mathcal{O}_{qu}^{(1)}$ and
$\mathcal{O}_{qu}^{(8)}$.
\begin{center}
    \begin{tabular}{|l|l|l|l|}
        \hline
        type &  & weight & $A_4$ invariant  \\ \hline
        $[(\overline{q}q)(\overline{u}u)]_{\mathbf{1}}$ &
        $[\overline{Q}_L\Gamma Q_L][\overline{U}_R\Gamma U_R]$ &
        $6_q+6_{qd}$ &
        $(Y^{(6)}_{\mathbf{r}_q}{Y^{(6)}_{\mathbf{r}'_q}}^*)
        [\overline{Q}_L\Gamma Q_L][\overline{U}_R\Gamma U_R]$  \\ \hline
    \end{tabular}
\end{center}
The construction is divided into two parts: one involving the modular forms
and the other involving the SMEFT bilinears.

\paragraph{Modular forms:}

The weight-$6$ modular forms include the singlet $Y^{(6)}_{\mathbf{1}}$,
the non-trivial singlets $Y^{(6)}_{\mathbf{1'}}$ and
$Y^{(6)}_{\mathbf{1''}}$, and the two triplets $Y^{(6)}_{\mathbf{3}a}$ and
$Y^{(6)}_{\mathbf{3}b}$. Their tensor products decompose as follows:

\begin{itemize}
    \item Singlet $\mathbf{1}$:
    \begin{equation}
    \begin{aligned}
    & Y^{(6)}_{\mathbf{1}}(\tau_q)\,{Y^{(6)}_{\mathbf{1}}}^{*}(\tau_q), \\
    & \bigl[Y^{(6)}_{\mathbf{3}a}(\tau_q)
            {Y^{(6)}_{\mathbf{3}a}}^{*}(\tau_q)\bigr]_{\mathbf{1}}, \\
    & \bigl[Y^{(6)}_{\mathbf{3}a}(\tau_q)
            {Y^{(6)}_{\mathbf{3}b}}^{*}(\tau_q)\bigr]_{\mathbf{1}}, \\
    & \bigl[Y^{(6)}_{\mathbf{3}b}(\tau_q)
            {Y^{(6)}_{\mathbf{3}a}}^{*}(\tau_q)\bigr]_{\mathbf{1}}, \\
    & \bigl[Y^{(6)}_{\mathbf{3}b}(\tau_q)
            {Y^{(6)}_{\mathbf{3}b}}^{*}(\tau_q)\bigr]_{\mathbf{1}}.
    \end{aligned}
    \end{equation}
    (5 modular singlets of type $\mathbf{1}$.)

    \item Non-trivial singlet $\mathbf{1'}$:
    \begin{equation}
    \begin{aligned}
    & \bigl[Y^{(6)}_{\mathbf{3}a}(\tau_q)
            {Y^{(6)}_{\mathbf{3}a}}^{*}(\tau_q)\bigr]_{\mathbf{1'}}, \\
    & \bigl[Y^{(6)}_{\mathbf{3}a}(\tau_q)
            {Y^{(6)}_{\mathbf{3}b}}^{*}(\tau_q)\bigr]_{\mathbf{1'}}, \\
    & \bigl[Y^{(6)}_{\mathbf{3}b}(\tau_q)
            {Y^{(6)}_{\mathbf{3}a}}^{*}(\tau_q)\bigr]_{\mathbf{1'}}, \\
    & \bigl[Y^{(6)}_{\mathbf{3}b}(\tau_q)
            {Y^{(6)}_{\mathbf{3}b}}^{*}(\tau_q)\bigr]_{\mathbf{1'}}.
    \end{aligned}
    \end{equation}
    (4 modular non-trivial singlets of type $\mathbf{1'}$.)

    \item Non-trivial singlet $\mathbf{1''}$:
    \begin{equation}
    \begin{aligned}
    & \bigl[Y^{(6)}_{\mathbf{3}a}(\tau_q)
            {Y^{(6)}_{\mathbf{3}a}}^{*}(\tau_q)\bigr]_{\mathbf{1''}}, \\
    & \bigl[Y^{(6)}_{\mathbf{3}a}(\tau_q)
            {Y^{(6)}_{\mathbf{3}b}}^{*}(\tau_q)\bigr]_{\mathbf{1''}}, \\
    & \bigl[Y^{(6)}_{\mathbf{3}b}(\tau_q)
            {Y^{(6)}_{\mathbf{3}a}}^{*}(\tau_q)\bigr]_{\mathbf{1''}}, \\
    & \bigl[Y^{(6)}_{\mathbf{3}b}(\tau_q)
            {Y^{(6)}_{\mathbf{3}b}}^{*}(\tau_q)\bigr]_{\mathbf{1''}}.
    \end{aligned}
    \end{equation}
    (4 modular non-trivial singlets of type $\mathbf{1''}$.)

    \item Triplets $\mathbf{3}$:
    \begin{equation}
    \begin{aligned}
    & \bigl[Y^{(6)}_{\mathbf{1}}(\tau_q)
            {Y^{(6)}_{\mathbf{3}a}}^{*}(\tau_q)\bigr]_{\mathbf{3}},\quad
      \bigl[Y^{(6)}_{\mathbf{1}}(\tau_q)
            {Y^{(6)}_{\mathbf{3}b}}^{*}(\tau_q)\bigr]_{\mathbf{3}}, \\
    & \bigl[Y^{(6)}_{\mathbf{3}a}(\tau_q)
            {Y^{(6)}_{\mathbf{1}}}^{*}(\tau_q)\bigr]_{\mathbf{3}},\quad
      \bigl[Y^{(6)}_{\mathbf{3}b}(\tau_q)
            {Y^{(6)}_{\mathbf{1}}}^{*}(\tau_q)\bigr]_{\mathbf{3}}, \\
    & \bigl[Y^{(6)}_{\mathbf{3}a}(\tau_q)
            {Y^{(6)}_{\mathbf{3}b}}^{*}(\tau_q)\bigr]_{\mathbf{3}_s},\quad
      \bigl[Y^{(6)}_{\mathbf{3}b}(\tau_q)
            {Y^{(6)}_{\mathbf{3}a}}^{*}(\tau_q)\bigr]_{\mathbf{3}_s}, \\
    & \bigl[Y^{(6)}_{\mathbf{3}a}(\tau_q)
            {Y^{(6)}_{\mathbf{3}b}}^{*}(\tau_q)\bigr]_{\mathbf{3}_a},\quad
      \bigl[Y^{(6)}_{\mathbf{3}b}(\tau_q)
            {Y^{(6)}_{\mathbf{3}a}}^{*}(\tau_q)\bigr]_{\mathbf{3}_a}, \\
    & \bigl[Y^{(6)}_{\mathbf{3}a}(\tau_q)
            {Y^{(6)}_{\mathbf{3}a}}^{*}(\tau_q)\bigr]_{\mathbf{3}_s},\quad
      \bigl[Y^{(6)}_{\mathbf{3}b}(\tau_q)
            {Y^{(6)}_{\mathbf{3}b}}^{*}(\tau_q)\bigr]_{\mathbf{3}_s},\\
    & \bigl[Y^{(6)}_{\mathbf{3}a}(\tau_q)
            {Y^{(6)}_{\mathbf{3}a}}^{*}(\tau_q)\bigr]_{\mathbf{3}_a},\quad
      \bigl[Y^{(6)}_{\mathbf{3}b}(\tau_q)
            {Y^{(6)}_{\mathbf{3}b}}^{*}(\tau_q)\bigr]_{\mathbf{3}_a}.
    \end{aligned}
    \end{equation}
    In total there are $12$ independent modular triplets (including both
    symmetric $\mathbf{3}_s$ and antisymmetric $\mathbf{3}_a$).
\end{itemize}

\paragraph{SMEFT operators:}

The right-handed up-type bilinears are
\begin{equation}
\begin{aligned}
& (\overline{u}_R \Gamma u_R)_\mathbf{1}, \quad
  (\overline{u}_R \Gamma c_R)_{\mathbf{1'}}, \quad
  (\overline{u}_R \Gamma t_R)_{\mathbf{1''}}, \\
& (\overline{c}_R \Gamma c_R)_\mathbf{1}, \quad
  (\overline{c}_R \Gamma t_R)_{\mathbf{1'}}, \quad
  (\overline{c}_R \Gamma u_R)_{\mathbf{1''}}, \\
& (\overline{t}_R \Gamma t_R)_\mathbf{1}, \quad
  (\overline{t}_R \Gamma u_R)_{\mathbf{1'}}, \quad
  (\overline{t}_R \Gamma c_R)_{\mathbf{1''}},
\end{aligned}
\end{equation}
which we denote schematically by $(\overline{U}_R\Gamma U_R)_{\rho_R}$, with
$\rho_R\in\{\mathbf{1},\mathbf{1'},\mathbf{1''}\}$. The left-handed bilinear
is
\begin{equation}
(\overline{Q}_L \Gamma Q_L)_{\rho_L},\qquad
\rho_L\in\{\mathbf{1},\mathbf{1'},\mathbf{1''},\mathbf{3}_s,\mathbf{3}_a\}.
\end{equation}
A generic $A_4$-invariant structure has the form
\begin{equation}
\Bigl[(\overline{Q}_L \Gamma Q_L)_{\rho_L}\,
      (Y^{(6)}_{\mathbf{r}_q}{Y^{(6)}_{\mathbf{r}'_q}}^{*})_{\rho_M}\,
      (\overline{U}_R\Gamma U_R)_{\rho_R}\Bigr]_{\mathbf{1}},
\end{equation}
with $\rho_M$ the irrep of the modular coefficient. For a fixed right-handed
irrep $\rho_R$, the allowed $(\rho_L,\rho_M)$ are determined by the $A_4$
tensor rules.

For $\rho_R=\mathbf{1}$ one finds:
\begin{itemize}
  \item $\rho_L=\mathbf{1}$ with any of the $5$ modular singlets of type
        $\mathbf{1}$;
  \item $\rho_L=\mathbf{1'}$ with any of the $4$ modular non-trivial singlets
        of type $\mathbf{1''}$;
  \item $\rho_L=\mathbf{1''}$ with any of the $4$ modular non-trivial singlets
        of type $\mathbf{1'}$;
  \item $\rho_L=\mathbf{3}_s$ or $\mathbf{3}_a$ with any of the $12$ modular
        triplets, where for each modular triplet there are two possible
        contractions (symmetric/antisymmetric), summarized as $2\times 12$.
\end{itemize}
Thus, for $\rho_R=\mathbf{1}$ there are
\begin{equation}
5(\rho_L=\mathbf{1})
+4(\rho_L=\mathbf{1'})
+4(\rho_L=\mathbf{1''})
+2\times 12(\rho_L=\mathbf{3})
= 37
\end{equation}
independent $(\rho_L,\rho_M)$ combinations. By symmetry, the counting for
$\rho_R=\mathbf{1'}$ and $\rho_R=\mathbf{1''}$ is identical, so each also
gives $37$ possibilities:
\begin{equation}
N_{Q,Y}^{(\rho_R=\mathbf{1'})}
=N_{Q,Y}^{(\rho_R=\mathbf{1''})}
=37.
\end{equation}

For each choice of $\rho_R$ there are three independent flavor bilinears
(e.g.\ for $\rho_R=\mathbf{1}$, the three diagonal structures
$(\overline{u}_R\Gamma u_R)_\mathbf{1}$,
$(\overline{c}_R\Gamma c_R)_\mathbf{1}$,
$(\overline{t}_R\Gamma t_R)_\mathbf{1}$), giving a factor of $3$. Finally,
there are three one-dimensional irreps for the right-handed bilinear,
$\rho_R\in\{\mathbf{1},\mathbf{1'},\mathbf{1''}\}$.

Altogether, the total number of $A_4$-invariant SMEFT operators of each type
$\mathcal{O}_{qu}^{(1,8)}$ is
\begin{equation}
N_{qu} =
\underbrace{3}_{\rho_R=\mathbf{1},\mathbf{1'},\mathbf{1''}}
\times
\underbrace{37}_{\text{modular+left-handed sector}}
\times
\underbrace{3}_{f_u=u,c,t}
= 333.
\end{equation}

\item $\bigl[(\overline{q}q)(\overline{d}d)\bigr]_{\mathbf{1}}$: 
The corresponding SMEFT operators are $\mathcal{O}_{qd}^{(1)}$ and
$\mathcal{O}_{qd}^{(8)}$.
\begin{center}
    \begin{tabular}{|l|l|l|l|}
        \hline
        type &  & weight & $A_4$ invariant \\ \hline
        $[(\overline{q}q) (\overline{d}d)]_{\mathbf{1}}$ &
        $[\overline{Q}_L \Gamma Q_L][\overline{D}_R \Gamma D_R]$ &
        $2_q + 2_{qd}$ &
        $\bigl(Y^{(2)}_{\mathbf{r}_q}{Y^{(2)}_{\mathbf{r}'_q}}^*\bigr)
        [\overline{Q}_L \Gamma Q_L][\overline{D}_R \Gamma D_R]$ \\ \hline
    \end{tabular}
\end{center}
The construction again splits into a modular part and a SMEFT-bilinear part.

\paragraph{Modular forms:}

At weight $2$ we have the triplet $Y^{(2)}_{\mathbf{3}_q}$. Contracting a
modular form with its conjugate yields the singlet $\mathbf{1}$, the non-trivial
singlets $\mathbf{1'}$ and $\mathbf{1''}$, and the two triplets
$\mathbf{3}_s$ and $\mathbf{3}_a$. We denote these contractions by
\begin{equation}
(YY^\ast)_\rho \equiv \bigl[Y^{(2)}_{\mathbf{3}_q}\,{Y^{(2)}_{\mathbf{3}_q}}^\ast\bigr]_{\rho},
\qquad
\rho\in\{\mathbf{1},\mathbf{1'},\mathbf{1''},\mathbf{3}_s,\mathbf{3}_a\}.
\end{equation}

\paragraph{SMEFT operators:}

The right-handed down-type bilinears are
\begin{equation}
\begin{aligned}
& (\overline{d}_R \Gamma d_R)_\mathbf{1}, \quad
  (\overline{d}_R \Gamma s_R)_{\mathbf{1'}}, \quad
  (\overline{d}_R \Gamma b_R)_{\mathbf{1''}}, \\
& (\overline{s}_R \Gamma s_R)_\mathbf{1}, \quad
  (\overline{s}_R \Gamma b_R)_{\mathbf{1'}}, \quad
  (\overline{s}_R \Gamma d_R)_{\mathbf{1''}}, \\
& (\overline{b}_R \Gamma b_R)_\mathbf{1}, \quad
  (\overline{b}_R \Gamma d_R)_{\mathbf{1'}}, \quad
  (\overline{b}_R \Gamma s_R)_{\mathbf{1''}} ,
\end{aligned}
\end{equation}
while the left-handed bilinear decomposes as
\begin{equation}
(\overline{Q}_L \Gamma Q_L)_{\rho_Q},\qquad
\rho_Q\in\{\mathbf{1},\mathbf{1'},\mathbf{1''},\mathbf{3}_s,\mathbf{3}_a\}.
\end{equation}

Since $(\overline{D}_R\Gamma D_R)$ only appears in one-dimensional
representations, we first count the number of singlet contractions in the
combined modular--left-handed sector
\begin{equation}
(YY^\ast)\,(\overline{Q}_L\Gamma Q_L).
\end{equation}
For each chosen one-dimensional irrep
$\rho\in\{\mathbf{1},\mathbf{1'},\mathbf{1''}\}$, there are seven independent
ways to form an $A_4$ singlet from $(YY^\ast)_\rho$ and
$(\overline{Q}_L\Gamma Q_L)_{\rho_Q}$.
To make this explicit, consider the singlet case
$\rho=\mathbf{1}$; the seven structures can be written schematically as
\begin{equation}
\begin{aligned}
\text{(i)}\;& \bigl[(YY^\ast)_{\mathbf{1}}\,
               (\overline{Q}_L\Gamma Q_L)_{\mathbf{1}}\bigr]_{\mathbf{1}}, \\
\text{(ii)}\;& \bigl[(YY^\ast)_{\mathbf{1'}}\,
               (\overline{Q}_L\Gamma Q_L)_{\mathbf{1''}}\bigr]_{\mathbf{1}}, \\
\text{(iii)}\;& \bigl[(YY^\ast)_{\mathbf{1''}}\,
               (\overline{Q}_L\Gamma Q_L)_{\mathbf{1'}}\bigr]_{\mathbf{1}}, \\
\text{(iv)}\;& \bigl[(YY^\ast)_{\mathbf{3}_s}\,
               (\overline{Q}_L\Gamma Q_L)_{\mathbf{3}_s}\bigr]_{\mathbf{1}}, \\
\text{(v)}\;& \bigl[(YY^\ast)_{\mathbf{3}_s}\,
               (\overline{Q}_L\Gamma Q_L)_{\mathbf{3}_a}\bigr]_{\mathbf{1}}, \\
\text{(vi)}\;& \bigl[(YY^\ast)_{\mathbf{3}_a}\,
               (\overline{Q}_L\Gamma Q_L)_{\mathbf{3}_s}\bigr]_{\mathbf{1}}, \\
\text{(vii)}\;& \bigl[(YY^\ast)_{\mathbf{3}_a}\,
               (\overline{Q}_L\Gamma Q_L)_{\mathbf{3}_a}\bigr]_{\mathbf{1}} .
\end{aligned}
\end{equation}
The same counting holds for the non-trivial singlets
$\rho=\mathbf{1'}$ and $\rho=\mathbf{1''}$, giving seven independent
$(YY^\ast)(\overline{Q}_L\Gamma Q_L)$ singlets for each irrep.

We now attach the right-handed sector. For a fixed irrep $\rho$,
there are three flavor-diagonal down-type bilinears in that irrep.
For example, in the singlet case the allowed choices are
\begin{equation}
(\overline{d}_R\Gamma d_R)_{\mathbf{1}},\quad
(\overline{s}_R\Gamma s_R)_{\mathbf{1}},\quad
(\overline{b}_R\Gamma b_R)_{\mathbf{1}},
\end{equation}
so combining with the seven structures above yields
$3\times 7 = 21$ $A_4$-invariant flavor structures in $\rho=\mathbf{1}$.
The same reasoning applies to $\rho=\mathbf{1'}$ and $\rho=\mathbf{1''}$,
each contributing another $21$ structures. Therefore, the total number of
$A_4$-invariant operators of each type $\mathcal{O}_{qd}^{(1,8)}$ is
\begin{equation}
N_{qd}
=
\underbrace{3}_{\rho=\mathbf{1},\mathbf{1'},\mathbf{1''}}
\times
\underbrace{7}_{(YY^\ast)(\overline{Q}_L\Gamma Q_L)\ \text{singlets per }\rho}
\times
\underbrace{3}_{f_d=d,s,b}
= 63 .
\end{equation}
\item $\bigl[(\overline{\ell}\ell)(\overline{e}e)\bigr]_{\mathbf{1}}$: 
The corresponding SMEFT operator is $\mathcal{O}_{\ell e}$.
\begin{center}
    \begin{tabular}{|l|l|l|l|}
        \hline
        type &  & weight & $A_4$ invariant  \\ \hline
        $[(\overline{\ell}\ell)(\overline{e}e)]_{\mathbf{1}}$ &
        $[\overline{L}_L\Gamma L_L][\overline{E}_R\Gamma E_R]$ &
        $2_e+2_{ed}$ &
        $\bigl(Y^{(2)}_{\mathbf{r}_e}{Y^{(2)}_{\mathbf{r}'_e}}^*\bigr)
        [\overline{L}_L\Gamma L_L][\overline{E}_R\Gamma E_R]$  \\ \hline
    \end{tabular}
\end{center}
The construction is fully analogous to the $\mathcal{O}_{qd}$ case, with the
replacement
\begin{equation}
(Q_L,\; D_R,\; Y^{(2)}_{\mathbf{3}_q}(\tau_q))
\;\longrightarrow\;
(L_L,\; E_R,\; Y^{(2)}_{\mathbf{3}_e}(\tau_e)).
\end{equation}
We nevertheless spell out the counting explicitly for clarity.

\paragraph{Modular forms:}

At weight $2$ the lepton-sector modular form is the triplet
$Y^{(2)}_{\mathbf{3}_e}(\tau_e)$. Contracting it with its conjugate yields the singlet
$\mathbf{1}$, the non-trivial singlets $\mathbf{1'}$ and $\mathbf{1''}$, and the two triplets
$\mathbf{3}_s$ and $\mathbf{3}_a$. We denote these by
\begin{equation}
(YY^\ast)_\rho \equiv 
\bigl[Y^{(2)}_{\mathbf{3}_e}(\tau_e)\,{Y^{(2)}_{\mathbf{3}_e}}^\ast(\tau_e)\bigr]_{\rho},
\qquad
\rho\in\{\mathbf{1},\mathbf{1'},\mathbf{1''},\mathbf{3}_s,\mathbf{3}_a\}.
\end{equation}

\paragraph{SMEFT operators:}

The left-handed lepton bilinear decomposes as
\begin{equation}
(\overline{L}_L \Gamma L_L)_{\rho_L},\qquad
\rho_L\in\{\mathbf{1},\mathbf{1'},\mathbf{1''},\mathbf{3}_s,\mathbf{3}_a\},
\end{equation}
while the right-handed charged-lepton bilinears are
\begin{equation}
\begin{aligned}
& (\overline{e}_R \Gamma e_R)_\mathbf{1}, \quad
  (\overline{e}_R \Gamma \mu_R)_{\mathbf{1'}}, \quad
  (\overline{e}_R \Gamma \tau_R)_{\mathbf{1''}}, \\
& (\overline{\mu}_R \Gamma \mu_R)_\mathbf{1}, \quad
  (\overline{\mu}_R \Gamma \tau_R)_{\mathbf{1'}}, \quad
  (\overline{\mu}_R \Gamma e_R)_{\mathbf{1''}}, \\
& (\overline{\tau}_R \Gamma \tau_R)_\mathbf{1}, \quad
  (\overline{\tau}_R \Gamma e_R)_{\mathbf{1'}}, \quad
  (\overline{\tau}_R \Gamma \mu_R)_{\mathbf{1''}} .
\end{aligned}
\end{equation}

Since the right-handed bilinear contributes only one-dimensional irreps, we
first count singlet contractions in the modular--left-handed sector
\begin{equation}
(YY^\ast)\,(\overline{L}_L\Gamma L_L).
\end{equation}
For each one-dimensional irrep
$\rho\in\{\mathbf{1},\mathbf{1'},\mathbf{1''}\}$, there are seven independent
ways to form an $A_4$ singlet. In the singlet case
$\rho=\mathbf{1}$, these seven structures are
\begin{equation}
\begin{aligned}
\text{(i)}\;& \bigl[(YY^\ast)_{\mathbf{1}}\,
               (\overline{L}_L\Gamma L_L)_{\mathbf{1}}\bigr]_{\mathbf{1}}, \\
\text{(ii)}\;& \bigl[(YY^\ast)_{\mathbf{1'}}\,
               (\overline{L}_L\Gamma L_L)_{\mathbf{1''}}\bigr]_{\mathbf{1}}, \\
\text{(iii)}\;& \bigl[(YY^\ast)_{\mathbf{1''}}\,
               (\overline{L}_L\Gamma L_L)_{\mathbf{1'}}\bigr]_{\mathbf{1}}, \\
\text{(iv)}\;& \bigl[(YY^\ast)_{\mathbf{3}_s}\,
               (\overline{L}_L\Gamma L_L)_{\mathbf{3}_s}\bigr]_{\mathbf{1}}, \\
\text{(v)}\;& \bigl[(YY^\ast)_{\mathbf{3}_s}\,
               (\overline{L}_L\Gamma L_L)_{\mathbf{3}_a}\bigr]_{\mathbf{1}}, \\
\text{(vi)}\;& \bigl[(YY^\ast)_{\mathbf{3}_a}\,
               (\overline{L}_L\Gamma L_L)_{\mathbf{3}_s}\bigr]_{\mathbf{1}}, \\
\text{(vii)}\;& \bigl[(YY^\ast)_{\mathbf{3}_a}\,
               (\overline{L}_L\Gamma L_L)_{\mathbf{3}_a}\bigr]_{\mathbf{1}} .
\end{aligned}
\end{equation}
The same counting applies to the non-trivial singlets $\rho=\mathbf{1'}$ and $\rho=\mathbf{1''}$,
each providing seven independent singlets in the
$(YY^\ast)(\overline{L}_L\Gamma L_L)$ sector.

We then attach the right-handed charged-lepton bilinear. For each fixed
irrep $\rho$, there are three flavor-diagonal bilinears in that irrep. For
instance, in the singlet case these are
\begin{equation}
(\overline{e}_R\Gamma e_R)_{\mathbf{1}},\quad
(\overline{\mu}_R\Gamma \mu_R)_{\mathbf{1}},\quad
(\overline{\tau}_R\Gamma \tau_R)_{\mathbf{1}} .
\end{equation}
Thus the number of $A_4$-invariant flavor structures in each
$\rho=\mathbf{1},\mathbf{1'},\mathbf{1''}$ channel is $3\times 7=21$.

Collecting the three one-dimensional irreps, the total number of
$A_4$-invariant operators of each type $\mathcal{O}_{\ell e}$ is
\begin{equation}
N_{\ell e}
=
\underbrace{3}_{\rho=\mathbf{1},\mathbf{1'},\mathbf{1''}}
\times
\underbrace{7}_{(YY^\ast)(\overline{L}_L\Gamma L_L)\ \text{singlets per }\rho}
\times
\underbrace{3}_{f_e=e,\mu,\tau}
= 63 .
\end{equation}
\end{enumerate}

\item $(\overline{L}R)(\overline{L}R)+\text{h.c.}$: 
The structure $(\overline{L}R)(\overline{L}R)$ gives rise to two types of
operators. They can be organized as follows:
\begin{itemize}
    \item \textbf{Operators invariant under $A_{4e}\otimes A_{4q}$.}
    \begin{equation}
    \left[(\overline{\ell}e)_\mathbf{1}\,(\overline{q}u)_\mathbf{1}\right]_{\mathbf{1}}.
    \end{equation}

    \item \textbf{Operators invariant under $A_{4q}$.}
    \begin{equation}
    \left[(\overline{q}u)\,(\overline{q}d)_\mathbf{1}\right]_{\mathbf{1}}.
    \end{equation}
\end{itemize}

\begin{enumerate}
%------------------------------------------------
\item $[(\overline{\ell}e)_\mathbf{1}(\overline{q}u)_\mathbf{1}]_{\mathbf{1}}
      +\text{h.c.}$: 
The corresponding SMEFT operators are
$\mathcal{O}_{\ell equ}^{(1)}+\text{h.c.}$ and
$\mathcal{O}_{\ell equ}^{(3)}+\text{h.c.}$.
\begin{center}
    \begin{tabular}{|l|l|l|l|}
        \hline
        type &  & weight & $A_4$ invariant  \\ \hline
        $(\overline{\ell}e)_\mathbf{1}(\overline{q}u)_\mathbf{1}$ &
        $[\overline{L}_L\Gamma E_R][\overline{Q}_L\Gamma U_R]$ &
        $2_{ed}+6_{qd}$ &
        $\bigl({Y_{\mathbf{r}_e}^{(2)}}^{*}{Y_{\mathbf{r}_q}^{(6)}}^{*}\bigr)
        [\overline{L}_L\Gamma E_R][\overline{Q}_L\Gamma U_R]$  \\ \hline
    \end{tabular}
\end{center}

The $A_4$-invariant contraction factorizes into a lepton part and a quark
part,
\begin{equation}
\bigl[{Y_{e}^{(2)}}^{*}
      (\overline{L}_L\Gamma E_R)_{\mathbf{3}}\bigr]_{\mathbf{1}}\;
\bigl[{Y_{q}^{(6)}}^{*}
      (\overline{Q}_L\Gamma U_R)_{\mathbf{3}}\bigr]_{\mathbf{1}} .
\end{equation}
For a fixed lepton flavor $f_e\in\{e,\mu,\tau\}$ and up-quark flavor
$f_u\in\{u,c,t\}$, each mixed bilinear
$(\overline{L}_L\Gamma E_R)$ and $(\overline{Q}_L\Gamma U_R)$ forms a unique
$A_4$ triplet.  
Hence, in each sector there is only \emph{one} singlet contraction between
the modular triplet and the bilinear triplet.

The counting is therefore:
\begin{itemize}
    \item $1_{\,\ell}$ from the unique contraction
    ${Y_{\mathbf{3}}^{(2)}}^{*}(\tau_e)\otimes(\overline{L}_L\Gamma E_R)_{\mathbf{3}}\to\mathbf{1}$;
    \item $2_{\,q}$ from the two independent weight $6$ quark triplets
    $Y^{(6)}_{\mathbf{3}a}(\tau_q)$ and $Y^{(6)}_{\mathbf{3}b}(\tau_q)$, each
    giving one singlet with $(\overline{Q}_L\Gamma U_R)_{\mathbf{3}}$;
    \item $9_{\,\text{flavor}}=3_{E_R}\times 3_{U_R}$ from choosing
    $(f_e,f_u)$.
\end{itemize}
Altogether, the number of independent flavor structures is
\begin{equation}
N_{\ell equ}
=
\underbrace{1_{\,\ell}}_{\text{lepton triplet contraction}}
\times
\underbrace{2_{\,q}}_{\text{two quark triplet modular forms}}
\times
\underbrace{9_{\,\text{flavor}}}_{3_{E_R}\times 3_{U_R}}
=18 .
\end{equation}

%------------------------------------------------
\item $[(\overline{q}u)(\overline{q}d)]_{\mathbf{1}}+\text{h.c.}$:
The corresponding SMEFT operators are
$\mathcal{O}_{quqd}^{(1)}+\text{h.c.}$ and
$\mathcal{O}_{quqd}^{(3)}+\text{h.c.}$.
\begin{center}
    \begin{tabular}{|l|l|l|l|}
        \hline
        type &  & weight & $A_4$ invariant  \\ \hline
        $[(\overline{q}u)(\overline{q}d)]_{\mathbf{1}}$ &
        $[\overline{Q}_L\Gamma U_R][\overline{Q}_L\Gamma D_R]$ &
        $8_{qd}$ &
        ${Y_{\mathbf{r}_q}^{(8)}}^{*}
        [\overline{Q}_L\Gamma U_R][\overline{Q}_L\Gamma D_R]$  \\ \hline
    \end{tabular}
\end{center}

The invariant has the schematic form
\begin{equation}
\bigl[{Y_{\mathbf{r}_q}^{(8)}}^{*}\,
      (\overline{Q}_L\Gamma U_R)_{\mathbf{3}}\,
      (\overline{Q}_L\Gamma D_R)_{\mathbf{3}}\bigr]_{\mathbf{1}} .
\end{equation}
For fixed flavors $f_u\in\{u,c,t\}$ and $f_d\in\{d,s,b\}$, the two mixed
bilinears are both $A_4$ triplets.
Thus, before inserting the modular form, there are
three one-dimensional channels ($\mathbf{1},\mathbf{1'},\mathbf{1''}$) and
two triplet channels ($\mathbf{3}_s,\mathbf{3}_a$).
At weight $8$, the quark modular sector contains two independent triplets
$Y^{(8)}_{\mathbf{3}a}(\tau_q)$ and $Y^{(8)}_{\mathbf{3}b}(\tau_q)$ and three
independent singlets with representations $\{\mathbf{1},\mathbf{1'},\mathbf{1''}\}$.
For the fermionic operator part, the bilinears $[\overline{Q}_L\Gamma U_R]$ and 
$[\overline{Q}_L\Gamma D_R]$ each transform as (generically different) $A_{4q}$
triplets. Therefore, their tensor product decomposes according to the $A_4$
Clebsch--Gordan rules as
\begin{equation}
\bigl[(\overline{q}\Gamma f_u)_{\mathbf{3}}\,
      (\overline{q}\Gamma f_d)_{\mathbf{3}}\bigr]_{\rho},
\qquad
\rho\in\{\mathbf{1},\mathbf{1'},\mathbf{1''},\mathbf{3}_s,\mathbf{3}_a\},
\end{equation}
where $f_u\in\{u,c,t\}$ and $f_d\in\{d,s,b\}$ denote the up- and down-type quark
flavors, respectively.

The counting is therefore:
\begin{itemize}
    \item $3_{\,\text{singlet}}$ from matching the three weight $8$ modular singlets
    $\mathbf{1},\mathbf{1'},\mathbf{1''}$ with the corresponding fermion-sector singlet
    channels $\mathbf{1},\mathbf{1'},\mathbf{1''}$;
    \item $2_{\,\text{triplet}}\times 2_{\,Y^{(8)}_{\mathbf{3}}}$ from the fermion-sector
    triplet channels $\mathbf{3}_{s,a}$ combined with the two weight $8$ triplet modular forms;
    \item $9_{\,\text{flavor}}=3_{U_R}\times 3_{D_R}$ from choosing
    $(f_u,f_d)$.
\end{itemize}
Hence
\begin{equation}
N_{quqd}
=
\underbrace{\bigl[3_{\,\text{singlet}}
+2_{\,\text{triplet}}\times 2_{\,Y^{(8)}_{\mathbf{3}}}\bigr]}_{=7\ \text{contractions per }(f_u,f_d)}
\times
\underbrace{9_{\,\text{flavor}}}_{3_{U_R}\times 3_{D_R}}
=63 .
\end{equation}
\end{enumerate}

%=========================================================
\item $(\overline{L}R)(\overline{R}L)+\text{h.c.}$: 
The structure $(\overline{L}R)(\overline{R}L)$ gives rise to one operator class:
\begin{itemize}
    \item Type1: \textbf{Operators invariant under $A_{4e}\otimes A_{4q}$.}
    \begin{equation}
    \left[(\overline{\ell}e)_\mathbf{1}\,(\overline{d}q)_\mathbf{1}\right]_{\mathbf{1}}.
    \end{equation}
\end{itemize}

\begin{enumerate}
\item $[(\overline{\ell}e)_\mathbf{1}(\overline{d}q)_\mathbf{1}]_{\mathbf{1}}
      +\text{h.c.}$:
The corresponding SMEFT operator is
$\mathcal{O}_{\ell edq}+\text{h.c.}$.
\begin{center}
    \begin{tabular}{|l|l|l|l|}
        \hline
        type &  & weight & $A_4$ invariant  \\ \hline
        $(\overline{\ell}e)_\mathbf{1}(\overline{d}q)_\mathbf{1}$ &
        $[\overline{L}_L\Gamma E_R][\overline{D}_R\Gamma Q_L]$ &
        $2_{ed}+2_{qd}$ &
        $\bigl({Y_{\mathbf{r}_e}^{(2)}}^{*}{Y_{\mathbf{r}_q}^{(2)}}^{*}\bigr)
        [\overline{L}_L\Gamma E_R][\overline{D}_R\Gamma Q_L]$ \\ \hline
    \end{tabular}
\end{center}

The $A_4$-invariant contraction again factorizes,
\begin{equation}
\bigl[{Y_{e}^{(2)}}^{*}
      (\overline{L}_L\Gamma E_R)_{\mathbf{3}}\bigr]_{\mathbf{1}}\;
\bigl[{Y_{q}^{(2)}}^{*}
      (\overline{D}_R\Gamma Q_L)_{\mathbf{3}}\bigr]_{\mathbf{1}} .
\end{equation}
For fixed flavors $f_e\in\{e,\mu,\tau\}$ and $f_d\in\{d,s,b\}$, each mixed
bilinear forms a unique triplet, and each triplet contracts with the
corresponding weight $2$ modular triplet in a unique way. Therefore:
\begin{itemize}
    \item $1_{\,\ell}$ from
    ${Y_{e}^{(2)}}^{*}\otimes(\overline{L}_L\Gamma E_R)_{\mathbf{3}}\to\mathbf{1}$;
    \item $1_{\,q}$ from
    ${Y_{q}^{(2)}}^{*}\otimes(\overline{D}_R\Gamma Q_L)_{\mathbf{3}}\to\mathbf{1}$;
    \item $9_{\,\text{flavor}}=3_{E_R}\times 3_{D_R}$ from choosing
    $(f_e,f_d)$.
\end{itemize}
Hence the number of independent flavor structures is
\begin{equation}
N_{\ell edq}
=
\underbrace{1_{\,\ell}}_{\text{lepton triplet contraction}}
\times
\underbrace{1_{\,q}}_{\text{quark triplet contraction}}
\times
\underbrace{9_{\,\text{flavor}}}_{3_{E_R}\times 3_{D_R}}
=9 .
\end{equation}
\end{enumerate}
\end{enumerate}
\end{itemize}

\section{Holomorphic approach in non-holomorphic modular forms}
\label{sec:Maabeta_construction}
In our previous setup, we assumed that the Yukawa couplings are given by holomorphic modular forms. This assumption can be interpreted as the low-energy effective description of an underlying SUSY ultraviolet completion. However, in general non-SUSY theories, the holomorphic condition on modular forms does not necessarily need to be imposed. As a result, it is natural to consider non-holomorphic modular forms, among which polyharmonic Maa\ss\ forms provide a particularly relevant example.
%In the following, we will present the explicit structure of polyharmonic Maaß forms and consider the model proposed in~\cite{qu2024nonholomorphicmodularflavorsymmetry}. 
%As an extension by using our constructing method, we aim to generate dimension-5 modular-invariant operators and part of the dimension-6 ones.
In the following, we present the explicit structure of polyharmonic Maa\ss\ forms and discuss the model proposed in~\cite{qu2024nonholomorphicmodularflavorsymmetry}. As an extension of our construction method, we aim to generate modular-invariant operators at dimension $5$, as well as leptonic part of those at dimension $6$.

\subsection{Benchmark Standard Model with polyharmonic Maa\ss\ forms}
The key difference for non-holomorphic modular forms is that the holomorphy
condition is no longer imposed. Instead, one considers
automorphic forms of weight $k$ that satisfy the weight-$k$ hyperbolic
Laplacian equation
\begin{equation}
    \Delta_k\,Y(\tau)=0,
\end{equation}
where $\Delta_k$ is the weight-$k$ hyperbolic Laplacian,
\begin{equation}
    \Delta_k
    =
    -y^2\left(\frac{\partial^2}{\partial x^2}+\frac{\partial^2}{\partial y^2}\right)
    +ik y\left(\frac{\partial}{\partial x}+i\frac{\partial}{\partial y}\right)
    =
    -4y^2\frac{\partial}{\partial \tau}\frac{\partial}{\partial \overline{\tau}}
    +2ik y\frac{\partial}{\partial \overline{\tau}}.
\end{equation}
Here $\tau=x+iy$ with $y=\mathrm{Im}\,\tau>0$. Imposing the growth condition
\begin{equation}
    Y(\tau)=\mathcal{O}(y^{\alpha})
    \qquad \text{as } y\to +\infty,\ \text{uniformly in } x,
\end{equation}
one obtains a polyharmonic Maa\ss\ form. Its Fourier expansion can be
written as
\begin{equation}
    Y(\tau)
    =
    \sum_{\substack{n \in \frac{1}{N}\mathbb{Z} \\ n \geqslant 0}}
    c^{+}(n)\,q^n
    +c^{-}(0)\,y^{1-k}
    +\sum_{\substack{n \in \frac{1}{N}\mathbb{Z} \\ n < 0}}
    c^{-}(n)\,\Gamma(1-k,-4\pi n y)\,q^n,
\end{equation}
where $q\equiv e^{2\pi i\tau}$. Here $\Gamma(s,z)$ is the incomplete gamma
function~\cite{book:Ono},
\begin{equation}
    \Gamma(s,z)=\int_{z}^{+\infty}e^{-t}\,t^{s-1}\,dt.
\end{equation}
The space of polyharmonic Maa\ss\ forms of integer weight $k$ and level $N$
has dimension $\dim \mathcal{PH}_k(\Gamma(N))$. We denote a basis by
$f_i(\tau)$, with $i=1,2,\ldots,\dim \mathcal{PH}_k(\Gamma(N))$. These
functions can be organized into irreducible multiplets of the finite modular
group $\Gamma'_N=SL(2,\mathbb{Z})/\Gamma(N)$ for general integer weights $k$,
and of $\Gamma_N=SL(2,\mathbb{Z})/\pm\Gamma(N)$ for even modular weights.
These features are analogous to the holomorphic case.

%In this work, we focus exclusively on the case $N = 3$ with even integer weight, corresponding to $\Gamma_3 \cong  A_4$. A key feature of the polyharmonic Maaß forms is that for weights $k \geq 4$, they coincide with the usual holomorphic modular forms. For lower weights, specifically when $k \leq 2$, the modular forms can always be arranged into a singlet $Y^{(k_Y)}_{\mathbf{1}}$ and a triplet $Y^{(k_Y)}_{\mathbf{3}}$ of $A_4$. The details are summarized in the Table~\ref{tab:PolyharmonicMaaßForms} below.
In this work, we focus exclusively on the case $N=3$ with even integer
weight, corresponding to $\Gamma_3 \cong A_4$. A key feature of polyharmonic Maa\ss\
forms is that, for weights $k\geq 4$, they coincide with the usual
holomorphic modular forms. For lower weights, in particular for $k\leq 2$,
the modular forms can always be organized into an $A_4$ singlet
$Y^{(k_Y)}_{\mathbf{1}}$ and an $A_4$ triplet $Y^{(k_Y)}_{\mathbf{3}}$. The
relevant information is summarized in Table~\ref{tab:PolyharmonicMaaßForms}.

\begin{table}[h]
    \centering
    \begin{tabular}{|c|c|}
    \hline
    \textbf{Weight} $k_Y$ & \textbf{Polyharmonic Maaß forms} $Y_{\mathbf{r}}^{(k_Y)}$ \\
    \hline
    $k_Y = -4$ & $Y_{\mathbf{1}}^{(-4)},\quad Y_{\mathbf{3}}^{(-4)}$ \\
    \hline
    $k_Y = -2$ & $Y_\mathbf{1}^{(-2)},\quad Y_\mathbf{3}^{(-2)}$ \\
    \hline
    $k_Y = 0$ & $Y_\mathbf{1}^{(0)},\quad Y_\mathbf{3}^{(0)}$ \\
    \hline
    $k_Y = 2$ & $Y_\mathbf{1}^{(2)},\quad Y_{\mathbf{3}}^{(2)}$ \\
    \hline
    $k_Y = 4$ & $Y_\mathbf{1}^{(4)},\quad Y_{\mathbf{1'}}^{(4)},\quad Y_{3}^{(4)}$ \\
    \hline
    $k_Y = 6$ & $Y_\mathbf{1}^{(6)},\quad Y_{\mathbf{3}a}^{(6)},\quad Y_{\mathbf{3}b}^{(6)}$ \\
    \hline
    \end{tabular}
    \caption{Polyharmonic Maaß forms $Y_{\mathbf{r}}^{(k_Y)}$ at various modular weights $k_Y$. }
    \label{tab:PolyharmonicMaaßForms}
    \end{table}

\begin{table}[H]
    \centering
    \begin{tabular}{ |c|c|c|c|c| } 
    \hline
    & $L_L$ & $(e^c_R, \mu^c_R, \tau^c_R)$ & $H$ & $Y^{(k_Y)}_{\mathbf{r}}(\tau_e)$ \\ 
    \hline
    $SU(2)$ & $2$ & $1$ & $2$ & $1$ \\ 
    \hline
    $A_4$ & ${\mathbf{3}}$ & $(\mathbf{1}, \mathbf{1}^{\prime\prime}, \mathbf{1}^{\prime})$ & ${\mathbf{1}}$ & ${\mathbf{r}}$ \\ 
    \hline
    $k$ & $-2$ & $(0, 2, 2)$ & $0$ & $k_Y$ \\ 
    \hline
    \end{tabular}
    \caption{Benchmark models for lepton sector.}
    \label{tab:Benchmark_lepton}
\end{table}
%In our model, when the modular forms are taken to be polyharmonic Maa$\upbeta$ forms,
%the kinetic terms can still be written as
%\begin{equation}
%    \mathcal{L}_{\rm kinetic}
%    \supset
%\frac{\Lambda_0^2}{\left(-i\tau + i\bar{\tau}\right)^2}\,
%D'_\mu \bar{\tau}\,D'^{\mu}\tau
%+
%    \sum_{\psi}
%    \frac{1}{\big[i(\overline{\tau} - \tau)\big]^{k}}\,
%    ( i\,{\psi}^\dagger \,\slashed{D}' \psi ).
%    \label{eq:non_kinetic}
%\end{equation}
In benchmark model, when the modular forms are taken to be polyharmonic Maa\ss\
forms, the fermion kinetic terms can still be written as
\begin{equation}
\mathcal{L}_{\rm kinetic}
\supset
\sum_{\psi}
\frac{1}{\left\langle i\overline{\tau} - i\tau \right\rangle^{k_\psi}}\,
\big[ i\,{\psi}^\dagger \,\slashed{D}' \psi \big]_{\mathbf{1}}\,.
\label{eq:non_holo_kinetic2}
\end{equation}
The covariant derivative still takes the form given in Eq.~\ref{eq:d}. 
%If we neglect the kinetic term of $\tau$, we may still perform the field redefinition in Eq.~\ref{eq:redefine} to bring the fermion kinetic terms to canonical form.
Since polyharmonic Maa{\ss} modular forms can carry zero or negative modular weights, unlike the additional kinetic terms in the holomorphic setup, the kinetic corrections here can be written as
\begin{align}
\Delta \mathcal{L}_{\rm kinetic}
=&
\sum_{\psi,k,\mathbf r,\mathbf r'}
\alpha_{\psi,k,\mathbf r,\mathbf r'}\,
\big\langle i\bar\tau-i\tau \big\rangle^{k-k_{\psi}}\,
\Big[\big(Y^{(k)}{Y^{(k)}}^{*}\big)_{\mathbf r}\,
\big(i\,\psi^\dagger \slashed{D}'\psi\big)_{\mathbf r'}\Big]_{\mathbf 1}
\;+\;\mathrm{h.c.}
\nonumber\\[4pt]
&+
\sum_{\psi,k,\mathbf r,\mathbf r'}
\beta_{\psi,k,\mathbf r,\mathbf r'}\,
\big\langle i\bar\tau-i\tau \big\rangle^{-k_{\psi}}\,
\Big[\big(Y^{(-k)}Y^{(k)}\big)_{\mathbf r}\,
\big(i\,\psi^\dagger \slashed{D}'\psi\big)_{\mathbf r'}\Big]_{\mathbf 1}
\;+\;\mathrm{h.c.},
\end{align}
where $k$ can take even non-positive integer values. Here we still keep only the structures associated with the SM renormalizable kinetic terms without modular forms in Eq.~\ref{eq:non_holo_kinetic2}.
Meanwhile, the Yukawa sector is now constructed from the
polyharmonic Maa\ss\ forms listed in Table~\ref{tab:PolyharmonicMaaßForms}. Concretely, we
adopt the modular-weight assignments of the benchmark model in Ref.~\cite{qu2024nonholomorphicmodularflavorsymmetry}
and focus on the lepton sector only. The Yukawa interactions can then be written as
\begin{equation}
\mathcal{L}_e
=
\alpha\bigl(E_1^c L Y_{\mathbf{3}}^{(-2)} H^*\bigr)_{\mathbf{1}}
+\beta\bigl(E_2^c L Y_{\mathbf{3}}^{(0)} H^*\bigr)_{\mathbf{1}}
+\gamma\bigl(E_3^c L Y_{\mathbf{3}}^{(0)} H^*\bigr)_{\mathbf{1}}
+\text{h.c.}\,.
\label{yukawa_nonholo}
\end{equation}

\subsection{Non-holomorphic dimension-5 case}

%In this benchmark model, we continue to assume that the approach in holomorphic case would hold. In analogy with our previous discussion, the setup that we do not include the dynamics of \(\tau\) nor the additional terms induced by powers of \(\mathrm{Im}\,\tau\) can still be regarded as MFV-like, in the sense that flavor symmetry breaking is entirely controlled by the modular parameter $\tau$. However, in the present case the building blocks are no longer given by $Y_{\mathbf{3}}^{(2)}$, but instead by the two modular triplets $Y_{\mathbf{3}}^{(-2)}$ and $Y_{\mathbf{3}}^{(0)}$ appearing in the Yukawa sector.
In this benchmark model, we continue to assume that the approach in holomorphic case still holds. In analogy with our previous discussion, the setup in which we neglect the
dynamics of $\tau$, as well as the additional terms constructed by powers of $\mathrm{Im}\,\tau$ with modular forms,
can still be regarded as MFV-like, in the sense that the breaking of the flavor symmetry
is entirely controlled by the modular parameter $\tau$. In the present case, however, the
building blocks are no longer given by $Y_{\mathbf{3}}^{(2)}$, but instead by the two modular
triplets $Y_{\mathbf{3}}^{(-2)}$ and $Y_{\mathbf{3}}^{(0)}$ that appear in the Yukawa sector.

%The crucial difference with respect to the holomorphic construction discussed above is that these two triplets are group-theoretically independent: since the tensor product of two non-holomorphic modular forms $Y_{\mathbf{r}}^{(k)}$ and $Y_{\mathbf{r}'}^{(k')}$, with modular weights $k$ and $k'$, does not in general satisfy the Laplacian condition~\cite{qu2024nonholomorphicmodularflavorsymmetry},
%\begin{equation}
%    \Delta_{k+k'}\bigl(Y_{\mathbf{r}}^{(k)} \otimes %Y_{\mathbf{r}'}^{(k')}\bigr)_{\mathbf{1,1',1'',3}}
%    \neq 0 \, ,
%\end{equation}
%the combination $Y_{\mathbf{r}}^{(k)} \otimes Y_{\mathbf{r}'}^{(k')}$ is not itself a modular form of weight $k+k'$, i.e.
%\begin{equation}
%    [Y_{\mathbf{r}}^{(k)} \otimes Y_{\mathbf{r}'}^{(k')}]_\mathbf{1,1',1'',3}
%    \notin M_{k+k'}(\Gamma(3)) \, .
%\end{equation}
%Consequently, non-holomorphic modular forms cannot, in general, be generated by taking tensor powers of a single seed, and must instead be treated as independent building blocks.
The crucial difference with respect to the holomorphic construction discussed above is that
these two triplets are group-theoretically independent. This is because the tensor product of
two non-holomorphic modular forms $Y_{\mathbf{r}}^{(k)}$ and $Y_{\mathbf{r}'}^{(k')}$, with modular
weights $k$ and $k'$, does not in general satisfy the Laplacian condition~\cite{qu2024nonholomorphicmodularflavorsymmetry}:
\begin{equation}
\Delta_{k+k'}\Bigl(\bigl[Y_{\mathbf{r}}^{(k)} \otimes Y_{\mathbf{r}'}^{(k')}\bigr]_{\rho}\Bigr)\neq 0\,,
\qquad
\rho\in\{\mathbf{1},\mathbf{1'},\mathbf{1''},\mathbf{3}\}\,.
\end{equation}
Therefore, the projected product $\bigl[Y_{\mathbf{r}}^{(k)} \otimes Y_{\mathbf{r}'}^{(k')}\bigr]_{\rho}$ is not itself a
modular form of weight $k+k'$, i.e.
\begin{equation}
\bigl[Y_{\mathbf{r}}^{(k)} \otimes Y_{\mathbf{r}'}^{(k')}\bigr]_{\rho}\notin \mathcal{PH}_{k+k'}(\Gamma(3))\,,
\qquad
\rho\in\{\mathbf{1},\mathbf{1'},\mathbf{1''},\mathbf{3}\}\,.
\end{equation}
Consequently, non-holomorphic modular forms cannot, in general, be generated by taking tensor products repeatedly from a single seed;
instead, they must be treated as independent building blocks.

Assuming that the effective theory is built from the building blocks
$\{Y_{\mathbf{3}}^{(-2)},\,Y_{\mathbf{3}}^{(0)}\}$, and that the special holomorphic structure
associated with $Y_{\mathbf{3}}^{(2)}$ is absent, the Weinberg operator can be written
schematically as
\begin{equation}
    -\mathcal{L}_w=\sum_{n}\frac{g_n}{\Lambda^2}[Y_{\mathbf{3}}^{(-2)}\otimes Y_{\mathbf{3}}^{(-2)}\otimes (Y_{\mathbf{3}}^{(0)})_{{\text{sym}}^{\otimes n}}]_{\mathbf{3}}\otimes (LLH^2)_{\mathbf{3}},
\end{equation}
and
\begin{equation}
    -\mathcal{L}'_w=\sum_{n}\frac{g'_n}{\Lambda^2}  [Y_{\mathbf{3}}^{(-2)}\otimes Y_{\mathbf{3}}^{(-2)}\otimes (Y_{\mathbf{3}}^{(0)})_{{\text{sym}}^{\otimes n}}]_{\mathbf{1,1',1''}}\otimes (LLH^2)_{\mathbf{1,1',1''}}.
\end{equation}
In this construction, $n$ can take any non-negative integer value, so an infinite tower of
terms is generated. If we treat all polyharmonic Maa\ss\ forms as independent building blocks, while still assuming that their holomorphic subsector is generated by the weight-$2$ triplet $Y^{(2)}_{\mathbf{3}}$, then the Weinberg operator can be written schematically as
\begin{equation}
    -\mathcal{L}_w
    =\sum_{k}\sum_{n}\frac{g_n}{\Lambda^2}
    \Bigl[Y_{\mathbf{3}}^{(-2+k)}\otimes Y_{\mathbf{3}}^{(-2-k)}\otimes (Y_{\mathbf{3}}^{(0)})_{{\text{sym}}^{\otimes n}}\Bigr]_{\mathbf{3}}
    \otimes (LLH^2)_{\mathbf{3}}\,,
\end{equation}
and
\begin{equation} -\mathcal{L}'_w=\sum_{k}\sum_{n}\frac{g'_n}{\Lambda^2} [Y_{\mathbf{3}}^{(-2+k)}\otimes Y_{\mathbf{3}}^{(-2-k)}\otimes (Y_{\mathbf{3}}^{(0)})_{{\text{sym}}^{\otimes n}}]_{\mathbf{1,1',1''}}\otimes (LLH^2)_{\mathbf{1,1',1''}},
\end{equation}
which still introduces an infinite set of redundant structures.

Moreover, for higher-dimensional operators whose total modular weight is negative, one is
forced to insert additional modular forms with positive weight in order to build
modular-invariant combinations. For example, for the operator $\mathcal{O}_{He}$ the
bilinear $(\overline{\mu}_R \Gamma \mu_R)_{\mathbf{1}}$ can be promoted to a modular-invariant
structure in several ways, such as
\begin{equation}
\begin{aligned}
& \bigl[Y^{(2)}_{\mathbf{1}}\,{Y^{(2)}_{\mathbf{1}}}^{*}\bigr]_{\mathbf{1}}\,
  (\overline{\mu}_R\Gamma \mu_R)_{\mathbf{1}},\\
& \bigl[Y^{(2)}_{\mathbf{3}}\,{Y^{(2)}_{\mathbf{3}}}^{*}\bigr]_{\mathbf{1}}\,
  (\overline{\mu}_R\Gamma \mu_R)_{\mathbf{1}}.
\end{aligned}
\end{equation}
Note that the modular forms introduced in this way do not appear in the Yukawa sector, and
thus represent genuinely new structures in the effective theory.
In the non-holomorphic case, we therefore need to impose stronger assumptions on the
allowed higher-dimensional operator basis.

%In contrast, in the holomorphic case discussed above, the combined approach from holomorphic modular forms implies that higher-dimensional operators can always be organized in the form
%\begin{equation}
%    [\mathcal{O}(D,\phi,Y,Y^*)]_{\mathbf{1}}    =    \bigl[\,Y^{(k_Y)}_{s}\otimes {Y_{s}^{*}}^{(k_Y')} \otimes \mathcal{O}(D,\phi)\,\bigr]_{\mathbf{1}} \,.
%    \label{tensorproduct}
%\end{equation}
%\textbf{This assumption in the construction in non-holomorphic case does not follows that all flavour structures arise from the leading-order Yukawa terms. Otherwise, it means any comformal factor compensation be provided entirely by independent modular forms. So in non-holomorphic case we treat this rule as minimal assumption to avoid introducing large number of SMEFT operators due to the non-holomorphic modular forms' structure of the construction.}
In contrast, in the holomorphic case discussed above, the holomorphic modular-form construction implies that higher-dimensional operators can always be organized in the form
\begin{equation}
    [\mathcal{O}(D,\phi,Y,Y^*)]_{\mathbf{1}}
    =
    \bigl[\,Y^{(k_Y)}_{s}\otimes {Y_{s}^{*}}^{(k_Y')} \otimes \mathcal{O}(D,\phi)\,\bigr]_{\mathbf{1}} \,.
    \label{tensorproduct}
\end{equation}
In the non-holomorphic case, adopting Eq.~\eqref{tensorproduct} as an organizing principle does not imply that all flavour structures are generated by the leading-order Yukawa sector. Rather, the required automorphy-factor compensation can be provided by additional, a priori independent non-holomorphic modular forms. We therefore treat Eq.~\eqref{tensorproduct} as a minimal working assumption, which avoids an uncontrolled proliferation of independent SMEFT operators induced by the non-holomorphic modular-form building blocks. Otherwise, the number of independent modular-invariant structures would become infinite.

%\textbf{In the construction of Eq.~\eqref{tensorproduct}, we think that, when the modulus $\tau$ is treated as a field, the $D$ operator and the $\xi$ operator~\cite{book:Ono,qu2024nonholomorphicmodularflavorsymmetry} (also see in Appendix~\ref{app:DXi}) acting on modular forms have an additional physical interpretation like EOM or IBP. And in the holomorphic case this property with our minimal assumption leads to all holomorphic modular forms being written as tensor products of the basic modular form $Y_{\mathbf{3}}^{(2)}$, which allows us to regard $Y_{\mathbf{3}}^{(2)}$ as a building block and finally obtain the structural form of this construction; however, this structural form is not restricted to being holomorphic or non-holomorphic, so we make such construction assumption in the non-holomorphic construction. In other words, in this construction, in the holomorphic case the $D$ and $\xi$ operators are realised by treating $Y_{\mathbf{3}}^{(2)}$ as the building block, while in the non-holomorphic case they may rely on a different construction, but the final structural form is the same.}
%\textbf{Even if such math operators does not lead to this construction case, we still assume that under certain constraints the non-holomorphic case follows this construction in order to introduce fewer operators; otherwise, the number of operators would become very large or even infinite.}

Motivated by this minimal assumption, the dimension-$5$ Weinberg operator can be constructed in close analogy with the holomorphic case. Using the Hilbert series in Eq.~\eqref{Hilbert_series2}, we obtain the following explicit contributions:
\begin{equation}
    \begin{aligned}
    & {Y^{(-4)}_{\mathbf{1}}}^{*}\, (h^\dagger)^2\, (l^\dagger)^2
    + {Y^{(-4)}_{\mathbf{3}}}^{*}\, (h^\dagger)^2\, (l^\dagger)^2
    + Y^{(0)}_{\mathbf{3}}\, {Y^{(-4)}_{\mathbf{1}}}^{*}\, (h^\dagger)^2\, (l^\dagger)^2
    + 5\, Y^{(0)}_{\mathbf{3}}\, {Y^{(-4)}_{\mathbf{3}}}^{*}\, (h^\dagger)^2\, (l^\dagger)^2 \\[8pt]
    & + Y^{(-4)}_{\mathbf{1}}\, h^2\, l^2
    + Y^{(-4)}_{\mathbf{1}}\, {Y^{(0)}_{\mathbf{3}}}^{*}\, h^2\, l^2
    + Y^{(-4)}_{\mathbf{3}}\, h^2\, l^2
    + 5\, Y^{(-4)}_{\mathbf{3}}\, {Y^{(0)}_{\mathbf{3}}}^{*}\, h^2\, l^2 \, .
    \end{aligned}
\end{equation}
However, once non-holomorphic modular forms are included, the conjugates of the two weight-$0$ modular forms, $Y^{(0)}_{\mathbf{1}}$ and $Y^{(0)}_{\mathbf{3}}$, are identified with the original forms. For the singlet $Y^{(0)}_{\mathbf{1}}=1$, this is trivial.
For the triplet $Y^{(0)}_{\mathbf{3}}=(Y^{(0)}_1,\,Y^{(0)}_2,\,Y^{(0)}_3)^{T}$, one readily checks that
\begin{equation}
{Y^{(0)}_{1}}^{*} = Y^{(0)}_{1}, \qquad
{Y^{(0)}_{2}}^{*} = Y^{(0)}_{3}, \qquad
{Y^{(0)}_{3}}^{*} = Y^{(0)}_{2},
\end{equation}
so that
\begin{equation}
{Y^{(0)}_{\mathbf{3}}}^{*}
=
\begin{pmatrix}
{Y^{(0)}_{1}}^{*} \\
{Y^{(0)}_{3}}^{*} \\
{Y^{(0)}_{2}}^{*}
\end{pmatrix}
=
\begin{pmatrix}
Y^{(0)}_{1} \\
Y^{(0)}_{2} \\
Y^{(0)}_{3}
\end{pmatrix}
= Y^{(0)}_{\mathbf{3}} \, .
\end{equation}

%Since $Y^{(0)}_{\mathbf{3}}$ is its conjugate, it can only appear on its own in the construction in Eq.~(\ref{tensorproduct}), because it can itself play the role of both $Y^{(k_Y)}_{s}$ and ${Y_{s}^{*}}^{(k_Y')}$. It is therefore natural to conclude that $Y^{(0)}_{\mathbf{3}}$ should not generate new Yukawa structures through products with other modular forms. In the Hilbert series, the contributions that must be removed in accordance with this requirement are
%\begin{equation}
%    {Y^{(-4)}_{\mathbf{1}}}\,{Y^{(0)}_{\mathbf{3}}}^{*}\, h^2 l^2
%    + 5\, {Y^{(-4)}_{\mathbf{3}}}\,{Y^{(0)}_{\mathbf{3}}}^{*}\, h^2 l^2
%    + \text{h.c.}\, .
%    \label{eq:removed_terms}
%\end{equation}
%Under this requirement, the Weinberg operator can be built from the following flavor structures:
%\begin{equation}
%    \begin{aligned}
%        & Y^{(-4)}_{\mathbf{3}}\;[(L H)(L H)]_{\mathbf{3}} ,\\
%        & Y^{(-4)}_{\mathbf{1}}\;[(L H)(L H)]_{\mathbf{1}} .
%    \end{aligned}
%\end{equation}
%The same construction applies to the conjugate Weinberg terms. This result coincides with the Weinberg operator employed in Ref.~\cite{qu2024nonholomorphicmodularflavorsymmetry}. In the following, we continue to employ the tensor-product construction of Eq.~\eqref{tensorproduct} in order to build the dimension-6 operators in the benchmark model.
Since $Y^{(0)}_{\mathbf{3}}$ is self-conjugate, it can itself play the role of both $Y^{(k_Y)}_{s}$ and ${Y_{s}^{*}}^{(k_Y')}$.
Therefore it is natural to require that $Y^{(0)}_{\mathbf{3}}$ does not generate new Yukawa structures when multiplied by other modular forms with our minimal construction.

In the Hilbert series, the contributions that must be removed in order to implement this requirement are
\begin{equation}
{Y^{(-4)}_{\mathbf{1}}}\,{Y^{(0)}_{\mathbf{3}}}^{*}\, h^2 l^2
+ 5\, {Y^{(-4)}_{\mathbf{3}}}\,{Y^{(0)}_{\mathbf{3}}}^{*}\, h^2 l^2
+ \text{h.c.}\, .
\label{eq:removed_terms}
\end{equation}

With these terms removed, the Weinberg operator can be built from the two flavor structures
\begin{equation}
\begin{aligned}
& Y^{(-4)}_{\mathbf{3}}\;[(L H)(L H)]_{\mathbf{3}} ,\\
& Y^{(-4)}_{\mathbf{1}}\;[(L H)(L H)]_{\mathbf{1}} .
\end{aligned}
\end{equation}
The result agrees with the minimal Weinberg operator used in Ref.~\cite{qu2024nonholomorphicmodularflavorsymmetry}. In the following, we continue to adopt the tensor-product construction of Eq.~\eqref{tensorproduct} when building the dimension-$6$ operators in the benchmark model.

\subsection{Non-holomorphic dimension-6 case}

In this subsection, we continue to classify the dimension-6 operators
according to the number of fermion fields, following the method described
in Sec.~\ref{sec:dim-6}. As before, we restrict our attention to
lepton-number-conserving operators.

\begin{itemize}
    \item \textbf{Bilinear structures}

    Bilinear operators are grouped according to their modular-invariant
    flavor structures:

    \begin{itemize}
        \item Operators of the type \(\overline{L}R + \text{h.c.}\):

        \begin{equation}
            (\overline{L}e)_{\mathbf{1}} \, .
        \end{equation}

        As an illustrative example, we consider the SMEFT operator
        \(Q_{eH}\). There are three possibilities:

        \begin{itemize}
            \item \([\overline{L}_L \Gamma e_R]\) with modular weight
            \(-2_d\). The corresponding term in the Hilbert series is
            \begin{equation}
                e \, h^2 \, h^\dagger \, L^\dagger \,
                {Y^{(-2)}_{\mathbf{3}}}^{*} \, .
            \end{equation}
            The associated \(A_4\)-invariant structure can be written as
            \begin{equation}
                {Y^{(-2)}_{\mathbf{3}}}^{*}\,
                [\overline{L}_L \Gamma e_R]_{\mathbf{3}} \, .
            \end{equation}

            \item \([\overline{L}_L \Gamma \mu_R]\) with modular weight
            \(0\). The Hilbert-series contribution is
            \begin{equation}
                h^2 \, h^\dagger \, L^\dagger \, \mu \,
                Y^{(0)}_{\mathbf{3}} \, ,
            \end{equation}
            leading to the structure
            \begin{equation}
                Y^{(0)}_{\mathbf{3}}\,
                [\overline{L}_L \Gamma \mu_R]_{\mathbf{3}} \, .
            \end{equation}

            \item \([\overline{L}_L \Gamma \tau_R]\) with modular weight
            \(0\). The Hilbert-series term is
            \begin{equation}
                h^2 \, h^\dagger \, L^\dagger \, \tau \,
                Y^{(0)}_{\mathbf{3}} \, ,
            \end{equation}
            and its flavor structure is analogous to the
            \([\overline{L}_L \Gamma \mu_R]\) case.
        \end{itemize}

        \item Operators of the type \(\overline{R}R\):

        \begin{equation}
            (\overline{e}e)_{\mathbf{1}} \, .
        \end{equation}

        As an example, we consider the SMEFT operator \(Q_{He}\).
        There are five distinct cases:

        \begin{itemize}
            \item \([\overline{e}_R \Gamma e_R]\) with modular weight
            \(0\). The corresponding term in the Hilbert series is
            \begin{equation}
                e \, e^\dagger \, h \, h^\dagger \, D \, ,
            \end{equation}
            which gives the simple flavor structure
            \begin{equation}
                [\overline{e}_R \Gamma e_R]_{\mathbf{1}} \, .
            \end{equation}

            \item \([\overline{\mu}_R \Gamma \mu_R]\) with modular weight
            \(-2 + (-2_d)\). The Hilbert-series contribution reads
            \begin{equation}
                h \, h^\dagger \, \mu \, \mu^\dagger \, D \,
                Y^{(2)}_{\mathbf{1}}\, {Y^{(2)}_{\mathbf{1}}}^{*}
                + h \, h^\dagger \, \mu \, \mu^\dagger \, D \,
                Y^{(2)}_{\mathbf{3}}\, {Y^{(2)}_{\mathbf{3}}}^{*} \, .
            \end{equation}
            The corresponding modular- and flavor-invariant combinations are
            \begin{equation}
                \begin{aligned}
                    &\bigl[ Y^{(2)}_{\mathbf{1}} \,
                    {Y^{(2)}_{\mathbf{1}}}^{*} \bigr]_{\mathbf{1}}\,
                    [\overline{\mu}_R \Gamma \mu_R]_{\mathbf{1}} \, ,\\[4pt]
                    &\bigl[ Y^{(2)}_{\mathbf{3}} \,
                    {Y^{(2)}_{\mathbf{3}}}^{*} \bigr]_{\mathbf{1}}\,
                    [\overline{\mu}_R \Gamma \mu_R]_{\mathbf{1}} \, .
                \end{aligned}
            \end{equation}

            \item \([\overline{\tau}_R \Gamma \tau_R]\) with modular weight
            \(-2 + (-2_d)\). The Hilbert-series terms are
            \begin{equation}
                h \, h^\dagger \, D \, \tau \, \tau^\dagger \,
                Y^{(2)}_{\mathbf{1}}\, {Y^{(2)}_{\mathbf{1}}}^{*}
                + h \, h^\dagger \, D \, \tau \, \tau^\dagger \,
                Y^{(2)}_{\mathbf{3}}\, {Y^{(2)}_{\mathbf{3}}}^{*} \, ,
            \end{equation}
            and their invariant contractions follow exactly the same pattern
            as in the \([\overline{\mu}_R \Gamma \mu_R]\) case.

            \item \([\overline{\mu}_R \Gamma \tau_R] + \text{h.c.}\) with
            modular weight \(-2 + (-2_d)\). The corresponding Hilbert-series
            term is
            \begin{equation}
                h \, h^\dagger \, \mu^\dagger \, D \, \tau \,
                Y^{(2)}_{\mathbf{3}}\, {Y^{(2)}_{\mathbf{3}}}^{*}
                + \text{h.c.}
            \end{equation}
            For \([\overline{\mu}_R \Gamma \tau_R]\), one possible
            invariant structure is
            \begin{equation}
                \bigl[ Y^{(2)}_{\mathbf{3}}\, {Y^{(2)}_{\mathbf{3}}}^{*}
                \bigr]_{\mathbf{1''}}\,
                [\overline{\mu}_R \Gamma \tau_R]_{\mathbf{1'}} \, ,
            \end{equation}
            while its conjugate term takes the form
            \begin{equation}
                \bigl[ Y^{(2)}_{\mathbf{3}}\, {Y^{(2)}_{\mathbf{3}}}^{*}
                \bigr]_{\mathbf{1'}}\,
                [\overline{\tau}_R \Gamma \mu_R]_{\mathbf{1''}} \, .
            \end{equation}
        \end{itemize}

        \item Operators of the type \(\overline{L}L\):

        \begin{equation}
            (\overline{L}L)_{\mathbf{1}} \, .
        \end{equation}

        As a representative example, we consider the SMEFT operator
        \(Q_{Hl}^{(1)}\). The Hilbert series for this class of operators
        can be written as
        \begin{equation}
            \begin{aligned}
            & h \, h^\dagger \, L \, L^\dagger \, D \,
              {Y^{(-2)}_{\mathbf{1}}}^{*} \, Y^{(-2)}_{\mathbf{1}}
            + 2 \, h \, h^\dagger \, L \, L^\dagger \, D \,
              {Y^{(-2)}_{\mathbf{3}}}^{*} \, Y^{(-2)}_{\mathbf{1}} \\
            & + 2 \, h \, h^\dagger \, L \, L^\dagger \, D \,
              {Y^{(-2)}_{\mathbf{1}}}^{*} \, Y^{(-2)}_{\mathbf{3}}
            + 7 \, h \, h^\dagger \, L \, L^\dagger \, D \,
              {Y^{(-2)}_{\mathbf{3}}}^{*} \, Y^{(-2)}_{\mathbf{3}} \, .
            \end{aligned}
        \end{equation}
        These contributions can be organized into explicit
        \(A_4\)-invariant structures as
        \begin{equation}
            \begin{aligned}
                \left[{Y^{(-2)}_\mathbf{1}}^*  Y^{(-2)}_\mathbf{1}\right]\left[\overline{L}_L\Gamma L_L\right]_{({\mathbf{1}},{\mathbf{1}})} ,&
                \left[{Y^{(-2)}_\mathbf{1}}^*  Y^{(-2)}_\mathbf{3}\right]\left[\overline{L}_L\Gamma L_L\right]_{({{\mathbf{3}},\mathbf{3}_\mathbf{s}})} ,\;\\
                \left[{Y^{(-2)}_\mathbf{1}}^*  Y^{(-2)}_\mathbf{3}\right]\left[\overline{L}_L\Gamma L_L\right]_{({{\mathbf{3}},\mathbf{3}_\mathbf{a}})} ,&
                \left[{Y^{(-2)}_\mathbf{3}}^*  Y^{(-2)}_\mathbf{1}\right]\left[\overline{L}_L\Gamma L_L\right]_{({{\mathbf{3}},\mathbf{3}_\mathbf{s}})} ,\;\\
                \left[{Y^{(-2)}_\mathbf{3}}^*  Y^{(-2)}_\mathbf{1}\right]\left[\overline{L}_L\Gamma L_L\right]_{({{\mathbf{3}},\mathbf{3}_\mathbf{a}})} ,&
                \left[{Y^{(-2)}_\mathbf{3}}^*  Y^{(-2)}_\mathbf{3}\right]\left[\overline{L}_L\Gamma L_L\right]_{({{\mathbf{1}},\mathbf{1}})} ,\\
                \left[{Y^{(-2)}_\mathbf{3}}^*  Y^{(-2)}_\mathbf{3}\right]\left[\overline{L}_L\Gamma L_L\right]_{({{\mathbf{1'}},\mathbf{1''}})} ,&
                \left[{Y^{(-2)}_\mathbf{3}}^*  Y^{(-2)}_\mathbf{3}\right]\left[\overline{L}_L\Gamma L_L\right]_{({{\mathbf{1''}},\mathbf{1'}})} ,\\
                \left[{Y^{(-2)}_\mathbf{3}}^*  Y^{(-2)}_\mathbf{3}\right]\left[\overline{L}_L\Gamma L_L\right]_{({{\mathbf{3}_\mathbf{s}},\mathbf{3}_\mathbf{s}})} ,\;&
                \left[{Y^{(-2)}_\mathbf{3}}^*  Y^{(-2)}_\mathbf{3}\right]\left[\overline{L}_L\Gamma L_L\right]_{({{\mathbf{3}_\mathbf{s}},\mathbf{3}_\mathbf{a}})} ,\\
                \left[{Y^{(-2)}_\mathbf{3}}^*  Y^{(-2)}_\mathbf{3}\right]\left[\overline{L}_L\Gamma L_L\right]_{({{\mathbf{3}_\mathbf{a}},\mathbf{3}_\mathbf{s}})} ,\;&
                \left[{Y^{(-2)}_\mathbf{3}}^*  Y^{(-2)}_\mathbf{3}\right]\left[\overline{L}_L\Gamma L_L\right]_{({{\mathbf{3}_\mathbf{a}},\mathbf{3}_\mathbf{a}})} .
            \end{aligned}
        \end{equation}   
    \end{itemize}

    \item \textbf{All Quadrilinear Structures}:

Quadrilinear operators are classified in complete analogy, according to
their flavor and modular structures:

\begin{itemize}
    \item Operators of the type \((\overline{L}L)(\overline{L}L)\):
    \begin{equation}
        \bigl[(\overline{L}L)(\overline{L}L)\bigr]_{\mathbf{1}} \, .
    \end{equation}
    In this case, the Hilbert series contains the terms
    \begin{equation}
        5\, L^2 (L^\dagger)^2 \,{Y^{(-4)}_{\mathbf{1}}}^{*} Y^{(-4)}_{\mathbf{1}}
        + 10\, L^2 (L^\dagger)^2 \,{Y^{(-4)}_{\mathbf{3}}}^{*} Y^{(-4)}_{\mathbf{1}}
        + 10\, L^2 (L^\dagger)^2 \,{Y^{(-4)}_{\mathbf{1}}}^{*} Y^{(-4)}_{\mathbf{3}}
        + 35\, L^2 (L^\dagger)^2 \,{Y^{(-4)}_{\mathbf{3}}}^{*} Y^{(-4)}_{\mathbf{3}} \, .
    \end{equation}
    The explicit construction of
    \([\overline{L}_L \Gamma L_L][\overline{L}_L \Gamma L_L]\)
    has already been discussed in Sec.~\ref{sec:dim-6}. In particular,
    for each singlet representation \(\mathbf{1}\), \(\mathbf{1'}\),
    \(\mathbf{1''}\) there are five distinct contractions, while for the
    triplet representation there are ten.

    On the modular side, the combination
    \({Y^{(-4)}_{\mathbf{1}}}^{*} Y^{(-4)}_{\mathbf{1}}\) can only form
    a singlet when tensored with
    \(\bigl[[\overline{L}_L \Gamma L_L] \otimes [\overline{L}_L \Gamma L_L]\bigr]_{\mathbf{1}}\).
    In contrast, \({Y^{(-4)}_{\mathbf{3}}}^{*} Y^{(-4)}_{\mathbf{1}}\) and
    \({Y^{(-4)}_{\mathbf{1}}}^{*} Y^{(-4)}_{\mathbf{3}}\) can form singlets
    only when contracted with
    \(\bigl[[\overline{L}_L \Gamma L_L] \otimes [\overline{L}_L \Gamma L_L]\bigr]_{\mathbf{3}}\),
    and there are ten triplet structures in each case. Finally, the
    tensor product
    \({Y^{(-4)}_{\mathbf{3}}}^{*} \otimes Y^{(-4)}_{\mathbf{3}}\) has the
    same decomposition pattern as
    \({Y^{(4)}_{\mathbf{3}}}^{*} \otimes Y^{(4)}_{\mathbf{3}}\), leading to
    35 possible contractions, exactly as in the holomorphic construction.

    \item Operators of the type \((\overline{R}R)(\overline{R}R)\):
    \begin{equation}
        \bigl[(\overline{e}e)(\overline{e}e)\bigr]_{\mathbf{1}} \, .
    \end{equation}
    Using the condition~\eqref{eq:invariant_condition}, we separate
    the operators according to the modular weights \(k_Y\) and \(k'_Y\).

    \begin{itemize}
        \item \(\boldsymbol{k_Y = 0}\) and \(\boldsymbol{k'_Y = 0}\).  

        The relevant SMEFT operator is
        \([\overline{e}_R \Gamma e_R]_{\mathbf{1}}
         [\overline{e}_R \Gamma e_R]_{\mathbf{1}}\).
        The Hilbert series yields
        \begin{equation}
            e^2 (e^\dagger)^2 \, ,
        \end{equation}
        corresponding to the invariant structure
        \begin{equation}
            [\overline{e}_R \Gamma e_R]_{\mathbf{1}}\,
            [\overline{e}_R \Gamma e_R]_{\mathbf{1}} \, .
        \end{equation}

        \item \(\boldsymbol{k_Y = 2}\) and \(\boldsymbol{k'_Y = 2}\).  

        In this case, one right-handed field in \(\overline{E}_R\) must be
        \(\overline{\mu}_R\) or \(\overline{\tau}_R\), while the other is
        \(\overline{e}_R\); the same flavor assignment applies to \(E_R\).
        The Hilbert series contains
        \begin{equation}
        \begin{aligned}
            & e\, e^\dagger \,\mu\, \tau^\dagger
              \,Y^{(2)}_{\mathbf{3}}\, {Y^{(2)}_{\mathbf{3}}}^{*}
            + e\, e^\dagger \,\mu\, \mu^\dagger
              \,Y^{(2)}_{\mathbf{1}}\, {Y^{(2)}_{\mathbf{1}}}^{*}
            + e\, e^\dagger \,\mu\, \mu^\dagger
              \,Y^{(2)}_{\mathbf{3}}\, {Y^{(2)}_{\mathbf{3}}}^{*} \\
            & + e\, e^\dagger \,\tau\, \tau^\dagger
              \,Y^{(2)}_{\mathbf{1}}\, {Y^{(2)}_{\mathbf{1}}}^{*}
            + e\, e^\dagger \,\tau\, \tau^\dagger
              \,Y^{(2)}_{\mathbf{3}}\, {Y^{(2)}_{\mathbf{3}}}^{*}
            + e\, e^\dagger \,\mu^\dagger \tau
              \,Y^{(2)}_{\mathbf{3}}\, {Y^{(2)}_{\mathbf{3}}}^{*} \, .
        \end{aligned}
        \end{equation}
        The corresponding invariant structures can be written as
        \begin{equation}
        \begin{aligned}
            &\bigl[ Y^{(2)}_{\mathbf{3}} {Y^{(2)}_{\mathbf{3}}}^{*} \bigr]_{\mathbf{1'}}\,
             \bigl[(\overline{e}_R \Gamma e_R)_{\mathbf{1}}
                   (\overline{\tau}_R \Gamma \mu_R)_{\mathbf{1''}}\bigr]_{\mathbf{1''}} ,\\
            &\bigl[ Y^{(2)}_{\mathbf{3}} {Y^{(2)}_{\mathbf{3}}}^{*} \bigr]_{\mathbf{1''}}\,
             \bigl[(\overline{e}_R \Gamma e_R)_{\mathbf{1}}
                   (\overline{\mu}_R \Gamma \tau_R)_{\mathbf{1'}}\bigr]_{\mathbf{1'}} ,\\
            &\bigl[ Y^{(2)}_{\mathbf{1}} {Y^{(2)}_{\mathbf{1}}}^{*} \bigr]_{\mathbf{1}}\,
             \bigl[(\overline{e}_R \Gamma e_R)_{\mathbf{1}}
                   (\overline{\mu}_R \Gamma \mu_R)_{\mathbf{1}}\bigr]_{\mathbf{1}} ,\\
            &\bigl[ Y^{(2)}_{\mathbf{3}} {Y^{(2)}_{\mathbf{3}}}^{*} \bigr]_{\mathbf{1}}\,
             \bigl[(\overline{e}_R \Gamma e_R)_{\mathbf{1}}
                   (\overline{\mu}_R \Gamma \mu_R)_{\mathbf{1}}\bigr]_{\mathbf{1}} ,\\
            &\bigl[ Y^{(2)}_{\mathbf{1}} {Y^{(2)}_{\mathbf{1}}}^{*} \bigr]_{\mathbf{1}}\,
             \bigl[(\overline{e}_R \Gamma e_R)_{\mathbf{1}}
                   (\overline{\tau}_R \Gamma \tau_R)_{\mathbf{1}}\bigr]_{\mathbf{1}} ,\\
            &\bigl[ Y^{(2)}_{\mathbf{3}} {Y^{(2)}_{\mathbf{3}}}^{*} \bigr]_{\mathbf{1}}\,
             \bigl[(\overline{e}_R \Gamma e_R)_{\mathbf{1}}
                   (\overline{\tau}_R \Gamma \tau_R)_{\mathbf{1}}\bigr]_{\mathbf{1}} .
        \end{aligned}
        \end{equation}

        \item \(\boldsymbol{k_Y = 4}\) and \(\boldsymbol{k'_Y = 2}\).  

        Here one right-handed field must be \(e_R\), while the remaining
        fields carry flavors different from \(e_R\) and \(\overline{e}_R\).
        The Hilbert series gives
        \begin{equation}
        \begin{aligned}
            & e\, \mu\, (\mu^\dagger)^2\,
              Y^{(4)}_{\mathbf{1'}}\, {Y^{(2)}_{\mathbf{1}}}^{*}
            + e\, \mu\, (\mu^\dagger)^2\,
              Y^{(4)}_{\mathbf{3}}\, {Y^{(2)}_{\mathbf{3}}}^{*}
            + e\, (\mu^\dagger)^2 \tau\,
              Y^{(4)}_{\mathbf{1}}\, {Y^{(2)}_{\mathbf{1}}}^{*}
            + e\, (\mu^\dagger)^2 \tau\,
              Y^{(4)}_{\mathbf{3}}\, {Y^{(2)}_{\mathbf{3}}}^{*} \\
            & + e\, \mu\, (\tau^\dagger)^2\,
              Y^{(4)}_{\mathbf{1}}\, {Y^{(2)}_{\mathbf{1}}}^{*}
            + e\, \mu\, (\tau^\dagger)^2\,
              Y^{(4)}_{\mathbf{3}}\, {Y^{(2)}_{\mathbf{3}}}^{*}
            + e\, \tau\, (\tau^\dagger)^2\,
              Y^{(4)}_{\mathbf{3}}\, {Y^{(2)}_{\mathbf{3}}}^{*} \, .
        \end{aligned}
        \end{equation}
        The corresponding invariant contractions are
        \begin{equation}
        \begin{aligned}
            &\bigl[ Y^{(4)}_{\mathbf{1'}} {Y^{(2)}_{\mathbf{1}}}^{*} \bigr]_{\mathbf{1'}}\,
             \bigl[(\overline{\mu}_R \Gamma \mu_R)_{\mathbf{1}}
                   (\overline{\mu}_R \Gamma e_R)_{\mathbf{1''}}\bigr]_{\mathbf{1''}} ,\\
            &\bigl[ Y^{(4)}_{\mathbf{3}} {Y^{(2)}_{\mathbf{3}}}^{*} \bigr]_{\mathbf{1'}}\,
             \bigl[(\overline{\mu}_R \Gamma \mu_R)_{\mathbf{1}}
                   (\overline{\mu}_R \Gamma e_R)_{\mathbf{1''}}\bigr]_{\mathbf{1''}} ,\\
            &\bigl[ Y^{(4)}_{\mathbf{1}} {Y^{(2)}_{\mathbf{1}}}^{*} \bigr]_{\mathbf{1}}\,
             \bigl[(\overline{\mu}_R \Gamma e_R)_{\mathbf{1''}}
                   (\overline{\mu}_R \Gamma \tau_R)_{\mathbf{1'}}\bigr]_{\mathbf{1}} ,\\
            &\bigl[ Y^{(4)}_{\mathbf{3}} {Y^{(2)}_{\mathbf{3}}}^{*} \bigr]_{\mathbf{1}}\,
             \bigl[(\overline{\mu}_R \Gamma e_R)_{\mathbf{1''}}
                   (\overline{\mu}_R \Gamma \tau_R)_{\mathbf{1'}}\bigr]_{\mathbf{1}} ,\\
            &\bigl[ Y^{(4)}_{\mathbf{1}} {Y^{(2)}_{\mathbf{1}}}^{*} \bigr]_{\mathbf{1}}\,
             \bigl[(\overline{\tau}_R \Gamma e_R)_{\mathbf{1'}}
                   (\overline{\tau}_R \Gamma \mu_R)_{\mathbf{1''}}\bigr]_{\mathbf{1}} ,\\
            &\bigl[ Y^{(4)}_{\mathbf{3}} {Y^{(2)}_{\mathbf{3}}}^{*} \bigr]_{\mathbf{1}}\,
             \bigl[(\overline{\tau}_R \Gamma e_R)_{\mathbf{1'}}
                   (\overline{\tau}_R \Gamma \mu_R)_{\mathbf{1''}}\bigr]_{\mathbf{1}} ,\\
            &\bigl[ Y^{(4)}_{\mathbf{3}} {Y^{(2)}_{\mathbf{3}}}^{*} \bigr]_{\mathbf{1''}}\,
             \bigl[(\overline{\tau}_R \Gamma \tau_R)_{\mathbf{1}}
                   (\overline{\tau}_R \Gamma e_R)_{\mathbf{1'}}\bigr]_{\mathbf{1}} .
        \end{aligned}
        \end{equation}

        \item \(\boldsymbol{k_Y = 2}\) and \(\boldsymbol{k'_Y = 4}\).  

        This case is simply the complex conjugate of the previous one,
        \(k_Y = 4\), \(k'_Y = 2\).

        \item \(\boldsymbol{k_Y = 4}\) and \(\boldsymbol{k'_Y = 4}\).  

        Here none of the right-handed fields may carry flavor \(e_R\) or
        \(\overline{e}_R\). The Hilbert series reads
        \begin{equation}
        \begin{aligned}
            & \mu\, (\mu^\dagger)^2 \tau\,
              Y^{(4)}_{\mathbf{1'}} {Y^{(4)}_{\mathbf{1'}}}^{*}
            + \mu\, (\mu^\dagger)^2 \tau\,
              Y^{(4)}_{\mathbf{3}} {Y^{(4)}_{\mathbf{3}}}^{*} \\
            & + \mu^2 \mu^\dagger \tau^\dagger\,
              Y^{(4)}_{\mathbf{1'}} {Y^{(4)}_{\mathbf{1}}}^{*}
            + \mu^2 \mu^\dagger \tau^\dagger\,
              Y^{(4)}_{\mathbf{1}} {Y^{(4)}_{\mathbf{1'}}}^{*}
            + \mu^2 \mu^\dagger \tau^\dagger\,
              Y^{(4)}_{\mathbf{3}} {Y^{(4)}_{\mathbf{3}}}^{*} \\
            & + \mu \tau (\tau^\dagger)^2\,
              Y^{(4)}_{\mathbf{1'}} {Y^{(4)}_{\mathbf{1}}}^{*}
            + \mu \tau (\tau^\dagger)^2\,
              Y^{(4)}_{\mathbf{1}} {Y^{(4)}_{\mathbf{1'}}}^{*}
            + \mu \tau (\tau^\dagger)^2\,
              Y^{(4)}_{\mathbf{3}} {Y^{(4)}_{\mathbf{3}}}^{*} \\
            & + \mu^\dagger \tau^2 \tau^\dagger\,
              Y^{(4)}_{\mathbf{1'}} {Y^{(4)}_{\mathbf{1'}}}^{*}
            + \mu^\dagger \tau^2 \tau^\dagger\,
              Y^{(4)}_{\mathbf{3}} {Y^{(4)}_{\mathbf{3}}}^{*} \\
            & + (\mu^\dagger)^2 \tau^2\,
              Y^{(4)}_{\mathbf{1'}} {Y^{(4)}_{\mathbf{1}}}^{*}
            + (\mu^\dagger)^2 \tau^2\,
              Y^{(4)}_{\mathbf{1}} {Y^{(4)}_{\mathbf{1'}}}^{*}
            + (\mu^\dagger)^2 \tau^2\,
              Y^{(4)}_{\mathbf{3}} {Y^{(4)}_{\mathbf{3}}}^{*} \\
            & + \mu^2 (\tau^\dagger)^2\,
              Y^{(4)}_{\mathbf{1'}} {Y^{(4)}_{\mathbf{1'}}}^{*}
            + \mu^2 (\tau^\dagger)^2\,
              Y^{(4)}_{\mathbf{3}} {Y^{(4)}_{\mathbf{3}}}^{*} \\
            & + \mu \mu^\dagger \tau \tau^\dagger\,
              Y^{(4)}_{\mathbf{1}} {Y^{(4)}_{\mathbf{1}}}^{*}
            + \mu \mu^\dagger \tau \tau^\dagger\,
              Y^{(4)}_{\mathbf{3}} {Y^{(4)}_{\mathbf{3}}}^{*} \, .
        \end{aligned}
        \end{equation}
        The corresponding invariant structures are
        \begin{equation}
        \begin{aligned}
            &\bigl[ Y^{(4)}_{\mathbf{1'}} {Y^{(4)}_{\mathbf{1'}}}^{*} \bigr]_{\mathbf{1''}}\,
             \bigl[(\overline{\mu}_R \Gamma \mu_R)_{\mathbf{1}}
                   (\overline{\mu}_R \Gamma \tau_R)_{\mathbf{1'}}\bigr]_{\mathbf{1'}} ,\\
            &\bigl[ Y^{(4)}_{\mathbf{3}} {Y^{(4)}_{\mathbf{3}}}^{*} \bigr]_{\mathbf{1''}}\,
             \bigl[(\overline{\mu}_R \Gamma \mu_R)_{\mathbf{1}}
                   (\overline{\mu}_R \Gamma \tau_R)_{\mathbf{1'}}\bigr]_{\mathbf{1'}} ,\\
            &\bigl[ Y^{(4)}_{\mathbf{1'}} {Y^{(4)}_{\mathbf{1}}}^{*} \bigr]_{\mathbf{1'}}\,
             \bigl[(\overline{\mu}_R \Gamma \mu_R)_{\mathbf{1}}
                   (\overline{\tau}_R \Gamma \mu_R)_{\mathbf{1''}}\bigr]_{\mathbf{1''}} ,\\
            &\bigl[ Y^{(4)}_{\mathbf{1}} {Y^{(4)}_{\mathbf{1'}}}^{*} \bigr]_{\mathbf{1'}}\,
             \bigl[(\overline{\mu}_R \Gamma \mu_R)_{\mathbf{1}}
                   (\overline{\tau}_R \Gamma \mu_R)_{\mathbf{1''}}\bigr]_{\mathbf{1''}} ,\\
            &\bigl[ Y^{(4)}_{\mathbf{3}} {Y^{(4)}_{\mathbf{3}}}^{*} \bigr]_{\mathbf{1'}}\,
             \bigl[(\overline{\mu}_R \Gamma \mu_R)_{\mathbf{1}}
                   (\overline{\tau}_R \Gamma \mu_R)_{\mathbf{1''}}\bigr]_{\mathbf{1''}} ,\\
            &\bigl[ Y^{(4)}_{\mathbf{1'}} {Y^{(4)}_{\mathbf{1}}}^{*} \bigr]_{\mathbf{1'}}\,
             \bigl[(\overline{\tau}_R \Gamma \tau_R)_{\mathbf{1}}
                   (\overline{\tau}_R \Gamma \mu_R)_{\mathbf{1''}}\bigr]_{\mathbf{1''}} ,\\
            &\bigl[ Y^{(4)}_{\mathbf{1}} {Y^{(4)}_{\mathbf{1'}}}^{*} \bigr]_{\mathbf{1'}}\,
             \bigl[(\overline{\tau}_R \Gamma \tau_R)_{\mathbf{1}}
                   (\overline{\tau}_R \Gamma \mu_R)_{\mathbf{1''}}\bigr]_{\mathbf{1''}} ,\\
            &\bigl[ Y^{(4)}_{\mathbf{3}} {Y^{(4)}_{\mathbf{3}}}^{*} \bigr]_{\mathbf{1'}}\,
             \bigl[(\overline{\tau}_R \Gamma \tau_R)_{\mathbf{1}}
                   (\overline{\tau}_R \Gamma \mu_R)_{\mathbf{1''}}\bigr]_{\mathbf{1''}} ,\\
            &\bigl[ Y^{(4)}_{\mathbf{1'}} {Y^{(4)}_{\mathbf{1'}}}^{*} \bigr]_{\mathbf{1''}}\,
             \bigl[(\overline{\tau}_R \Gamma \tau_R)_{\mathbf{1}}
                   (\overline{\mu}_R \Gamma \tau_R)_{\mathbf{1'}}\bigr]_{\mathbf{1'}} ,\\
            &\bigl[ Y^{(4)}_{\mathbf{3}} {Y^{(4)}_{\mathbf{3}}}^{*} \bigr]_{\mathbf{1''}}\,
             \bigl[(\overline{\tau}_R \Gamma \tau_R)_{\mathbf{1}}
                   (\overline{\mu}_R \Gamma \tau_R)_{\mathbf{1'}}\bigr]_{\mathbf{1'}} ,\\
            &\bigl[ Y^{(4)}_{\mathbf{1'}} {Y^{(4)}_{\mathbf{1'}}}^{*} \bigr]_{\mathbf{1''}}\,
             \bigl[(\overline{\tau}_R \Gamma \mu_R)_{\mathbf{1''}}
                   (\overline{\tau}_R \Gamma \mu_R)_{\mathbf{1''}}\bigr]_{\mathbf{1'}} ,\\
            &\bigl[ Y^{(4)}_{\mathbf{3}} {Y^{(4)}_{\mathbf{3}}}^{*} \bigr]_{\mathbf{1''}}\,
             \bigl[(\overline{\tau}_R \Gamma \mu_R)_{\mathbf{1''}}
                   (\overline{\tau}_R \Gamma \mu_R)_{\mathbf{1''}}\bigr]_{\mathbf{1'}} ,\\
            &\bigl[ Y^{(4)}_{\mathbf{1'}} {Y^{(4)}_{\mathbf{1}}}^{*} \bigr]_{\mathbf{1'}}\,
             \bigl[(\overline{\mu}_R \Gamma \tau_R)_{\mathbf{1'}}
                   (\overline{\mu}_R \Gamma \tau_R)_{\mathbf{1'}}\bigr]_{\mathbf{1''}} ,\\
            &\bigl[ Y^{(4)}_{\mathbf{1}} {Y^{(4)}_{\mathbf{1'}}}^{*} \bigr]_{\mathbf{1'}}\,
             \bigl[(\overline{\mu}_R \Gamma \tau_R)_{\mathbf{1'}}
                   (\overline{\mu}_R \Gamma \tau_R)_{\mathbf{1'}}\bigr]_{\mathbf{1''}} ,\\
            &\bigl[ Y^{(4)}_{\mathbf{3}} {Y^{(4)}_{\mathbf{3}}}^{*} \bigr]_{\mathbf{1'}}\,
             \bigl[(\overline{\mu}_R \Gamma \tau_R)_{\mathbf{1'}}
                   (\overline{\mu}_R \Gamma \tau_R)_{\mathbf{1'}}\bigr]_{\mathbf{1''}} ,\\
            &\bigl[ Y^{(4)}_{\mathbf{1}} {Y^{(4)}_{\mathbf{1}}}^{*} \bigr]_{\mathbf{1}}\,
             \bigl[(\overline{\mu}_R \Gamma \mu_R)_{\mathbf{1}}
                   (\overline{\tau}_R \Gamma \tau_R)_{\mathbf{1}}\bigr]_{\mathbf{1}} ,\\
            &\bigl[ Y^{(4)}_{\mathbf{3}} {Y^{(4)}_{\mathbf{3}}}^{*} \bigr]_{\mathbf{1}}\,
             \bigl[(\overline{\mu}_R \Gamma \mu_R)_{\mathbf{1}}
                   (\overline{\tau}_R \Gamma \tau_R)_{\mathbf{1}}\bigr]_{\mathbf{1}} .
        \end{aligned}
        \end{equation}
    \end{itemize}

    \item Operators involving \((\overline{L}L)(\overline{R}R)\):
    \begin{equation}
        \bigl[(\overline{l}l)(\overline{e}e)\bigr]_{\mathbf{1}} \, .
    \end{equation}
    We can separate these operators into nine cases, according to the flavor
    assignments of the right-handed fields:
    \begin{itemize}
        \item \(e = 1\), \(e^\dagger = 1\):\\[2pt]
        The Hilbert series gives
        \begin{equation}
            e \, e^\dagger \, l \, l^\dagger \, {Y^{(-2)}_{\mathbf{1}}}^{*} \, Y^{(-2)}_{\mathbf{1}}
            + 2\, e \, e^\dagger \, l \, l^\dagger \, {Y^{(-2)}_{\mathbf{3}}}^{*} \, Y^{(-2)}_{\mathbf{1}}
            + 2\, e \, e^\dagger \, l \, l^\dagger \, {Y^{(-2)}_{\mathbf{1}}}^{*} \, Y^{(-2)}_{\mathbf{3}}
            + 7\, e \, e^\dagger \, l \, l^\dagger \, {Y^{(-2)}_{\mathbf{3}}}^{*} \, Y^{(-2)}_{\mathbf{3}} \, .
        \end{equation}
        The corresponding \(A_4\)-invariant combinations can be written as
        \begin{equation}
        \begin{aligned}
            \bigl[{Y^{(-2)}_{\mathbf{1}}}^{*} Y^{(-2)}_{\mathbf{1}}\bigr]_{\mathbf{1}}\,
            &\bigl[(\overline{L}_L \Gamma L_L)_{\mathbf{1}}(\overline{e}_R \Gamma e_R)\bigr]_{\mathbf{1}} ,\\
            \bigl[{Y^{(-2)}_{\mathbf{3}}}^{*} Y^{(-2)}_{\mathbf{1}}\bigr]_{\mathbf{3}}\,
            &\bigl[(\overline{L}_L \Gamma L_L)_{\mathbf{3}_{\mathrm{s}},\mathbf{3}_{\mathrm{a}}}
                   (\overline{e}_R \Gamma e_R)\bigr]_{\mathbf{3}_{\mathrm{s}},\mathbf{3}_{\mathrm{a}}} ,\\
            \bigl[{Y^{(-2)}_{\mathbf{1}}}^{*} Y^{(-2)}_{\mathbf{3}}\bigr]_{\mathbf{3}}\,
            &\bigl[(\overline{L}_L \Gamma L_L)_{\mathbf{3}_{\mathrm{s}},\mathbf{3}_{\mathrm{a}}}
                   (\overline{e}_R \Gamma e_R)\bigr]_{\mathbf{3}_{\mathrm{s}},\mathbf{3}_{\mathrm{a}}} ,\\
            \bigl[{Y^{(-2)}_{\mathbf{3}}}^{*} Y^{(-2)}_{\mathbf{3}}\bigr]_{\mathbf{1}}\,
            &\bigl[(\overline{L}_L \Gamma L_L)_{\mathbf{1}}
                   (\overline{e}_R \Gamma e_R)\bigr]_{\mathbf{1}} ,\\
            \bigl[{Y^{(-2)}_{\mathbf{3}}}^{*} Y^{(-2)}_{\mathbf{3}}\bigr]_{\mathbf{1''}}\,
            &\bigl[(\overline{L}_L \Gamma L_L)_{\mathbf{1'}}
                   (\overline{e}_R \Gamma e_R)\bigr]_{\mathbf{1'}} ,\\
            \bigl[{Y^{(-2)}_{\mathbf{3}}}^{*} Y^{(-2)}_{\mathbf{3}}\bigr]_{\mathbf{1'}}\,
            &\bigl[(\overline{L}_L \Gamma L_L)_{\mathbf{1''}}
                   (\overline{e}_R \Gamma e_R)\bigr]_{\mathbf{1''}} ,\\
            \bigl[{Y^{(-2)}_{\mathbf{3}}}^{*} Y^{(-2)}_{\mathbf{3}}\bigr]_{\mathbf{3}_{\mathrm{s}}}\,
            &\bigl[(\overline{L}_L \Gamma L_L)_{\mathbf{3}_{\mathrm{s}}}
                   (\overline{e}_R \Gamma e_R)\bigr]_{\mathbf{3}_{\mathrm{s}}} ,\\
            \bigl[{Y^{(-2)}_{\mathbf{3}}}^{*} Y^{(-2)}_{\mathbf{3}}\bigr]_{\mathbf{3}_{\mathrm{a}}}\,
            &\bigl[(\overline{L}_L \Gamma L_L)_{\mathbf{3}_{\mathrm{s}}}
                   (\overline{e}_R \Gamma e_R)\bigr]_{\mathbf{3}_{\mathrm{s}}} ,\\
            \bigl[{Y^{(-2)}_{\mathbf{3}}}^{*} Y^{(-2)}_{\mathbf{3}}\bigr]_{\mathbf{3}_{\mathrm{s}}}\,
            &\bigl[(\overline{L}_L \Gamma L_L)_{\mathbf{3}_{\mathrm{a}}}
                   (\overline{e}_R \Gamma e_R)\bigr]_{\mathbf{3}_{\mathrm{a}}} ,\\
            \bigl[{Y^{(-2)}_{\mathbf{3}}}^{*} Y^{(-2)}_{\mathbf{3}}\bigr]_{\mathbf{3}_{\mathrm{a}}}\,
            &\bigl[(\overline{L}_L \Gamma L_L)_{\mathbf{3}_{\mathrm{a}}}
                   (\overline{e}_R \Gamma e_R)\bigr]_{\mathbf{3}_{\mathrm{a}}} .
        \end{aligned}
        \end{equation}

        \item \(e = 1\), \(\mu^\dagger = 1\):\\[2pt]
        The Hilbert series gives
        \begin{equation}
            e \, l \, l^\dagger \, \mu^\dagger \, {Y^{(-2)}_{\mathbf{1}}}^{*}
            + 2\, e \, l \, l^\dagger \, \mu^\dagger \, {Y^{(-2)}_{\mathbf{3}}}^{*} \, .
        \end{equation}
        The corresponding invariant operators are
        \begin{equation}
        \begin{aligned}
            \bigl[{Y^{(-2)}_{\mathbf{1}}}^{*}\bigr]_{\mathbf{1}}\,
            &\bigl[(\overline{L}_L \Gamma L_L)_{\mathbf{1'}}
                   (\overline{\mu}_R \Gamma e_R)\bigr]_{\mathbf{1}} ,\\
            \bigl[{Y^{(-2)}_{\mathbf{3}}}^{*}\bigr]_{\mathbf{3}}\,
            &\bigl[(\overline{L}_L \Gamma L_L)_{\mathbf{3}_{\mathrm{s}},\mathbf{3}_{\mathrm{a}}}
                   (\overline{\mu}_R \Gamma e_R)\bigr]_{\mathbf{3}_{\mathrm{s}},\mathbf{3}_{\mathrm{a}}} .
        \end{aligned}
        \end{equation}

        \item \(e = 1\), \(\tau^\dagger = 1\):\\[2pt]
        The Hilbert series gives
        \begin{equation}
            e \, l \, l^\dagger \, \tau^\dagger \, {Y^{(-2)}_{\mathbf{1}}}^{*}
            + 2\, e \, l \, l^\dagger \, \tau^\dagger \, {Y^{(-2)}_{\mathbf{3}}}^{*} \, .
        \end{equation}
        The invariant combinations may be written as
        \begin{equation}
        \begin{aligned}
            \bigl[{Y^{(-2)}_{\mathbf{1}}}^{*}\bigr]_{\mathbf{1}}\,
            &\bigl[(\overline{L}_L \Gamma L_L)_{\mathbf{1''}}
                   (\overline{\tau}_R \Gamma e_R)\bigr]_{\mathbf{1}} ,\\
            \bigl[{Y^{(-2)}_{\mathbf{3}}}^{*}\bigr]_{\mathbf{3}}\,
            &\bigl[(\overline{L}_L \Gamma L_L)_{\mathbf{3}_{\mathrm{s}},\mathbf{3}_{\mathrm{a}}}
                   (\overline{\tau}_R \Gamma e_R)\bigr]_{\mathbf{3}_{\mathrm{s}},\mathbf{3}_{\mathrm{a}}} .
        \end{aligned}
        \end{equation}

        \item \(\mu = 1\), \(e^\dagger = 1\):\\[2pt]
        This case is simply the complex conjugate of the case
        \(e = 1\), \(\mu^\dagger = 1\).

        \item \(\mu = 1\), \(\mu^\dagger = 1\):\\[2pt]
        The Hilbert series gives
        \begin{equation}
            l \, l^\dagger \, \mu \, \mu^\dagger \, D
            + 2\, l \, l^\dagger \, \mu \, \mu^\dagger \, Y^{(0)}_{\mathbf{3}}
            + 2\, l \, l^\dagger \, \mu \, \mu^\dagger \, {Y^{(0)}_{\mathbf{3}}}^{*} \, .
        \end{equation}
        The corresponding invariants can be summarized as
        \begin{equation}
        \begin{aligned}
            &\bigl[(\overline{L}_L \Gamma L_L)_{\mathbf{1}}
                   (\overline{\mu}_R \Gamma \mu_R)\bigr]_{\mathbf{1}} ,\\
            &\bigl[(\overline{L}_L \Gamma L_L)_{\mathbf{3}_{\mathrm{s}},\mathbf{3}_{\mathrm{a}}}
                   (\overline{\mu}_R \Gamma \mu_R)\bigr]_{\mathbf{3}_{\mathrm{s}},\mathbf{3}_{\mathrm{a}}}
              \, Y^{(0)}_{\mathbf{3}} ,\\
            &\bigl[(\overline{L}_L \Gamma L_L)_{\mathbf{3}_{\mathrm{s}},\mathbf{3}_{\mathrm{a}}}
                   (\overline{\mu}_R \Gamma \mu_R)\bigr]_{\mathbf{3}_{\mathrm{s}},\mathbf{3}_{\mathrm{a}}}
              \, {Y^{(0)}_{\mathbf{3}}}^{*} .
        \end{aligned}
        \end{equation}

        \item \(\mu = 1\), \(\tau^\dagger = 1\):\\[2pt]
        The Hilbert series gives
        \begin{equation}
            l \, l^\dagger \, \mu \, \tau^\dagger
            + 2\, l \, l^\dagger \, \mu \, \tau^\dagger \, Y^{(0)}_{\mathbf{3}}
            + 2\, l \, l^\dagger \, \mu \, \tau^\dagger \, {Y^{(0)}_{\mathbf{3}}}^{*} \, .
        \end{equation}
        The corresponding invariants are
        \begin{equation}
        \begin{aligned}
            &\bigl[(\overline{L}_L \Gamma L_L)_{\mathbf{1'}}
                   (\overline{\tau}_R \Gamma \mu_R)\bigr]_{\mathbf{1}} ,\\
            &\bigl[(\overline{L}_L \Gamma L_L)_{\mathbf{3}_{\mathrm{s}},\mathbf{3}_{\mathrm{a}}}
                   (\overline{\tau}_R \Gamma \mu_R)\bigr]_{\mathbf{3}_{\mathrm{s}},\mathbf{3}_{\mathrm{a}}}
              \, Y^{(0)}_{\mathbf{3}} ,\\
            &\bigl[(\overline{L}_L \Gamma L_L)_{\mathbf{3}_{\mathrm{s}},\mathbf{3}_{\mathrm{a}}}
                   (\overline{\tau}_R \Gamma \mu_R)\bigr]_{\mathbf{3}_{\mathrm{s}},\mathbf{3}_{\mathrm{a}}}
              \, {Y^{(0)}_{\mathbf{3}}}^{*} .
        \end{aligned}
        \end{equation}

        \item \(\tau = 1\), \(e^\dagger = 1\):\\[2pt]
        This case is the complex conjugate of the case
        \(e = 1\), \(\tau^\dagger = 1\).

        \item \(\tau = 1\), \(\mu^\dagger = 1\):\\[2pt]
        This case is the complex conjugate of the case
        \(\mu = 1\), \(\tau^\dagger = 1\).

        \item \(\tau = 1\), \(\tau^\dagger = 1\):\\[2pt]
        The construction rules are identical to those in the case
        \(\mu = 1\), \(\mu^\dagger = 1\), with the flavor index
        replacement \(\mu \rightarrow \tau\).
    \end{itemize}

\end{itemize}

\end{itemize}

\section{Conclusion}
\label{sec:conclusion}
%In our work, we first reviewed the Standard Model flavor structure, including the origin of fermion mass matrices and the CKM/PMNS mixings. As one possible solution to the flavor problem, we recalled the MFV implementation in SMEFT, where the only sources of flavor and CP violation in all higher-dimensional operators are the SM Yukawa matrices $Y_u$, $Y_d$, and $Y_e$. In parallel, we also reviewed the traditional discrete flavor-symmetry framework, and introduced flavor symmetry as an economical realization of flavor breaking, in which Yukawa couplings are provided by modular forms rather than flavon alignments. Building on this, we explicitly constructed a modular-symmetric Standard Model by taking the full modular flavor group to be $A_4^{(q)}\otimes A_4^{(e)}$, assigning left-handed fermions to $A_4$ triplets and right-handed fermions to inequivalent singlets, and writing renormalizable Yukawa interactions.
%\textbf{In holomorphic case, we make a minimal assumption like MFV that leading-order Yukawa is the only source to build flavor structure, so that we neglect the dynamical effects of $\tau$ and $\mathrm{Im}\,\tau$, and modular forms serve as elementary spurion building blocks. By doing so, flavor breaking is controlled solely by the modulus VEV and only by modular forms in an MFV-like manner. Meanwhile only with this minimal choice infinite number of operators would not be introduced.} This modular-symmetric SM setup provides the reference framework for constructing modular-invariant SMEFT operators using the Hilbert-series method.

In our work, we first reviewed the Standard Model flavor structure, including the origin of fermion mass matrices and the CKM/PMNS mixings. As one possible solution to the flavor problem, we recalled the MFV implementation in SMEFT, in which the SM Yukawa matrices $Y_u$, $Y_d$, and $Y_e$ provide the only sources of flavor and CP violation in all higher-dimensional operators. In parallel, we also reviewed the traditional discrete flavor-symmetry framework, and introduced modular flavor symmetry as an economical realization of flavor breaking, where Yukawa couplings are furnished by modular forms rather than flavon alignments. Building on this, we followed the constructions in Ref.~\cite{Kobayashi_2022} and constructed a modular-symmetric Standard Model by taking the full modular flavor group to be $A_4^{(q)}\otimes A_4^{(e)}$, assigning left-handed fermions to $A_4$ triplets and right-handed fermions to inequivalent singlets, and writing renormalizable Yukawa interactions.
The resulting holomorphic Yukawa couplings admit an interpretation as the low-energy imprint of a modular-invariant $\mathcal{N}=1$ SUSY UV completion, where they descend from holomorphic superpotential interactions.
In the holomorphic case, we adopted an MFV-like assumption that the leading-order Yukawa sector constitutes the only source of flavor breaking relevant for constructing higher-dimensional operators.
We restricted the operator construction to a finite set of holomorphic modular-form multiplets and did not treat $\mathrm{Im}(\tau)$ as an independent building block. Instead, we took the simplest form of the kinetic terms and allowed $\mathrm{Im}(\tau)$ to appear there, as required by modular invariance, while excluding any $\mathrm{Im}(\tau)$-dependent structures from the higher-dimensional operator basis. This prescription yielded a finite operator tower, even though $\mathrm{Im}(\tau)$ can enter the kinetic sector.
This modular-symmetric SM setup provides the reference framework for constructing modular-invariant SMEFT operators using the Hilbert-series method.

In our setup, the breaking of $A_{4}^{(q)}\otimes A_{4}^{(e)}$ is entirely controlled by the vacuum expectation values $\langle\tau_q\rangle$ and $\langle\tau_e\rangle$. Their effects in higher-dimensional operators are encoded solely through the corresponding modular forms. Accordingly, we took the modular forms appearing in the renormalizable Yukawa sector as the fundamental spurions of the EFT, and built all higher-dimensional operator structures from them. In the holomorphic $N=3$ case, all higher-weight modular forms are generated from the fundamental weight-$2$ triplet $Y^{(2)}_{\mathbf{3}}$ via symmetric tensor products, so that $Y^{(2)}_{\mathbf{3}}$ serves as the unique modular-form spurion. Using this property, we constructed the modular Hilbert series and verified that the resulting operator basis was organized in close analogy with the Yukawa sector, which significantly simplified the classification.
The resulting modular-invariant SMEFT contained six independent dimension-$5$ operators, $2961$ dimension-$6$ operators, and $360$ dimension-$7$ operators. The explicit dimension-$5$ operators were summarized in Table~\ref{tab:dim5_A4}, while the dimension-$6$ results were collected in Tables~\ref{tab:dim6_boson_bilinear} and~\ref{tab:classes4-9}. We constructed all dimension-$5$ operators and all baryon- and lepton-number conserving dimension-$6$ operators. In Appendix~\ref{dim_7}, we further discussed the structure of bilinear dimension-$7$ operators without field-strength tensors.

We then extended the construction to non-holomorphic modular forms. Unlike the minimal assumption adopted in the holomorphic case, we imposed a constructive assumption in the non-holomorphic setting to avoid an infinite proliferation of operators. For higher-dimensional operators, we assumed that the required automorphy-factor compensation was provided solely by a single modular form, or together with a complex conjugate form if necessary. 
%This organizing assumption was essential to prevent an uncontrolled proliferation of independent operator structures, which would otherwise become infinite. 
This is because non-holomorphic modular forms were not closed under multiplication. Under this assumption, higher-dimensional operators could still be organized in the familiar tensor-product form, 
$
\bigl[Y_{\mathbf{r}}^{(k_Y)}\,{Y_{\mathbf{r}'}^{(k_Y')}}^{*}\,\mathcal{O}\bigr]_{\mathbf{1}}
$,
as in the holomorphic case. The assumption was also consistent with the fact that, at sufficiently large weight, in particular for weights $k\ge 4$ in the $N=3$ case, the relevant polyharmonic Maa$\upbeta$ forms coincided with ordinary holomorphic modular forms. Within the benchmark Standard Model setup with Maa\ss\ forms, this reproduced the Weinberg-operator structure of Ref.~\cite{qu2024nonholomorphicmodularflavorsymmetry} and allowed us to derive the complete set of dimension-$6$ operators in the same non-holomorphic modular setting.

%With our minimal assumption, both the holomorphic and the non-holomorphic settings lead to a finite set of modular-invariant SMEFT operators. This assumption also provides a systematic way to organize modular forms together with the SM field content. In particular, even if the SM Lagrangian is taken to be non-holomorphic, one may still employ holomorphic modular forms and regard a small set of fundamental holomorphic forms as the basic building blocks. The reason is that, in higher-dimensional operators, such holomorphic building blocks extend in a controlled manner and can compensate the loss of covariance induced by non-holomorphic structures. This mechanism is not available if one starts directly from genuinely non-holomorphic modular objects, such as Maa\ss\ forms. In this sense, holomorphic modular forms impose stronger constraints on operator construction than the non-holomorphic setting.

In the holomorphic part of this work we performed a detailed comparison with the modular-invariant SMEFT operator construction of Ref.~\cite{Kobayashi_2022}. Our Hilbert-series method automatically accounted for the algebraic relations among modular forms, typically in the Weinberg operator and for four-fermion current--current operators
$(\overline{L}_L\Gamma L_L)(\overline{L}_L\Gamma L_L)$, and thus yielded an independent and complete operator basis. A concrete illustration of the resulting redundancies was given in Sec.~\ref{dimension-W}, where we used the Weinberg operator as an explicit example.
After removing redundant operators, the corresponding Hilbert-series expressions are given in
Eq.~\eqref{eq:MSSMHS5} for the Weinberg operator,
Eq.~\eqref{barllcc} for $(\overline{\ell}\,\ell)(\overline{\ell}\,\ell)$,
and Eq.~\eqref{barqqcc} for $(\overline{q}\,q)(\overline{q}\,q)$.
The other operators, for all bilinear operators discussed there, as well as for the four-fermion operators built purely from right-handed fields and their conjugates, the total multiplicities agreed with our results. The apparent difference concerned how modular forms were embedded into the SMEFT operators. 
Moreover, our Hilbert series allows us to extract the operator structures explicitly in the form
$
\bigl[\,Y_{\mathbf{r}}^{(k_Y)}\,{Y_{\mathbf{r}'}^{(k_Y')}}^{*}\,\mathcal{O}\,\bigr]_{\mathbf{1}}\,,
$
without having to introduce couplings that involve multiple insertions of $Y^{(2)}_{\mathbf{3}}$ and its complex conjugate.
In the SMEFT we can classify operators as holomorphic or non-holomorphic by their chiral Lorentz structure. 
Holomorphic operators are built from the components $\{\,X^-_{\mu\nu},\,\bar R,\,L\,\}$~\cite{Alonso_2014}, where the complex (anti-)self-dual decomposition of field strengths is defined by
\begin{equation}
  X_{\mu\nu}^{\pm}\equiv \frac12\left(X_{\mu\nu}\mp i\,\widetilde X_{\mu\nu}\right)\,,
  \qquad \widetilde X_{\mu\nu}^{\pm}=\pm i\,X_{\mu\nu}^{\pm}\,,
\end{equation}
with
\begin{equation}
  \widetilde X_{\mu\nu}\equiv \frac12\,\epsilon_{\mu\nu\rho\sigma}\,X^{\rho\sigma}\,,
  \qquad X_{\mu\nu}\in\{G_{\mu\nu}^A,\,W_{\mu\nu}^I,\,B_{\mu\nu}\},
\end{equation}
and scalar fields such as the Higgs do not enter the holomorphy classification.
%Concerning Higgs insertions, we adopt the practical convention that holomorphic operators involve Higgs fields only through linear insertions of $H^\dagger$ and $\widetilde H^\dagger$ (with $\widetilde H \equiv i\sigma_2 H^{*}$), and contain no Hermitian Higgs bilinear $H^\dagger H$.%
%\footnote{
%Throughout this work, operators containing at least one insertion of the Hermitian bilinear $H^\dagger H$ are treated as non-holomorphic.
%This convention is motivated by the supersymmetric analogy: Hermitian bilinears naturally resemble K\"ahler/$D$-term structures rather than superpotential-like F-terms.
%Nevertheless, such Higgs-dependent structures may still be generated effectively after integrating out heavy states, even when the renormalizable UV interactions are purely holomorphic; this possibility is part of our matching discussion.}
With this criterion, the unique dimension-five operator is the holomorphic Weinberg operator $(LH)(LH)$. 
At dimension six, the fermionic operators can be organized schematically as~\footnote{Ref.~\cite{Alonso_2014} classifies the Yukawa-like class $\psi^2H^3$ as non-holomorphic, since these operators can be removed in favor of non-holomorphic structures using the EOM. In contrast, we focus on the chirality-based operator structure and therefore treat $\psi^2H^3$ as holomorphic in our classification.}
\begin{align}
\mathcal{O}_{h} \subset \Bigl\{
  & \psi^{2} X H,\ \psi^{2}H^{3},\ (\overline{L}R)(\overline{L}R),\ 
  \mathcal{O}_{qqq},\ \mathcal{O}_{duu}
\Bigr\}, \\
\mathcal{O}_{n} \subset \Bigl\{
  & \psi^{2}H^{2}D,\ (\overline{L}L)(\overline{L}L),\ (\overline{R}R)(\overline{R}R),\ 
  (\overline{L}L)(\overline{R}R),\ (\overline{L}R)(\overline{R}L)+\text{h.c.},\ 
  \mathcal{O}_{duq},\ \mathcal{O}_{qqu}
\Bigr\},
\end{align}
where we omit the bosonic classes in this summary.

In our assumption that the SM Yukawa couplings are furnished by holomorphic modular forms, together with an MFV-like organizing principle, modular-invariant SMEFT operators are constructed by dressing an underlying SMEFT operator structure with appropriate (anti-)holomorphic modular-form factors. 
Accordingly, whether an operator is completed with purely holomorphic modular-form factors or with non-holomorphic mixed factors built from holomorphic modular forms and their complex conjugates is governed primarily by whether the underlying SMEFT structure is holomorphic or non-holomorphic in the above chirality-based sense.
Within this MFV-like setup, the matching of modular-invariant SMEFT operators can be interpreted consistently in terms of a modular-invariant SUSY UV completion, subject to two minimal restrictions on admissible matching channels:
\begin{enumerate}
  \item \textbf{Minimal $\tau$-dependence.}
  The modulus $\tau$ enters the EFT only through (anti-)holomorphic modular forms.
  \item \textbf{$\tau$-independent heavy masses.}
  Heavy-mediator mass parameters are taken to be $\tau$-independent constants.
\end{enumerate}
These conditions confine $\tau$-dependence to interaction couplings through holomorphic modular forms (and, when required, their complex conjugates), and exclude modulus-dependent propagator denominators such as $1/M(\tau)^2$.
They also motivate discarding kinetic-mediated contributions whenever modular covariance forces explicit $\mathrm{Im}(\tau)$-dependent prefactors to appear in covariant kinetic structures, since such prefactors would propagate into Wilson coefficients and lie outside our minimal operator class. For holomorphic operators, the required holomorphic modular forms typically originate only from F-term structures in a SUSY UV completion, whereas D-term structures in the SUSY UV generally introduce non-holomorphic modular factors, preventing the matched EFT operator from remaining holomorphic.
%A more detailed discussion and an explicit dictionary to superspace diagnostics are collected in Appendix~\ref{app:XXX}.

A representative non-holomorphic modular-invariant operator is
\begin{equation}
\Bigl[
  Y^{(4)}_{\mathbf r_e}(\tau)\,Y^{(4)\,*}_{\mathbf r_e}(\tau)\,
  \bigl[\overline{L}_L\Gamma L_L\bigr]\,
  \bigl[\overline{L}_L\Gamma L_L\bigr]
\Bigr]_{\mathbf{1}}\,.
\end{equation}
One SUSY-compatible UV realization is a Zee-type holomorphic trilinear coupling involving a heavy scalar mediator $S$ that is a modular singlet,
\begin{equation}
\bigl[
  Y^{(4)}_{\mathbf r}(\tau)\,
  (L^T C i\sigma_2 L)_{\mathbf r'}\,S
\bigr]_{\mathbf 1},
\end{equation}
so that integrating out $S$ generates the above four-fermion structure at low energies. 
Alternatively, a current--current structure can be produced by integrating out a heavy vector state $X_\mu$ coupled to lepton currents, schematically from an interaction of the form
\begin{equation}
\Bigl[
  (Y^{(2)}Y^{(2)\,*})_{\mathbf r}\,
  \bigl(\overline{L}_L\gamma^\mu L_L\,X_\mu\bigr)_{\mathbf r'}
\Bigr]_{\mathbf 1}.
\end{equation}
While the stringy factorization ansatz of Ref.~\cite{Kobayashi_2022} naturally accommodates superpotential-type trilinear couplings, and hence holomorphic modular-form dependence in the underlying Yukawa interactions. 
However, it does not fix a generic light--light--heavy current coupling of the form $\overline{L}_L\gamma^\mu L_L X_\mu$. 
Therefore, the operator construction in their paper cannot be obtained from the D-term sector of a modular SUSY UV completion.
In our MFV-like setup, any such current coupling is required not to introduce new independent flavor spurions; its flavor tensor must be expandable in modular-form combinations. 

In summary, our MFV-like assumption can be motivated not only from the F-term sector of a modular SUSY UV completion, but also provides a general restriction on admissible modular SUSY D-term structures. This restriction avoids the appearance of infinite kinetic terms of the kind discussed in Ref.~\cite{Chen_2020}.
Within the modular-invariant SMEFT, holomorphic operators are typically generated only from F-term structures in modular SUSY UV completions in tree-level matching, whereas non-holomorphic operators can generally arise from either F-term or D-term structures. A more detailed discussion is provided in Appendix~\ref{app:susy_realization_summary}.

Finally, this MFV-like assumption still suffers from several shortcomings. First, it relies on simplifying assumptions neglecting the kinetic term of $\tau$ and omitting additional contributions involving $\mathrm{Im}\,\tau$ together with modular forms. Second, the resulting operator count is not dramatically smaller than in the conventional SM case, so many free parameters remain in phenomenological applications. For comparison, at dimension six the number of invariant operators with three generations is $3045$, whereas our result is only slightly smaller. Moreover, the number of baryon-number--conserving operators in our setup is $2733$, which is still more than the $894$ baryon-number--conserving operators in the dimension-6 MFV-SMEFT. In this sense, the modular constraints considered here are not yet sufficiently restrictive. Addressing this issue requires additional constraints within the present framework.

%约束太少
%一种新的考虑非全纯模形式的思考方法，即并不是直接去构造非全纯模形式，而是将全纯模形式及其共轭当做building block来贡献非全纯的部分。
%与文章2112.00493在bilinear部分构成是完全等价的。对于四费米子部分如前文所说，2112.00493的ansatz可以认为是Y^{(2)}_{\mathbf{3}}当成building building block时一种特殊的basis选取，而这种basis选取无法显示的消除constrain还有appendix\label{app:forms}中的冗余。

%从MFV的思想中，给了yukawa更基本的结构，也就是利用模形式，还有LO的待定系数，以更精确的拟合yukawa couplings。

%规范变换的底空间(时空本身)邻域内也是不变的，比如x与x+\delta x，在规范变换下都是底空间不变的，但是对于fix point \tau_0来说，它的邻域是变换的
%模对称变换：$\Lambda_\tau\rightarrow \Lambda_{\gamma\tau}=\frac{1}{c\tau+d}\Lambda_\tau=\frac{1}{|c\tau+d|}e^{i\theta}\Lambda_\tau$
%weight 来源于SL(2,R)而非SL(2,Z) 也就是说如果要在物理上引入模形式，为什么会有SL(2,R)到SL(2,Z)这个连续群的分离实现？这个实现怎么物理上或者几何上理解？ 在SL(2,Z)下的SM，在SL(2,R)中又是什么？从SL(2,R)到SL(2,Z)的过度是怎么体现的？

\section*{acknowledgments}
    We would like to thank Gui-Jun Ding and Ye-Ling Zhou for valuable comments on the draft. This work is supported by the National Science Foundation of China under Grants No. 12347105, No. 12375099 and No. 12447101, and the National Key Research and Development Program of China Grant No. 2020YFC2201501, No. 2021YFA0718304.

% ====== 正文章节结束 ======
\clearpage
\begin{appendices}

\section{Flavor construction in general \texorpdfstring{$A_4$}{A4} flavor symmetry}
\subsection{Tensor product rule of \texorpdfstring{$A_4$}{A4} group}
\label{app:tensor_product}
In the discrete flavor group $A_4$, the tensor products of irreducible representations decompose as
\begin{equation}
\begin{aligned}
&\mathbf{1}\otimes\mathbf{1} \;=\; \mathbf{1}, 
\qquad \mathbf{1'}\otimes \mathbf{1''} \;=\; \mathbf{1},\\
&\mathbf{1'}\otimes\mathbf{1'} \;=\; \mathbf{1''}, 
\qquad \mathbf{1''}\otimes \mathbf{1''} \;=\; \mathbf{1},\\
&\mathbf{1}\otimes \mathbf{3} \;=\; \mathbf{3},\quad 
 \mathbf{1'}\otimes \mathbf{3} \;=\; \mathbf{3},\quad 
 \mathbf{1''}\otimes \mathbf{3} \;=\; \mathbf{3},\\
&\mathbf{3}\otimes \mathbf{3} \;=\; 
 \mathbf{1}\oplus \mathbf{1'}\oplus \mathbf{1''}\oplus 
 \mathbf{3}_{\mathbf{s}}\oplus \mathbf{3}_{\mathbf{a}} \, .
\end{aligned}
\end{equation}
Here $\mathbf{3}_{\mathbf{s}}$ and $\mathbf{3}_{\mathbf{a}}$ denote, respectively, the symmetric and antisymmetric components in 
$\mathbf{3}\otimes\mathbf{3}$. 

Specifically, the group $A_{4}$ admits the presentation
\begin{equation}
    S^{2}=(ST)^{3}=T^{3}=e,
    \label{eq:a4}
\end{equation}
and it can be realized as a semidirect product of the residual subgroups
\begin{equation}
    G_{\ell}\simeq \mathbb{Z}_{3}=\{ e,T,T^2 \},\qquad 
    G_{\nu}\simeq \mathbb{Z}_{2}\times \mathbb{Z}_{2}=\{ e,S, TST^2,T^2ST\},
\quad\text{so that}\quad
A_{4}\simeq G_{\nu}\rtimes G_{\ell}.
\end{equation}

%More concretely, we label the four vertices of a regular tetrahedron by $1,2,3,4$. The generator $S$ is the permutation $(12)(34)$, corresponding to a $180^\circ$ rotation about the axis through the midpoints of the opposite edges connecting vertices $1$–$2$ and $3$–$4$. The generator $T$ is the permutation $(123)$, corresponding to a $120^\circ$ rotation about the axis through vertex $4$ and the center of the opposite face $(123)$. Using the generators $S$ and $T$ and the group relations in Eq.~(\ref{eq:a4}), the group $A_4$ has 12 elements, which we can take to be
%\[
%\{ e,\, S,\, TST^{2},\, T^{2}ST,\, T,\, TS,\, ST,\, STS,\, T^{2},\, ST^{2},\, T^{2}S,\, TST \}.
%\]
%These 12 elements fall into four conjugacy classes,
%\[
%C_{1} = \{ e \},\quad 
%C_{2} = \{ S,\, TST^{2},\, T^{2}ST \},\quad
%C_{3} = \{ T,\, TS,\, ST,\, STS \},\quad
%C_{4} = \{ T^{2},\, ST^{2},\, T^{2}S,\, TST \}.
%\]
%Since $A_4$ has four conjugacy classes, it has four inequivalent irreducible representations. By the standard sum–of–squares formula for the dimensions $d_i$ of the irreducible representations,
%\[
%\sum_{i=1}^{4} d_i^2 = |A_4| = 12,
%\]
%with $d_i \in \mathbb{Z}_{>0}$. The only possibility is
%\[
%(d_1,d_2,d_3,d_4) = (1,1,1,3),
%\]
%so $A_4$ has three one-dimensional representations and one three-dimensional representation. Using the orthogonality relations for characters, we obtain the representation character in Table~\ref{tab:A4-character-table}.

More concretely, we label the four vertices of a regular tetrahedron by $1,2,3,4$. 
The generator $S$ is the permutation $(12)(34)$, corresponding to 
a $180^\circ$ rotation about the axis through the midpoints of the opposite 
edges connecting vertices $1$--$2$ and $3$--$4$. The generator $T$ is the 
permutation $(123)$, corresponding to a $120^\circ$ rotation about the axis 
through vertex $4$ and the center of the opposite face $(123)$.
Using the generators $S$ and $T$ and the group relations in Eq.~(\ref{eq:a4}), 
the group $A_{4}$ has 12 elements, which can be chosen as
\begin{equation}
\nonumber
\{ e,\, S,\, TST^{2},\, T^{2}ST,\, T,\, TS,\, ST,\, STS,\, T^{2},\, ST^{2},\, T^{2}S,\, TST \}.
\end{equation}
These 12 elements fall into four conjugacy classes,
\begin{equation}
\nonumber
\begin{aligned}
C_{1} &= \{ e \},\quad 
C_{2} = \{ S,\, TST^{2},\, T^{2}ST \},\\
C_{3} &= \{ T,\, TS,\, ST,\, STS \},\quad
C_{4} = \{ T^{2},\, ST^{2},\, T^{2}S,\, TST \}.
\end{aligned}
\end{equation}
Since $A_{4}$ has four conjugacy classes, it has four inequivalent irreducible representations. 
By the standard sum--of--squares formula for the dimensions $d_i$ of the irreducible representations,
\begin{equation}
\nonumber
\sum_{i=1}^{4} d_i^2 = |A_4| = 12,
\end{equation}
with $d_i \in \mathbb{Z}_{>0}$. The only possibility is
\begin{equation}
\nonumber
(d_1,d_2,d_3,d_4) = (1,1,1,3),
\end{equation}
so $A_4$ has three one-dimensional representations and one three-dimensional representation.
Using the character orthogonality relations, we obtain the character table shown in Table~\ref{tab:A4-character-table}.

In the three--dimensional irreducible representation of $A_{4}$, the
generators $S$ and $T$ can be written in two convenient bases.

\paragraph{(i) Eigenbasis of $S$.}
With eigenvalues $1,-1,-1$ for $S$, the generators are represented by
\begin{equation}
    T=\begin{pmatrix}
0 & 1 & 0\\
0 & 0 & 1\\
1 & 0 & 0
\end{pmatrix},
\qquad
S=\begin{pmatrix}
1 & 0 & 0\\
0 & -1 & 0\\
0 & 0 & -1
\end{pmatrix}.
\end{equation}

\paragraph{(ii) Eigenbasis of $T$.}
Let $\omega=e^{2\pi i/3}$. In the basis where
$T'=\mathrm{diag}(1,\omega,\omega^{2})$, the generators read
\begin{equation}
    T'=\begin{pmatrix}
1 & 0 & 0\\
0 & \omega & 0\\
0 & 0 & \omega^{2}
\end{pmatrix},
\qquad
S'=\frac{1}{3}\begin{pmatrix}
-1 & 2 & 2\\
2 & -1 & 2\\
2 & 2 & -1
\end{pmatrix}.
\end{equation}

The two bases are related by the unitary matrix
\begin{equation}
\nonumber
P=\frac{1}{\sqrt{3}}
\begin{pmatrix}
1 & 1 & 1\\
1 & \omega & \omega^{2}\\
1 & \omega^{2} & \omega
\end{pmatrix},
\end{equation}
such that
\begin{equation}
\nonumber
T' = P^{-1}TP, \qquad S' = P^{-1}SP .
\end{equation}
In the eigenbasis of $S$, denote two $A_{4}$ triplets by
$a^{T}=(a_{1},a_{2},a_{3})$ and $b^{T}=(b_{1},b_{2},b_{3})$, with
$\omega=e^{2\pi i/3}$. In this basis, the tensor-product decomposition
$\mathbf{3}\otimes\mathbf{3}=\mathbf{1}\oplus\mathbf{1'}\oplus\mathbf{1''}\oplus
\mathbf{3}_{\mathbf{s}}\oplus\mathbf{3}_{\mathbf{a}}$ reads
\begin{equation}
\begin{aligned}
\mathbf{1}\;&\equiv\;(ab)
   \;=\; a_{1}b_{1}+a_{2}b_{2}+a_{3}b_{3},\\
\mathbf{1'}\;&\equiv\;(ab)^{\prime}
   \;=\; a_{1}b_{1}+\omega^{2}a_{2}b_{2}+\omega\,a_{3}b_{3},\\
\mathbf{1''}\;&\equiv\;(ab)^{\prime\prime}
   \;=\; a_{1}b_{1}+\omega\,a_{2}b_{2}+\omega^{2}a_{3}b_{3},\\
\mathbf{3}_{\mathbf{s}}\;&\equiv\;(ab)_{\mathbf{3}_{\mathbf{s}}}
   \;=\;
   \begin{pmatrix}
     a_{3}b_{2}+a_{2}b_{3}\\[2pt]
     a_{1}b_{3}+a_{3}b_{1}\\[2pt]
     a_{2}b_{1}+a_{1}b_{2}
   \end{pmatrix},
\qquad
\mathbf{3}_{\mathbf{a}}\;\equiv\;(ab)_{\mathbf{3}_{\mathbf{a}}}
   \;=\;
   \begin{pmatrix}
     a_{3}b_{2}-a_{2}b_{3}\\[2pt]
     a_{1}b_{3}-a_{3}b_{1}\\[2pt]
     a_{2}b_{1}-a_{1}b_{2}
   \end{pmatrix}.
\end{aligned}
\label{eq:A4_CG_Sbasis}
\end{equation}
% --- A4 tensor products in S- and T-eigenbases ---

In the eigenbasis of $S$, the singlet contractions of two
triplets can be written in matrix form as
\begin{equation}
  (ab)_{\mathbf{1}} \equiv a^{T} G_{1}\, b,\qquad
  (ab)_{\mathbf{1}'} \equiv a^{T} G_{1'}\, b,\qquad
  (ab)_{\mathbf{1}''} \equiv a^{T} G_{1''}\, b .
\end{equation}
Here the coefficient matrices are
\begin{equation}
  G_{1}=\mathrm{diag}(1,1,1),\qquad
  G_{1'}=\mathrm{diag}(1,\omega^{2},\omega),\qquad
  G_{1''}=\mathrm{diag}(1,\omega,\omega^{2}),
\end{equation}
with $\omega\equiv e^{2\pi i/3}$.
The $S$- and $T$-eigenbases are related by the change of basis
\begin{equation}
  a'_i = P^{-1}_{ij} a_j,\qquad b'_i = P^{-1}_{ij} b_j.
\end{equation}

Using this transformation, the singlet contractions can be recast in the
eigenbasis of $T$. In the following, $(a_1,a_2,a_3)$ and $(b_1,b_2,b_3)$
denote the components in the $T$-eigenbasis:
\begin{equation}
  (ab)_{\mathbf{1}} \;=\; a_1 b_1 + a_3 b_2 + a_2 b_3 ,
\end{equation}
\begin{equation}
  (ab)_{\mathbf{1}'} \;=\; a_3 b_3 + a_1 b_2 + a_2 b_1 ,
\end{equation}
\begin{equation}
  (ab)_{\mathbf{1}''} \;=\; a_2 b_2 + a_3 b_1 + a_1 b_3 .
\end{equation}

For the triplet products, we first express the contractions in the $S$-basis
and then apply the change of basis above to obtain their forms in the
$T$-basis. The symmetric and antisymmetric triplets are
\begin{equation}
  (ab)_{\mathbf{3}_{\mathbf{s}}}
  =
  \begin{pmatrix}
    2 a_1 b_1 - a_2 b_3 - a_3 b_2\\[2pt]
    2 a_3 b_3 - a_1 b_2 - a_2 b_1\\[2pt]
    2 a_2 b_2 - a_3 b_1 - a_1 b_3
  \end{pmatrix},
\end{equation}
\begin{equation}
  (ab)_{\mathbf{3}_{\mathbf{a}}}
  =
  \begin{pmatrix}
    a_3 b_2 - a_2 b_3\\[2pt]
    a_2 b_1 - a_1 b_2\\[2pt]
    a_1 b_3 - a_3 b_1
  \end{pmatrix}.
\end{equation}
These five tensor-product rules are the ones adopted in this work.

\subsection{Flavor construction in a generic \texorpdfstring{$A_4$}{A4} flavor symmetry}
\label{flavor_model}
For a given fermion species (e.g.\ the up-type quarks), one may perform a flavor rotation on the flavor triplet $\Psi_u=(u, c, t)^T$ as
\begin{equation}
\Psi_u \;\to\; U\,\Psi_u,\qquad U\in U(3),
\label{eq:weakbasisU3}
\end{equation}
without changing $\mathcal L_{\rm kinetic}+\mathcal L_{\rm gauge}$. All such related
field triplets define different weak bases (or flavor bases). Once Yukawa
couplings are switched on, the up-type mass matrix $M_u$ is in general non-diagonal in a
generic weak basis and is diagonalized by a biunitary transformation,
\begin{equation}
U_L^{u\dagger} M_u U_R^u = M_u^{\rm diag}\equiv \mathrm{diag}(m_u,m_c,m_t).
\label{eq:mubidiag}
\end{equation}
This transformation maps the weak basis to the mass basis.
When a non-Abelian discrete flavor symmetry is imposed, the weak-basis freedom is
restricted by the representation assignments. As a concrete example, consider an
$A_4$ flavor symmetry where the three up-type quark flavors are assigned to the
one-dimensional irreducible representations
\begin{equation}
u_R \sim \mathbf{1},\qquad c_R \sim \mathbf{1}',\qquad t_R \sim \mathbf{1}'' ,
\label{eq:A4assign}
\end{equation}
and assemble them into a reducible flavor triplet
$\Psi_u=(u,\,c,\,t)^T$. In the standard $A_4$ presentation with generators $S$ and $T$
satisfying $S^2=(ST)^3=T^3=e$, the action of $T$ on these one-dimensional irreps is
\begin{equation}
\mathbf{1}:~T=1,\qquad
\mathbf{1}':~T=\omega,\qquad
\mathbf{1}'':~T=\omega^2,\qquad \omega=e^{2\pi i/3},
\end{equation}
so that in the triplet space one has the diagonal representation matrix
\begin{equation}
\rho_u(T)=\mathrm{diag}(1,\omega,\omega^2).
\label{eq:Tdiag}
\end{equation}
A weak basis where $\rho_u(T)$ takes the diagonal form in Eq.~\eqref{eq:Tdiag} is referred to
as the $T$-diagonal basis; choosing the assignments in Eq.~\eqref{eq:A4assign} is therefore
equivalent to fixing a $T$-diagonal weak basis.

Under a general weak-basis change in Eq.~\eqref{eq:weakbasisU3}, the representation matrix of
$T$ transforms by similarity,
\begin{equation}
\rho_u(T)\;\to\;\rho_u'(T)=U\,\rho_u(T)\,U^\dagger .
\end{equation}
Requiring that the irrep assignments in Eq.~\eqref{eq:A4assign} remain valid in the transformed
basis is equivalent to demanding $\rho_u'(T)=\rho_u(T)$, i.e.
$U\,\rho_u(T)=\rho_u(T)\,U$. Writing this condition in components gives, for
$i\neq j$,
\begin{equation}
\bigl(\rho_u(T)_{jj}-\rho_u(T)_{ii}\bigr)\,U_{ij}=0 .
\end{equation}
Since the eigenvalues of $\rho_u(T)$ are $\{1,\omega,\omega^2\}$ and are all distinct, we have
$U_{ij}=0$ for $i\neq j$. Hence the only allowed weak-basis transformations that
preserve the assignments in Eq.~\eqref{eq:A4assign} are rephasings,
\begin{equation}
U=\mathrm{diag}(e^{i\alpha_1},e^{i\alpha_2},e^{i\alpha_3}),
\label{eq:Udiagphase}
\end{equation}
so the original $U(3)$ freedom collapses to independent phase redefinitions.

In a generic $A_4$ flavor framework with the lepton mixing matrix approximated by the tri-bimaximal (TB) form in Eq.~\eqref{eq:TB}, the lepton-sector Yukawa interactions can be implemented by introducing $A_4$-triplet flavons $\varphi_T$ and $\varphi_S$ (both transforming as $\mathbf{3}$ of $A_4$) and an $A_4$-trivial singlet $\xi$. Schematically, one may write~\cite{Altarelli_2005,Altarelli_2006}
\begin{equation}
    \mathcal{L}
= y_e\, e^{c}\,(\varphi_T\, L)\,\frac{h_d}{\Lambda}
+ y_\mu\, \mu^{c}\,(\varphi_T\, L)^{\prime}\,\frac{h_d}{\Lambda}
+ y_\tau\, \tau^{c}\,(\varphi_T\, L)^{\prime\prime}\,\frac{h_d}{\Lambda}
+ \frac{x_a}{\Lambda^{2}}\,\xi\,(L h_u\, L h_u)
+ \frac{x_b}{\Lambda^{2}}\,(\varphi_S\, L h_u\, L h_u)
+ \text{h.c.} + \ldots,
\end{equation}
where $\Lambda$ is the cutoff scale, and the leptonic and scalar fields transform under $A_4$ as
$\{e^c,h_u,h_d\} \sim \mathbf{1}$, $\mu^c\sim \mathbf{1}^{\prime\prime}$, $\tau^c\sim \mathbf{1}^\prime$, and $L\sim \mathbf{3}$.
The symbols $(\cdots)_{\mathbf{1},\mathbf{1'},\mathbf{1''}}$ denote $A_4$ contractions into the corresponding singlet representations.
Specifically, we adopt the eigenbasis of the $A_4$ generator $S$ for all triplet fields. In this basis, the flavon components $\varphi_{T i}$ and $\varphi_{S i}$, as well as the singlet $\xi$, are taken to be \emph{real} scalar fields.

For the one-dimensional irreducible representations,
each field $e^c$, $h_u$, $h_d$, $\mu^c$, and $\tau^c$
is an ordinary complex scalar or spin-$\tfrac{1}{2}$ field (as appropriate) that additionally transforms according to its assigned $A_4$ irrep.

The left-handed $SU(2)_L$ lepton doublets are
\begin{equation}
    L_\ell \;=\;
\begin{pmatrix}
\ell_L\\ \nu_{\ell}
\end{pmatrix},\qquad \ell=e,\mu,\tau,
\end{equation}
and are assembled into an $A_4$ triplet in flavor space,
\begin{equation}
    L \;=\;
\begin{pmatrix}
L_e\\ L_\mu\\ L_\tau
\end{pmatrix}
\sim \mathbf{3},
\end{equation}
so that the three flavors $(L_e,L_\mu,L_\tau)$ transform into one another under the triplet representation of $A_4$.
With these field representations, once the Higgs fields develop their VEVs and the flavons acquire vacuum expectation values, the electroweak gauge symmetry
$SU(2)_L\times U(1)_Y$ and the $A_4$ flavor symmetry are spontaneously broken.
We assume the following vacuum alignment for the flavons,
\begin{equation}
  \langle \varphi_T \rangle = (v_T,\,0,\,0),\qquad
  \langle \varphi_S \rangle = (v_S,\,v_S,\,v_S),\qquad
  \langle \xi \rangle = u,
\end{equation}
and the electroweak Higgs doublets acquire VEVs
\begin{equation}
  \langle h_u\rangle = v_u,\qquad
  \langle h_d\rangle = v_d .
\end{equation}

% Charged--lepton masses
After symmetry breaking, the Yukawa terms take the form
\begin{equation}
\begin{aligned}
\mathcal{L}_{\ell}
&= v_d\,\frac{v_T}{\Lambda}\!\left( y_e\, e^{c} e \;+\; y_\mu\, \mu^{c} \mu \;+\; y_\tau\, \tau^{c} \tau \right)
\\[2pt]
&\quad +\, x_a\, v_u^{2}\,\frac{u}{\Lambda^{2}}\!\left( \nu_e \nu_e \;+\; 2\,\nu_\mu \nu_\tau \right)
\\[2pt]
&\quad +\, x_b\, v_u^{2}\,\frac{2 v_S}{3\Lambda^{2}}
\!\left( \nu_e \nu_e \;+\; \nu_\mu \nu_\mu \;+\; \nu_\tau \nu_\tau
 \;-\; \nu_e \nu_\mu \;-\; \nu_\mu \nu_\tau \;-\; \nu_\tau \nu_e \right)
\;+\; \text{h.c.} \;+\; \cdots \, .
\end{aligned}
\end{equation}
In this setup, the charged-lepton and neutrino mass matrices take the form:
\begin{equation}
m_\ell \;=\; v_d\,\frac{v_T}{\Lambda}\!
\begin{pmatrix}
y_e & 0 & 0\\[2pt]
0 & y_\mu & 0\\[2pt]
0 & 0 & y_\tau
\end{pmatrix},
\qquad
m_\nu \;=\; \frac{v_u^{2}}{\Lambda}\!
\begin{pmatrix}
a+\tfrac{2}{3}b & -\tfrac{1}{3}b & -\tfrac{1}{3}b\\[4pt]
-\tfrac{1}{3}b & \tfrac{2}{3}b & a-\tfrac{1}{3}b\\[4pt]
-\tfrac{1}{3}b & a-\tfrac{1}{3}b & \tfrac{2}{3}b
\end{pmatrix},
\end{equation}
with
\begin{equation}
a \equiv 2x_a\,\frac{u}{\Lambda},\qquad
b \equiv 2x_b\,\frac{v_S}{\Lambda}\, .
\end{equation}
In this setup the neutrino mass matrix is diagonalized by the tri-bimaximal matrix
$U_{\mathrm{TB}}$ in Eq.~\eqref{eq:TB}, namely
\begin{equation}
  U_{\mathrm{TB}}^{T}\, m_\nu \, U_{\mathrm{TB}}
  \;=\; \frac{v_u^{2}}{\Lambda}\,
  \mathrm{diag}\!\bigl(m_1=a+b,\; m_2=a,\; m_3=-a+b\bigr)\, .
\end{equation}

In particular, when the flavon acquires the alignment
$
\langle \varphi_T \rangle = (v_T,\,0,\,0)\,,
$
the set of group elements that leaves it invariant is
$\{e,\,T,\,T^2\}\simeq \mathbb{Z}_3$.
By contrast, for the alignment
$
\langle \varphi_S \rangle = (v_S,\,v_S,\,v_S)\,,
$
the invariant set is $\{e,\,S\}\simeq \mathbb{Z}_2$.
Thus, this choice of vacuum alignments induces different residual symmetries in the
charged-lepton and neutrino sectors ($\mathbb{Z}_3$ and $\mathbb{Z}_2$, respectively),
which is precisely the pattern compatible with tri-bimaximal models.

A distinctive feature of the $A_4$ setup is that its three inequivalent singlet
representations map uniquely to the three charged-lepton flavors. If one interchanges the
assignments of $\mu^c$ and $\tau^c$, i.e.\ takes
$e^c\sim\mathbf{1}$, $\mu^c\sim\mathbf{1'}$, $\tau^c\sim\mathbf{1''}$,
then after electroweak and flavor breaking, the charged-lepton Yukawa terms read
\begin{equation}
  \mathcal{L}_\ell
  \;=\;
  v_T\,\frac{v_d}{\Lambda}\,
  \bigl(y_e\,e^c e \;+\; y_\mu\,\mu^c \tau \;+\; y_\tau\,\tau^c \mu \bigr) \,,
\end{equation}
which correspond to the mass matrix
\begin{equation}
  m'_\ell
  \;=\;
  v_T\,\frac{v_d}{\Lambda}
  \begin{pmatrix}
    y_e & 0   & 0 \\
    0   & 0   & y_\tau \\
    0   & y_\mu & 0
  \end{pmatrix},
  \qquad
  m'_\ell m_\ell^{\prime\dagger}
  \;=\;
  v_T^{\,2}\,\frac{v_d^{\,2}}{\Lambda^{2}}
  \begin{pmatrix}
    |y_e|^2 & 0 & 0\\
    0 & |y_\tau|^2 & 0\\
    0 & 0 & |y_\mu|^2
  \end{pmatrix}.
\end{equation}

Consider first the diagonalization by multiplying on the \emph{right} with the
permutation matrix
\begin{equation}
  P \;=\;
  \begin{pmatrix}
    1 & 0 & 0\\
    0 & 0 & 1\\
    0 & 1 & 0
  \end{pmatrix}.
\end{equation}
This gives
\begin{equation}
  m'_{\ell,\mathrm{diag}}
  \;=\;
  m'_\ell P
  \;=\;
  v_T\,\frac{v_d}{\Lambda}
  \begin{pmatrix}
    y_e & 0 & 0\\
    0   & y_\tau & 0\\
    0   & 0 & y_\mu
  \end{pmatrix}.
\end{equation}
While the PMNS matrix is unaffected by this column permutation,
one should note that in this option the parameter $y_\mu$ actually furnishes the
$\tau$ mass and $y_\tau$ the $\mu$ mass (the right-handed fields are permuted).

Alternatively, diagonalizing by multiplying on the left with
$P^\dagger$ yields
\begin{equation}
  m'_{\ell,\mathrm{diag}}
  \;=\;
  P^\dagger m'_\ell
  \;=\;
  v_T\,\frac{v_d}{\Lambda}
  \begin{pmatrix}
    y_e & 0 & 0\\
    0   & y_\mu & 0\\
    0   & 0 & y_\tau
  \end{pmatrix},
\end{equation}
which realizes the conventional identification
$m_e\propto y_e$, $m_\mu\propto y_\mu$, and $m_\tau\propto y_\tau$.
Although the second option restores the conventional identification of the
charged-lepton masses, the PMNS matrix is correspondingly permuted. In
particular,
\begin{equation}
  U'_{\mathrm{HPS}}
  \;=\;
  P^\dagger\,U_{\mathrm{HPS}}
  \;=\;
  \begin{pmatrix}
    \sqrt{\tfrac{2}{3}} & \tfrac{1}{\sqrt{3}} & 0\\
    -\tfrac{1}{\sqrt{6}} & \tfrac{1}{\sqrt{3}} & \tfrac{1}{\sqrt{2}}\\
    -\tfrac{1}{\sqrt{6}} & \tfrac{1}{\sqrt{3}} & -\tfrac{1}{\sqrt{2}}
  \end{pmatrix}.
\end{equation}
Therefore, with the correct assignment, the Yukawa coefficients align with the
charged-lepton mass terms without spoiling the tri-bimaximal (TBM) approximation.

\section{Mathematical Properties of \texorpdfstring{$A_4$}{A4} Modular Forms}
\subsection{Lattices and \texorpdfstring{$SL(2,\mathbb{Z})$}{SL(2,Z)} basis changes}
A two--dimensional lattice can be written as
\begin{equation}
\Lambda = \mathbb{Z} e_1 \oplus \mathbb{Z} e_2 ,
\end{equation}
where $e_1$ and $e_2$ are two linearly independent basis vectors. In a
standard coordinate choice one may take $e_1=(1,0)^T$ and $e_2=(0,1)^T$,
while more generally, for a torus with periods $\omega_{1,2}$ one may take
$e_1=(\omega_1,0)^T$ and $e_2=(0,\omega_2)^T$.

The modular group $SL(2,\mathbb{Z})$ consists of integer matrices with unit
determinant,
\begin{equation}
SL(2,\mathbb{Z})=
\left\{
g=
\begin{pmatrix}
a & b\\
c & d
\end{pmatrix}
\ \Big|\ 
a,b,c,d\in\mathbb{Z},\ ad-bc=1
\right\}.
\end{equation}
It acts on the lattice basis by linear recombination,
\begin{equation}
g
\begin{pmatrix}
e_1 \\ e_2
\end{pmatrix}
=
\begin{pmatrix}
a & b\\
c & d
\end{pmatrix}
\begin{pmatrix}
e_1 \\ e_2
\end{pmatrix}
=
\begin{pmatrix}
e_1' \\ e_2'
\end{pmatrix}
=
\begin{pmatrix}
a e_1 + b e_2\\
c e_1 + d e_2
\end{pmatrix}.
\end{equation}
This change of basis has three important geometric properties:
\begin{enumerate}
\item \textbf{Area preservation.} The oriented area spanned by $(e_1,e_2)$ is
preserved because
\begin{equation}
\det
\begin{pmatrix}
a & b\\
c & d
\end{pmatrix}
=1 .
\end{equation}

\item \textbf{Orientation preservation.} Since $\det g>0$, the ordered pair
$(e_1',e_2')$ keeps the same (right--handed or left--handed) orientation as
$(e_1,e_2)$.

\item \textbf{The lattice itself is unchanged.} Although the basis changes,
the set of lattice points is the same:
\begin{equation}
\Lambda' = \mathbb{Z} e_1' \oplus \mathbb{Z} e_2'
         = \mathbb{Z}(a e_1 + b e_2)\oplus \mathbb{Z}(c e_1 + d e_2)
         = \Lambda .
\end{equation}
Thus $SL(2,\mathbb{Z})$ relates different descriptions of the same lattice.
\end{enumerate}

Property 3 can be understood equivalently as keeping the basis fixed but
transforming the integer coordinates of lattice vectors. Any lattice vector
can be decomposed as
\begin{equation}
v = (e_1,e_2)
\begin{pmatrix}
v_1\\
v_2
\end{pmatrix},
\qquad v_1,v_2\in\mathbb{Z}.
\end{equation}
Under a basis change $(e_1,e_2)\to(e_1',e_2')$ with $e'_i=g e_i$, the same
geometric vector $v$ is represented by new coordinates $(v_1',v_2')$:
\begin{equation}
(e_1',e_2')
\begin{pmatrix}
v_1'\\
v_2'
\end{pmatrix}
=
(e_1,e_2)
\begin{pmatrix}
v_1\\
v_2
\end{pmatrix}.
\end{equation}
Using $(e_1',e_2')=(e_1,e_2) g^T$, one finds
\begin{equation}
\begin{pmatrix}
v_1'\\
v_2'
\end{pmatrix}
=
(g^T)^{-1}
\begin{pmatrix}
v_1\\
v_2
\end{pmatrix},
\qquad (g^T)^{-1}\in SL(2,\mathbb{Z}).
\end{equation}
Since $(g^T)^{-1}$ is again an $SL(2,\mathbb{Z})$ matrix, we may relabel it
as $g$ and write the coordinate action simply as
\begin{equation}
\begin{pmatrix}
v_1'\\
v_2'
\end{pmatrix}
=
g
\begin{pmatrix}
v_1\\
v_2
\end{pmatrix},
\qquad g\in SL(2,\mathbb{Z}).
\end{equation}

The linear transformations of integer coordinates can be grouped by the
value of $\mathrm{tr}\,g$:
\begin{enumerate}
\item \textbf{Finite-order case (elliptic elements): discrete rotations around lattice points.}
Typical generators are
\begin{equation}
\mathrm{tr}\,g=-2:\ 
g=
\begin{pmatrix}
-1 & 0\\
0 & -1
\end{pmatrix}
\quad\Rightarrow\quad
(v_1',v_2')=(-v_1,-v_2)
\ \text{(rotation by $180^\circ$)},
\end{equation}
\begin{equation}
\mathrm{tr}\,g=-1:\ 
g=
\begin{pmatrix}
0 & -1\\
1 & -1
\end{pmatrix}
\ \text{(counterclockwise rotation by $120^\circ$)},
\end{equation}
\begin{equation}
\mathrm{tr}\,g=0:\ 
g=
\begin{pmatrix}
0 & -1\\
1 & 0
\end{pmatrix}
\ \text{(counterclockwise rotation by $90^\circ$)},
\end{equation}
\begin{equation}
\mathrm{tr}\,g=1:\ 
g=
\begin{pmatrix}
0 & -1\\
1 & 1
\end{pmatrix}
\ \text{(counterclockwise rotation by $60^\circ$)}.
\end{equation}
The transpose $g^T$ generates the corresponding rotation in the opposite
direction.

\item \textbf{Parabolic case ($|\mathrm{tr}\,g|=2$ with $g\neq \pm I$): lattice shear/translation.}
A standard family is
\begin{equation}
T^N=
\begin{pmatrix}
1 & N\\
0 & 1
\end{pmatrix},
\qquad
T^N
\begin{pmatrix}
v_1\\
v_2
\end{pmatrix}
=
\begin{pmatrix}
v_1+N v_2\\
v_2
\end{pmatrix},
\end{equation}
with inverse
\begin{equation}
(T^N)^{-1}=
\begin{pmatrix}
1 & -N\\
0 & 1
\end{pmatrix}.
\end{equation}

\item \textbf{Hyperbolic case ($|\mathrm{tr}\,g|>2$): stretching/compression.}
These transformations rescale lattice directions (one expanded and the other
contracted) while preserving the lattice area.
\end{enumerate}

As an explicit illustration, take an integer vector
\begin{equation}
v=
\begin{pmatrix}
v_1\\
v_2
\end{pmatrix}
=
\begin{pmatrix}
1\\
5
\end{pmatrix},
\end{equation}
and act with the $\mathrm{tr}\,g=-1$ matrix
\begin{equation}
g=
\begin{pmatrix}
0 & -1\\
1 & -1
\end{pmatrix},
\end{equation}
corresponding to a counterclockwise $120^\circ$ rotation. Then
\begin{equation}
v' = g v =
\begin{pmatrix}
-5\\
-4
\end{pmatrix}.
\end{equation}

For the hexagonal (or $A_2$) lattice, the inner product in the basis
$(1,0)^T,(0,1)^T$ is encoded by the metric
\begin{equation}
G=
\begin{pmatrix}
1 & -\tfrac12\\
-\tfrac12 & 1
\end{pmatrix}.
\end{equation}
For the matrix $g$ above, one finds that it preserves this metric,
\begin{equation}
g^T G g = G,
\end{equation}
so the norm is invariant:
\begin{equation}
\langle v',v'\rangle
= v^T g^T G g v
= v^T G v
= \langle v,v\rangle
= 21.
\end{equation}
The angle between $v$ and $v'$ follows from
\begin{equation}
\cos\theta
=
\frac{\langle v',v\rangle}
{\sqrt{\langle v,v\rangle\,\langle v',v'\rangle}}
=
\frac{-21/2}{21}
=
-\frac12,
\end{equation}
hence $\theta=120^\circ$, consistent with the geometric interpretation.

As discussed above, an element
$
g=\begin{pmatrix} a & b \\ c & d \end{pmatrix}\in SL(2,\mathbb{Z})
$
acts on a rank--two lattice either as a change of basis or as an active
transformation of lattice vectors. We now specialize to the complex torus,
for which the lattice can be parameterized by a complex modulus $\tau$.

Consider the lattice
\begin{equation}
\Lambda_\tau = \mathbb{Z}\,\omega_1 + \mathbb{Z}\,\omega_2 \subset \mathbb{C},
\end{equation}
and choose a convenient basis such that $\omega_2=1$ and $\omega_1=\tau$.
Equivalently, the period vector can be represented as
\begin{equation}
\begin{pmatrix}
\omega_1\\
\omega_2
\end{pmatrix}
=
\begin{pmatrix}
\tau\\
1
\end{pmatrix},
\end{equation}
so that $\Lambda_\tau=\mathbb{Z}\tau+\mathbb{Z}$. Acting with
$g\in SL(2,\mathbb{Z})$ on the periods gives
\begin{equation}
\begin{pmatrix} a & b \\ c & d \end{pmatrix}
\begin{pmatrix} \tau \\ 1 \end{pmatrix}
=
\begin{pmatrix} a\tau+b \\ c\tau+d \end{pmatrix}
=
(c\tau+d)
\begin{pmatrix} \tau' \\ 1 \end{pmatrix},
\qquad
\tau'=\frac{a\tau+b}{c\tau+d}.
\end{equation}
Therefore the transformed lattice can be expressed as
\begin{equation}
\Lambda_{\tau'}=\mathbb{Z}\tau'+\mathbb{Z}
\;\simeq\;
(c\tau+d)^{-1}\Lambda_\tau,
\end{equation}
i.e.\ $\Lambda_{\tau'}$ differs from $\Lambda_\tau$ only by a global complex
rescaling by $(c\tau+d)^{-1}$. The corresponding complex tori
\begin{equation}
T_\tau=\mathbb{C}/\Lambda_\tau,
\qquad
T_{\tau'}=\mathbb{C}/\Lambda_{\tau'}
\end{equation}
are thus conformally equivalent: the modular transformation does not change
the intrinsic shape of the torus, but it rescales all lattice vectors by the
overall factor $(c\tau+d)^{-1}$.

This observation explains the appearance of the automorphy factor in modular
symmetry. If $\tau$ is taken as the unique parameter describing the torus,
then any quantity that is sensitive to the overall scale of lattice vectors
must transform covariantly under $SL(2,\mathbb{Z})$. Concretely, for an
object $\Phi(\tau)$ with modular weight $k$, modular invariance requires
\begin{equation}
\Phi(\tau)\;\longrightarrow\;\Phi(\tau')
=(c\tau+d)^{-k}\,\Phi(\tau),
\end{equation}
so that the rescaling of the lattice is compensated by the factor
$(c\tau+d)^{-k}$. Although the discussion above was phrased in terms of lattice
basis vectors, the same covariance applies to any vector on the torus, in
accordance with property~3 in the previous subsection.

\subsection{Ring structure of \texorpdfstring{$A_4$}{A4} modular forms}
\label{app:forms}

The graded ring of holomorphic modular forms of level $\Gamma(3)$ is
generated by $Y_i$ modulo a single quadratic relation,
\begin{equation}\label{eq:QuadRel}
(YY)_{\mathbf{1}''}=Y_2^2+2\,Y_1Y_3 \;=\; 0 .
\end{equation}
Every modular form of weight $2n$ can be expressed as a homogeneous polynomial of total
degree $n$ in the variables $Y_i$, reduced using Eq.~\eqref{eq:QuadRel}.

\paragraph{Weight 2.} There is only the triplet
\begin{equation}
M_2 \cong \mathbf 3,\qquad Y^{(2)}_{\mathbf 3}=(Y_1,Y_2,Y_3).
\end{equation}

\paragraph{Weight 4.} From \(\mathrm{Sym}^2(\mathbf 3)\) and \eqref{eq:QuadRel}:
\begin{align}
M_4 \cong \mathbf 1 \oplus \mathbf 1' \oplus \mathbf 3,\qquad
\begin{cases}
Y^{(4)}_{\mathbf 1}   = Y_1^2 + 2Y_2Y_3,\\
Y^{(4)}_{\mathbf 1'}  = Y_3^2 + 2Y_1Y_2,\\
Y^{(4)}_{\mathbf 1''} = Y_2^2 + 2Y_1Y_3 \equiv 0,\\
Y^{(4)}_{\mathbf 3}   = \bigl(Y_1^2-Y_2Y_3,\; Y_3^2-Y_1Y_2,\; Y_2^2-Y_1Y_3\bigr).
\end{cases}
\end{align}

\paragraph{Weight 6.} From \(\mathrm{Sym}^3(\mathbf 3)\) and \eqref{eq:QuadRel}:
\begin{align}
M_6 \cong \mathbf 1 \oplus \mathbf 3 \oplus \mathbf 3,\qquad
\begin{cases}
Y^{(6)}_{\mathbf 1} = Y_1^3 + Y_2^3 + Y_3^3 - 3Y_1Y_2Y_3,\\[2pt]
Y^{(6)}_{\mathbf 3a} =
\bigl(Y_1^3+2Y_1Y_2Y_3,\; Y_1^2Y_2+2Y_2^2Y_3,\; Y_1^2Y_3+2Y_3^2Y_2\bigr),\\[2pt]
Y^{(6)}_{\mathbf 3b} =
\bigl(Y_3^3+2Y_1Y_2Y_3,\; Y_3^2Y_1+2Y_1^2Y_2,\; Y_3^2Y_2+2Y_2^2Y_1\bigr),\\[2pt]
Y^{(6)}_{\mathbf 3c} =
\bigl(Y_2^3+2Y_1Y_2Y_3,\; Y_2^2Y_3+2Y_3^2Y_1,\; Y_2^2Y_1+2Y_1^2Y_3\bigr)\equiv 0.
\end{cases}
\end{align}
For holomorphic vector-valued forms $F\in\mathcal{M}_k(\rho)$ and
$G\in\mathcal{M}_{k'}(\rho')$, their pointwise product transforms with
weight $k+k'$ and representation $\rho\otimes\rho'$.

It is convenient to collect all pairwise products via the linear map
\begin{equation}\label{eq:multmap}
\mu_{k,k'}:\ \mathcal{M}_k \otimes \mathcal{M}_{k'} \longrightarrow \mathcal{M}_{k+k'}.
\end{equation}
In general, $\mu_{k,k'}$ is not injective, and typically
$\dim(\mathcal{M}_k)\,\dim(\mathcal{M}_{k'})>\dim(\mathcal{M}_{k+k'})$.
Hence many product tensors lie in $\ker\mu_{k,k'}$ and should be rewritten as
linear combinations of a basis of $\mathcal{M}_{k+k'}$.
Keeping both the original products and the corresponding linear combinations
produces an overcomplete, redundant set; in practice one expresses
results in terms of a chosen basis of $\mathcal{M}_{k+k'}$.

For example, consider the case $\mathcal{M}_2 \otimes \mathcal{M}_{4}$.
Let $Y\equiv Y^{(2)}_{\mathbf{3}}=(Y_1,Y_2,Y_3)$ and
\begin{equation}
X\equiv Y^{(4)}_{\mathbf{3}}=(X_1,X_2,X_3)=
\bigl(Y_1^2-Y_2Y_3,\; Y_3^2-Y_1Y_2,\; Y_2^2-Y_1Y_3\bigr).
\end{equation}
All identities below are understood as equalities in the quotient ring by the constrain
Eq.~\eqref{eq:QuadRel}.

\paragraph{(i) $\mathbf{3}\otimes \mathbf{1} \to \mathbf{3}$.}
\begin{equation}
Y^{(4)}_{\mathbf{1}}\cdot Y \;=\; Y^{(6)}_{\mathbf{3a}}.
\end{equation}

\paragraph{(ii) $\mathbf{3}\otimes \mathbf{1}' \to \mathbf{3}$.}
Using $\mathbf{1}'\otimes\mathbf{3}\cong\mathbf{3}$ as the fixed cyclic permutation
$(v_1,v_2,v_3)\mapsto (v_2,v_3,v_1)$, one has
\begin{equation}
Y^{(4)}_{\mathbf{1}'}\cdot Y \;\cong\; Y^{(6)}_{\mathbf{3b}}.
\end{equation}

\paragraph{(iii) $\mathbf{3}\otimes \mathbf{3} \to \mathbf{1}\oplus \mathbf{1}'\oplus \mathbf{1}''\oplus \mathbf{3}\oplus \mathbf{3}$.}
Define the symmetric and antisymmetric triplets via the tensor products in the eigenbasis of $T$:
\begin{equation}
\begin{aligned}
V_s=(Y\otimes X)_{\mathbf{3}_{\mathbf{s}}}&=
\begin{pmatrix}
2Y_1X_1-Y_2X_3-Y_3X_2\\
2Y_3X_3-Y_1X_2-Y_2X_1\\
2Y_2X_2-Y_1X_3-Y_3X_1
\end{pmatrix},\qquad
V_a=(Y\otimes X)_{\mathbf{3}_{\mathbf{a}}}=
\begin{pmatrix}
Y_2X_3-Y_3X_2\\
Y_1X_2-Y_2X_1\\
Y_3X_1-Y_1X_3
\end{pmatrix}.
\end{aligned}
\end{equation}
Then one has the explicit componentwise identities
\begin{equation}
\begin{aligned}
(Y\otimes X)_{\mathbf{1}} &= Y^{(6)}_{\mathbf{1}},\\
(Y\otimes X)_{\mathbf{1}'} &= 0,\qquad (Y\otimes X)_{\mathbf{1}''}=0,\\[3pt]
V_s &= 2\,Y^{(6)}_{\mathbf{3a}} - Y^{(6)}_{\mathbf{3b}},\\
V_a &= -\,Y^{(6)}_{\mathbf{3b}} + Y^{(6)}_{\mathbf{3c}}
      \;=\; -\,Y^{(6)}_{\mathbf{3b}}\quad\text{since } Y^{(6)}_{\mathbf{3c}}\equiv 0.
\end{aligned}
\end{equation}
Thus every irrep component of
$Y^{(2)}_{\mathbf{3}}\otimes Y^{(4)}_{\mathbf{1},\mathbf{1}',\mathbf{3}}$
is a linear combination of the weight-$6$ basis
$\{Y^{(6)}_{\mathbf{1}},Y^{(6)}_{\mathbf{3a}},Y^{(6)}_{\mathbf{3b}}\}$;
the $\mathbf{1}'$ and $\mathbf{1}''$ projections vanish identically.

If one replaces holomorphic modular forms by (harmonic) Maa{\ss} forms,
the pointwise product still transforms with weight $k+k'$, but it
generically fails to satisfy the Laplace equation $\Delta_{k+k'}(fg)=0$.
Hence closure under products and the finite-basis reductions above are
specific to the holomorphic setting.

\subsection{Constructing rules for \texorpdfstring{$A_4$}{A4} high weight modular forms with \texorpdfstring{$k>2$}{k2}}
For convenience, we divide the even modular weights into two congruence classes,
$k = 4m$ and $k = 4m + 2$, with $m\in\mathbb{Z}_{>0}$. Before imposing the constraint in
Eq.~\eqref{eq:modular_constrain}, the corresponding dimensions are
\begin{equation}
\nonumber
\frac{(2m + 1)(2m + 2)}{2}
\quad\text{and}\quad
\frac{(2m + 2)(2m + 3)}{2} \, .
\end{equation}
For $k = 4m$, any singlet modular form can be written as
\begin{equation}
\nonumber
\bigl(Y_{\mathbf{1}}^{(4)}\bigr)^a
\bigl(Y_{\mathbf{1}'}^{(4)}\bigr)^b
\bigl(Y_{\mathbf{1}''}^{(4)}\bigr)^c \, ,
\end{equation}
while a triplet takes the form
\begin{equation}
\nonumber
\bigl(Y_{\mathbf{1}}^{(4)}\bigr)^\alpha
\bigl(Y_{\mathbf{1}'}^{(4)}\bigr)^\beta
\bigl(Y_{\mathbf{1}''}^{(4)}\bigr)^\gamma
Y^{(4)}_{\mathbf{3}} \, .
\end{equation}
Here all exponents are non-negative integers and satisfy the weight-counting
relations
\begin{equation}
    \begin{aligned}
        4a + 4b + 4c &= 4m \,,\\
        4\alpha + 4\beta + 4\gamma + 4 &= 4m \, .
    \end{aligned}
    \label{weight_4m_relation}
\end{equation}
In this case, the total number of possible combinations of $(a,b,c)$
(corresponding to singlet structures) is $m(m+1)/2$, while the number of
possible combinations of $(\alpha,\beta,\gamma)$ (corresponding to triplet
structures) is $(m+1)(m+2)/2$. The total dimension can be cross-checked against the
general counting and reads
\begin{equation}
    3 \cdot \frac{m(m+1)}{2}
    + \frac{(m+1)(m+2)}{2}
    = \frac{(2m + 2)(2m + 1)}{2} \, .
\end{equation}

For $k = 4m + 2$, each singlet can be written as a monomial of the form
\begin{equation}
\nonumber
  \bigl(Y_{\mathbf{1}}^{(4)}\bigr)^a
  \bigl(Y_{\mathbf{1}'}^{(4)}\bigr)^b
  \bigl(Y_{\mathbf{1}''}^{(4)}\bigr)^c
  \bigl(Y^{(2)}_{\mathbf{3}} \otimes Y^{(4)}_{\mathbf{3}}\bigr)_{\mathbf{1}} \,,
\end{equation}
while each triplet can be written as
\begin{equation}
\nonumber
  \bigl(Y_{\mathbf{1}}^{(4)}\bigr)^\alpha
  \bigl(Y_{\mathbf{1}'}^{(4)}\bigr)^\beta
  \bigl(Y_{\mathbf{1}''}^{(4)}\bigr)^\gamma
  Y^{(2)}_{\mathbf{3}} \,.
\end{equation}
The corresponding weight-counting relations read
\begin{equation}
    \begin{aligned}
        4a + 4b + 4c + 6 &= 4m + 2 \,,\\
        4\alpha + 4\beta + 4\gamma + 2 &= 4m + 2 \,.
    \end{aligned}
    \label{weight_4m+2_relation}
\end{equation}
The number of singlet structures is $m(m+1)/2$, while the number of triplet
structures is $(m+1)(m+2)/2$. Consequently, the total dimension is
\begin{equation}
    \frac{m(m+1)}{2} + 3 \cdot \frac{(m+1)(m+2)}{2}
    = \frac{(2m+2)(2m+3)}{2} \,,
\end{equation}
in agreement with the general counting formula.

Now impose the constraint in Eq.~\eqref{eq:modular_constrain}, namely
$Y_{\mathbf{1}''}^{(4)} = 0$. In the tensor construction, this amounts to removing all
structures that contain at least one factor of $Y_{\mathbf{1}''}^{(4)}$. In terms of
the exponents, this means $c \geq 1$ or $\gamma \geq 1$. To count such contributions, we define
shifted variables $c' = c - 1$ and $\gamma' = \gamma - 1$. The relations
in Eq.~\eqref{weight_4m_relation} and Eq.~\eqref{weight_4m+2_relation} then become
\begin{equation}
\nonumber
    \begin{aligned}
        4a + 4b + 4c' &= 4(m-1) \,,\\
        4\alpha + 4\beta + 4\gamma' + 4 &= 4(m-1) \,,
    \end{aligned}
\end{equation}
and
\begin{equation}
\nonumber
    \begin{aligned}
        4a + 4b + 4c' + 6 &= 4(m-1) + 2 \,,\\
        4\alpha + 4\beta + 4\gamma' + 2 &= 4(m-1) + 2 \,.
    \end{aligned}
\end{equation}
This shows that the total dimension of all terms involving
$Y_{\mathbf{1}''}^{(4)}$ coincides with the dimension of modular forms
with weight decreased by $4$. Of course, when $c = 0$ or $\gamma = 0$,
the corresponding relations in terms of $c'$ or $\gamma'$ should not be included.

In terms of the generic weight parameter $k = 2n$, the dimension formula
\begin{equation}
\nonumber
    \frac{(n+2)(n+1)}{2} - (2n+1)
    = \frac{(n-2+2)(n-2+1)}{2}
\end{equation}
simply expresses the fact that the difference between the unconstrained
and constrained dimensions at weight $k = 2n$ is equal to the dimension
of the modular forms at weight $k = 2n - 4$. Moreover, from the explicit
construction of symmetric tensors and the corresponding cancellation, we
see that, for modular weight $k = 2n$, all symmetric tensors built from
$(n-2)$ copies of $Y_{\mathbf{3}}^{(2)}$ must have their tensor products
with $Y_{\mathbf{1}''}^{(4)}$ removed.

As an explicit example, consider modular forms of weight $k = 8$, i.e.\
$Y^{(8)}_{\mathbf{r}}$ with $m = 2$. The total dimension is
\begin{equation}
\nonumber
  \frac{(2m+2)(2m+1)}{2}
  = \frac{(2\cdot 2 + 2)(2\cdot 2 + 1)}{2}
  = 15
  = 6 + 3 \cdot 3 \,,
\end{equation}
which corresponds to six singlets,
\begin{equation}
\nonumber
\begin{aligned}
  &Y_{\mathbf{1}}^{(4)} Y_{\mathbf{1}}^{(4)},\qquad
  Y_{\mathbf{1}}^{(4)} Y_{\mathbf{1}'}^{(4)},\qquad
  Y_{\mathbf{1}}^{(4)} Y_{\mathbf{1}''}^{(4)},\\
  &Y_{\mathbf{1}'}^{(4)} Y_{\mathbf{1}'}^{(4)},\qquad
  Y_{\mathbf{1}'}^{(4)} Y_{\mathbf{1}''}^{(4)},\qquad
  Y_{\mathbf{1}''}^{(4)} Y_{\mathbf{1}''}^{(4)} \,,
\end{aligned}
\end{equation}
and three triplets,
\begin{equation}
\nonumber
  Y^{(8)}_{\mathbf{3}a}
  = Y_{\mathbf{1}}^{(4)} Y_{\mathbf{3}}^{(4)},\quad
  Y^{(8)}_{\mathbf{3}b}
  = Y_{\mathbf{1}'}^{(4)} Y_{\mathbf{3}}^{(4)},\quad
  Y^{(8)}_{\mathbf{3}c}
  = Y_{\mathbf{1}''}^{(4)} Y_{\mathbf{3}}^{(4)} \, .
\end{equation}
Under the constraint in Eq.~\eqref{eq:modular_constrain}, the contributions
involving $Y_{\mathbf{1}''}^{(4)}$ must be removed. In particular, the
singlet combinations
\begin{equation}
\nonumber
  Y_{\mathbf{1}}^{(4)} Y_{\mathbf{1}''}^{(4)},\quad
  Y_{\mathbf{1}'}^{(4)} Y_{\mathbf{1}''}^{(4)},\quad
  Y_{\mathbf{1}''}^{(4)} Y_{\mathbf{1}''}^{(4)}
\end{equation}
and the triplet
\begin{equation}
\nonumber
  Y^{(8)}_{\mathbf{3}c}
  = Y_{\mathbf{1}''}^{(4)} Y_{\mathbf{3}}^{(4)}
\end{equation}
are precisely of the form
$
  \bigl(Y_{\mathbf{3}}^{(2)} \otimes Y_{\mathbf{3}}^{(2)}\bigr)_{\mathrm{sym}}
  \otimes Y_{\mathbf{1}''}^{(4)}
$
and are therefore eliminated by the constraint.

\paragraph{Examples.}
We record the explicit modular forms used in our analysis.

\medskip
\noindent\textbf{Weight 4.}
From the symmetric square of the triplet $Y^{(2)}_{\mathbf{3}}$,
\begin{equation}
    \begin{aligned}
        &Y^{(4)}_{\mathbf{1}}   = \bigl(Y^{(2)}_{\mathbf{3}} \otimes Y^{(2)}_{\mathbf{3}}\bigr)_{\mathbf{1}},\\
        &Y^{(4)}_{\mathbf{1'}}  = \bigl(Y^{(2)}_{\mathbf{3}} \otimes Y^{(2)}_{\mathbf{3}}\bigr)_{\mathbf{1'}},\\
        &Y^{(4)}_{\mathbf{1''}} = \bigl(Y^{(2)}_{\mathbf{3}} \otimes Y^{(2)}_{\mathbf{3}}\bigr)_{\mathbf{1''}}
           \;\overset{\eqref{eq:modular_constrain}}{=}\; 0,\\
        &Y^{(4)}_{\mathbf{3}}   = \bigl(Y^{(2)}_{\mathbf{3}} \otimes Y^{(2)}_{\mathbf{3}}\bigr)_{\mathbf{3}} \, .
    \end{aligned}
\end{equation}

\medskip
\noindent\textbf{Weight 6.}
Using the exponent relations in Eq.~\eqref{weight_4m+2_relation} with $m=1$,
\begin{equation}
\nonumber
\begin{aligned}
    4a + 4b + 4c &= 0,\\
    4\alpha + 4\beta + 4\gamma &= 4 \, ,
\end{aligned}
\end{equation}
whose solutions are
\begin{equation}
\nonumber
    a=b=c=0,
    \qquad
    (\alpha,\beta,\gamma)\in\{(1,0,0),(0,1,0),(0,0,1)\}.
\end{equation}
The resulting modular forms are
\begin{equation}
    \begin{aligned}
        &Y^{(6)}_{\mathbf{1}}  = \bigl(Y^{(2)}_{\mathbf{3}}\otimes Y^{(4)}_{\mathbf{3}}\bigr)_{\mathbf{1}},\\
        &Y^{(6)}_{\mathbf{3}a} = \bigl(Y^{(2)}_{\mathbf{3}}\otimes Y^{(4)}_{\mathbf{1}}\bigr)_{\mathbf{3}},\\
        &Y^{(6)}_{\mathbf{3}b} = \bigl(Y^{(2)}_{\mathbf{3}}\otimes Y^{(4)}_{\mathbf{1'}}\bigr)_{\mathbf{3}},\\
        &Y^{(6)}_{\mathbf{3}c} = \bigl(Y^{(2)}_{\mathbf{3}}\otimes Y^{(4)}_{\mathbf{1''}}\bigr)_{\mathbf{3}}
           \;\overset{\eqref{eq:modular_constrain}}{=}\; 0 \, .
    \end{aligned}
\end{equation}

\medskip
\noindent\textbf{Weight 8.}
Using the relations in Eq.~\eqref{weight_4m_relation} with $m=2$,
\begin{equation}
\nonumber
    \begin{aligned}
        4a + 4b + 4c &= 8,\\
        4\alpha + 4\beta + 4\gamma &= 4 \, ,
    \end{aligned}
\end{equation}
with solutions
\begin{equation}
\nonumber
    \begin{aligned}
        &(a,b,c)\in\{(2,0,0),(0,2,0),(0,0,2),(1,1,0),(1,0,1),(0,1,1)\},\\
        &(\alpha,\beta,\gamma)\in\{(1,0,0),(0,1,0),(0,0,1)\}.
    \end{aligned}
\end{equation}
Thus there are six singlets and three triplets before imposing the constraint.
They are
\begin{equation}
    \begin{aligned}
        &Y^{(8)}_{\mathbf{1}a}   = \bigl(Y^{(4)}_{\mathbf{1}}\otimes Y^{(4)}_{\mathbf{1}}\bigr)_{\mathbf{1}},\\
        &Y^{(8)}_{\mathbf{1''}a} = \bigl(Y^{(4)}_{\mathbf{1'}}\otimes Y^{(4)}_{\mathbf{1'}}\bigr)_{\mathbf{1''}},\\
        &Y^{(8)}_{\mathbf{1'}a}  = \bigl(Y^{(4)}_{\mathbf{1''}}\otimes Y^{(4)}_{\mathbf{1''}}\bigr)_{\mathbf{1'}}
            \;\overset{\eqref{eq:modular_constrain}}{=}\; 0,\\
        &Y^{(8)}_{\mathbf{1'}b}  = \bigl(Y^{(4)}_{\mathbf{1}}\otimes Y^{(4)}_{\mathbf{1'}}\bigr)_{\mathbf{1'}},\\
        &Y^{(8)}_{\mathbf{1''}b} = \bigl(Y^{(4)}_{\mathbf{1}}\otimes Y^{(4)}_{\mathbf{1''}}\bigr)_{\mathbf{1''}}
            \;\overset{\eqref{eq:modular_constrain}}{=}\; 0,\\
        &Y^{(8)}_{\mathbf{1}b}   = \bigl(Y^{(4)}_{\mathbf{1'}}\otimes Y^{(4)}_{\mathbf{1''}}\bigr)_{\mathbf{1}}
            \;\overset{\eqref{eq:modular_constrain}}{=}\; 0,\\
        &Y^{(8)}_{\mathbf{3}a}   = \bigl(Y^{(4)}_{\mathbf{3}}\otimes Y^{(4)}_{\mathbf{1}}\bigr)_{\mathbf{3}},\\
        &Y^{(8)}_{\mathbf{3}b}   = \bigl(Y^{(4)}_{\mathbf{3}}\otimes Y^{(4)}_{\mathbf{1'}}\bigr)_{\mathbf{3}},\\
        &Y^{(8)}_{\mathbf{3}c}   = \bigl(Y^{(4)}_{\mathbf{3}}\otimes Y^{(4)}_{\mathbf{1''}}\bigr)_{\mathbf{3}}
            \;\overset{\eqref{eq:modular_constrain}}{=}\; 0 \, .
    \end{aligned}
\end{equation}
In particular, the four structures containing $Y^{(4)}_{\mathbf{1''}}$ vanish upon
imposing the constraint in Eq.~\eqref{eq:modular_constrain}.

\section{SMEFT Operators}  % Appendix A
\subsection{Dimension-6 SMEFT Operators}
Here the dimension-6 SMEFT operators are adopted from the Ref.~\cite{Grzadkowski:2010es}, and only the baryon-number and lepton-number conserving fermionic operators are presented. 

\begin{center}

\begin{minipage}[t]{4cm}

    \renewcommand{\arraystretch}{1.5}
    \begin{tabular}[t]{c|c}
    \multicolumn{2}{c}{$1: \psi^2H^3 + \hbox{h.c.}$} \\
    \hline
    $\mathcal{O}_{eH}$           & $(H^\dag H)(\bar l^p e_r H)$ \\
    $\mathcal{O}_{uH}$          & $(H^\dag H)(\bar q^p u_r \widetilde H )$ \\
    $\mathcal{O}_{dH}$           & $(H^\dag H)(\bar q^p d_r H)$\\
    \hline
    \end{tabular}
\end{minipage}
\quad
\begin{minipage}[t]{5.2cm}
    \renewcommand{\arraystretch}{1.5}
    \begin{tabular}[t]{c|c}
    \multicolumn{2}{c}{$2:\psi^2 XH+\hbox{h.c.}$} \\
    \hline
    $\mathcal{O}_{eW}$      & $(\bar l^p \sigma^{\mu\nu} e_r) \tau^I H W_{\mu\nu}^I$ \\
    $\mathcal{O}_{eB}$        & $(\bar l^p \sigma^{\mu\nu} e_r) H B_{\mu\nu}$ \\
    $\mathcal{O}_{uG}$        & $(\bar q^p \sigma^{\mu\nu} \lambda^A u_r) \widetilde H \, G_{\mu\nu}^A$ \\
    $\mathcal{O}_{uW}$        & $(\bar q^p \sigma^{\mu\nu} u_r) \tau^I \widetilde H \, W_{\mu\nu}^I$ \\
    $\mathcal{O}_{uB}$        & $(\bar q^p \sigma^{\mu\nu} u_r) \widetilde H \, B_{\mu\nu}$ \\
    $\mathcal{O}_{dG}$        & $(\bar q^p \sigma^{\mu\nu} \lambda^A d_r) H\, G_{\mu\nu}^A$ \\
    $\mathcal{O}_{dW}$         & $(\bar q^p \sigma^{\mu\nu} d_r) \tau^I H\, W_{\mu\nu}^I$ \\
    $\mathcal{O}_{dB}$        & $(\bar q^p \sigma^{\mu\nu} d_r) H\, B_{\mu\nu}$ \\
\hline    
\end{tabular}
\end{minipage}
\quad
\begin{minipage}[t]{5.4cm}
    \renewcommand{\arraystretch}{1.5}
    \begin{tabular}[t]{c|c}
    \multicolumn{2}{c}{$3:\psi^2H^2 D$} \\
    \hline
    $\mathcal{O}_{H l}^{(1)}$      & $(H^\dag i\overleftrightarrow{D}_\mu H)(\bar l^p \gamma^\mu l_r)$\\
    $\mathcal{O}_{H l}^{(3)}$      & $(H^\dag i\overleftrightarrow{D}^I_\mu H)(\bar l^p \tau^I \gamma^\mu l_r)$\\
    $\mathcal{O}_{H e}$            & $(H^\dag i\overleftrightarrow{D}_\mu H)(\bar e^p \gamma^\mu e_r)$\\
    $\mathcal{O}_{H q}^{(1)}$      & $(H^\dag i\overleftrightarrow{D}_\mu H)(\bar q^p \gamma^\mu q_r)$\\
    $\mathcal{O}_{H q}^{(3)}$      & $(H^\dag i\overleftrightarrow{D}^I_\mu H)(\bar q^p \tau^I \gamma^\mu q_r)$\\
    $\mathcal{O}_{H u}$            & $(H^\dag i\overleftrightarrow{D}_\mu H)(\bar u^p \gamma^\mu u_r)$\\
    $\mathcal{O}_{H d}$            & $(H^\dag i\overleftrightarrow{D}_\mu H)(\bar d^p \gamma^\mu d_r)$\\
    $\mathcal{O}_{H u d}$          & $i(\widetilde H ^\dag D_\mu H)(\bar u^p \gamma^\mu d_r)$\\
    \hline
    \end{tabular}
\end{minipage}

\vspace{0.25cm}

\begin{minipage}[t]{4.75cm}
    \renewcommand{\arraystretch}{1.5}
    \begin{tabular}[t]{c|c}
    \multicolumn{2}{c}{$4: (\bar LL)(\bar LL)$} \\
    \hline
    $\mathcal{O}_{ll}$        & $(\bar l^p \gamma_\mu l_r)(\bar l^s \gamma^\mu l_t)$ \\
    $\mathcal{O}_{qq}^{(1)}$  & $(\bar q^p \gamma_\mu q_r)(\bar q^s \gamma^\mu q_t)$ \\
    $\mathcal{O}_{qq}^{(3)}$  & $(\bar q^p \gamma_\mu \tau^I q_r)(\bar q^s \gamma^\mu \tau^I q_t)$ \\
    $\mathcal{O}_{lq}^{(1)}$                & $(\bar q^p\gamma^\mu q_r)(\bar l^s \gamma_\mu l_t)$ \\
    $\mathcal{O}_{lq}^{(3)}$                & $(\bar q^p \gamma^\mu \tau^I q_r)(\bar l^s \gamma_\mu \tau^I l_t)$ \\
    \hline
    \end{tabular}
\end{minipage}
\quad
\begin{minipage}[t]{5.25cm}
    \renewcommand{\arraystretch}{1.5}
    \begin{tabular}[t]{c|c}
    \multicolumn{2}{c}{$4:(\bar RR)(\bar RR)$} \\
    \hline
    $\mathcal{O}_{ee}$               & $(\bar e^p \gamma_\mu e_r)(\bar e^s \gamma^\mu e_t)$ \\
    $\mathcal{O}_{uu}$        & $(\bar u^p \gamma_\mu u_r)(\bar u^s \gamma^\mu u_t)$ \\
    $\mathcal{O}_{dd}$        & $(\bar d^p \gamma_\mu d_r)(\bar d^s \gamma^\mu d_t)$ \\
    $\mathcal{O}_{eu}$                      & $(\bar u^p \gamma^\mu u_r)(\bar e^s \gamma_\mu e_t)$ \\
    $\mathcal{O}_{ed}$                      & $(\bar d^p\gamma^\mu d_r)(\bar e^s \gamma_\mu e_t)$ \\
    $\mathcal{O}_{ud}^{(1)}$                & $(\bar u^p \gamma_\mu u_r)(\bar d^s \gamma^\mu d_t)$ \\
    $\mathcal{O}_{ud}^{(8)}$                & $(\bar u^p \gamma_\mu \lambda^A u_r)(\bar d^s \gamma^\mu \lambda^A d_t)$ \\
    \hline
    \end{tabular}
\end{minipage}
\quad
\begin{minipage}[t]{5cm}
    \renewcommand{\arraystretch}{1.5}
    \begin{tabular}[t]{c|c}
    \multicolumn{2}{c}{$4:(\bar LL)(\bar RR)$} \\
    \hline
    $\mathcal{O}_{le}$               & $(\bar l^p \gamma_\mu l_r)(\bar e^s \gamma^\mu e_t)$ \\
    $\mathcal{O}_{lu}$               & $(\bar u^p \gamma^\mu u_r)(\bar l^s \gamma_\mu l_t)$ \\
    $\mathcal{O}_{ld}$               & $(\bar d^p \gamma^\mu d_r)(\bar l^s \gamma_\mu l_t)$ \\
    $\mathcal{O}_{qe}$               & $(\bar q^p \gamma_\mu q_r)(\bar e^s \gamma^\mu e_t)$ \\
    $\mathcal{O}_{qu}^{(1)}$         & $(\bar q^p \gamma_\mu q_r)(\bar u^s \gamma^\mu u_t)$ \\
    $\mathcal{O}_{qu}^{(8)}$         & $(\bar q^p \gamma_\mu \lambda^A q_r)(\bar u^s \gamma^\mu \lambda^A u_t)$ \\
    $\mathcal{O}_{qd}^{(1)}$ & $(\bar q^p \gamma_\mu q_r)(\bar d^s \gamma^\mu d_t)$ \\
    $\mathcal{O}_{qd}^{(8)}$ & $(\bar q^p \gamma_\mu \lambda^A q_r)(\bar d^s \gamma^\mu \lambda^A d_t)$\\
    \hline
    \end{tabular}
\end{minipage}

\vspace{0.25cm}

\begin{minipage}[t]{3.75cm}
    \renewcommand{\arraystretch}{1.5}
    \begin{tabular}[t]{c|c}
    \multicolumn{2}{c}{$4:(\bar LR)(\bar RL)+\hbox{h.c.}$} \\
    \hline
    $\mathcal{O}_{ledq}$ & $(\bar d^p q_{jr})(\bar l^{is} e_t)$ \\
    \hline
    \end{tabular}
\end{minipage}
\quad
\begin{minipage}[t]{5.5cm}
    \renewcommand{\arraystretch}{1.5}
    \begin{tabular}[t]{c|c}
    \multicolumn{2}{c}{$4:(\bar LR)(\bar L R)+\hbox{h.c.}$} \\
    \hline
    $\mathcal{O}_{quqd}^{(1)}$ & $(\bar q^{jp} u_r) \epsilon_{jk} (\bar q^{ks} d_t)$ \\
    $\mathcal{O}_{quqd}^{(8)}$ & $(\bar q^{jp} \lambda^A u_r) \epsilon_{jk} (\bar q^{ks} \lambda^A d_t)$ \\
    $\mathcal{O}_{lequ}^{(1)}$ & $(\bar q^{kp} u_r)\epsilon_{jk}(\bar l^{js} e_t)  $ \\
    $\mathcal{O}_{lequ}^{(3)}$ & $ (\bar q^{kp} \sigma^{\mu\nu} u_r)\epsilon_{jk} (\bar l^{js} \sigma_{\mu\nu} e_t)$ \\
    \hline
    \end{tabular}
\end{minipage}
    
\end{center}

\subsection{Holomorphy and SUSY UV matching for modular-invariant SMEFT operators}
\label{app:susy_realization_summary}

In general $\mathcal N=1$ SUSY, superspace holomorphy refers to whether an interaction originates from a chiral superspace integral (an $F$-term) rather than a full superspace integral (a $D$-term). 
Here $\Phi$ denotes a chiral superfield and $V$ the gauge vector superfield. 
Superpotential couplings and gauge--kinetic terms take the schematic form
\begin{equation}
  \int d^2\theta\, W(\Phi)\,,
  \qquad
  \int d^2\theta\, f_{ab}(\Phi)\, W^{a\alpha} W^b_{\alpha}\,,
  \label{eq:app_Fterms}
\end{equation}
where $W(\Phi)$ and $f_{ab}(\Phi)$ are holomorphic functions of chiral superfields $\Phi$.  
By contrast, K\"ahler and more general kinetic structures arise from
\begin{equation}
  \int d^4\theta\, K(\Phi,\bar\Phi;V)\,,
  \label{eq:app_Dterms}
\end{equation}
with $K$ a real function of $\Phi$ and $\Phi^\dagger$. 
Specifically, the $\mathcal N=1$ SUSY action can be written as
\begin{equation}
S \;=\;
\int d^4x\, d^2\theta\, d^2\bar\theta\;
K\!\left(\Phi,\bar\Phi;V\right)
\;+\;
\left[
\int d^4x\, d^2\theta\;
\left(
W(\Phi)
+\frac{1}{4}\, f_{ab}(\Phi)\, W^{a\alpha} W^{b}{}_{\alpha}
\right)
\;+\; \text{h.c.}
\right].
\end{equation}
In a modular-invariant setup, the modulus $\tau$ enters F-terms by holomorphic modular forms, while D-terms are generically non-holomorphic and may depend on both $\tau$ and $\tau^*$. For a renormalizable SUSY action, $f_{ab}$ is taken to be constant, and the superpotential can be written schematically as
\begin{equation}
W(\Phi)=Y(\tau)^{k_W}\,\varphi^{(I_1)}\cdots \varphi^{(I_n)}\,,
\qquad
k_W=\sum_{a=1}^{n} k_{I_a}\,.
\end{equation}
For the K\"ahler potential, we have
\begin{equation}
K(\Phi,\bar\Phi,V)
=\frac{1}{\left\langle -i\tau+i\bar\tau \right\rangle^{k_K}}
\,
K(\varphi,\bar\varphi,V)
\;+\Delta K\;,
\end{equation}
in which
\begin{equation}
  \Delta K \;=\;
  \sum_{k,\mathbf r,\mathbf r'}\,
  \alpha_{k,\mathbf r,\mathbf r'}\;
  \big\langle i\bar\tau - i\tau \big\rangle^{\,k-k_{K}}\,
  \Big[\big(Y^{(k)}{Y^*}^{(k)}\big)_{\mathbf r}\;
  K(\varphi,\bar\varphi,V)_{\mathbf r'}\Big]_{\mathbf 1}\,,
\end{equation}
where $k$ is any even integer satisfying $k\ge 2$.

Here, we focus directly on the modular SUSY UV matching mechanisms that generate the modular-invariant SMEFT operators discussed in the main text, and we use the F- and D-term language only to label the UV origin of the relevant couplings.  
%This perspective is more closely aligned with the practical matching question: which types of heavy supermultiplet exchange (tree or loop) can generate a given SMEFT class, and what modular dependence is expected in the Wilson coefficients.

We impose the MFV-like assumptions used in the conclusions that $\tau$-dependence in the EFT is restricted to (anti-)holomorphic modular forms, and heavy mediator masses are taken to be $\tau$-independent constants.  Accordingly, the modular dressing of an SMEFT operator structure $\mathcal O$ can be written schematically as
\begin{equation}
  \mathcal O \ \longrightarrow\ 
  \bigl[\,Y^{(k)}_{\mathbf r}(\tau)\, Y^{(k')*}_{\mathbf r'}(\tau)\, \mathcal O\,\bigr]_{\mathbf 1}\,,
  \label{eq:app_dressing}
\end{equation}
with the tensor product taken in flavor space and projected to the singlet.  In particular, these assumptions exclude modulus-dependent propagator denominators such as $1/M(\tau)^2$.  They also motivate discarding kinetic-mediated contributions whenever modular covariance forces explicit $\mathrm{Im}(\tau)$-dependent prefactors to appear in covariant kinetic structures, since such prefactors would propagate into Wilson coefficients and lie outside the minimal dressing class~\eqref{eq:app_dressing}. 
Under these restriction, the SUSY UV kinetic term in modular-invariant SUSY UV completion is uniquely constrained to take the form
\begin{equation}
  K_{\rm UV}=\sum_{\mathbf r,\mathbf r'}\,
  \beta_{\mathbf r,\mathbf r'}\;\Big[\big(Y^{(k_K)}{Y^*}^{(k_K)}\big)_{\mathbf r}\;
  K(\varphi,\bar\varphi,V)_{\mathbf r'}\Big]_{\mathbf 1}\,.
\end{equation}
%In this setup, we discuss which common tree-level and loop-level exchange mechanisms in a modular-invariant SUSY UV completion can generate these modular-invariant SMEFT operators as a extension of Ref.~\cite{de_Blas_2018}. A compact dictionary is provided in Table~\ref{tab:warsaw_susy}.

In this setup, we discuss which common tree-level and loop-level exchange mechanisms in a modular-invariant SUSY UV completion can generate these modular-invariant SMEFT operators
Our organization follows the tree-level UV/IR dictionary in Refs.~\cite{de_Blas_2018,Li:2022abx}. 
A compact summary is collected in Table~\ref{tab:warsaw_susy}.

\paragraph{Two-fermion classes.}
At the level of SUSY UV matching, the Weinberg operator $(LH)(LH)$ is the standard example generated at tree level by exchanging a heavy chiral supermultiplet through holomorphic superpotential couplings (seesaw-type mediators), i.e.\ via the exchange of its scalar or fermionic component in components language. In this case the modular dependence of the Wilson coefficient is inherited from the holomorphic modular-form dependence of the superpotential couplings, while the heavy masses are $\tau$-independent by assumption.

The two-fermion classes $\psi^2H^3$ and $\psi^2H^2D$ can also be generated at tree level.  
The holomorphic class $\psi^2H^3$  is most naturally induced by integrating out heavy scalars through holomorphic superpotential couplings.  
By contrast, the non-holomorphic class $\psi^2H^2D$  admits two standard tree-level realizations: heavy-vector exchange via current couplings, which is rooted in UV kinetic/K\"ahler structures and is therefore ``D-term-like'' in origin; and heavy vectorlike-fermion exchange, corresponding to exchanging the fermionic component of a heavy (vectorlike) chiral supermultiplet through holomorphic superpotential couplings, i.e.\ a UV F-term origin.  
Under our MFV-like assumption, matching contributions from both the F-term and D-term sectors are allowed,
and for D-term--induced operators the $\tau$-dependence of the Wilson coefficients typically appears in the mixed form
$Y(\tau)\,Y^{*}(\tau)$.

Finally, dipole operators $\psi^2 X H$ are chirality-holomorphic. 
In renormalizable SUSY UV completions, they are typically generated only at one loop. 
From the UV perspective, the matching necessarily requires holomorphic superpotential Yukawa/trilinear couplings (F-term) together with gauge interactions (D-term/kinetic), and the heavy states running in the loop typically include charged/colored heavy fermions and heavy scalars (and their superpartners). 
Here the $D$-term/kinetic vertices should couple to modular factors in a holomorphic way; otherwise the matching would introduce intrinsically non-holomorphic couplings, which would in turn require a nontrivial (and model-dependent) choice of modular-weight assignments to restore modular covariance.
Under our minimal modular-spurion setup, the $\tau$-dependence of the resulting Wilson coefficients is carried by modular forms appearing in the superpotential couplings and gauge interactions.

\paragraph{Four-fermion classes.}
For baryon-number conserving four-fermion operators, a useful UV-level distinction is between current--current structures and operators of the $(\overline{L}R)(\overline{R}L)+\mathrm{h.c.}$ type.
Current--current operators such as $(\overline{L}L)(\overline{L}L)$, $(\overline{R}R)(\overline{R}R)$ and $(\overline{L}L)(\overline{R}R)$ are readily generated at tree level by heavy-vector exchange through couplings to fermion currents; from the UV viewpoint these couplings are naturally rooted in kinetic/K\"ahler structures and are therefore D-term-like in origin. 
Operators of the $(\overline{L}R)(\overline{R}L)+\mathrm{h.c.}$ type can also arise from heavy-vector exchange, possibly after Fierz rearrangements, depending on the mediator quantum numbers.
Both types of non-holomorphic four-fermion classes may also be produced by exchanging heavy chiral multiplets through holomorphic superpotential interactions (typically via heavy-scalar exchange), in which case the underlying UV couplings are holomorphic but the resulting four-fermion operator appears in the low-energy EFT with a non-holomorphic structure.
These are precisely the cases in which mixed modular dependence of the form $Y(\tau)Y^*(\tau)$ is expected and is consistent with Eq.~\eqref{eq:app_dressing}.

By contrast, the schematic class $(\overline{L}R)(\overline{L}R)$ (such as $\mathcal O^{(1,8)}_{quqd}$ and $\mathcal O^{(1,3)}_{lequ}$) is holomorphic in the chirality-based classification.  
In SUSY-UV interpretation, generating this structure most naturally requires a tree-level heavy-scalar exchange. 
Equivalently, in a supersymmetric completion this corresponds to exchanging the scalar component of a heavy chiral supermultiplet, with the required renormalizable trilinear couplings arising from holomorphic superpotential interactions (an F-term origin). 
Under our minimal modular-spurion setup, the modulus dependence in these trilinear couplings is carried only by holomorphic modular forms, so that the resulting Wilson coefficient inherits a purely holomorphic modular-form factor.

Meanwhile, the dimension-six baryon-number violating (BNV) operators can be accommodated within the same SUSY-UV matching logic, but with baryon-number violating superpotential couplings.
Concretely, their minimal realization most economically proceeds via a tree-level heavy-scalar exchange, so that the effective four-fermion interaction arises after integrating out the exchanged scalar.
Equivalently, the required renormalizable trilinear vertices originate from holomorphic superpotential interactions (an F-term origin).
Under our minimal modular-spurion assumptions, any $\tau$-dependence enters only through holomorphic modular forms appearing in these superpotential trilinears (with $\tau$-independent mediator masses), and the resulting Wilson coefficients inherit the corresponding holomorphic modular-form factors.
\newpage

\begin{table}[!htbp]
  \centering
  \small
  \setlength{\tabcolsep}{6pt}
  \renewcommand{\arraystretch}{1.25}
  \begin{tabular}{|>{\centering\arraybackslash}p{4.05cm}
                  |>{\centering\arraybackslash}p{3.55cm}
                  |>{\centering\arraybackslash}p{6.05cm}|}
    \hline
    \textbf{Warsaw class / fermionic structure}
    & \textbf{modular-invariant SMEFT holomorphy}
    & \textbf{modular SUSY UV mechanism realization}
    \\
    \hline 

    \textbf{Dim-5:} $(LH)(LH)$ (Weinberg)
    & holomorphic
    & \textbf{Tree level:} \textbf{UV F-term (heavy scalar/fermion)}
    %heavy chiral supermultiplet exchange (seesaw-type), with holomorphic modular-form couplings and $\tau$-independent mediator masses.
    \\
    \hline

    \textbf{Dim-6:} $\psi^2 X H$ (dipoles)
    & holomorphic
    & \textbf{One loop:} \textbf{One loop:} \textbf{UV D-term (gauge vertices)}\footnotemark + \textbf{UV F-term (Yukawa insertion)}
    \\
    \hline

    \textbf{Dim-6:} $\psi^2 H^3$
    & holomorphic
    & \textbf{Tree level:} 
    %typically generated by integrating out heavy chiral/scalar multiplets through 
    \textbf{UV F-term (heavy scalar)} 
    %(superpotential) couplings
    %, inducing the corresponding non-holomorphic two-fermion class in the low-energy EFT, consistent with mixed modular factors $Y(\tau)Y^*(\tau)$ under Eq.~\eqref{eq:app_dressing}.
    \\
    \hline

    \textbf{Dim-6:} $\psi^2 H^2 D$
    & non-holomorphic
    & \textbf{Tree level:} \textbf{UV D-term (heavy vector) or/and \newline UV F-term (vectorlike fermion)}
    %generated by \textbf{(i) UV D-term} heavy vector exchange (current-type) and/or \textbf{(ii) UV F-term} heavy chiral/scalar exchange inducing the derivative/current structure in the EFT; modular dependence compatible with $Y(\tau)Y^*(\tau)$ under the minimal dressing.
    \\
    \hline

    \textbf{Dim-6:} $(\overline{L}L)(\overline{L}L)$
    & non-holomorphic
    & \textbf{Tree level:} 
    \textbf{UV D-term (heavy vector) or/and \newline UV F-term (heavy scalar)}
    %\textbf{(i) UV D-term} heavy vector exchange (current--current), or \textbf{(ii) UV F-term} heavy chiral/scalar exchange (two holomorphic superpotential trilinears) yielding a four-fermion operator with mixed modular dependence consistent with $Y(\tau)Y^*(\tau)$.
    \\
    \hline

    \textbf{Dim-6:} $(\overline{R}R)(\overline{R}R)$
    & non-holomorphic
    & \textbf{Tree level:} 
    \textbf{UV D-term (heavy vector) or/and \newline UV F-term (heavy scalar)}
    %\textbf{(i) UV D-term} heavy vector exchange, or \textbf{(ii) UV F-term} heavy chiral/scalar exchange; $\tau$-dependence organized by modular-form combinations (typically $YY^*$) with $\tau$-independent mediator masses.
    \\
    \hline

    \textbf{Dim-6:} $(\overline{L}L)(\overline{R}R)$
    & non-holomorphic
    & \textbf{Tree level:} 
    \textbf{UV D-term (heavy vector) or/and \newline UV F-term (heavy scalar)}
    %\textbf{(i) UV D-term} heavy vector exchange, or \textbf{(ii) UV F-term} heavy chiral/scalar exchange; modular dependence restricted to modular-form combinations (typically $YY^*$) in our minimal setup.
    \\
    \hline

    \textbf{Dim-6:} $(\overline{L}R)(\overline{R}L)$ + h.c.
    & non-holomorphic
    & \textbf{Tree level:}
    \textbf{UV D-term (heavy vector) or/and \newline UV F-term (heavy scalar)}
    %\textbf{(i) UV D-term} heavy vector exchange (after Fierz rearrangements), or \textbf{(ii) UV F-term} heavy chiral/scalar exchange; in either case Wilson coefficients can depend on $\tau$ only through modular-form combinations such as $YY^*$ under the minimal rule.
    \\
    \hline

    \textbf{Dim-6:} $(\overline{L}R)(\overline{L}R)$
   % \newline (e.g.\ $\mathcal O^{(1,8)}_{quqd}$, $\mathcal O^{(1,3)}_{lequ}$)
    & holomorphic
    & \textbf{Tree level:}
    \textbf{UV F-term (heavy scalar)}
    %\textbf{Model-dependent / typically non-manifest:} not generated by the simplest manifest tree-level exchanges in a generic SUSY UV completion; when present it typically requires additional UV structure and/or non-manifest effects (e.g.\ SUSY-breaking insertions and/or loop thresholds, or integrating out incomplete supermultiplets). Any allowed $\tau$-dependence is still restricted to holomorphic modular forms in the relevant UV couplings, with $\tau$-independent mediator masses.
    \\
    \hline

    \textbf{Dim-6:} $\mathcal O_{duq}$,\ $\mathcal O_{qqu}$
    & non-holomorphic
    & \textbf{Tree level:}
    \textbf{UV F-term (heavy scalar/fermion)}
    %\textbf{Model-dependent:} can be generated at tree level by specific heavy scalar/chiral mediators (\textbf{UV F-term} exchange) or by heavy vectors (\textbf{UV D-term} exchange). In our minimal framework, Wilson coefficients may depend on $\tau$ only through modular-form combinations (typically $YY^*$) and heavy masses are $\tau$-independent.
    \\
    \hline

    \textbf{Dim-6:} $\mathcal O_{qqq}$,\ $\mathcal O_{duu}$
    & holomorphic
    & \textbf{Tree level:}
    \textbf{UV F-term (heavy scalar/fermion)}
    %\textbf{Tree level (if allowed):} generated by \textbf{UV F-term} baryon-number violating superpotential couplings (RPV-type) followed by heavy scalar/chiral exchange (often sfermion exchange), yielding a four-fermion operator. Modular dependence enters holomorphically in the trilinear couplings; heavy masses $\tau$-independent.
    \\
    \hline

  \end{tabular}
  \caption{Summary of modular SUSY UV matching mechanisms for the fermionic (dimension-five and dimension-six) operator classes in the modular-invariant SMEFT.}
  \label{tab:warsaw_susy}
\end{table}
\footnotetext{The modular factors in the $D$-term/gauge vertices are taken to be holomorphic, which shoulde be model dependent; otherwise non-holomorphic couplings would enter the matching.}

\subsection{Some dimension-7 modular invariant operators}
\label{dim_7}
At dimension-$7$, we present the modular-invariant operator structures corresponding to the bilinear SMEFT operators that do not involve field-strength tensors, namely the classes $\phi^2H^4$, $\phi^2H^3D$, and $\phi^2H^2D^2$.
\begin{itemize}
  \item $\phi^2H^4$ and $\phi^2H^2D^2$: \\
  The associated operators are
  \begin{equation}
    \mathcal{O}_{LH} = \epsilon_{ij} \epsilon_{mn} (L^i_L C L_L^m) H^j H^n (H^\dagger H),
  \end{equation}
  \begin{equation}
    \mathcal{O}_{LHD}^{(1)} = \epsilon_{ij} \epsilon_{mn} L^i_L C (D^\mu L^j_L) H^m (D_\mu H^n),
  \end{equation}
  \begin{equation}
    \mathcal{O}_{LHD}^{(2)} = \epsilon_{im} \epsilon_{jn} L^i_L C (D^\mu L^j_L) H^m (D_\mu H^n).
  \end{equation}
  For these classes, the corresponding modular-invariant structures obtained from the Hilbert series are
  \begin{equation}
  \begin{aligned}
    &h^3 \, h^\dagger \, l^2 \, Y^{(4)}_{\mathbf{1}}(\tau_e)
    + h^3 \, h^\dagger \, l^2 \, Y^{(4)}_{\mathbf{1'}}(\tau_e)
    + h^3 \, h^\dagger \, l^2 \, Y^{(4)}_{\mathbf{3}}(\tau_e)\\
    &\quad + 2\, h^2 \, l^2 \, D^2 \, Y^{(4)}_{\mathbf{1}}(\tau_e)
    + 2\, h^2 \, l^2 \, D^2 \, Y^{(4)}_{\mathbf{1'}}(\tau_e)
    + 2\, h^2 \, l^2 \, D^2 \, Y^{(4)}_{\mathbf{3}}(\tau_e).
  \end{aligned}
  \end{equation}
  One can see that this structure is almost identical to the construction of the dimension-$5$ Weinberg operator. The last three terms, each with coefficient $2$, correspond to the two distinct operator structures that arise from the $\phi^2 H^2 D^2$ class. Following the construction shown in Eq.~\eqref{eq:Weinberg}, their explicit forms can be written down in a straightforward way. For instance, the operator
  \begin{equation}
  \nonumber
    \mathcal{O}_{LHD}^{(1)} = \epsilon_{ij} \epsilon_{mn}
    L_L^i C (D^\mu L_L^j) H^m (D_\mu H^n)
  \end{equation}
  corresponds to the $A_{4e}$-invariant structures
  \begin{equation}
  \begin{aligned}
    &\left[\left[\epsilon_{ij} \epsilon_{mn} L^i_L C (D^\mu L^j_L) H^m (D_\mu H^n)\right]_{\mathbf{3}} \otimes Y^{(4)}_{\mathbf{3}}(\tau_e)\right]_{\mathbf{1}},\\
    &\left[\left[\epsilon_{ij} \epsilon_{mn} L^i_L C (D^\mu L^j_L) H^m (D_\mu H^n)\right]_{\mathbf{1}} \otimes Y^{(4)}_{\mathbf{1}}(\tau_e)\right]_{\mathbf{1}},\\
    &\left[\left[\epsilon_{ij} \epsilon_{mn} L^i_L C (D^\mu L^j_L) H^m (D_\mu H^n)\right]_{\mathbf{1''}} \otimes Y^{(4)}_{\mathbf{1'}}(\tau_e)\right]_{\mathbf{1}}.
  \end{aligned}
  \end{equation}

  \item $\phi^2H^3D$: \\
  The associated operator is
  \begin{equation}
    \mathcal{O}_{LHDe} = \epsilon_{ij} \epsilon_{mn} (L^i_L C \gamma_\mu E_R) H^j H^m D^\mu H^n .
  \end{equation}
  The corresponding modular-invariant structures obtained from the Hilbert series are
  \begin{equation}
  \begin{aligned}
    &e \, h^3 \, l \, D \, Y^{(2)}_{\mathbf{3}}(\tau_e)
    + h^3 \, l \, \mu \, D \, Y^{(2)}_{\mathbf{3}}(\tau_e)
    + h^3 \, l \, D \, \tau \, Y^{(2)}_{\mathbf{3}}(\tau_e) .
  \end{aligned}
  \end{equation}
  Similarly, these terms directly lead to the corresponding invariant operator structures,
  \begin{equation}
  \begin{aligned}
    &\left[\left[\epsilon_{ij} \epsilon_{mn} (L^i_L C \gamma_\mu e_R) H^j H^m D^\mu H^n\right]_{\mathbf{3}}\otimes Y^{(2)}_{\mathbf{3}}(\tau_e)\right]_{\mathbf{1}},\\
    &\left[\left[\epsilon_{ij} \epsilon_{mn} (L^i_L C \gamma_\mu \mu_R) H^j H^m D^\mu H^n\right]_{\mathbf{3}}\otimes Y^{(2)}_{\mathbf{3}}(\tau_e)\right]_{\mathbf{1}},\\
    &\left[\left[\epsilon_{ij} \epsilon_{mn} (L^i_L C \gamma_\mu \tau_R) H^j H^m D^\mu H^n\right]_{\mathbf{3}}\otimes Y^{(2)}_{\mathbf{3}}(\tau_e)\right]_{\mathbf{1}}.
  \end{aligned}
  \end{equation}
\end{itemize}

\section{Non-holomorphic modular forms in \texorpdfstring{$A_4$}{A4} modular symmetry}
\label{app:nonholoN3}

In this appendix we summarize the ingredients needed for non-holomorphic (polyharmonic Maa{\ss}) modular forms at level $N=3$. We first collect the basic differential operators and their modular properties, and then list explicit $q$-expansions for the multiplets with weights $k<2$. For $k>2$, all structures coincide with the holomorphic case and are summarized in Appendix~\ref{app:forms}.
At weight $k=2$, the polyharmonic Maa{\ss} sector contains the $A_4$ singlet $\widehat{E}_2(\tau)$ in addition to the triplet $Y^{(2)}_{\mathbf{3}}$; we denote this singlet by $Y^{(2)}_{\mathbf{1}}\equiv \widehat{E}_2(\tau)$.
\subsection{Differential operators \texorpdfstring{$D$}{D} and \texorpdfstring{$\xi$}{xi}}
\label{app:DXi}

We adopt the standard Maa{\ss}--Shimura operators
\begin{equation}
  D \;=\; \frac{1}{2\pi i}\,\frac{\partial}{\partial\tau}\,,
  \qquad
  \xi_k \;=\; 2 i\, y^{k}\,\frac{\partial}{\partial\overline{\tau}}\,,
  \qquad y = \mathrm{Im}\,\tau\, .
  \label{eq:DXI_def_app}
\end{equation}
Let $Y(\tau)$ be a polyharmonic Maa{\ss} form of weight $k$ and level $N$. Then
\begin{equation}
\begin{aligned}
\bigl(D^{1-k}Y\bigr)(\gamma\tau) &= (c\tau+d)^{2-k}\,\bigl(D^{1-k}Y\bigr)(\tau),\\
\bigl(\xi_k Y\bigr)(\gamma\tau) &= (c\tau+d)^{2-k}\,\bigl(\xi_k Y\bigr)(\tau),
\end{aligned}
\qquad
\gamma=\begin{pmatrix}a&b\\ c&d\end{pmatrix}\in\Gamma(N).
\label{eq:DXI_modularity_app}
\end{equation}
Hence both $D^{\,1-k}$ and $\xi_k$ map (polyharmonic) weight-$k$ forms to weight-$2-k$ modular forms.
Using the Fourier--Whittaker basis ($q=e^{2\pi i\tau}$), one has for $n>0$:
\begin{equation}
  D^{\,1-k}(q^n)=n^{\,1-k}q^n,\qquad
  D^{\,1-k}\!\left(y^{1-k}\right)=(-4\pi)^{k-1}(1-k)!,\qquad
  D^{\,1-k}\!\bigl(\Gamma(1-k,4\pi n y)\,q^{-n}\bigr)=0,
  \label{eq:D_on_basis_app}
\end{equation}
and
\begin{equation}
  \xi_k(q^n)=0,\qquad
  \xi_k\!\left(y^{1-k}\right)=1-k,\qquad
  \xi_k\!\bigl(\Gamma(1-k,4\pi n y)\,q^{-n}\bigr)=-(4\pi n)^{1-k}q^{-n}.
  \label{eq:xi_on_basis_app}
\end{equation}
Because level-$N$ holomorphic modular forms do not exist at negative weight, nontrivial polyharmonic Maa{\ss} forms necessarily satisfy $k\le 2$.

\subsection{Level \texorpdfstring{$A_4$}{A4}: explicit series for \texorpdfstring{$k\le 2$}{k 2}}
\label{app:N3_series}
The non-holomorphic multiplets of weights $k=0,-2,-4$ can be obtained by acting with $D$ and $\xi_k$ on holomorphic forms of higher weight. Below we collect convenient representatives of their $q$-expansions; dots indicate higher-order terms that follow the same pattern.

\paragraph{Weight $k=0$ triplet.}
There exists a non-holomorphic weight-$0$ triplet $Y^{(0)}_{\mathbf{3}}=(Y^{(0)}_{\mathbf{3},1},Y^{(0)}_{\mathbf{3},2},Y^{(0)}_{\mathbf{3},3})^{T}$ and the trivial singlet $Y^{(0)}_{\mathbf{1}}=1$ such that
\begin{equation}
  \xi_0\!\bigl(Y^{(0)}_{\mathbf{3}}\bigr)=\Omega\,Y^{(2)}_{\mathbf{3}},
  \qquad
  D\!\bigl(Y^{(0)}_{\mathbf{3}}\bigr)=-\frac{1}{4\pi}\,Y^{(2)}_{\mathbf{3}},
  \label{eq:k0_relation_app}
\end{equation}
where $\Omega$ is an intertwiner implementing the equivalence between the real triplet and its complex conjugate.
A convenient choice of $q$-expansions is
\begin{align}
Y^{(0)}_{\mathbf{3},1} \;=\;&
 y
 - \frac{3\,e^{-4\pi y}}{\pi\,q}
 - \frac{9\,e^{-8\pi y}}{2\pi\,q^{2}}
 - \frac{e^{-12\pi y}}{\pi\,q^{3}}
 - \frac{21\,e^{-16\pi y}}{4\pi\,q^{4}}
 - \frac{18\,e^{-20\pi y}}{5\pi\,q^{5}}
 - \frac{3\,e^{-24\pi y}}{2\pi\,q^{6}}
 + \cdots \notag\\
&\hspace{1em}
 - \frac{9\log 3}{4\pi}
 - \frac{3q}{\pi}
 - \frac{9q^{2}}{2\pi}
 - \frac{q^{3}}{\pi}
 - \frac{21q^{4}}{4\pi}
 - \frac{18q^{5}}{5\pi}
 - \frac{3q^{6}}{2\pi}
 + \cdots \,,
\\[6pt]
Y^{(0)}_{\mathbf{3},2} \;=\;&
 \frac{27\,q^{1/3} e^{\pi y/3}}{\pi}
 \left(
   \frac{e^{-3\pi y}}{4q}
 + \frac{e^{-7\pi y}}{5q^{2}}
 + \frac{5\,e^{-11\pi y}}{16q^{3}}
 + \frac{2\,e^{-15\pi y}}{11q^{4}}
 + \frac{2\,e^{-19\pi y}}{7q^{5}}
 + \frac{4\,e^{-23\pi y}}{17q^{6}}
 + \cdots
 \right) \notag\\
&\hspace{1em}
 + \frac{9\,q^{1/3}}{2\pi}
 \left(
   1
 + \frac{7q}{4}
 + \frac{8q^{2}}{7}
 + \frac{9q^{3}}{5}
 + \frac{14q^{4}}{13}
 + \frac{31q^{5}}{16}
 + \frac{20q^{6}}{19}
 + \cdots
 \right)\!,
\\[6pt]
Y^{(0)}_{\mathbf{3},3} \;=\;&
 \frac{9\,q^{2/3} e^{2\pi y/3}}{2\pi}
 \left(
   \frac{e^{-2\pi y}}{q}
 + \frac{7\,e^{-6\pi y}}{4q^{2}}
 + \frac{8\,e^{-10\pi y}}{7q^{3}}
 + \frac{9\,e^{-14\pi y}}{5q^{4}}
 + \frac{14\,e^{-18\pi y}}{13q^{5}}
 + \frac{31\,e^{-22\pi y}}{16q^{6}}
 + \cdots
 \right) \notag\\
&\hspace{1em}
 + \frac{27\,q^{2/3}}{\pi}
 \left(
   \frac{1}{4}
 + \frac{q}{5}
 + \frac{5q^{2}}{16}
 + \frac{2q^{3}}{11}
 + \frac{2q^{4}}{7}
 + \frac{9q^{5}}{17}
 + \frac{21q^{6}}{20}
 + \cdots
 \right)\!.
\end{align}

\paragraph{Weight $k=-2$ multiplets from weight $4$.}
Acting with $\xi_{-2}$ (or $D^{3}$) on weight-$4$ holomorphic forms yields
\begin{equation}
  \xi_{-2}\!\bigl(Y_{\mathbf{1}}^{(-2)}\bigr)=Y_{\mathbf{1}}^{(4)},\qquad
  \xi_{-2}\!\bigl(Y_{\mathbf{3}}^{(-2)}\bigr)=\Omega\,Y_{\mathbf{3}}^{(4)},
  \qquad
  D^{3}\!\bigl(Y_{\mathbf{1}}^{(-2)}\bigr)=-\frac{2}{(4\pi)^{3}}Y_{\mathbf{1}}^{(4)},\qquad
  D^{3}\!\bigl(Y_{\mathbf{3}}^{(-2)}\bigr)=-\frac{2}{(4\pi)^{3}}Y_{\mathbf{3}}^{(4)}.
  \label{eq:kminus2_map_app}
\end{equation}
The corresponding $q$-series start as
\begin{align}
Y^{(-2)}_{\mathbf{1}}(\tau)
= \;& \frac{y^{3}}{3}
 - \frac{15\,\Gamma(3,4\pi y)}{4\pi^{3} q}
 - \frac{135\,\Gamma(3,8\pi y)}{32\pi^{3} q^{2}}
 - \frac{35\,\Gamma(3,12\pi y)}{9\pi^{3} q^{3}}
 + \cdots \notag\\
&\;
 - \frac{\pi\,\zeta(3)}{12\,\zeta(4)}
 - \frac{15\,q}{2\pi^{3}}
 - \frac{135\,q^{2}}{16\pi^{3}}
 - \frac{70\,q^{3}}{9\pi^{3}}
 - \frac{1095\,q^{4}}{128\pi^{3}}
 - \frac{189\,q^{5}}{25\pi^{3}}
 - \frac{35\,q^{6}}{4\pi^{3}}
 + \cdots .
\end{align}
\begin{align}
Y^{(-2)}_{\mathbf{3},1}(\tau)
= \;& \frac{y^{3}}{3}
 + \frac{21\,\Gamma(3,4\pi y)}{16\pi^{3} q}
 + \frac{189\,\Gamma(3,8\pi y)}{128\pi^{3} q^{2}}
 + \frac{169\,\Gamma(3,12\pi y)}{144\pi^{3} q^{3}}
 + \frac{1533\,\Gamma(3,16\pi y)}{1024\pi^{3} q^{4}}
 + \cdots \notag\\
&\;
 + \frac{\pi\,\zeta(3)}{40\,\zeta(4)}
 + \frac{21\,q}{8\pi^{3}}
 + \frac{189\,q^{2}}{64\pi^{3}}
 + \frac{169\,q^{3}}{72\pi^{3}}
 + \frac{1533\,q^{4}}{512\pi^{3}}
 + \frac{1323\,q^{5}}{500\pi^{3}}
 + \frac{169\,q^{6}}{64\pi^{3}}
 + \cdots ,
\\[6pt]
Y^{(-2)}_{\mathbf{3},2}(\tau)
= \;& -\,\frac{729\,q^{1/3}}{16\pi^{3}}
\Biggl(
    \frac{\Gamma(3,8\pi y/3)}{16\,q}
  + \frac{7\,\Gamma(3,20\pi y/3)}{125\,q^{2}}
  + \frac{65\,\Gamma(3,32\pi y/3)}{1024\,q^{3}}
  + \frac{74\,\Gamma(3,44\pi y/3)}{1331\,q^{4}}
  + \cdots
\Biggr)\notag\\[2pt]
& -\,\frac{81\,q^{1/3}}{16\pi^{3}}
\left(
  1 + \frac{73\,q}{64} + \frac{344\,q^{2}}{343} + \frac{567\,q^{3}}{500}
  + \frac{20198\,q^{4}}{2197} + \frac{4681\,q^{5}}{4096} + \cdots
\right),
\\[6pt]
Y^{(-2)}_{\mathbf{3},3}(\tau)
= \;& -\,\frac{81\,q^{2/3}}{32\pi^{3}}
\Biggl(
  \frac{\Gamma(3,4\pi y/3)}{q}
 + \frac{73\,\Gamma(3,16\pi y/3)}{64\,q^{2}}
 + \frac{344\,\Gamma(3,28\pi y/3)}{343\,q^{3}}
 + \frac{567\,\Gamma(3,40\pi y/3)}{500\,q^{4}}
 + \cdots
\Biggr)\notag\\[2pt]
&\;
 - \frac{729\,q^{2/3}}{8\pi^{3}}
\left(
  \frac{1}{16} + \frac{7\,q}{125} + \frac{65\,q^{2}}{1024} + \frac{74\,q^{3}}{1331}
  + \cdots
\right) .
\end{align}

\paragraph{Weight $k=-4$ multiplets from weight $6$.}
Similarly, from weight-$6$ holomorphic forms one obtains
\begin{equation}
  \xi_{-4}\!\bigl(Y_{\mathbf{1}}^{(-4)}\bigr)=Y_{\mathbf{1}}^{(6)},\qquad
  \xi_{-4}\!\bigl(Y_{\mathbf{3}}^{(-4)}\bigr)=\Omega\,Y_{\mathbf{3}}^{(6)},
  \qquad
  D^{5}\!\bigl(Y_{\mathbf{1}}^{(-4)}\bigr)=-\frac{24}{(4\pi)^{5}}Y_{\mathbf{1}}^{(6)},\qquad
  D^{5}\!\bigl(Y_{\mathbf{3}}^{(-4)}\bigr)=-\frac{24}{(4\pi)^{5}}Y_{\mathbf{3}}^{(6)}.
  \label{eq:kminus4_map_app}
\end{equation}
A convenient set of $q$-expansions is
\begin{align}
Y^{(-4)}_{\mathbf{1}}(\tau)
= \;& \frac{y^{5}}{5}
 + \frac{63\,\Gamma(5,4\pi y)}{128\,\pi^{5}\,q}
 + \frac{2079\,\Gamma(5,8\pi y)}{4096\,\pi^{5}\,q^{2}}
 + \frac{427\,\Gamma(5,12\pi y)}{864\,\pi^{5}\,q^{3}}
 + \frac{66591\,\Gamma(5,16\pi y)}{131072\,\pi^{5}\,q^{4}}
 + \cdots \notag\\
&\;
 + \frac{\pi\,\zeta(5)}{80\,\zeta(6)}
 + \frac{189\,q}{16\,\pi^{5}}
 + \frac{6237\,q^{2}}{512\,\pi^{5}}
 + \frac{427\,q^{3}}{36\,\pi^{5}}
 + \frac{199773\,q^{4}}{16384\,\pi^{5}}
 + \cdots \, ,
\end{align}
\begin{align}
Y^{(-4)}_{\mathbf{3},1}(\tau)
= \;& \frac{y^{5}}{5}
 - \frac{549}{3328\,\pi^{5}}
 \Biggl(
   \frac{\Gamma(5,4\pi y)}{q}
 + \frac{33\,\Gamma(5,8\pi y)}{32\,q^{2}}
 + \frac{14641\,\Gamma(5,12\pi y)}{14823\,q^{3}}
 + \frac{1057\,\Gamma(5,16\pi y)}{1024\,q^{4}}
 + \cdots
 \Biggr) \notag\\
&\;
 - \frac{3\pi\,\zeta(5)}{728\,\zeta(6)}
 - \frac{1647}{416\,\pi^{5}}
 \Biggl(
   q + \frac{33 q^{2}}{32}
   + \frac{14641 q^{3}}{14823}
   + \frac{1057 q^{4}}{1024}
   + \frac{3126 q^{5}}{3125}
   + \cdots
 \Biggr) \, ,
\\[6pt]
Y^{(-4)}_{\mathbf{3},2}(\tau)
= \;& \frac{72171\,q^{1/3}}{212992\,\pi^{5}}
 \Biggl(
   \frac{\Gamma(5,8\pi y/3)}{q}
 + \frac{33344\,\Gamma(5,20\pi y/3)}{34375\,q^{2}}
 + \frac{1025\,\Gamma(5,32\pi y/3)}{1024\,q^{3}}
 + \cdots
 \Biggr) \notag\\
&\;
 + \frac{6561\,q^{1/3}}{832\,\pi^{5}}
 \left(
   1 + \frac{1057 q}{1024}
     + \frac{16808 q^{2}}{16807}
     + \frac{51579 q^{3}}{50000}
     + \frac{371294 q^{4}}{371293}
     + \cdots
 \right) \, ,
\\[6pt]
Y^{(-4)}_{\mathbf{3},3}(\tau)
= \;& \frac{2187\,q^{2/3}}{6656\,\pi^{5}}
 \Biggl(
   \frac{\Gamma(5,4\pi y/3)}{q}
 + \frac{1057\,\Gamma(5,16\pi y/3)}{1024\,q^{2}}
 + \frac{16808\,\Gamma(5,28\pi y/3)}{16807\,q^{3}}
 + \cdots
 \Biggr) \notag\\
&\;
 + \frac{216513\,q^{2/3}}{26624\,\pi^{5}}
 \left(
   1 + \frac{33344 q}{34375}
     + \frac{1025 q^{2}}{1024}
     + \frac{1717888 q^{3}}{1771561}
     + \frac{16808 q^{4}}{16807}
     + \cdots
 \right) .
\end{align}

%标准模型Yukawa部分，flavor symmetry东西 sector的东西，一般U_35 如何motivate一般A_4，怎么加modular  把标准模型flavor加进去
%把个数列出来
%标准模型加右手中微子

\end{appendices}

\newpage
% 参考文献（如果有 .bib）
\bibliography{ref}

\providecommand{\href}[2]{#2}\begingroup\raggedright\begin{thebibliography}{100}

\bibitem{Weinberg:1979sa}
S.~Weinberg, \emph{{Baryon and Lepton Nonconserving Processes}},
  \href{http://dx.doi.org/10.1103/PhysRevLett.43.1566}{\emph{Phys. Rev. Lett.}
  {\bf 43} (1979) 1566--1570}.

\bibitem{Buchmuller:1985jz}
W.~Buchmuller and D.~Wyler, \emph{{Effective Lagrangian Analysis of New
  Interactions and Flavor Conservation}},
  \href{http://dx.doi.org/10.1016/0550-3213(86)90262-2}{\emph{Nucl. Phys. B}
  {\bf 268} (1986) 621--653}.

\bibitem{Grzadkowski:2010es}
B.~Grzadkowski, M.~Iskrzynski, M.~Misiak and J.~Rosiek, \emph{{Dimension-Six
  Terms in the Standard Model Lagrangian}},
  \href{http://dx.doi.org/10.1007/JHEP10(2010)085}{\emph{JHEP} {\bf 10} (2010)
  085}, [\href{https://arxiv.org/abs/1008.4884}{{\tt 1008.4884}}].

\bibitem{Lehman_2014}
L.~Lehman, \emph{Extending the standard model effective field theory with the
  complete set of dimension-7 operators},
  \href{http://dx.doi.org/10.1103/physrevd.90.125023}{\emph{Physical Review D}
  {\bf 90} (Dec., 2014) }.

\bibitem{Li:2020gnx}
H.-L. Li, Z.~Ren, J.~Shu, M.-L. Xiao, J.-H. Yu and Y.-H. Zheng, \emph{{Complete
  set of dimension-eight operators in the standard model effective field
  theory}}, \href{http://dx.doi.org/10.1103/PhysRevD.104.015026}{\emph{Phys.
  Rev. D} {\bf 104} (2021) 015026},
  [\href{https://arxiv.org/abs/2005.00008}{{\tt 2005.00008}}].

\bibitem{Li:2020xlh}
H.-L. Li, Z.~Ren, M.-L. Xiao, J.-H. Yu and Y.-H. Zheng, \emph{{Complete set of
  dimension-nine operators in the standard model effective field theory}},
  \href{http://dx.doi.org/10.1103/PhysRevD.104.015025}{\emph{Phys. Rev. D} {\bf
  104} (2021) 015025}, [\href{https://arxiv.org/abs/2007.07899}{{\tt
  2007.07899}}].

\bibitem{Li:2022tec}
H.-L. Li, Z.~Ren, M.-L. Xiao, J.-H. Yu and Y.-H. Zheng, \emph{{Operators for
  generic effective field theory at any dimension: on-shell amplitude basis
  construction}}, \href{http://dx.doi.org/10.1007/JHEP04(2022)140}{\emph{JHEP}
  {\bf 04} (2022) 140}, [\href{https://arxiv.org/abs/2201.04639}{{\tt
  2201.04639}}].

\bibitem{Lehman:2015via}
L.~Lehman and A.~Martin, \emph{{Hilbert Series for Constructing Lagrangians:
  expanding the phenomenologist's toolbox}},
  \href{http://dx.doi.org/10.1103/PhysRevD.91.105014}{\emph{Phys. Rev. D} {\bf
  91} (2015) 105014}, [\href{https://arxiv.org/abs/1503.07537}{{\tt
  1503.07537}}].

\bibitem{Lehman:2015coa}
L.~Lehman and A.~Martin, \emph{{Low-derivative operators of the Standard Model
  effective field theory via Hilbert series methods}},
  \href{http://dx.doi.org/10.1007/JHEP02(2016)081}{\emph{JHEP} {\bf 02} (2016)
  081}, [\href{https://arxiv.org/abs/1510.00372}{{\tt 1510.00372}}].

\bibitem{Henning:2015daa}
B.~Henning, X.~Lu, T.~Melia and H.~Murayama, \emph{{Hilbert series and operator
  bases with derivatives in effective field theories}},
  \href{http://dx.doi.org/10.1007/s00220-015-2518-2}{\emph{Commun. Math. Phys.}
  {\bf 347} (2016) 363--388}, [\href{https://arxiv.org/abs/1507.07240}{{\tt
  1507.07240}}].

\bibitem{Henning:2017fpj}
B.~Henning, X.~Lu, T.~Melia and H.~Murayama, \emph{{Operator bases,
  $S$-matrices, and their partition functions}},
  \href{http://dx.doi.org/10.1007/JHEP10(2017)199}{\emph{JHEP} {\bf 10} (2017)
  199}, [\href{https://arxiv.org/abs/1706.08520}{{\tt 1706.08520}}].

\bibitem{Marinissen:2020jmb}
C.~B. Marinissen, R.~Rahn and W.~J. Waalewijn, \emph{{..., 83106786, 114382724,
  1509048322, 2343463290, 27410087742, ... efficient Hilbert series for
  effective theories}},
  \href{http://dx.doi.org/10.1016/j.physletb.2020.135632}{\emph{Phys. Lett. B}
  {\bf 808} (2020) 135632}, [\href{https://arxiv.org/abs/2004.09521}{{\tt
  2004.09521}}].

\bibitem{Henning:2015alf}
B.~Henning, X.~Lu, T.~Melia and H.~Murayama, \emph{{2, 84, 30, 993, 560, 15456,
  11962, 261485, ...: Higher dimension operators in the SM EFT}},
  \href{http://dx.doi.org/10.1007/JHEP08(2017)016}{\emph{JHEP} {\bf 08} (2017)
  016}, [\href{https://arxiv.org/abs/1512.03433}{{\tt 1512.03433}}].

\bibitem{Sun:2022aag}
H.~Sun, Y.-N. Wang and J.-H. Yu, \emph{{Hilbert Series and Operator Counting on
  the Higgs Effective Field Theory}},
  \href{https://arxiv.org/abs/2211.11598}{{\tt 2211.11598}}.

\bibitem{Kondo:2022wcw}
D.~Kondo, H.~Murayama and R.~Okabe, \emph{{23, 381, 6242, 103268, 1743183,
  {\textellipsis} : Hilbert series for CP-violating operators in SMEFT}},
  \href{http://dx.doi.org/10.1007/JHEP03(2023)107}{\emph{JHEP} {\bf 03} (2023)
  107}, [\href{https://arxiv.org/abs/2212.02413}{{\tt 2212.02413}}].

\bibitem{Cal__2023}
S.~Calò, C.~Marinissen and R.~Rahn, \emph{Discrete symmetries and efficient
  counting of operators},
  \href{http://dx.doi.org/10.1007/jhep05(2023)215}{\emph{Journal of High Energy
  Physics} {\bf 2023} (May, 2023) }.

\bibitem{Chivukula:1987py}
R.~S. Chivukula and H.~Georgi, \emph{Composite technicolor standard model},
  \href{http://dx.doi.org/10.1016/0370-2693(87)90713-1}{\emph{Phys. Lett. B}
  {\bf 188} (1987) 99--104}.

\bibitem{DAmbrosio:2002vsn}
G.~D'Ambrosio, G.~F. Giudice, G.~Isidori and A.~Strumia, \emph{{Minimal flavor
  violation: An Effective field theory approach}},
  \href{http://dx.doi.org/10.1016/S0550-3213(02)00836-2}{\emph{Nucl. Phys. B}
  {\bf 645} (2002) 155--187}, [\href{https://arxiv.org/abs/hep-ph/0207036}{{\tt
  hep-ph/0207036}}].

\bibitem{Bonnefoy:2020yee}
Q.~Bonnefoy, E.~Gendy and C.~Grojean, \emph{{Positivity bounds on Minimal
  Flavor Violation}},
  \href{http://dx.doi.org/10.1007/JHEP04(2021)115}{\emph{JHEP} {\bf 04} (2021)
  115}, [\href{https://arxiv.org/abs/2011.12855}{{\tt 2011.12855}}].

\bibitem{Aoude:2020dwv}
R.~Aoude, T.~Hurth, S.~Renner and W.~Shepherd, \emph{{The impact of flavour
  data on global fits of the MFV SMEFT}},
  \href{http://dx.doi.org/10.1007/JHEP12(2020)113}{\emph{JHEP} {\bf 12} (2020)
  113}, [\href{https://arxiv.org/abs/2003.05432}{{\tt 2003.05432}}].

\bibitem{Bruggisser:2021duo}
S.~Bruggisser, R.~Sch\"afer, D.~van Dyk and S.~Westhoff, \emph{{The Flavor of
  UV Physics}}, \href{http://dx.doi.org/10.1007/JHEP05(2021)257}{\emph{JHEP}
  {\bf 05} (2021) 257}, [\href{https://arxiv.org/abs/2101.07273}{{\tt
  2101.07273}}].

\bibitem{Kobayashi:2021uam}
T.~Kobayashi and H.~Otsuka, \emph{{On stringy origin of minimal flavor
  violation}},
  \href{http://dx.doi.org/10.1140/epjc/s10052-022-09986-4}{\emph{Eur. Phys. J.
  C} {\bf 82} (2022) 25}, [\href{https://arxiv.org/abs/2108.02700}{{\tt
  2108.02700}}].

\bibitem{Bruggisser:2022rhb}
S.~Bruggisser, D.~van Dyk and S.~Westhoff, \emph{{Resolving the flavor
  structure in the MFV-SMEFT}},
  \href{http://dx.doi.org/10.1007/JHEP02(2023)225}{\emph{JHEP} {\bf 02} (2023)
  225}, [\href{https://arxiv.org/abs/2212.02532}{{\tt 2212.02532}}].

\bibitem{Bartocci:2023nvp}
R.~Bartocci, A.~Biek\"otter and T.~Hurth, \emph{{A global analysis of the SMEFT
  under the minimal MFV assumption}},
  \href{http://dx.doi.org/10.1007/JHEP05(2024)074}{\emph{JHEP} {\bf 05} (2024)
  074}, [\href{https://arxiv.org/abs/2311.04963}{{\tt 2311.04963}}].

\bibitem{Bartocci:2024fgj}
R.~Bartocci, \emph{{Global analysis of the $U(3)^5$ symmetric SMEFT}},  in
  \emph{{58th Rencontres de Moriond on Electroweak Interactions and Unified
  Theories}}, 5, 2024.
\newblock \href{https://arxiv.org/abs/2405.10101}{{\tt 2405.10101}}.

\bibitem{sun2025flavorcpsymmetriesstandard}
H.~Sun and J.-H. Yu, \emph{Flavor and cp symmetries in the standard model
  effective field theory},  2025.

\bibitem{Ding_2025}
G.-J. Ding and J.~W. Valle, \emph{The symmetry approach to quark and lepton
  masses and mixing},
  \href{http://dx.doi.org/10.1016/j.physrep.2024.12.005}{\emph{Physics Reports}
  {\bf 1109} (Mar., 2025) 1–105}.

\bibitem{Altarelli_2010}
G.~Altarelli and F.~Feruglio, \emph{Discrete flavor symmetries and models of
  neutrino mixing},
  \href{http://dx.doi.org/10.1103/revmodphys.82.2701}{\emph{Reviews of Modern
  Physics} {\bf 82} (Sept., 2010) 2701–2729}.

\bibitem{bartlett2005categoricalaspectstopologicalquantum}
B.~H. Bartlett, \emph{Categorical aspects of topological quantum field
  theories},  2005.

\bibitem{Altarelli_2005}
G.~Altarelli and F.~Feruglio, \emph{Tri-bimaximal neutrino mixing from discrete
  symmetry in extra dimensions},
  \href{http://dx.doi.org/10.1016/j.nuclphysb.2005.05.005}{\emph{Nuclear
  Physics B} {\bf 720} (Aug., 2005) 64–88}.

\bibitem{ma2004nonabeliandiscretefamilysymmetries}
E.~Ma, \emph{Non-abelian discrete family symmetries of leptons and quarks},
  2004.

\bibitem{Ishimori_2010}
H.~Ishimori, T.~Kobayashi, H.~Ohki, Y.~Shimizu, H.~Okada and M.~Tanimoto,
  \emph{Non-abelian discrete symmetries in particle physics},
  \href{http://dx.doi.org/10.1143/ptps.183.1}{\emph{Progress of Theoretical
  Physics Supplement} {\bf 183} (2010) 1–163}.

\bibitem{King_2014}
S.~F. King, A.~Merle, S.~Morisi, Y.~Shimizu and M.~Tanimoto, \emph{Neutrino
  mass and mixing: from theory to experiment},
  \href{http://dx.doi.org/10.1088/1367-2630/16/4/045018}{\emph{New Journal of
  Physics} {\bf 16} (Apr., 2014) 045018}.

\bibitem{Harrison_2002}
P.~Harrison, D.~Perkins and W.~Scott, \emph{Tri-bimaximal mixing and the
  neutrino oscillation data},
  \href{http://dx.doi.org/10.1016/s0370-2693(02)01336-9}{\emph{Physics Letters
  B} {\bf 530} (Mar., 2002) 167–173}.

\bibitem{Harrison1_2002}
P.~Harrison and W.~Scott, \emph{Symmetries and generalisations of tri-bimaximal
  neutrino mixing},
  \href{http://dx.doi.org/10.1016/s0370-2693(02)01753-7}{\emph{Physics Letters
  B} {\bf 535} (May, 2002) 163–169}.

\bibitem{Xing_2002}
Z.-z. Xing, \emph{Nearly tri-bimaximal neutrino mixing and cp violation},
  \href{http://dx.doi.org/10.1016/s0370-2693(02)01649-0}{\emph{Physics Letters
  B} {\bf 533} (May, 2002) 85–93}.

\bibitem{harrison2004statustribimaximalneutrinomixing}
P.~F. Harrison and W.~G. Scott, \emph{Status of tri/bi-maximal neutrino
  mixing},  2004.

\bibitem{Harrison_2004}
P.~Harrison and W.~Scott, \emph{The simplest neutrino mass matrix},
  \href{http://dx.doi.org/10.1016/j.physletb.2004.05.039}{\emph{Physics Letters
  B} {\bf 594} (Aug., 2004) 324–332}.

\bibitem{Babu_2003}
K.~Babu, E.~Ma and J.~Valle, \emph{Underlying a4 symmetry for the neutrino mass
  matrix and the quark mixing matrix},
  \href{http://dx.doi.org/10.1016/s0370-2693(02)03153-2}{\emph{Physics Letters
  B} {\bf 552} (Jan., 2003) 207–213}.

\bibitem{Altarelli_2006}
G.~Altarelli and F.~Feruglio, \emph{Tri-bimaximal neutrino mixing, and the
  modular symmetry},
  \href{http://dx.doi.org/10.1016/j.nuclphysb.2006.02.015}{\emph{Nuclear
  Physics B} {\bf 741} (May, 2006) 215–235}.

\bibitem{Ding:2023htn}
G.-J. Ding and S.~F. King, \emph{{Neutrino mass and mixing with modular
  symmetry}}, \href{http://dx.doi.org/10.1088/1361-6633/ad52a3}{\emph{Rept.
  Prog. Phys.} {\bf 87} (2024) 084201},
  [\href{https://arxiv.org/abs/2311.09282}{{\tt 2311.09282}}].

\bibitem{Chen_2020}
M.-C. Chen, S.~Ramos-Sánchez and M.~Ratz, \emph{A note on the predictions of
  models with modular flavor symmetries},
  \href{http://dx.doi.org/10.1016/j.physletb.2019.135153}{\emph{Physics Letters
  B} {\bf 801} (Feb., 2020) 135153}.

\bibitem{Liu:2020akv}
X.-G. Liu, C.-Y. Yao and G.-J. Ding, \emph{{Modular invariant quark and lepton
  models in double covering of $S_4$ modular group}},
  \href{http://dx.doi.org/10.1103/PhysRevD.103.056013}{\emph{Phys. Rev. D} {\bf
  103} (2021) 056013}, [\href{https://arxiv.org/abs/2006.10722}{{\tt
  2006.10722}}].

\bibitem{Yao:2020qyy}
C.-Y. Yao, J.-N. Lu and G.-J. Ding, \emph{{Modular Invariant $A_{4}$ Models for
  Quarks and Leptons with Generalized CP Symmetry}},
  \href{http://dx.doi.org/10.1007/JHEP05(2021)102}{\emph{JHEP} {\bf 05} (2021)
  102}, [\href{https://arxiv.org/abs/2012.13390}{{\tt 2012.13390}}].

\bibitem{Behera:2025tpj}
M.~K. Behera, P.~Ittisamai, C.~Pongkitivanichkul and P.~Uttayarat,
  \emph{{Phenomenology of Inverse Seesaw Using $S_3$ Modular Symmetry}},
  \href{https://arxiv.org/abs/2504.12954}{{\tt 2504.12954}}.

\bibitem{Behera:2024ark}
M.~K. Behera, P.~Ittisamai, C.~Pongkitivanichkul and P.~Uttayarat,
  \emph{{Neutrino phenomenology in the modular S3 seesaw model}},
  \href{http://dx.doi.org/10.1103/PhysRevD.110.035004}{\emph{Phys. Rev. D} {\bf
  110} (2024) 035004}, [\href{https://arxiv.org/abs/2403.00593}{{\tt
  2403.00593}}].

\bibitem{Meloni:2023aru}
D.~Meloni and M.~Parriciatu, \emph{{A simplest modular S$_{3}$ model for
  leptons}}, \href{http://dx.doi.org/10.1007/JHEP09(2023)043}{\emph{JHEP} {\bf
  09} (2023) 043}, [\href{https://arxiv.org/abs/2306.09028}{{\tt 2306.09028}}].

\bibitem{Marciano_2024}
S.~Marciano, D.~Meloni and M.~Parriciatu, \emph{Minimal seesaw and leptogenesis
  with the smallest modular finite group},
  \href{http://dx.doi.org/10.1007/jhep05(2024)020}{\emph{Journal of High Energy
  Physics} {\bf 2024} (May, 2024) }.

\bibitem{behera2024neutrinophenomenologymodulars3}
M.~K. Behera, P.~Ittisamai, C.~Pongkitivanichkul and P.~Uttayarat,
  \emph{Neutrino phenomenology in the modular $s_3$ seesaw model},  2024.

\bibitem{Meloni_2023}
D.~Meloni and M.~Parriciatu, \emph{A simplest modular s3 model for leptons},
  \href{http://dx.doi.org/10.1007/jhep09(2023)043}{\emph{Journal of High Energy
  Physics} {\bf 2023} (Sept., 2023) }.

\bibitem{mishra2020neutrinomixingleptogenesismodular}
S.~Mishra, \emph{Neutrino mixing and leptogenesis with modular $s_3$ symmetry
  in the framework of type iii seesaw},  2020.

\bibitem{Okada_2019}
H.~Okada and Y.~Orikasa, \emph{{Modular $S_{3}$ symmetric radiative seesaw
  model}}, \href{http://dx.doi.org/10.1103/PhysRevD.100.115037}{\emph{Physical
  Review D} {\bf 100} (Dec., 2019) 115037}.

\bibitem{Kobayashi_2019}
T.~Kobayashi, Y.~Shimizu, K.~Takagi, M.~Tanimoto, T.~H. Tatsuishi and
  H.~Uchida, \emph{Finite modular subgroups for fermion mass matrices and
  baryon/lepton number violation},
  \href{http://dx.doi.org/10.1016/j.physletb.2019.05.034}{\emph{Physics Letters
  B} {\bf 794} (July, 2019) 114–121}.

\bibitem{nomura2023texturezerosrealizationthreeloop}
T.~Nomura, H.~Okada and H.~Otsuka, \emph{Texture zeros realization in a
  three-loop radiative neutrino mass model from modular $a_4$ symmetry},  2023.

\bibitem{Kumar_2024}
R.~Kumar, P.~Mishra, M.~K. Behera, R.~Mohanta and R.~Srivastava,
  \emph{Predictions from scoto-seesaw with a4 modular symmetry},
  \href{http://dx.doi.org/10.1016/j.physletb.2024.138635}{\emph{Physics Letters
  B} {\bf 853} (June, 2024) 138635}.

\bibitem{gogoi2023leptogenesisdarkmatterminimal}
J.~Gogoi, L.~Sarma and M.~K. Das, \emph{Leptogenesis and dark matter in minimal
  inverse seesaw using $a_4$ modular symmetry},  2023.

\bibitem{mishra2023exploringmodelsmodularsymmetry}
P.~Mishra, M.~K. Behera, P.~Panda, M.~Ghosh and R.~Mohanta, \emph{Exploring
  models with modular symmetry in neutrino oscillation experiments},  2023.

\bibitem{Ding:2022bzs}
G.-J. Ding, S.~F. King, J.-N. Lu and B.-Y. Qu, \emph{{Leptogenesis in SO(10)
  models with A$_{4}$ modular symmetry}},
  \href{http://dx.doi.org/10.1007/JHEP10(2022)071}{\emph{JHEP} {\bf 10} (2022)
  071}, [\href{https://arxiv.org/abs/2206.14675}{{\tt 2206.14675}}].

\bibitem{Chen:2021zty}
P.~Chen, G.-J. Ding and S.~F. King, \emph{{SU(5) GUTs with A$_{4}$ modular
  symmetry}}, \href{http://dx.doi.org/10.1007/JHEP04(2021)239}{\emph{JHEP} {\bf
  04} (2021) 239}, [\href{https://arxiv.org/abs/2101.12724}{{\tt 2101.12724}}].

\bibitem{nomura2023quarkleptonmodelflavor}
T.~Nomura and H.~Okada, \emph{Quark and lepton model with flavor specific dark
  matter and muon $g-2$ in modular $a_4$ and hidden $u(1)$ symmetries},  2023.

\bibitem{kim2023fermilatgevexcessmuon}
J.~Kim and H.~Okada, \emph{Fermi-lat gev excess and muon $g-2$ in a modular
  $a_4$ symmetry},  2023.

\bibitem{devi2023retrievingtexturezeros31}
M.~R. Devi, \emph{Retrieving texture zeros in 3+1 active-sterile neutrino
  framework under the action of $a_4$ modular-invariants},  2023.

\bibitem{Ding:2021eva}
G.-J. Ding, S.~F. King and J.-N. Lu, \emph{{SO(10) models with A$_{4}$ modular
  symmetry}}, \href{http://dx.doi.org/10.1007/JHEP11(2021)007}{\emph{JHEP} {\bf
  11} (2021) 007}, [\href{https://arxiv.org/abs/2108.09655}{{\tt 2108.09655}}].

\bibitem{dasgupta2021diracradiativeneutrinomass}
A.~Dasgupta, T.~Nomura, H.~Okada, O.~Popov and M.~Tanimoto, \emph{Dirac
  radiative neutrino mass with modular symmetry and leptogenesis},  2021.

\bibitem{Nomura_2020}
T.~Nomura, H.~Okada and O.~Popov, \emph{A modular a4 symmetric scotogenic
  model}, \href{http://dx.doi.org/10.1016/j.physletb.2020.135294}{\emph{Physics
  Letters B} {\bf 803} (Apr., 2020) 135294}.

\bibitem{pathak2025neutrinomassgenesisscotoinverse}
G.~Pathak, P.~Das and M.~K. Das, \emph{Neutrino mass genesis in scoto-inverse
  seesaw with modular $a_4$},  2025.

\bibitem{Wang_2020}
X.~Wang, \emph{Lepton flavor mixing and cp violation in the minimal type-(i+ii)
  seesaw model with a modular a4 symmetry},
  \href{http://dx.doi.org/10.1016/j.nuclphysb.2020.115105}{\emph{Nuclear
  Physics B} {\bf 957} (Aug., 2020) 115105}.

\bibitem{nomura2024novelapproachradiativelinear}
T.~Nomura and H.~Okada, \emph{A more novel approach of radiative linear seesaw
  in a modular $a_4$ symmetry},  2024.

\bibitem{Kashav_2021}
M.~Kashav and S.~Verma, \emph{Broken scaling neutrino mass matrix and
  leptogenesis based on a4 modular invariance},
  \href{http://dx.doi.org/10.1007/jhep09(2021)100}{\emph{Journal of High Energy
  Physics} {\bf 2021} (Sept., 2021) }.

\bibitem{Kashav_2023}
M.~Kashav and S.~Verma, \emph{On minimal realization of topological lorentz
  structures with one-loop seesaw extensions in a4 modular symmetry},
  \href{http://dx.doi.org/10.1088/1475-7516/2023/03/010}{\emph{Journal of
  Cosmology and Astroparticle Physics} {\bf 2023} (Mar., 2023) 010}.

\bibitem{Kobayashi_2020}
T.~Kobayashi, T.~Nomura and T.~Shimomura, \emph{{Type {II} seesaw models with
  modular $A_{4}$ symmetry}},
  \href{http://dx.doi.org/10.1103/PhysRevD.102.035019}{\emph{Physical Review D}
  {\bf 102} (Aug., 2020) 035019}.

\bibitem{nomura2019inverseseesawmodela4modular}
T.~Nomura, H.~Okada and S.~Patra, \emph{An inverse seesaw model with
  $a_4$-modular symmetry},  2019.

\bibitem{behera2020modulara4symmetricscotogenic}
M.~K. Behera, S.~Singirala, S.~Mishra and R.~Mohanta, \emph{A modular $a_4$
  symmetric scotogenic model for neutrino mass and dark matter},  2020.

\bibitem{Kobayashi1_2018}
T.~Kobayashi, N.~Omoto, Y.~Shimizu, K.~Takagi, M.~Tanimoto and T.~H. Tatsuishi,
  \emph{Modular a4 invariance and neutrino mixing},
  \href{http://dx.doi.org/10.1007/jhep11(2018)196}{\emph{Journal of High Energy
  Physics} {\bf 2018} (Nov., 2018) }.

\bibitem{Kobayashi_2018}
T.~Kobayashi, K.~Tanaka and T.~H. Tatsuishi, \emph{Neutrino mixing from finite
  modular groups},
  \href{http://dx.doi.org/10.1103/physrevd.98.016004}{\emph{Physical Review D}
  {\bf 98} (July, 2018) }.

\bibitem{Ding:2019gof}
G.-J. Ding, S.~F. King, X.-G. Liu and J.-N. Lu, \emph{{Modular S$_{4}$ and
  A$_{4}$ symmetries and their fixed points: new predictive examples of lepton
  mixing}}, \href{http://dx.doi.org/10.1007/JHEP12(2019)030}{\emph{JHEP} {\bf
  12} (2019) 030}, [\href{https://arxiv.org/abs/1910.03460}{{\tt 1910.03460}}].

\bibitem{Penedo2_2019}
J.~Penedo and S.~Petcov, \emph{Lepton masses and mixing from modular s4
  symmetry},
  \href{http://dx.doi.org/10.1016/j.nuclphysb.2018.12.016}{\emph{Nuclear
  Physics B} {\bf 939} (Feb., 2019) 292–307}.

\bibitem{Novichkov2_2019}
P.~P. Novichkov, J.~T. Penedo, S.~T. Petcov and A.~V. Titov, \emph{Modular s4
  models of lepton masses and mixing},
  \href{http://dx.doi.org/10.1007/jhep04(2019)005}{\emph{Journal of High Energy
  Physics} {\bf 2019} (Apr., 2019) }.

\bibitem{Kobayashi2_2019}
T.~Kobayashi, Y.~Shimizu, K.~Takagi, M.~Tanimoto and T.~H. Tatsuishi,
  \emph{$a_4$ lepton flavor model and modulus stabilization from $s_4$ modular
  symmetry},
  \href{http://dx.doi.org/10.1103/PhysRevD.100.115045}{\emph{Physical Review D}
  {\bf 100} (dec, 2019) 115045}.

\bibitem{Liu_2021}
X.-G. Liu, C.-Y. Yao and G.-J. Ding, \emph{Modular invariant quark and lepton
  models in double covering of the modular group $s_4$},
  \href{http://dx.doi.org/10.1103/PhysRevD.103.056013}{\emph{Physical Review D}
  {\bf 103} (mar, 2021) 056013}.

\bibitem{varzielas2023quarksmodulars4cusp}
I.~de~Medeiros~Varzielas, M.~Levy, J.~T. Penedo and S.~T. Petcov, \emph{Quarks
  at the modular $s_4$ cusp},  2023.

\bibitem{King_2020}
S.~F. King and Y.-L. Zhou, \emph{Trimaximal tm$_1$ mixing with two modular
  $s_4$ groups},
  \href{http://dx.doi.org/10.1103/PhysRevD.101.015001}{\emph{Physical Review D}
  {\bf 101} (jan, 2020) 015001}.

\bibitem{Ding:2021zbg}
G.-J. Ding, S.~F. King and C.-Y. Yao, \emph{{Modular $S_4\times SU(5)$ GUT}},
  \href{http://dx.doi.org/10.1103/PhysRevD.104.055034}{\emph{Phys. Rev. D} {\bf
  104} (2021) 055034}, [\href{https://arxiv.org/abs/2103.16311}{{\tt
  2103.16311}}].

\bibitem{Ding:2024inn}
G.-J. Ding, J.-N. Lu, S.~T. Petcov and B.-Y. Qu, \emph{{Non-holomorphic modular
  S$_{4}$ lepton flavour models}},
  \href{http://dx.doi.org/10.1007/JHEP01(2025)191}{\emph{JHEP} {\bf 01} (2025)
  191}, [\href{https://arxiv.org/abs/2408.15988}{{\tt 2408.15988}}].

\bibitem{Penedo_2019}
J.~Penedo and S.~Petcov, \emph{Lepton masses and mixing from modular s4
  symmetry},
  \href{http://dx.doi.org/10.1016/j.nuclphysb.2018.12.016}{\emph{Nuclear
  Physics B} {\bf 939} (Feb., 2019) 292–307}.

\bibitem{Ding:2019xna}
G.-J. Ding, S.~F. King and X.-G. Liu, \emph{{Neutrino mass and mixing with
  $A_5$ modular symmetry}},
  \href{http://dx.doi.org/10.1103/PhysRevD.100.115005}{\emph{Phys. Rev. D} {\bf
  100} (2019) 115005}, [\href{https://arxiv.org/abs/1903.12588}{{\tt
  1903.12588}}].

\bibitem{Yao:2020zml}
C.-Y. Yao, X.-G. Liu and G.-J. Ding, \emph{{Fermion masses and mixing from the
  double cover and metaplectic cover of the $A_5$ modular group}},
  \href{http://dx.doi.org/10.1103/PhysRevD.103.095013}{\emph{Phys. Rev. D} {\bf
  103} (2021) 095013}, [\href{https://arxiv.org/abs/2011.03501}{{\tt
  2011.03501}}].

\bibitem{Novichkov_2019}
P.~P. Novichkov, J.~T. Penedo, S.~T. Petcov and A.~V. Titov, \emph{Modular a5
  symmetry for flavour model building},
  \href{http://dx.doi.org/10.1007/jhep04(2019)174}{\emph{Journal of High Energy
  Physics} {\bf 2019} (Apr., 2019) }.

\bibitem{Kobayashi_2022}
T.~Kobayashi, H.~Otsuka, M.~Tanimoto and K.~Yamamoto, \emph{Modular symmetry in
  the smeft},
  \href{http://dx.doi.org/10.1103/physrevd.105.055022}{\emph{Physical Review D}
  {\bf 105} (Mar., 2022) }.

\bibitem{qu2024nonholomorphicmodularflavorsymmetry}
B.-Y. Qu and G.-J. Ding, \emph{Non-holomorphic modular flavor symmetry},  2024.

\bibitem{Pontecorvo:1957qd}
B.~Pontecorvo, \emph{{Inverse Beta Processes and Nonconservation of Lepton
  Charge}}, {\emph{Sov. Phys. JETP} {\bf 7} (1958) 172--173}.

\bibitem{Maki:1962mu}
Z.~Maki, M.~Nakagawa and S.~Sakata, \emph{{Remarks on the unified model of
  elementary particles}},
  \href{http://dx.doi.org/10.1143/PTP.28.870}{\emph{Prog. Theor. Phys.} {\bf
  28} (1962) 870--880}.

\bibitem{feruglio2017neutrinomassesmodularforms}
F.~Feruglio, \emph{Are neutrino masses modular forms?},  2017.

\bibitem{Gerard1983FermionMS}
J.~M. Gerard, \emph{Fermion mass spectrum insu(2)l×u(1)}, {\emph{Zeitschrift
  f{\"u}r Physik C Particles and Fields} {\bf 18} (1983) 145--154}.

\bibitem{ParticleDataGroup:2024cfk}
{\scshape Particle Data Group} collaboration, S.~Navas et~al., \emph{{Review of
  particle physics}},
  \href{http://dx.doi.org/10.1103/PhysRevD.110.030001}{\emph{Phys. Rev. D} {\bf
  110} (2024) 030001}.

\bibitem{Isidori:2010kg}
G.~Isidori, Y.~Nir and G.~Perez, \emph{{Flavor Physics Constraints for Physics
  Beyond the Standard Model}},
  \href{http://dx.doi.org/10.1146/annurev.nucl.012809.104534}{\emph{Ann. Rev.
  Nucl. Part. Sci.} {\bf 60} (2010) 355},
  [\href{https://arxiv.org/abs/1002.0900}{{\tt 1002.0900}}].

\bibitem{Faroughy:2020ina}
D.~A. Faroughy, G.~Isidori, F.~Wilsch and K.~Yamamoto, \emph{{Flavour
  symmetries in the SMEFT}},
  \href{http://dx.doi.org/10.1007/JHEP08(2020)166}{\emph{JHEP} {\bf 08} (2020)
  166}, [\href{https://arxiv.org/abs/2005.05366}{{\tt 2005.05366}}].

\bibitem{okada2021unificationquarkleptonflavors}
H.~Okada and M.~Tanimoto, \emph{Towards unification of quark and lepton flavors
  in $a_4$ modular invariance},  2021.

\bibitem{Feng:2007ur}
B.~Feng, A.~Hanany and Y.-H. He, \emph{{Counting gauge invariants: The
  Plethystic program}},
  \href{http://dx.doi.org/10.1088/1126-6708/2007/03/090}{\emph{JHEP} {\bf 03}
  (2007) 090}, [\href{https://arxiv.org/abs/hep-th/0701063}{{\tt
  hep-th/0701063}}].

\bibitem{Okada_2021}
H.~Okada and M.~Tanimoto, \emph{Modular invariant flavor model of $a_4$ and
  hierarchical structures at nearby fixed points},
  \href{http://dx.doi.org/10.1103/PhysRevD.103.015005}{\emph{Physical Review D}
  {\bf 103} (Jan., 2021) }.

\bibitem{book:Ono}
K.~Bringmann, A.~Folsom, K.~Ono and L.~Rolen, \emph{Harmonic Maass Forms and
  Mock Modular Forms: Theory and Applications}, vol.~64 of \emph{Colloquium
  Publications}.
\newblock American Mathematical Society, 2017.

\bibitem{Alonso_2014}
R.~Alonso, E.~E. Jenkins and A.~V. Manohar, \emph{Holomorphy without
  supersymmetry in the standard model effective field theory},
  \href{http://dx.doi.org/10.1016/j.physletb.2014.10.045}{\emph{Physics Letters
  B} {\bf 739} (Dec., 2014) 95–98}.

\bibitem{de_Blas_2018}
J.~de~Blas, J.~C. Criado, M.~Pérez-Victoria and J.~Santiago, \emph{Effective
  description of general extensions of the standard model: the complete
  tree-level dictionary},
  \href{http://dx.doi.org/10.1007/jhep03(2018)109}{\emph{Journal of High Energy
  Physics} {\bf 2018} (Mar., 2018) }.

\bibitem{Li:2022abx}
H.-L. Li, Y.-H. Ni, M.-L. Xiao and J.-H. Yu, \emph{{The bottom-up EFT: complete
  UV resonances of the SMEFT operators}},
  \href{http://dx.doi.org/10.1007/JHEP11(2022)170}{\emph{JHEP} {\bf 11} (2022)
  170}, [\href{https://arxiv.org/abs/2204.03660}{{\tt 2204.03660}}].

\end{thebibliography}\endgroup

\end{document}